\documentclass[a4paper,12pt,twoside]{book}
\usepackage[utf8]{inputenc}
\usepackage[german,main=english]{babel}
\usepackage{latexsym}
\usepackage{amssymb}
\usepackage{physics}
\usepackage{amsmath}
\usepackage{bm}
\usepackage{amsfonts}
\usepackage{tabls}
\usepackage{bbold}
\usepackage{color}
\usepackage{slashed}
\usepackage{appendix}
\usepackage{graphicx}
\usepackage{blindtext}
\usepackage{setspace}
\usepackage{caption}
\usepackage{subcaption}
\usepackage{relsize}
\usepackage[bottom,symbol]{footmisc}
\usepackage[a4paper,top=4cm,bottom=4.5cm,left=2.5cm,right=2.5cm]{geometry}
\usepackage{tocbibind}
\usepackage{fancyhdr} 
\usepackage{parskip}
\usepackage{titlesec}
\titlespacing*{\section}{0pt}{16pt}{11pt}
\titlespacing*{\subsection}{0pt}{11pt}{8pt}
\titlespacing*{\subsubsection}{0pt}{8pt}{6pt}
\makeatletter
\newcommand\footnoteref[1]{\protected@xdef\@thefnmark{\ref{#1}}\@footnotemark}
\makeatother
\usepackage[sort&compress,square,numbers]{natbib}
\usepackage[colorlinks = true,
            linkcolor = blue,
            urlcolor  = blue,
            citecolor = blue,
            anchorcolor = blue,
            filecolor = blue,
            pdftex,
            bookmarks,
            hypertexnames=false,
            debug]{hyperref}
\usepackage{bookmark}
\definecolor{ao(english)}{rgb}{0.0, 0.5, 0.0}

\usepackage{subfiles}
\newcommand{\up}{\uparrow}
\newcommand{\down}{\downarrow}
\newcommand{\eps}{\varepsilon}
\newcommand{\bk}{\mathbf{k}}
\newcommand{\bq}{\mathbf{q}}
\newcommand{\bkq}{{\mathbf{k}+\mathbf{q}}}
\newcommand{\mbk}{{-\mathbf{k}}}
\newcommand{\gmk}{g^-_\mathbf{k}}
\newcommand{\hmk}{h^-_\mathbf{k}}
\newcommand{\hpk}{h^+_\mathbf{k}}
\newcommand{\emk}{e^-_\mathbf{k}}
\newcommand{\epk}{e^+_\mathbf{k}}
\newcommand{\bQ}{\mathbf{Q}}
\newcommand{\ellnot}{{-\ell}}
\newcommand{\PauliMat}{\sigma}
\renewcommand{\L}{\Lambda}
\newcommand{\Lini}{{\Lambda_\mathrm{ini}}}
\newcommand{\Lfin}{{\Lambda_\mathrm{fin}}}
\newcommand{\deL}{\partial_\L}
\newcommand{\Lamc}{{\L_c}}
\newcommand\shortdots{\hbox to 1em{.\hss.\hss.}}
\newcommand{\rndup}[1]{\left\lceil\frac{{#1}}{2}\right\rceil}
\newcommand{\rnddo}[1]{\left\lfloor\frac{{#1}}{2}\right\rfloor}
\newcommand{\rndupnotwo}[1]{\left\lceil{#1}\right\rceil}
\newcommand{\rnddonotwo}[1]{\left\lfloor{#1}\right\rfloor}

\newcommand{\phx}{\overline{ph}}
\newcommand{\bzero}{\mathbf{0}}
\newcommand{\kx}{\mathrm{k}_x}
\newcommand{\ky}{\mathrm{k}_y}
\newcommand{\bs}[1]{\mathbf{#1}}
\newcommand{\chit}{\widetilde{\chi}}
\newcommand{\Gammat}{\widetilde{\Gamma}}
\newcommand{\psibar}{\overline{\psi}}
\newcommand{\msml}[1]{{\mathsmaller{\mathsmaller{\mathsmaller{#1}}}}}
\newcommand{\smsqr}{\msml{\square}}
\newcommand{\munu}{{\mu\nu}}
\newcommand{\dmu}{\partial_\mu}
\newcommand{\dnu}{\partial_\nu}
\newcommand{\Luv}{{\Lambda_\mathrm{uv}}}
\newcommand{\br}{\mathbf{r}}
\newcommand{\cA}{\mathcal{A}}
\newcommand{\cB}{\mathcal{B}}

\newcommand{\cJ}{\mathcal{J}}

\newcommand{\cP}{\mathcal{P}}
\newcommand{\cR}{\mathcal{R}}
\newcommand{\cS}{\mathcal{S}}
\renewcommand{\a}{a}
\renewcommand{\b}{b}

\title{}
\author{}
\date{}

\begin{document}
\begin{titlepage}

{\setstretch{2.0}
%
%
%
\begin{center}
\textcolor{black}{
{\Huge{\bf Long-range order, bosonic fluctuations, and pseudogap in strongly correlated electron systems}}\\
}
\end{center}}

{\setstretch{1.4}
\vspace{20mm} \par \noindent

\begin{center}
Von der Fakult\"at Mathematik und Physik der Universit\"at Stuttgart\\
zur Erlangung der W\"urde eines Doktors der\\ 
Naturwissenschaften (Dr.~rer.~nat.) genehmigte Abhandlung
\end{center}

} 

\vspace{12mm} 
\begin{center}
vorgelegt von
\end{center}

\vspace{3mm}
\begin{center}\textcolor{black}{
{\LARGE{\bf Pietro Maria Bonetti}}\\
}\end{center}

\vspace{3mm} 
\begin{center}
aus Aosta, Italien
\end{center}

\vspace{6mm}
\begin{center}
\begin{tabular}{rl}

Hauptberichter:& \vspace{2mm}{Prof.~Dr.~Walter Metzner}\\
Mitberichterin:& \vspace{2mm}{Prof.~Dr.~Maria Daghofer}\\
             & \vspace{2mm}{Prof.~Dr.~Sabine Andergassen}\\
Pr\"ufungsvorsitzender:& \vspace{2mm}{Prof.~Dr.~Sebastian Loth}\\

\end{tabular}
\end{center}
\vspace{10mm}
\begin{center}
\begin{tabular}{rl}
    Tag der Einreichung: &19. Juli 2022\\
    Tag der m\"undlichen Pr\"ufung: &23. September 2022\\ 
    \vspace{2mm}
\end{tabular}
\end{center}

\vspace{6mm}
\begin{center}
{\Large{\bf
Max-Planck-Institut f\"ur Festk\"orperforschung \\
Universit\"at Stuttgart\\ \vspace{3mm} 2022
\vspace{2mm}\\
}}
\end{center}

\end{titlepage}

\clearpage{\pagestyle{empty}\cleardoublepage}
\frontmatter
\begin{titlepage}                                                           \thispagestyle{empty}                                                       
\topmargin=6.5cm                                                            
\raggedleft                                                                 
\emph{
Non chiederci la parola che squadri da ogni lato\\
l'animo nostro informe, e a lettere di fuoco\\
lo dichiari e risplenda come un croco\\
Perduto in mezzo a un polveroso prato.\\
\bigskip
Ah l'uomo che se ne va sicuro,\\
agli altri ed a se stesso amico,\\
e l'ombra sua non cura che la canicola\\
stampa sopra uno scalcinato muro!\\
\bigskip
Non domandarci la formula che mondi possa aprirti\\
s\`i qualche storta sillaba e secca come un ramo.\\
Codesto solo oggi possiamo dirti,\\
ci\`o che non siamo, ci\`o che non vogliamo.\\
}
\vspace{1cm}

\emph{Ossi di Seppia.} Eugenio Montale

\end{titlepage}
\clearpage{\pagestyle{empty}\cleardoublepage}        
\chapter*{Abstract}
\addcontentsline{toc}{chapter}{Abstract}

    \rhead[\fancyplain{}{\bfseries Abstract}]{\fancyplain{}{\bfseries\thepage}}
    \lhead[\fancyplain{}{\bfseries\thepage}]{\fancyplain{}{\bfseries Abstract}}
    Cuprate high transition temperature superconductors are of fundamental interest in condensed matter physics due to their rich phase diagram, where several phases and regimes compete and coexist with each other. Among the models proposed to describe the physics of the electrons in the copper-oxide planes, the two-dimensional Hubbard model has gained the most popularity. Despite its apparent simplicity, the search for an approximate solution able to capture all its phases still represents a challenging problem. Mean-field theory often represents a good starting point to describe ordered phases, but, in order to capture several physical features, it is necessary to analyze fluctuations of the order parameter. 
    
    Among the various methods proposed to treat the Hubbard model, in this thesis we focus on the moderate coupling functional renormalization group (fRG) and its combination with the dynamical mean-field theory (DMFT), which extends it to strong coupling. We deal with the problem of identifying bosonic fluctuations in the vertex function, describing the effective interaction between two electrons in the many-body medium, which exhibits an intricate dependence on momenta and frequencies already at moderate coupling. In the symmetric phase, when no symmetry of the model is broken, the goal is achieved by employing the recently introduced single-boson exchange decomposition. This decomposition allows for a clear and physically intuitive parametrization of the vertex function in terms of processes involving the exchange of a single boson, describing a collective excitation, and a residual part. Moreover, the single-boson exchange decomposition allows for a \emph{substantial} reduction of the computational complexity of the vertex function. We also reformulate the previously introduced combination of fRG with mean-field theory, designed to access symmetry broken phases, by explicitly introducing a bosonic field. This reformulation is proven to be equivalent to the "purely fermionic" approach, but it represents a convenient starting point on top of which one can include order parameter fluctuations while keeping the full, non-simplified, frequency and momentum dependence of the vertex. 
    
    A widely discussed and challenging problem is the emergence of a \emph{pseudogap} in the Hubbard model and its relation to the pseudogap regime observed in the cuprates. In this thesis we assume this phase to be characterized by strong magnetic fluctuations. Following previous works, we fractionalize the electron in a chargon, carrying the electron's charge, and a charge neutral spinon, which represents local fluctuations of the spin orientation. The chargons undergo N\'eel or spiral magnetic order below a density-dependent transition temperature $T^*$. Charge transport coefficients are only weakly affected by directional fluctuations of the spin orientation, so that in their computation one can consider only chargon degrees of freedom. We perform a DMFT computation of the magnetic order parameter for fermions (that can be interpreted as chargons) displaying spiral magnetic ordering, starting from the two-dimensional Hubbard model. The magnetic order leads to a Fermi surface reconstruction. We compute DC charge transport properties by combining the renormalized band structure as obtained from the DMFT with a phenomenological scattering rate. We obtain a pronounced drop of the longitudinal conductivity and the Hall number in a narrow doping regime below a critical doping $p^*$, above which magnetic order disappears, in agreement with recent transport measurements for cuprate superconductors in high magnetic fields in the pseudogap regime. 
    
    Directional fluctuations of the spin orientation are described by a non-linear sigma model. We derive formulas for the non-linear sigma model parameters, namely the spin stiffnesses, by expanding the inverse of the transverse order parameter susceptibilities in powers of momentum and frequency, and we prove via local \emph{Ward identities} that this approach is equivalent to the computation of the system's response to a fictitious SU(2) gauge field. At finite electron or hole doping, the chargons form small Fermi surfaces, which can induce \emph{Landau} damping of the Goldstone modes of the magnetic state, which for low energies coincide with the directional fluctuations of the spins. A spiral magnetic state is host to three Goldstone modes, two of which correspond to out-of-plane fluctuations, and one to in-plane fluctuations of the spins. The decay rate of the in-plane mode is found to be of the order of its excitation energy, while the decay rate of the out-of-plane modes is smaller so that these modes are asymptotically stable. We also perform a computation of the chargon order parameter in the pseudogap regime. We employ a \emph{renormalized} Hartree-Fock theory, using effective interactions extracted from a fRG flow. The spin stiffnesses are computed through the response to a fictitious SU(2) gauge field. Spinon fluctuations prevent long-range ordering of the electrons at any finite temperature but, at least in the weak coupling regime, not in the ground state. The phase where the chargon degrees of freedom are magnetically ordered shares many features with the pseudogap regime observed in high-$T_c$ cuprates: a strong reduction of the charge carrier density, a spin gap, Fermi arcs, and electronic nematicity.

\clearpage{\pagestyle{empty}\cleardoublepage}
\chapter*{Deutsche Zusammenfassung}
\addcontentsline{toc}{chapter}{Deutsche Zusammenfassung}

    \rhead[\fancyplain{}{\bfseries Deutsche Zusammenfassung}]{\fancyplain{}{\bfseries\thepage}}
    \lhead[\fancyplain{}{\bfseries\thepage}]{\fancyplain{}{\bfseries Deutsche Zusammenfassung}}
    Kuprat-Supraleiter mit hoher Sprungtemperatur sind aufgrund ihres reichhaltigen Phasen\-diagramms mit mehreren miteinander konkurrierenden und koexistierenden Phasen von grundlegendem Interesse für die Physik der kondensierten Materie. Unter den Modellen, die zur Beschreibung der Physik der Elektronen in den Kupfer-Oxid-Ebenen vorgeschlagen wurden, hat das zweidimensionale Hubbard-Modell die größte Popularität erlangt. Trotz seiner scheinbaren Einfachheit ist die Suche nach einer Lösung, die alle Phasen erfassen kann, immer noch ein schwieriges Problem. Die Molekularfeldtheorie stellt oft einen guten Ausgangspunkt für die Beschreibung geordneter Phasen dar. Trotzdem müssen die Fluktuationen des Ordnungs\-parameters analysiert werden, um bestimmte physikalische Eigenschaften zu erfassen. 
    
    Unter den zahlreichen Methoden, die zur Behandlung des Hubbard-Modells vorgeschlagen wurden, konzentrieren wir uns in dieser Arbeit auf die funktionale Renormierungsgruppentheorie (fRG) mit moderater Kopplung und ihre Kombination mit der dynamischen Molekular\-feld\-theorie (DMFT), die ihre Anwendbarkeit auf starke Kopplung ausdehnt. Wir befassen uns mit dem Problem der Identifizierung bosonischer Fluktuationen in der Vertexfunktion, die die effektive Wechselwirkung zwischen zwei Elektronen im Vielteilchenmedium beschreibt, und bereits bei mäßiger Kopplung eine komplizierte Abhängigkeit von Impulsen und Frequenzen aufweist. In der symmetrischen Phase, wenn keine Symmetrie des Modells gebrochen ist, wird das Ziel durch die kürzlich eingeführte Ein-Bosonen-Austauschzerlegung (single-boson exchange decomposition) erreicht. Diese Zerlegung ermöglicht eine klare und physikalisch intuitive Parametrisierung der Vertexfunktion in Form von Prozessen, die den Austausch eines einzelnen Bosons, das eine kollektive Anregung beschreibt, beinhalten, sowie einen Restteil. Darüber hinaus reduziert die Zerlegung in Einzelbosonen-Austauschprozesse erheblich die Rechen\-komplexität der Vertexfunktion. Zur Beschreibung der symmetriegebrochenen Phasen formulieren wir auch die zuvor eingeführte Kombination der fRG mit der Molekularfeldtheorie neu, indem wir explizit ein bosonisches Feld einführen. Diese Neuformulierung erweist sich als äquivalent zum "rein fermionischen" Ansatz, stellt aber einen bequemeren Ausgangspunkt dar zur Einbeziehung von Ordnungsparameterfluktuationen, wobei man die vollständige, nicht vereinfachte Frequenz- und Impulsabhängigkeit der Vertexfunktion beibehält. 
    
    Ein viel diskutiertes und schwieriges Problem ist das Auftreten eines \emph{Pseudogaps} im Hubbard-Modell und seine Beziehung zum in den Kupraten beobachteten Pseudogapbereich. In dieser Arbeit gehen wir davon aus, dass diese Phase durch starke magnetische Fluktuationen gekenn\-zeichnet ist. In Anlehnung an frühere Arbeiten fraktionieren wir das Elektron in ein Chargon, das die Ladung des Elektrons trägt, und ein ladungsneutrales Spinon, das die Fluktuationen der Spinorientierung repräsentiert. Die Chargons nehmen unterhalb einer dichteabhängigen Übergangstemperatur $T^*$ eine N\'eel- oder spiralförmige magnetische Ordnung an. Die Ladungstransportkoeffizienten werden nur schwach von Fluktuationen der Spin\-orientierung be\-einflusst, sodass man zu ihrer Berechnung nur die Freiheitsgrade der Ladungen berücksichtigen muss. Wir führen eine DMFT-Berechnung des magnetischen Ordnungs\-parameters für Fermionen (die als Chargons interpretiert werden können) durch. Hier zeigt sich ausgehend von dem zweidimensionalen Hubbard-Modell eine spiralförmige magnetische Ordnung, welche zu einer Rekonstruktion der Fermi-Fläche führt. Wir berechnen die Gleichstrom\--Ladungs\-transport\-eigenschaften, indem wir die renormierte Bandstruktur, wie sie sich aus der DMFT ergibt, mit einer phänomenologischen Zerfallsrate kombinieren.  Wir erhalten einen ausgeprägten Abfall der longitudinalen Leitfähigkeit und der Hall-Zahl in einem engen Dotierungsbereich unterhalb einer kritischen Dotierung $p^*$, oberhalb derer die magnetische Ordnung verschwindet. Dies ist in Übereinstimmung mit Transportmessungen für Kuprat-Supraleiter in hohen Magnetfeldern im Pseudogapbereich. 
    
    Fluktuationen der Spinorientierung werden durch ein nichtlineares Sigma\--Modell be\-schrieben. Wir leiten Formeln für die Parameter des nichtlinearen Sigma-Modells ab, nämlich die Spinsteifigkeiten, indem wir den Kehrwert der transversalen Suszeptibilitäten des Ordnungsparameters im Bereich spiralförmiger magnetischer Ordnung entwickeln. Wir zeigen durch lokale \emph{Ward-Identitäten}, dass dieser Ansatz äquivalent zur Berechnung der Reaktion des Systems auf ein fiktives SU(2)-Eichfeld ist. Bei endlicher Elektronen- oder Lochdotierung bilden die Chargons kleine Fermi-Flächen, die eine \emph{Landau}-Dämpfung der Goldstone-Moden des magnetischen Zustands hervorrufen können. Diese Moden sind die niederenergetische Richtungsfluktuationen der Spins. Ein spiralförmiger magnetischer Zustand beherbergt drei Goldstone-Moden, von denen zwei Moden den Fluktuationen außerhalb der Ebene (out-of-plane Mode) und eine Mode den Fluktuationen innerhalb der Ebene der Spiralordnung der Spins entsprechen (in-plane Mode). Die Zerfallsrate der in-plane-Mode liegt in der Größenordnung ihrer Anregungsenergie, während die Zerfallsrate der out-of-plane-Moden kleiner ist, so dass diese Moden asymptotisch stabil sind. Wir führen eine Berechnung des Chargon\--Ordnungs\-parameters im Pseudogapbereich durch. Dazu verwenden wir eine \emph{renormierte} Hartree-Fock-Theorie mit effektiven Wechselwirkungen, die aus einem fRG-Fluss extrahiert werden. Die Spinsteifigkeiten werden anhand der Reaktion auf ein fiktives SU(2)-Eichfeld berechnet. Spinon-Fluktuationen verhindern eine lang\-reich\-weitige Ordnung der Elektronen bei jeder endlichen Temperatur, in \"Ubereinstimmung mit dem Mermin-Wagner-Theorem, jedoch nicht im Grundzustand. Die Phase, in der die Chargon-Freiheitsgrade magnetisch geordnet sind, weist viele Gemeinsamkeiten mit dem in Kupraten beobachteten Pseudogapbereich auf: eine starke Reduzierung der Ladungsträgerdichte, eine Spin-Lücke, Fermi-Bögen und elektronische Nematizität.

\clearpage{\pagestyle{empty}\cleardoublepage}
%
{\hypersetup{linkcolor=black}
\rhead[\fancyplain{}{\bfseries\leftmark}]{\fancyplain{}{\bfseries\thepage}}
\lhead[\fancyplain{}{\bfseries\thepage}]{\fancyplain{}{\bfseries CONTENTS}}
\tableofcontents
\cleardoublepage}
%
\mainmatter
\addtocontents{toc}{\setcounter{tocdepth}{1}}
\phantomsection
\addcontentsline{toc}{chapter}{Introduction}
\setcounter{page}{1}

    \rhead[\fancyplain{}{\bfseries Introduction}]{\fancyplain{}{\bfseries\thepage}}
    \lhead[\fancyplain{}{\bfseries\thepage}]{\fancyplain{}{\bfseries Introduction}}
    \chapter*{Introduction}
    \label{chap: intro}

    \section*{Context and motivation}
\begin{figure}[t]
        \centering
        \includegraphics[width=0.275\textwidth]{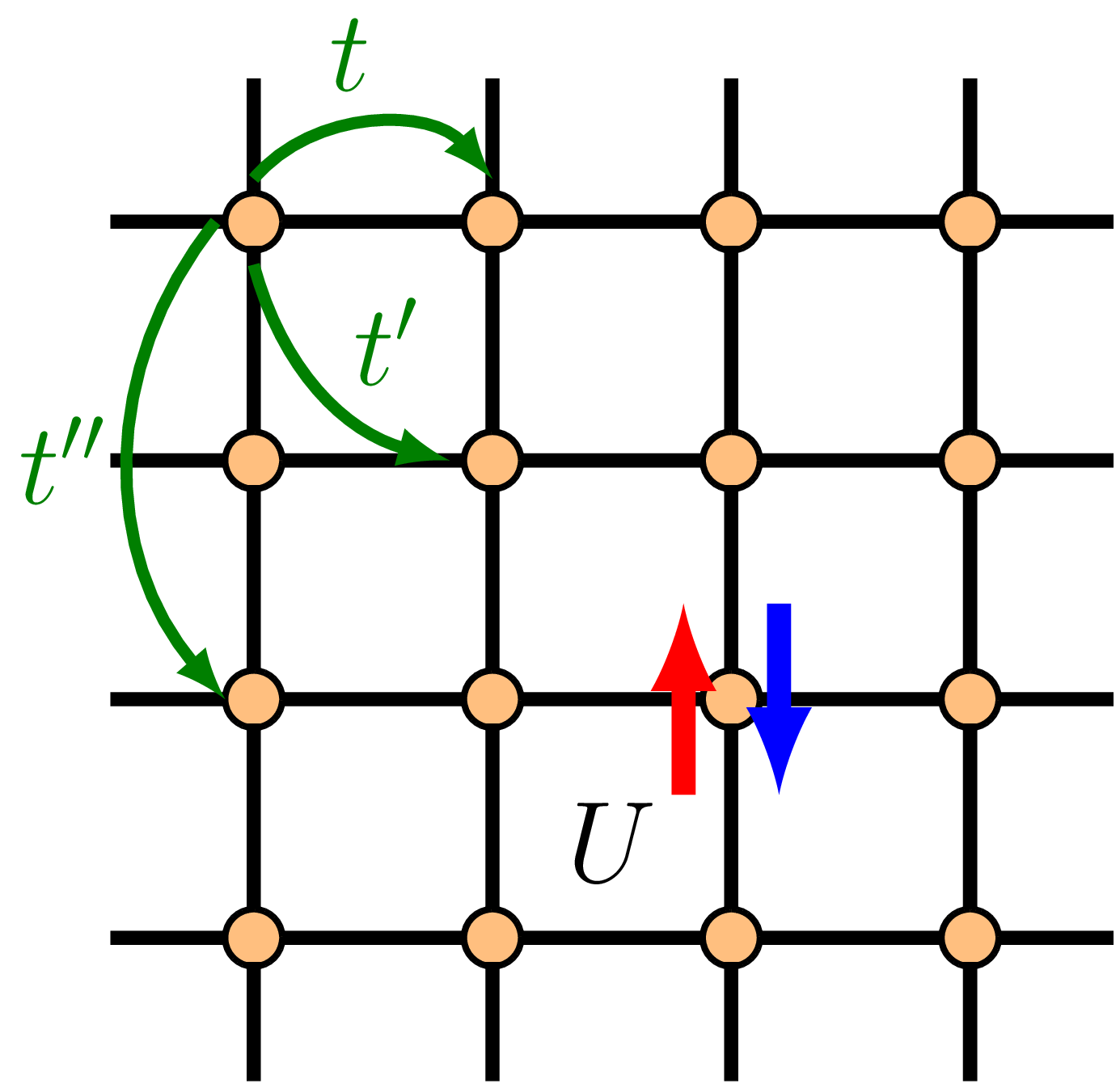}
        \caption{Pictorial representation of the Hubbard model on a square lattice. Here we consider hopping amplitudes $t$, $t'$ and $t^{\prime\prime}$ between nearest, next-to-nearest and third-neighboring sites, respectively. The onsite repulsive interaction only acts between opposite spin electrons, as, due to the Pauli principle, equal spin electrons can never occupy the same site.}
        \label{fig_intro: hubbard}
    \end{figure}

    Since the discovery of high-temperature superconductivity in copper-oxide compounds in the late 1980's~\cite{Bednorz1986}, the strongly correlated electron problem has gained considerable attention among condensed matter theorists. In fact, the conduction electrons lying within the stacked Cu-O planes, where the relevant physics is expected to take place, strongly interact with each other. These strong correlations produce a very rich phase diagram spanned by chemical doping and temperature~\cite{Keimer2015}: while the undoped compounds are antiferromagnetic Mott insulators, chemical substitution weakens the magnetic correlations and produces a so-called "superconducting dome" centered around optimal doping. Aside from these phases, many others have been found to coexist and compete with them, such as charge- and spin-density waves~\cite{Comin2014}, pseudogap~\cite{Proust2019}, and strange metal~\cite{Hussey2008}. 
    From a theoretical perspective, the early experiments on the cuprate materials immediately stimulated the search for a model able to describe at least some of the many competing phases. 
    In 1987, Anderson proposed the single-band two-dimensional Hubbard model to describe the electrons moving in the copper-oxide planes~\cite{Anderson1987}. Despite the real materials exhibiting several bands with complex structures, Zhang and Rice suggested that the Cu-O hybridization produces a singlet whose propagation through the lattice can be described by a single band model~\cite{Zhang1988}. While some other models have been proposed, such as the $t$-$J$ one~\cite{Zhang1988}, describing the motion of holes in a Heisenberg antiferromagnet and corresponding to the strong coupling limit of the Hubbard model~\cite{Chao1977,Chao1978}, or a more complex three-band model~\cite{Emery1987,Emery1988}, the Hubbard model has gained the most popularity because of its (apparent) simplicity.
    The model has been originally introduced by Hubbard~\cite{Hubbard1963,Hubbard1964}, Kanamori~\cite{Kanamori1963}, and Gutzwiller~\cite{Gutzwiller1963}, to describe correlation phenomena in three-dimensional systems with partially filled $d$- and $f$-bands. It consists of spin-$\frac{1}{2}$ electrons moving on a square lattice, with quantum mechanical hopping amplitudes $t_{jj'}$ between the sites labeled as $j$ and $j'$ and experiencing an on-site repulsive interaction $U$ (see Fig.~\ref{fig_intro: hubbard}).

\begin{figure}[t]
        \centering
        \includegraphics[width=0.5\textwidth]{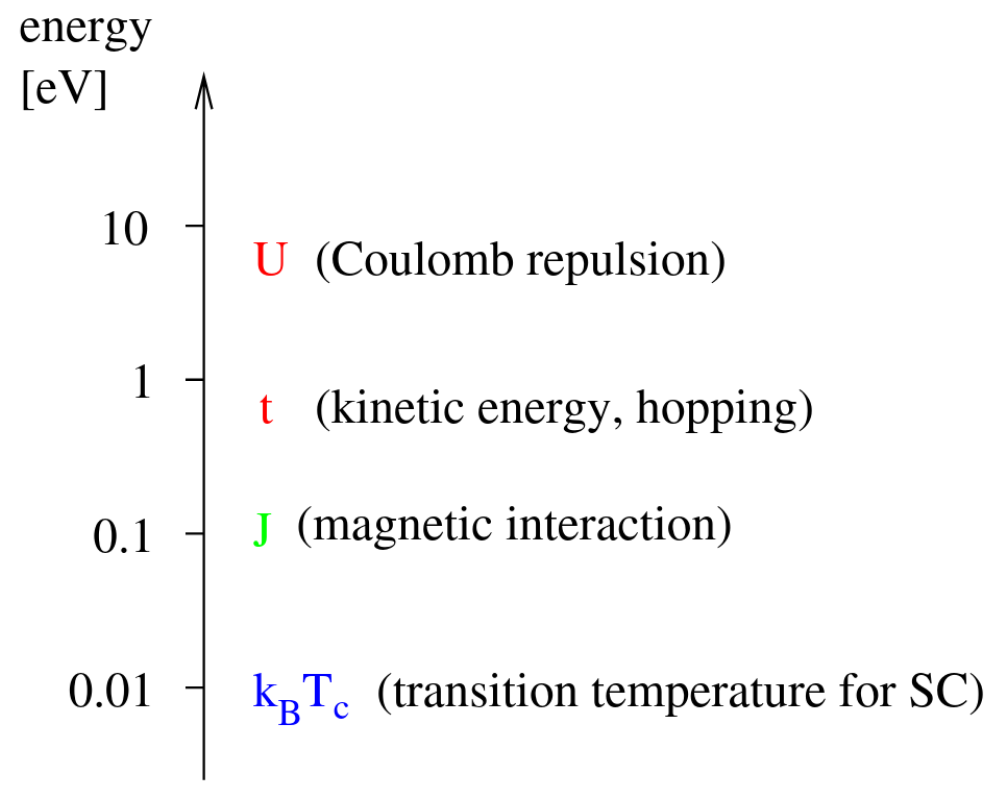}
        \caption{Hierarchy of energy scales in cuprate superconductors. Taken from Ref.~\cite{Metzner2012}.}
        \label{fig_intro: hubbard scales}
    \end{figure}

    Despite its apparent simplicity, the competition of different energy scales (Fig.~\ref{fig_intro: hubbard scales}) in the Hubbard model leads to various phases, some of which are still far from being understood~\cite{Qin2021}. One of the key ingredients is the competition between the localization energy scale $U$ and the kinetic energy (given by the bandwith $D=8t$) that instead tends to delocalize the electrons. This gives rise, at half filling, that is, when the single band is half occupied, to the celebrated Mott metal-to-insulator (MIT) transition. At weak coupling the system is in a metallic phase, characterized by itinerant electrons. Above a given critical value of the onsite repulsion $U$, the energy gained by localizing becomes lower than that of the metal, realizing a correlated (Mott) insulator. To capture this kind of physics was one of the early successes of the dynamical mean-field theory (DMFT)~\cite{Metzner1989,Georges1992,Georges1996}. 
    
    Another important energy scale is given by the antiferromagnetic exchange coupling $J$. Indeed, the half-filled Hubbard model with only nearest neighbor hopping amplitude $t$
    at strong coupling can be mapped onto the antiferromagnetic Heisenberg model with coupling constant $J=4t^2/U$, where the electron spins are the only degrees of freedom, as the charge fluctuations get frozen out. Therefore, the ground state at half filling is a N\'eel antiferromagnet. This is true not only at strong, but also at weak coupling and for finite hopping amplitudes to sites further than nearest neighbors. Indeed, a crossover takes place by varying the interaction strength. At small $U$, the instability to antiferromagnetism is driven by the Fermi surface (FS) geometry, with the wave vector $\bQ=(\pi/a,\pi/a)$ being a nesting vector (where $a$ is the lattice spacing), that is, it maps some points (hot spots) of the FS onto other points on the FS. In the particular case of zero hopping amplitudes beyond nearest neighbors (sometimes called pure Hubbard model), the nesting becomes perfect, with every point on the FS being a hot spot, implying that even infinitesimally small values of the coupling $U$ produce an antiferromagnetic state. In the more general case a minimal interaction strength $U_c$ is required to destabilize the paramagnetic phase. The state characterized by magnetic order occurring on top of a metallic state goes under the name of Slater antiferromagnet~\cite{Slater1951}. Differently, at strong coupling, local moments form on the lattice sites due to the freezing of charge fluctuations, which order antiferromagnetically, too~\cite{AuerbachBook1994,GebhardBook1997}. At intermediate coupling, the system is in a state that is something in between the two limits. In the pure Hubbard model case, a canonical particle-hole transformation~\cite{Micnas1990} maps the repulsive half-filled Hubbard model onto the attractive one, in which the crossover mentioned above becomes the BCS-BEC crossover~\cite{Eagles1969,Leggett1980,Nozieres1985}, describing the evolution from a weakly-coupled superconductor formed by loosely bound Cooper pairs, to a strongly coupled one, where the electrons tightly bind, forming bosonic particles which undergo Bose-Einstein condensation\footnote{Actually, in the pure attractive Hubbard model at half filling, the charge density wave and superconducting order parameters combine together to form an order parameter with SU(2) symmetry, which is the equivalent of the magnetization in the repulsive model.}. 

    Upon small electron or hole doping, the antiferromagnetic order gets weakened but may survive, giving rise to an \emph{itinerant antiferromagnet}, with small Fermi surfaces consisting of hole or electron pockets. Depending on the parameters, doping makes the N\'eel antiferromagnetic state unstable, with the spins rearranging in order to maximize the charge carrier kinetic energy, leading to an \emph{incommensurate} spiral magnet with ordering wave vector $\bQ\neq(\pi/a,\pi/a)$. The transition from a N\'{e}el to a spiral incommensurate antiferromagnet has been found not only at weak~\cite{Igoshev2010,Fresard1991,Chubukov1995}, but also at strong~\cite{Vilardi2018,Bonetti2020_I} coupling, as well as in the $t$-$J$ model~\cite{Shraiman1989,Kotov2004}, which describes the large-$U$ limit of the doped Hubbard model. At a given doping value, the (incommensurate) antiferromagnetic order finally ends, leaving room for other phases. At \emph{finite temperature}, long-range antiferromagnetic ordering is prevented by the Mermin-Wagner theorem~\cite{Mermin1966}, but strong magnetic fluctuations survive, leaving their signature in the electron spectrum~\cite{Borejsza2004,Scheurer2018}.  
    
\begin{figure}[t]
        \centering
        \includegraphics[width=0.65\textwidth]{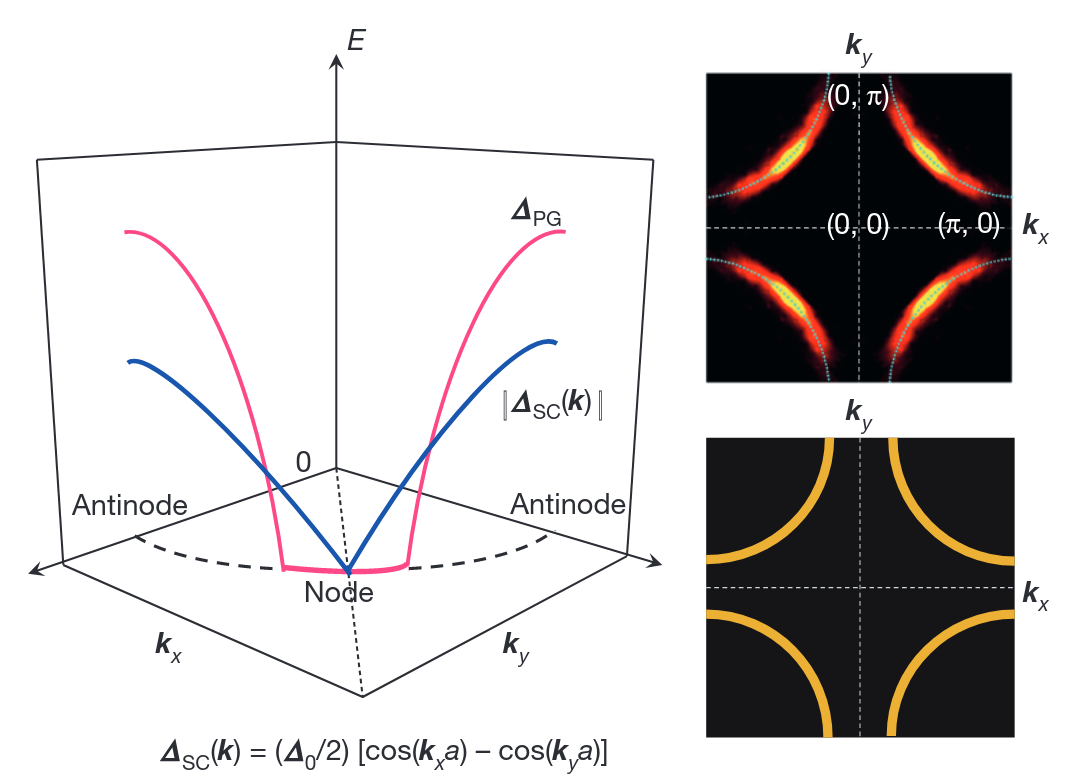}
        \caption{Left panel: superconducting $d$-wave gap and pseudogap as functions of the lattice momentum $\bk$. While the superconducting gap vanishes in a single $\bk$-point, the pseudogap is zero over a finite portion of the bare Fermi surface. Right panel: pseudogap spectral function (top) exhibiting the characteristic Fermi arcs, and spectral function without pseudogap (bottom), marked by a large Fermi surface. Taken from Ref.~\cite{Keimer2015}.}
        \label{fig_intro: pseudogap exp}
    \end{figure}

    At finite doping, magnetic fluctuations generate an effective attractive interaction between the electrons, eventually leading to an instability towards a $d$-wave superconducting state, characterized by a gap that vanishes at the nodal points of the underlying Fermi surface (see left panel of Fig.~\ref{fig_intro: pseudogap exp}). At least in the weak coupling limit, the presence of a superconducting state has to be expected, because, as pointed out by Kohn and Luttinger~\cite{Kohn1965}, as long as a sharp Fermi surface is present, every kind of (weak) interaction produces an attraction in a certain angular momentum channel, causing the onset of superconductivity. In other words, the Cooper instability always occurs in a Fermi liquid as soon as the interactions are turned on. At weak and moderate coupling, several methods have found $d$-wave superconducting phases and/or instabilities coexisting and competing with (incommensurate) antiferromagnetic ones. Among these methods, we list the fluctuation exchange approximation (FLEX)~\cite{Bickers1989,Bickers2004}, and the functional renormalization group (fRG)~\cite{Halboth2000,Halboth2000_PRL,Zanchi1998,Zanchi2000,Friederich2011,Husemann2009,Husemann2012}. 
    
    The FLEX approximation consists of a decoupling of the fluctuating magnetic and pairing channels, describing the $d$-wave pairing instability as a spin fluctuation mediated mechanism. On the other hand, the fRG~\cite{Metzner2012}, based on an exact flow equation~\cite{Wetterich1993,Berges2002}, provides an \emph{unbiased} treatment of \emph{all} the competing channels (including, for example, also charge fluctuations). The unavoidable truncation of the hierarchy of the flow equations, however, limits the applicability of this method to weak-to-moderate coupling values. Important progress has been made in this direction by replacing the \emph{bare} initial conditions with a converged DMFT solution~\cite{Taranto2014}, therefore "boosting"~\cite{Metzner2014} the fRG to strong coupling. One of the most challenging issues of the so-called DMF\textsuperscript{2}RG (DMFT+fRG) approach is the frequency dependence of the vertex function, representing the effective interaction felt by two electrons in the many-body medium, which has to be fully retained to properly capture strong coupling effects~\cite{Vilardi2019}. Similarly to the fRG, the parquet approximation~\cite{Zheleznyak1997,Eckhardt2020} (PA) treats all fluctuations on equal footing. 
    Self-consistent parquet equations are hard to converge numerically, and this has prevented their application to physically relevant parameter regimes so far. A notable advancement in this direction has been brought by the development of the multiloop fRG, that, by means of an improved truncation of the exact flow equations, controlled by a parameter $\ell$ counting the number of loops present in the flow diagrams, has been shown to become equivalent to the PA in the limit $\ell\to\infty$~\cite{Kugler2018_PRB,Kugler2018_PRL}. 

    Aside from antiferromagnetism and superconductivity, the Hubbard model is host to other intriguing phases, which have also been experimentally observed. One of those is the \emph{pseudogap} phase, characterized by the suppression of spectral weight at the antinodal points of the Fermi surface, forming so-called \emph{Fermi arcs} (see Fig.~\ref{fig_intro: pseudogap exp}). A full theoretical understanding of the mechanisms behind this behavior is still lacking, even though several works with various numerical methods~\cite{Vilk1997,Schaefer2015,VanLoon2018_II,Eckhardt2020,Simkovic2020,Hille2020_II} have found a considerable suppression of the spectral function close to the antinodal points in the Hubbard model. In all these works, the momentum-selective insulating nature of the computed self-energy seems to arise from strong antiferromagnetic fluctuations~\cite{Gunnarsson2015}. This can be described, at least in the weak coupling regime, by plugging in a Fock-like diagram for the self-energy an Ornstein-Zernike formula for the spin susceptibility, that is (in imaginary frequencies),
    \begin{equation*}
        \chi^m(\bq,\Omega)\simeq \frac{Z_m}{\Omega^2+c_s^2(\bq-\bQ)^2+(c_s\xi^{-1})^{2}}, 
    \end{equation*}
    with $c_s$ the spin wave velocity, $\bQ$ the antiferromagnetic ordering wave vector, $\xi$ the magnetic correlation length, and $Z_m$ a constant. According to the analysis carried out by Vilk and Tremblay~\cite{Vilk1997}, a gap opens at the antinodal points when $\xi\gg v_F/(\pi T)$, with $v_F$ the Fermi velocity and $T$ the temperature. More recent studies speculate that the pseudogap is connected to the onset of topological order in a fluctuating (that is, without long-range order) antiferromagnet~\cite{Sachdev2016,Chatterjee2017_II,Scheurer2018,Wu2018,Sachdev2018,Sachdev2019}.

    \begin{figure}[t]
        \centering
        \includegraphics[width=0.65\textwidth]{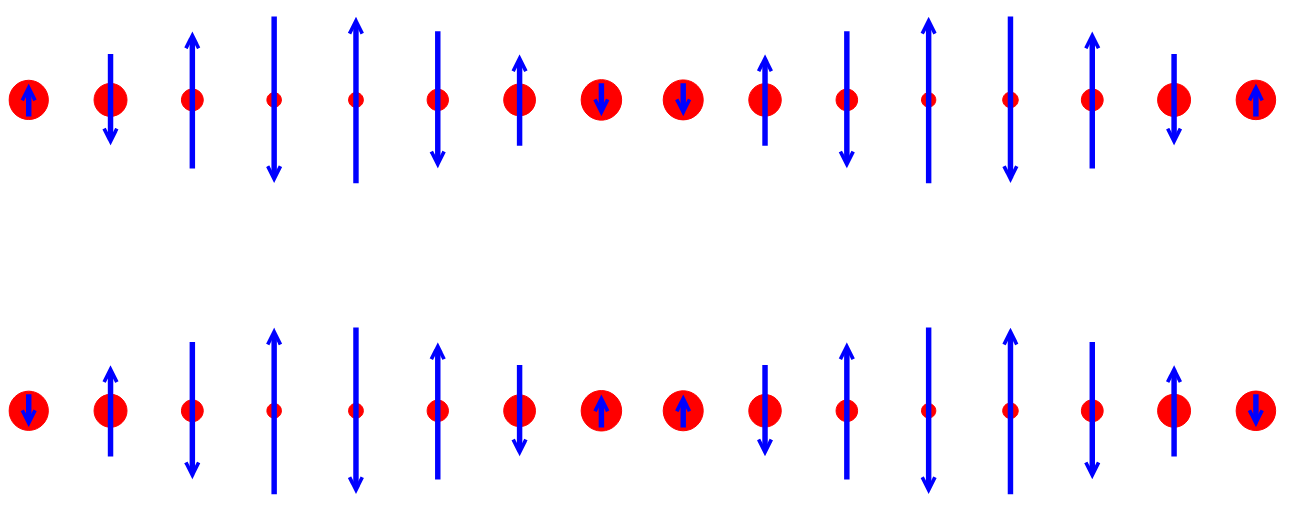}
        \caption{Pictorial representation of stripe order. The spins are ordered antiferromagnetically with a magnetization amplitude that gets modulated along one lattice direction. At the same time, the charge density also gets modulated, taking its maximum value where the magnetization density is minimal. Taken from Ref.~\cite{Tocchio2019}.}
        \label{fig_intro: stripes}
    \end{figure}
    Numerical calculations have also shown the emergence of a stripe phase, where the antiferromagnetic order parameter shows a modulation along one lattice direction, accompanied by a charge modulation (see Fig.~\ref{fig_intro: stripes}). Stripe order can be understood as an instability of a spiral phase~\cite{Shraiman1989,Dombre1990,Chubukov2021}, that is, a uniform incommensurate antiferromagnetic phase. It can also be viewed as the result of phase separation occurring in a hole-doped antiferromagnet~\cite{Schulz1989,Emery1990}. Stripe phases have been observed in several works, with methods starting from Hartree-Fock~\cite{Zaanen1989,Poilblanc1989}, up to the most recent density matrix renormalization group and quantum Monte Carlo studies of "Hubbard cylinders" at strong coupling~\cite{Zheng2017,Qin2020}. Stripe order is found to compete with other magnetic orders, such as uniform spiral magnetic phases~\cite{Shraiman1989}, as well as with $d$-wave superconductivity~\cite{White1999}.
    
    Among the phases listed above, the pseudogap remains one of the most puzzling ones. Most of its properties can be described by assuming some kind of magnetic order (often N\'eel or spiral) that causes a reconstruction of the large Fermi surface into smaller pockets and the appearance of Fermi arcs in the spectral function. However, no signature of static, long-range order has been found in experiments performed in this regime. In this thesis (see Chapter~\ref{chap: pseudogap} in particular), we theoretically model the pseudogap phase as a short-range ordered magnetic phase, where spin fluctuations prevent symmetry breaking at finite temperature (and they may do so even in the ground state), while many features of the long-range ordered state are retained, such as transport properties, superconductivity, and the spectral function. This is achieved by \emph{fractionalizing} the electron into a fermionic "chargon" and a charge neutral bosonic "spinon", carrying the spin quantum number of the original electron. In this way, one can assume magnetic order for the chargon degrees of freedom which gets eventually destroyed by the spinon fluctuations. 
    %
    \section*{Outline}
    This thesis is organized as it follows:
    \begin{itemize}
        \item In Chapter~\ref{chap: methods} we provide a short introduction of the main methods to approach the many body problem that we have used throughout this thesis. These are the functional renormalization group (fRG) and the dynamical mean-field theory (DMFT). In particular, we discuss various truncations of the fRG flow equations and the limitation of their validity to weak coupling. We finally present the usage of the DMFT as an initial condition of the fRG flow to access nonperturbative regimes.
        \item In Chapter~\ref{chap: spiral DMFT}, we present a study of transport coefficients across the transition between the pseudogap and the Fermi liquid phases of the cuprates. We model the pseudogap phase with long-range spiral magnetic order and perform nonperturbative computations in this regime via the DMFT. Subsequently, we extract an effective mean-field model, and using the formulas of Ref.~\cite{Mitscherling2018}, we compute the transport coefficients, which we can compare with the experimental results of Ref.~\cite{Badoux2016}.
        
        The results of this chapter have appeared in the publication:
        \begin{itemize}
            \item P.~M.~Bonetti\footnotemark[1], J.~Mitscherling\footnotemark[1]\footnotetext[1]{Equal contribution}, D.~Vilardi, and W.~Metzner, Charge carrier drop at the onset of pseudogap behavior in the two-dimensional Hubbard model, \href{https://link.aps.org/doi/10.1103/PhysRevB.101.165142}{Phys.~Rev.~B \textbf{101}, 165142 (2020)}.
        \end{itemize}
        \item In Chapter~\ref{chap: fRG+MF} we present the fRG+MF (mean-field) framework, introduced in Refs.~\cite{Wang2014,Yamase2016} that allows to continue the fRG flow into a spontaneously symmetry broken phase by means of a relatively simple truncation of the flow equations, that can be formally integrated, resulting into renormalized Hartree-Fock equations. 
        
        After presenting the general formalism, we apply the method to study the coexistence and competition of antiferromagnetism and superconductivity in the Hubbard model at weak coupling, by means of a state-of-the-art parametrization of the frequency dependence, thus methodologically improving the results of Ref.~\cite{Yamase2016}. We conclude the chapter by reformulating the fRG+MF equations in a mixed boson-fermion representation, where the explicit introduction of a bosonic field allows for a systematic inclusion of the collective fluctuations on top of the MF.
        
        The results of this chapter have appeared in the following publications:
        \begin{itemize}
            \item D.~Vilardi, P.~M.~Bonetti, and W.~Metzner, Dynamical functional renormalization group computation of order parameters and critical temperatures in the two-dimensional Hubbard model, \href{https://link.aps.org/doi/10.1103/PhysRevB.102.245128}{Phys.~Rev.~B \textbf{102}, 245128 (2020)}.
        
            \item P.~M.~Bonetti, Accessing the ordered phase of correlated Fermi systems:  Vertex bosonization and mean-field theory within the functional renormalization group, \href{https://link.aps.org/doi/10.1103/PhysRevB.102.235160}{Phys.~Rev.~B \textbf{102}, 235160 (2020)}.
        
        \end{itemize}
        \item In Chapter~\ref{chap: Bos Fluct Norm}, we present a reformulation of the fRG flow equations that exploits the \emph{single boson exchange } (SBE) representation of the two-particle vertex, introduced in Ref.~\cite{Krien2019_I}. The key idea of this parametrization is to represent the vertex in terms of processes each of which involves the exchange of a single boson, corresponding to a collective fluctuation, between two electrons, and a residual interaction. On the one hand, this decomposition offers numerical advantages, highly simplifying the computational complexity of the vertex function; on the other hand, it provides physical insight into the collective excitations of the correlated system. The chapter contains a formulation of the flow equations and results obtained by the application of this formalism to the Hubbard model at strong coupling, using the DMFT approximation as an initial condition of the fRG flow.
        
        The results of this chapter have appeared in:
        \begin{itemize}
            \item P.~M.~Bonetti,  A.~Toschi,  C.~Hille,  S.~Andergassen,  and  D.~Vilardi,  Single  boson exchange representation of the functional renormalization group for strongly interacting many-electron systems, \href{https://doi.org/10.1103/PhysRevResearch.4.013034}{Phys.~Rev.~Research \textbf{4}, 013034 (2022)}.
        \end{itemize}
        \item In Chapter~\ref{chap: low energy spiral}, we analyze the low-energy properties of magnons in an itinerant spiral magnet. In particular, we show that the \emph{complete} breaking of the SU(2) symmetry gives rise to three Goldstone modes. For each of these, we present a low energy expansion of the magnetic susceptibilities within the random phase approximation (RPA), and derive formulas for the spin stiffnesses and spectral weights. We also show that \emph{local Ward identities} enforce that these quantities can be alternatively computed from the response to a gauge field. Moreover, we analyze the size and the low-momentum and frequency dependence of the Landau damping of the Goldstone modes, due to their decay into particle-hole pairs. 
        
        The results of this chapter have appeared in:
        \begin{itemize}
            \item P.~M.~Bonetti, and W.~Metzner, Spin stiffness, spectral weight, and Landau damping of magnons in metallic spiral magnets, \href{https://link.aps.org/doi/10.1103/PhysRevB.105.134426}{Phys.~Rev.~B \textbf{105}, 134426 (2022)}.
            \item P.~M.~Bonetti, Local Ward identities for collective excitations in fermionic systems with spontaneously broken symmetries, \href{https://doi.org/10.48550/arXiv.2204.04132}{arXiv:2204.04132}, \textit{accepted in Physical Review B} (2022). 
        \end{itemize}
        \item In Chapter~\ref{chap: pseudogap}, we formulate a theory for the pseudogap phase in high-$T_c$ cuprates. This is achieved \emph{fractionalizing} the electron into a "chargon", carrying the original electron charge, and a charge neutral "spinon", which is a SU(2) matrix providing a time and space dependent local spin reference frame. We then consider a magnetically ordered state for the chargons where the Fermi surface gets reconstructed. Despite the chargons display long-range order, symmetry breaking at finite temperature is prevented by spinon fluctuations, in agreement with the Mermin-Wagner theorem. We subsequently derive an effective theory for the spinons integrating out the chargon degrees of freedom. The spinon dynamics is governed by a non-linear sigma model (NL$\sigma$M). By performing a large-$N$ expansion of the NL$\sigma$M derived from the two-dimensional Hubbard model at moderate coupling, we find a broad finite temperature pseudogap regime. At weak or moderate coupling $U$, however, spinon fluctuations are not strong enough to destroy magnetic long-range order in the ground state, except possibly near the edges of the pseudogap regime at large hole doping. The spectral functions in the hole doped pseudogap regime have the form of hole pockets with a truncated spectral weight on the backside, similar to the experimentally observed Fermi arcs. 
        The results of this chapter appear in
        \begin{itemize}
            \item P.~M.~Bonetti, and W.~Metzner, SU(2) gauge theory of the pseudogap phase in the two-dimensional Hubbard model, \href{https://doi.org/10.48550/arXiv.2207.00829}{arXiv:2207.00829} (2022).
        \end{itemize}
    \end{itemize}
    %
    

\cleardoublepage
    \rhead[\fancyplain{}{\bfseries Methods}]{\fancyplain{}{\bfseries\thepage}}
    \lhead[\fancyplain{}{\bfseries\thepage}]{\fancyplain{}{\bfseries Methods}}
    \chapter{Methods}
    \label{chap: methods}

    \section{Functional renormalization group (fRG)}
    The original idea of an exact flow equation for a generating functional dates back to Wetterich~\cite{Wetterich1993}, who derived it for a bosonic theory. Since then, the concept of a \emph{nonperturbative} renormalization group, that is, distinct from the \emph{perturbative} Wilsonian one~\cite{Wilson1975}, has been applied in many contexts, ranging from quantum gravity to statistical physics (see Ref.~\cite{Dupuis2021} for an overview). The first application of the Wetterich equation to correlated Fermi systems is due to Salmhofer and Honerkamp~\cite{Salmhofer2001}, in the context of the Hubbard model. 
    
    In this section, we present the functional renormalization group equations for the one-particle-irreducible (1PI) correlators of fermionic fields. The derivation closely follows Ref.~\cite{Metzner2012}, and we refer to it and to Refs.~\cite{Salmhofer1999,Berges2002,Kopietz2010} for further details.
        
    \subsection{Generating functionals}
    We start by defining the generating functional of \emph{connected} Green's functions as~\cite{NegeleOrland}
    \begin{equation}
        W\left[\eta,\overline{\eta}\right]=-\ln\int\!\mathcal{D}\Psi \mathcal{D}\overline{\Psi}\,
        e^{-\mathcal{S}\left[\Psi,\overline{\Psi}\right]+\left(\overline{\eta},\Psi\right)+\left(\overline{\Psi},\eta\right)},
        \label{eq_methods: W functional}
    \end{equation}
    where the symbol $(\overline{\eta},\Psi)$ is a shorthand for $\sum_{x}\overline{\eta}(x)\,\Psi(x)$, with $x$ a collective variable grouping a set of suitable quantum numbers and imaginary time or frequency. The bare action $\mathcal{S}$ typically consists of a noninteracting one-body term $\mathcal{S}_0$
    \begin{equation}
        \mathcal{S}_0\left[\Psi,\overline{\Psi}\right]=
        -\left(\overline{\Psi},G_0^{-1}\Psi\right),
    \end{equation}
    with $G_0$ the bare propagator, and a two-body interaction
    \begin{equation}
        \mathcal{S}_\mathrm{int}\left[\Psi,\overline{\Psi}\right]=
        \frac{1}{(2!)^2}
        \sum_{\substack{x_1',x_2',\\x_1,x_2}} \lambda(x_1',x_2';x_1,x_2)\, \overline{\Psi}(x_1')\overline{\Psi}(x_2')\Psi(x_2)\Psi(x_1),
        \label{eq_methods: Sint}
    \end{equation}
    with $\lambda$ describing the two-body potential. Deriving Eq.~\eqref{eq_methods: W functional} with respect to the source fields $\eta$ and $\overline{\eta}$, one can obtain the correlation functions corresponding to connected Feynman diagrams. In general, we define the connected $m$-particle Green's function as
    \begin{equation}
        G^{(2m)}(x_1,\shortdots,x_m,x_1',\shortdots,x_m')=(-1)^m
        \frac{\delta^{(2m)} W\left[\eta,\overline{\eta}\right]}{\delta\overline{\eta}(x_1)\shortdots\delta\overline{\eta}(x_m)\delta\eta(x_m')\shortdots\delta\eta(x_1')}\Bigg\rvert_{\eta,\overline{\eta}=0}.
        \label{eq_methods: m-particle Gfs}
    \end{equation}
    In particular, the $m=1$ case gives the interacting propagator.
    
    Another relevant functional is the so-called effective action, which generates all the 1PI correlators, that is, all correlators which cannot be divided into two distinct parts by removing a propagator line. It is defined as the Legendre transform of $W$
    \begin{equation}
        \Gamma\left[\psi,\overline{\psi}\right] = W\left[\eta,\overline{\eta}\right] + \left(\overline{\eta},\psi\right)+\left(\overline{\psi},\eta\right),
    \end{equation}
    where the fields $\psi$ and $\overline{\psi}$, represent the expectation values of the original fields $\Psi$ and $\overline{\Psi}$ in presence of the sources. They are related to $\eta$ and $\overline{\eta}$ via
    \begin{subequations}
        \begin{align}
            &\psi = - \frac{\delta W}{\delta\overline{\eta}},\\
            &\overline{\psi} = + \frac{\delta W}{\delta\eta},
        \end{align}
        \label{eq_methods: psi from G}
    \end{subequations}
    and the inverse relations read as 
    \begin{subequations}
        \begin{align}
            &\frac{\delta\Gamma}{\delta\psi} = -\overline{\eta},\\
            &\frac{\delta\Gamma}{\delta\overline{\psi}} = +\eta.
        \end{align}
    \end{subequations}
    Deriving $\Gamma$, one can obtain the 1PI $m$-particle correlators, that is,
    \begin{equation}
        \Gamma^{(2m)}(x_1,\shortdots,x_m,x_1',\shortdots,x_m')=
        \frac{\delta^{(2m)} \Gamma\left[\psi,\overline{\psi}\right]}{\delta\overline{\psi}(x_1')\shortdots\delta\overline{\psi}(x_m')\delta\psi(x_m)\shortdots\delta\psi(x_1)}\Bigg\rvert_{\psi,\overline{\psi}=0}.
    \end{equation}
    In particular the $m=1$ case gives the inverse interacting propagator,
    \begin{equation}
        \Gamma^{(2)} = G^{-1} = G_0^{-1} - \Sigma,
    \end{equation}
    with $\Sigma$ the self-energy, and the $m=2$ case the so-called two-particle vertex or effective interaction. It is possible to derive~\cite{NegeleOrland} a particular relation between the $W$ and $\Gamma$ functional, called reciprocity relation. It reads as
    \begin{equation}
        \mathbf{\Gamma}^{(2)}\left[\psi,\overline{\psi}\right]=\left(\mathbf{G}^{(2)}\left[\eta,\overline{\eta}\right]\right)^{-1},
        \label{eq_methods: reciprocity rel}
    \end{equation}
    with
    \begin{equation}
        \mathbf{G}^{(2)}\left[\eta,\overline{\eta}\right] = -
        \left(
        \begin{array}{cc}
            \frac{\delta^2W}{\delta\overline{\eta}(x)\delta\eta(x')} & -\frac{\delta^2W}{\delta\overline{\eta}(x)\delta\overline{\eta}(x')} \\
            -\frac{\delta^2W}{\delta\eta(x)\delta\eta(x')} & \frac{\delta^2W}{\delta\eta(x)\delta\overline{\eta}(x')}
        \end{array}
        \right),
    \end{equation}
    and
    \begin{equation}
        \mathbf{\Gamma}^{(2)}\left[\psi,\overline{\psi}\right] = 
        \left(
        \begin{array}{cc}
            \frac{\delta^2\Gamma}{\delta\overline{\psi}(x')\delta\psi(x)} & \frac{\delta^2\Gamma}{\delta\overline{\psi}(x')\delta\overline{\psi}(x)} \\
            \frac{\delta^2\Gamma}{\delta\psi(x')\delta\psi(x)} & \frac{\delta^2\Gamma}{\delta\psi(x')\delta\overline{\psi}(x)}
        \end{array}
        \right).
        \label{eq_methods: gamma2}
    \end{equation}
    \subsection{Derivation of the exact flow equation}
    For single band, translationally invariant systems, the bare propagator $G_0$ takes a simple form in momentum and imaginary frequency space:
    \begin{equation}
        G_0(\bk,\nu)= \frac{1}{i\nu -\xi_\bk},
    \end{equation}
    where $\nu$ is a fermionic Matusbara frequency, taking the value $(2n+1)\pi T$ ($n\in\mathbb{Z}$) at finite temperature $T$, and $\xi_\bk$ the band dispersion relative to the chemical potential $\mu$. At low temperatures $G_0$ exhibits a nearly singular structure at $\nu\sim 0$ and $\xi_\bk=0$, which highly influences the physics of the correlated system. This is a manifestation of the importance of the low energy excitations, that is, those close to the Fermi surface, at low temperatures. Therefore, one might be tempted to perform the integral in \eqref{eq_methods: W functional} step by step, including first the high energy modes and then, gradually, the low energy ones. This can be achieved regularizing the propagator via a scale-dependent function, that is,
    \begin{equation}
        G_0^\L(\bk,\nu)=\frac{\Theta^\L(\bk,\nu)}{i\nu -\xi_\bk},
    \end{equation}
    where $\Theta^\L(\bk,\nu)$ is a function that vanishes for $\nu\ll \L $ and/or $\xi_\bk\ll \L$ and tends to one for $\nu\gg\L$ and/or $\xi_\bk\gg\L$. In this way, one can define a scale-dependent action as
    \begin{equation}
        \mathcal{S}^\L\left[\Psi,\overline{\Psi}\right]=-\left(\overline{\Psi},Q_0^\L\Psi\right)+\mathcal{S}_\mathrm{int}\left[\Psi,\overline{\Psi}\right],
    \end{equation}
    with $Q_0^\L=(G_0^\L)^{-1}$, as well as a scale-dependent $ W$-functional
    \begin{equation}
         W^\L\left[\eta,\overline{\eta}\right]=-\ln\int\!\mathcal{D}\Psi \mathcal{D}\overline{\Psi}\,
        e^{-\mathcal{S}^\L\left[\Psi,\overline{\Psi}\right]+\left(\overline{\eta},\Psi\right)+\left(\overline{\Psi},\eta\right)}.
        \label{eq_methods: W functional Lambda}
    \end{equation}
    Differentiating Eq.~\eqref{eq_methods: W functional Lambda} with respect to $\L$, we obtain an exact flow equation for $ W$: 
    \begin{equation}
        \begin{split}
            \deL  W^\L=e^{ W^\L}\deL e^{- W^\L}
            &=e^{ W^\L}\int\!\mathcal{D}\Psi \mathcal{D}\overline{\Psi}\,\left(\overline{\Psi},\dot{Q}_0^\L\Psi\right) \,e^{-\mathcal{S}^\L\left[\Psi,\overline{\Psi}\right]+\left(\overline{\eta},\Psi\right)+\left(\overline{\Psi},\eta\right)}\\
            &=e^{ W^\L}\left(\frac{\delta}{\delta\eta},\dot{Q}_0^\L \frac{\delta}{\delta\overline{\eta}}\right)e^{- W^\L}\\
            &=\left(\frac{\delta W^\L}{\delta\eta},\dot{Q}_0^\L\frac{\delta W^\L}{\delta\overline{\eta}}\right)+\tr\left[\dot{Q}_0^\L\frac{\delta^2 W^\L}{\delta\overline{\eta}\delta\eta}\right],
        \end{split}
        \label{eq_methods: flow equation W}
    \end{equation}
    with $\dot{Q}_0^\L$ a shorthand for $\deL Q_0^\L$. Expanding $ W^\L$ in powers of the source fields, one can derive the flow equations for the connected Green's functions in Eq.~\eqref{eq_methods: m-particle Gfs}. Since 1PI correlators are easier to handle, we exploit the above result to derive a flow equation for the effective action functional $\Gamma^\L$:
    \begin{equation}
        \deL\Gamma^\L\left[\psi,\overline{\psi}\right]=
        \left(\deL\overline{\eta}^\L,\psi\right)+
        \left(\overline{\psi},\deL\eta^\L\right)+
        \deL W^\L\left[\eta^\L,\overline{\eta}^\L\right],
    \end{equation}
    where $\eta^\L$ and $\overline{\eta}^\L$ are solutions of the implicit equations
    \begin{subequations}
        \begin{align}
            &\psi =- \frac{\delta W^\L}{\delta\overline{\eta}},\\
            &\overline{\psi} = \frac{\delta W^\L}{\delta\eta}.
        \end{align}
    \end{subequations}
    Using the properties of the Legendre transform, we get
    \begin{equation}
        \deL\Gamma^\L\left[\psi,\overline{\psi}\right]=\deL W^\L\left[\eta^\L,\overline{\eta}^\L\right]\Big\rvert_{\eta^\L,\overline{\eta}^\L\,\mathrm{fixed}},
    \end{equation}
    that, combined with Eq.~\eqref{eq_methods: psi from G}, \eqref{eq_methods: reciprocity rel}, and \eqref{eq_methods: flow equation W} gives
    \begin{equation}
        \deL\Gamma^\L\left[\psi,\overline{\psi}\right]=
        -\left(\overline{\psi},\dot{Q}_0^\L\psi\right)-\frac{1}{2}\tr\left[\dot{\mathbf{Q}}_0^\L\left(\mathbf{\Gamma}^{(2)\L}\right)^{-1}\right],
        \label{eq_methods: Wetterich eq G0L}
    \end{equation}
    with $\mathbf{\Gamma}^{(2)\Lambda}$ the same as in Eq.~\eqref{eq_methods: gamma2}, and
    \begin{equation}
        \dot{\mathbf{Q}}_0^\L(x,x')=
        \left(
        \begin{array}{cc}
            \dot{Q}_0^\L(x,x') & 0 \\
            0 & -\dot{Q}_0^\L(x',x)
        \end{array}
        \right).
    \end{equation}
    Alternatively~\cite{Berges2002}, one can define the regularized bare propagators via a regulator $R^\L$:
    \begin{equation}
        G_0^\L=\frac{1}{G_0^{-1}-R^\L},
    \end{equation}    
    and introduce the concept of effective average action,
    \begin{equation}
        \Gamma_R^\L\left[\psi,\overline{\psi}\right]=\Gamma^\L\left[\psi,\overline{\psi}\right]-\left(\overline{\psi},R^\L\psi\right),
    \end{equation}
    so that the flow equation for $\Gamma_R^\L$ becomes
    \begin{equation}
        \begin{split}
            \deL\Gamma^\L_R\left[\psi,\overline{\psi}\right]=&-\frac{1}{2}\tr\left[\dot{\mathbf{R}}^\L\left(\mathbf{\Gamma}_R^{(2)\L}+\mathbf{R}^\L\right)^{-1}\right]\\
            =&-\frac{1}{2}\widetilde{\partial}_\L\tr\ln\left[\mathbf{\Gamma}_R^{(2)\L}+\mathbf{R}^\L\right],    
        \end{split}
        \label{eq_methods: Wetterich eq RL}
    \end{equation}
    with $\mathbf{R}^\L$ defined similarly to $\mathbf{Q}_0^\L$, and the symbol $\widetilde{\partial}_\L$ is defined as $\displaystyle{\widetilde{\partial}_\L=\dot{R}^\L\frac{\delta}{\delta R^\L}}$. 
    
    Eq.~\eqref{eq_methods: Wetterich eq G0L} (or \eqref{eq_methods: Wetterich eq RL}) is the so-called Wetterich equation and describes the exact evolution of the effective action functional. For the whole approach to make sense, it is necessary to completely remove the regularization of $G_0$ at the final scale $\L=\Lfin$, that is, $G_0^{\Lfin}=G_0$, so that at the final scale the scale-dependent effective action is the effective action of the many-body problem defined by action $\mathcal{S}$. Furthermore, like any other first order differential equation, Eq.~\eqref{eq_methods: Wetterich eq G0L} must be complemented with an initial condition at the initial scale $\Lini$. If we choose the function $\Theta^\L$ such that $G_0^{\Lini}= 0$, the integral in \eqref{eq_methods: W functional Lambda} is exactly given by the saddle point approximation, and Legendre transforming we get
    \begin{equation}
        \Gamma^{\Lini}\left[\psi,\overline{\psi}\right]=\mathcal{S}\left[\psi,\overline{\psi}\right]+\left(\overline{\psi},\left[Q_0^{\Lini}-G_0^{-1}\right]\psi\right)
        =\mathcal{S}^{\Lini}\left[\psi,\overline{\psi}\right],
    \end{equation}
    or, in terms of the effective average action,
    \begin{equation}
        \Gamma^{\Lini}_R\left[\psi,\overline{\psi}\right]=\mathcal{S}\left[\psi,\overline{\psi}\right].
        \label{eq_methods: fRG Gamma ini}
    \end{equation}
    \subsection{Expansion in the fields}
\begin{figure}[t]
        \centering
        \includegraphics[width=1.0\textwidth]{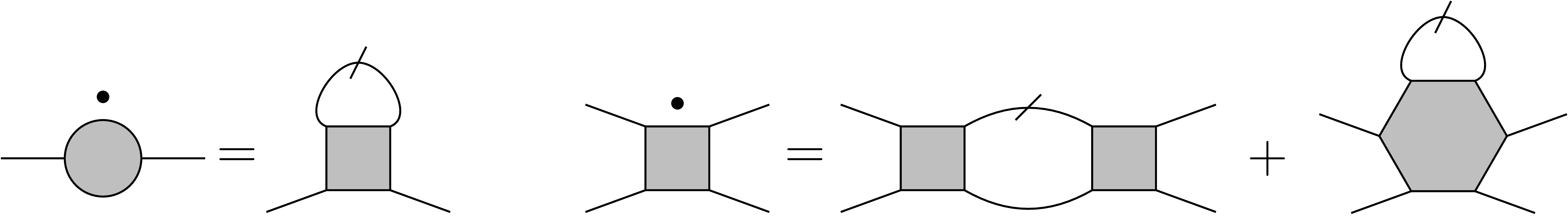}
        \caption{Schematic representation of the flow equations for the self-energy (left) and vertex (right). The ticked lines represent single-scale propagators, and the dots over the symbols scale derivatives.}
        \label{fig_methods: flow diagrams}
    \end{figure}

    A common approach to tackle Eq.~\eqref{eq_methods: Wetterich eq G0L} is to expand the effective action functional $\Gamma^\L$ in powers of the fields, where the coefficient of the $2m$-th power corresponds to the $m$ particle vertex up to a prefactor. We write $\mathbf{\Gamma}^{(2)\L}$ as 
    \begin{equation}
        \mathbf{\Gamma}^{(2)\L}\left[\psi,\overline{\psi}\right] = \left(\mathbf{G}^\L\right)^{-1} - \widetilde{\mathbf{\Sigma}}^\L\left[\psi,\overline{\psi}\right],
        \label{eq_methods: Gamma2=G-Sigma}
    \end{equation}
    where $\left(\mathbf{G}^\L\right)^{-1}$ is at the same time the field-independent part of $\mathbf{\Gamma}^{(2)\L}$ and the interacting propagator, and $\widetilde{\mathbf{\Sigma}}^\L$ vanishes for zero fields, that is, it is \emph{at least} quadratic in $\psi$, $\overline{\psi}$. We further notice that, as long as no pairing is present in the system, $\mathbf{G}^\L(x,x')$ can be expressed as $\mathrm{diag}\left(G^\L(x,x'),-G^\L(x',x)\right)$. Inserting \eqref{eq_methods: Gamma2=G-Sigma} into \eqref{eq_methods: Wetterich eq G0L} and writing
    \begin{equation}
        \left(\mathbf{\Gamma}^{(2)\L}\right)^{-1} 
        = \left(1-\mathbf{G}^\L \widetilde{\mathbf{\Sigma}}^\L \right)^{-1}\mathbf{G}^\L
        = \mathbf{G}^\L + \mathbf{G}^\L\widetilde{\mathbf{\Sigma}}^\L\mathbf{G}^\L+ \mathbf{G}^\L\widetilde{\mathbf{\Sigma}}^\L\mathbf{G}^\L\widetilde{\mathbf{\Sigma}}^\L\mathbf{G}^\L + \dots,
    \end{equation}
    we get
    \begin{equation}
        \begin{split}
            \deL\Gamma^\L\left[\psi,\overline{\psi}\right]=
            -\left(\overline{\psi},\dot{Q}_0^\L\psi\right)-\tr\left[\dot{Q}_0^\L G^\L\right]
            +\frac{1}{2}\tr\left[\mathbf{S}^\L\left(\widetilde{\mathbf{\Sigma}}^\L+\widetilde{\mathbf{\Sigma}}^\L\mathbf{G}^\L\widetilde{\mathbf{\Sigma}}^\L+\dots\right)\right],
        \end{split}
        \label{eq_methods: flow equation expanded in the fields}
    \end{equation}
    where we have defined a \emph{single-scale propagator} $\mathbf{S}^\L=-\mathbf{G}^\L\dot{\mathbf{Q}}_0^\L\mathbf{G}^\L=\widetilde{\partial}_\L\mathbf{G}^\L$, which, in a normal system, reads as $\mathbf{S}^\L(x,x')=\mathrm{diag}\left(S^\L(x,x'),-S^\L(x',x)\right)$, with $S^\L=\widetilde{\partial}_\L G^\L$. Here, the the symbol $\widetilde{\partial}_\L$ is intended as $\widetilde{\partial}_\L=\displaystyle{\dot{Q}_0^\L\frac{\delta}{\delta Q_0^\L}}=\displaystyle{\deL\left(\frac{1}{\Theta^\L}\right)\frac{\delta}{\delta(1/\Theta^\L)}}$.
    If we now write 
    \begin{equation}
        \begin{split}
            \Gamma^\L\left[\psi,\overline{\psi}\right] =
            &\Gamma^{(0)\L} 
            - \sum_{x,x'} \Gamma^{(2)\L}(x',x)\,\overline{\psi}(x')\psi(x) \\
            &+ \frac{1}{(2!)^2}\sum_{\substack{x_1',x_2',\\x_1,x_2}} \Gamma^{(4)\L}(x_1',x_2',x_1,x_2)\,\overline{\psi}(x_1')\overline{\psi}(x_2')\psi(x_2)\psi(x_1)\\
            &- \frac{1}{(3!)^2}\sum_{\substack{x_1',x_2',x_3',\\x_1,x_2,x_3}} \Gamma^{(6)\L}(x_1',x_2',x_3',x_1,x_2,x_3)\,\overline{\psi}(x_1')\overline{\psi}(x_2')\overline{\psi}(x_3')\psi(x_3)\psi(x_2)\psi(x_1) \\
            & + \dots,
        \end{split}
    \end{equation}
    and compare the coefficients in Eq.~\eqref{eq_methods: flow equation expanded in the fields}, we can derive the flow equations for all the different moments $\Gamma^{(2m)\L}$ of the effective action. We remark that, since we are dealing with fermions, all the $\Gamma^{(2m)\L}$ vertices are \emph{antisymmetric} under the exchange of a pair of primed or non-primed indices, that is
    \begin{equation}
        \begin{split}
            &\Gamma^{(2m)\L}(x_1',\shortdots,x_{\overline{i}}',\shortdots,x_{\overline{j}}',\shortdots, x_m',x_1,\shortdots,x_i,\shortdots,x_j,\shortdots, x_m)\\
            &=(-1)\Gamma^{(2m)\L}(x_1',\shortdots,x_{\overline{j}}',\shortdots,x_{\overline{i}}',\shortdots, x_m',x_1,\shortdots,x_i,\shortdots,x_j,\shortdots, x_m)\\
            &=(-1)\Gamma^{(2m)\L}(x_1',\shortdots,x_{\overline{i}}',\shortdots,x_{\overline{j}}',\shortdots, x_m',x_1,\shortdots,x_j,\shortdots,x_i,\shortdots, x_m).
        \end{split}
    \end{equation}
    The 0-th moment of the effetive action, $\Gamma^{(0)\L}$, is given by $T^{-1}\Omega^\L$, with $T$ the temperature and $\Omega^\L$ the grand canonical potential~\cite{Metzner2012}, so that we have
    \begin{equation}
        \deL\Omega^\L = -T\tr\left[\dot{Q}_0^\L G^\L\right]. 
    \end{equation}
    The flow equation for the 2nd moment reads as
    \begin{equation}
        \deL\Gamma^{(2)\L} = \dot{Q}_0^\L - \Tr\left[S^\L \Gamma^{(4)\L}\right].
    \end{equation}
    Noticing that $\Gamma^{(2)\L}=(G^\L)^{-1}-\Sigma^\L$, we can extract the flow equation for the self-energy:
    \begin{equation}
        \deL\Sigma^\L(x',x) = \sum_{y,y'} S^\L(y,y') \Gamma^{(4)\L}(x',y',x,y),
        \label{eq_methods: flow eq Sigma xx'}
    \end{equation}
    where its initial condition can be extracted from~\eqref{eq_methods: fRG Gamma ini}, and it reads as $\Sigma^\Lini(x',x)=0$.
    Similarly, one can derive the evolution equation for the two-particle vertex $\Gamma^{(4)\L}$:
    \begin{equation}
        \begin{split}
            \deL\Gamma^{(4)\L}(x_1',x_2',x_1,x_2) =\hskip 2cm&\\
             \sum_{\substack{y_1',y_2',\\y_1,y_2}}
                \bigg[P^\L(y_1',y_2',y_1,y_2)
                    \Big(
                        &+\Gamma^{(4)}(x_1',y_2',x_1,y_1)\Gamma^{(4)}(y_1',x_2',y_2,x_2)\\[-5mm]
                        &-\Gamma^{(4)}(x_1',y_1',y_2,x_2)\Gamma^{(4)}(y_2',x_2',x_1,y_1)\\
                        &-\frac{1}{2}\Gamma^{(4)}(x_1',x_2',y_1,y_2)\Gamma^{(4)}(y_1',y_2',x_1,x_2)        
                    \Big)
                \bigg]\\
                & \hskip -4.75cm -\sum_{y,y'}S^\L(y,y')\Gamma^{(6)\L}(x_1',x_2',y',x_1,x_2,y),
        \end{split}
        \label{eq_methods: flow eq vertex xx'}
    \end{equation}
    with
    \begin{equation}
        P^\L(y_1',y_2',y_1,y_2) = S^\L(y_1,y_1')G^\L(y_2,y_2') + S^\L(y_2,y_2')G^\L(y_1,y_1').
        \label{eq_methods: GS+SG}
    \end{equation}
    The initial condition for the two particle vertex, reads as $\Gamma^{(4)\Lini}=\lambda$, with $\lambda$ the bare two-particle interaction in Eq.~\eqref{eq_methods: Sint}.
    In Fig.~\ref{fig_methods: flow diagrams} a schematic representation of the flow equations for the self-energy and the two-particle vertex is shown.
    \subsection{Truncations}
    Inspecting Eqs.~\eqref{eq_methods: flow eq Sigma xx'} and, in particular \eqref{eq_methods: flow eq vertex xx'}, we notice that the flow equation for the self-energy requires the knowledge of the two-particle vertex, whose flow equation involves $\Sigma^\L$ (through $G^\L$ and $S^\L$), $\Gamma^{(4)\L}$, and $\Gamma^{(6)\L}$. Considering higher order terms, one can prove that the right hand side of the flow equation for $\Gamma^{(2\overline{m})\L}$ involves all the $\Gamma^{(2m)\L}$, with $m\leq\overline{m}+1$. This produces an infinite hierarchy of flow equations for the $m$-particle 1PI correlators that, for practical reasons, needs to be truncated at some given order. Since for most purposes the calculation of the self-energy and of the two-particle vertex is sufficient, the truncations often work as approximations on the three-particle vertex $\Gamma^{(6)\L}$. The simplest one could perform is the so-called \emph{1-loop} ($1\ell$) truncation, where the three-particle vertex is set to zero all along the flow, in this way, the last term in Eq.~\eqref{eq_methods: flow eq vertex xx'} can be discarded to compute the flow equation of the two-particle vertex. 
    
\begin{figure}[t]
        \centering
        \includegraphics[width=1.0\textwidth]{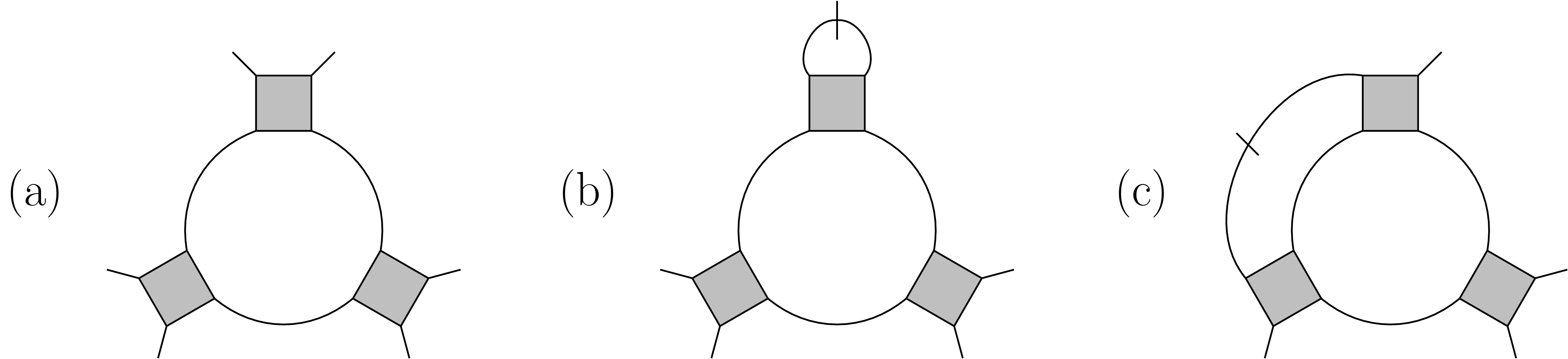}
        \caption{(a) Feynman diagram representing the approximate integration of the flow equation for $\Gamma^{(6)\L}$. (b-c) Contributions to the vertex flow equations with non-overlapping (b) and overlapping (c) loops. Here, the ticked lines represent single-scale propagators $S^\L$.}
        \label{fig_methods: integrated Gamma6}
    \end{figure}

    Alternatively, one can \emph{approximately} integrate the flow equation for $\Gamma^{(6)}$, obtaining the loop diagram schematically shown in panel (a) of Fig.~\ref{fig_methods: integrated Gamma6}, and insert it into the last term of the flow equation for the vertex. One can then classify the resulting terms into two classes depending on whether the corresponding diagram displays non-overlapping or overlapping loops (see (b) and (c) panels of Fig.~\ref{fig_methods: integrated Gamma6}). By considering only the former class, one can easily prove that these terms coming from $\Gamma^{(6)\L}$ can be reabsorbed into the first ones of Eq.~\eqref{eq_methods: flow eq vertex xx'} by replacing the single-scale propagator $S^\L$ with the full derivative of the Green's function $\deL G^\L = S^\L + G^\L (\deL \Sigma^\L) G^\L$ in Eq.~\eqref{eq_methods: GS+SG}, so that one can rewrite
    \begin{equation}
        P^\L(y_1',y_2',y_1,y_2) = \deL \left[G^\L(y_1,y_1')G^\L(y_2,y_2')\right].
        \label{eq_methods: P Katanin}
    \end{equation}
    This approximation is known under the name of Katanin scheme~\cite{Katanin2004} and, when considering only one of the first three terms in Eq.~\eqref{eq_methods: flow eq vertex xx'} (with $P^\Lambda$ as in Eq.~\eqref{eq_methods: P Katanin}), becomes equivalent to a \emph{Hartree-Fock} approximation for the self-energy, combined with a ladder resummation for the vertex~\cite{Salmhofer2004}.
    The more involved inclusion of diagrams with \emph{both} non-overlapping and overlapping loops leads to the \emph{2-loop} ($2\ell$) truncation, introduced by Eberlein~\cite{Eberlein2014}.
    
    Finally, Kugler and von Delft~\cite{Kugler2018_PRL,Kugler2018_PRB} have recently developed the so-called \emph{multiloop approximation}, which systematically and approximately takes into account contributions from higher order 1PI vertices in the fashion of a loop expansion. They also proved that in the limit of infinite loops this truncation becomes equivalent to the parquet approximation~\cite{Roulet1969,Bickers2004}, based on a diagrammatic approach, rather than on a flow equation. 
    
    In the context of statistical physics, where one mainly deals with bosonic rather than fermionic fields, other \emph{nonperturbative} truncations are possible. One can, for example, write the effective action as a one-body term (propagator) plus a local potential that only depends on the absolute value of the field, and then compute the flow of these two terms. In this way, one is able to include contributions from vertices with an arbitrary number of external legs. This approximation goes under the name of \emph{local potential approximation} (LPA). For a more detailed discussion on the LPA and its extensions, see Ref.~\cite{Berges2002} and references therein.
    \subsection{Vertex flow equation}
    \label{subs_methods: vertex flow eq}
\begin{figure}
        \centering
        \includegraphics[width=0.9\textwidth]{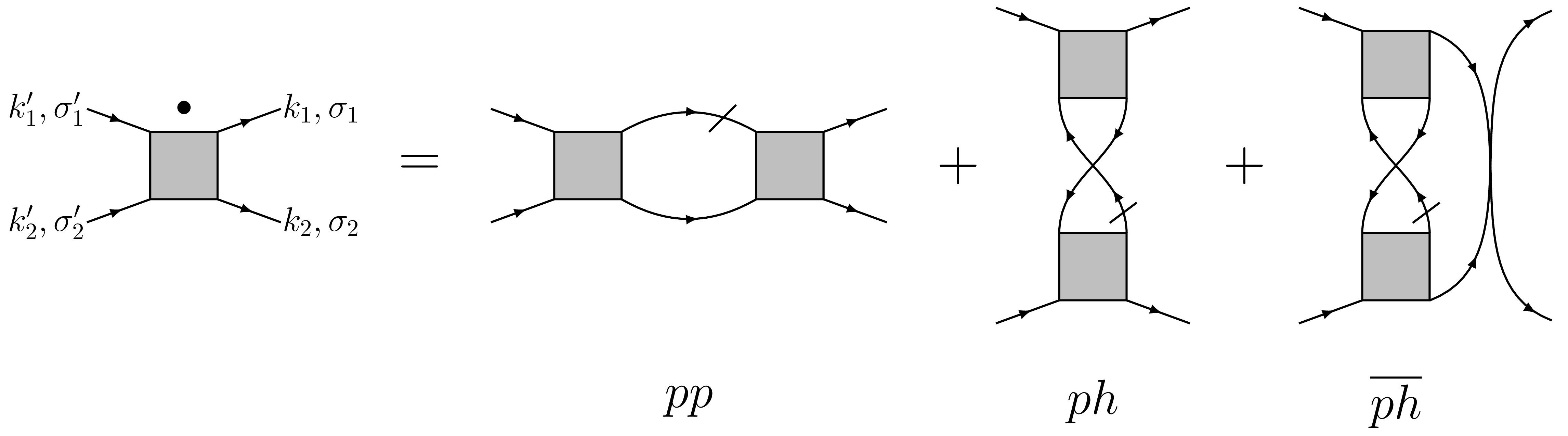}
        \caption{Schematic representation of the first three terms in Eq.~\eqref{eq_methods: flow eq vertex xx'}, also referred to as $pp$, $ph$, and $\overline{ph}$ channels (see text). For each channel, another diagram with the "tick" on the other internal fermionic line exists.}
        \label{fig_methods: diagrams flow vertex}
    \end{figure}

    We now turn our attention to the first three terms of Eq.~\eqref{eq_methods: flow eq vertex xx'} and neglect the contribution from the three-particle vertex, in a $1\ell$ approximation. Following the order of Eq.~\eqref{eq_methods: flow eq vertex xx'}, we call them particle-hole ($ph$), particle-hole-crossed ($\phx$) and particle-particle ($pp$) channels, respectively. In Fig.~\ref{fig_methods: diagrams flow vertex} we show a diagrammatic representation of each term. 

    If we now consider a rotationally and translationally invariant system of spin-$\frac{1}{2}$ fermions, we can choose the set of quantum numbers and imaginary frequency as $x=\{k,\sigma\}$, where $k=(\bk,\nu)$ is a collective variable encoding the spatial momentum and the frequency, and $\sigma=\up,\down$ is the spin projection. Under these assumptions, the propagator reads as 
    \begin{equation}
        G^\L(x',x)= G^\L_{\sigma'\sigma}(k',k) = G^\L(k)\delta_{\sigma\sigma'}\delta(k-k'),
    \end{equation}
    and a similar relation holds for $S^\L$. Analogously, we can write the two-particle vertex as 
    \begin{equation}
        \Gamma^{(4)\L}(x_1',x_2',x_1,x_2) = \Gamma^{(4)\L}_{\sigma_1'\sigma_2'\sigma_1^{\phantom{'}}\sigma_2^{\phantom{'}}}(k_1',k_2',k_1)\,\delta(k_1'+k_2'-k_1-k_2),
        \label{eq_methods: V en mom cons}
    \end{equation}
    where spin rotation invariance constrains the dependency of the vertex on spin-projections
    \begin{equation}
        \Gamma^{(4)\L}_{\sigma_1'\sigma_2'\sigma_1^{\phantom{'}}\sigma_2^{\phantom{'}}}(k_1',k_2',k_1)=
        V^\L(k_1',k_2',k_1)\delta_{\sigma_1'\sigma_1^{\phantom{'}}}\delta_{\sigma_2'\sigma_2^{\phantom{'}}}
        +\overline{V}^\L(k_1',k_2',k_1)\delta_{\sigma_1'\sigma_2^{\phantom{'}}}\delta_{\sigma_2'\sigma_1^{\phantom{'}}}.
        \label{eq_methods: V SU(2) inv}
    \end{equation}
    Finally, the antisymmetry properties of $\Gamma^{(4)\L}$ enforce $\overline{V}^\L(k_1',k_2',k_1)=-V^\L(k_2',k_1',k_1)$. Inserting \eqref{eq_methods: V en mom cons} and \eqref{eq_methods: V SU(2) inv} into \eqref{eq_methods: flow eq vertex xx'}, and with some straightforward calculations, we obtain a flow equation for $V^\L=\Gamma^{(4)\L}_{\up\down\up\down
    }$
    \begin{equation}
        \deL V^\L(k_1',k_2',k_1) = \mathcal{T}_{ph}^\L(k_1',k_2',k_1)+\mathcal{T}_{\phx}^\L(k_1',k_2',k_1)+\mathcal{T}_{pp}^\L(k_1',k_2',k_1).
        \label{eq_methods: flow Eq V k}
    \end{equation}
    From now on, we define the symbol $\int_{k=(\bk,\nu)}=T\sum_\nu\int\frac{d^d\bk}{(2\pi)^d}$ as the sum over the Matsubara frequencies and an integral over the spatial momentum, which can be either unbounded, for continuum systems, or, for lattice systems, a Brillouin zone momentum. In the case of zero temperature, the sum $T\sum_\nu$ is replaced by an integral. The particle-hole, particle-hole-crossed, and particle-particle contributions in Eq.~\eqref{eq_methods: flow Eq V k} have been defined as
    \begin{subequations}
        \begin{align}
            &\mathcal{T}_{ph}^\L(k_1',k_2',k_1)
            =\int_p P^\L(p,p+k_1-k_1') \big[2V^\L(k_1',p+k_1-k_1',k_1)V^\L(p,k_2',p+k_1-k_1') \nonumber\\[-2mm]
                &\hskip6.5cm-V^\L(p+k_1-k_1',k_1',k_1)V^\L(p,k_2',p+k_1-k_1') \\
                &\hskip6.5cm-V^\L(k_1',p+k_1-k_1',k_1)V^\L(k_2',p,p+k_1-k_1') \nonumber
            \big],\\
            &\mathcal{T}_{\phx}^\L(k_1',k_2',k_1)
            =-\int_p P^\L(p,p+k_2'-k_1) 
                V^\L(k_1',p+k_2'-k_1,p)V^\L(p,k_2',k_1),\\
            &\mathcal{T}_{pp}^\L(k_1',k_2',k_1)
            =-\int_p P^\L(p,k_1'+k_2'-p) 
                V^\L(k_1',k_2',p)V^\L(p,k_1'+k_2'-p,k_1),
        \end{align}
        \label{eq_methods: Tph Tphx Tpp}
    \end{subequations}
    respectively, and $P^\Lambda(p,p')$ reads as
    \begin{equation}
        P^\L(p,p') = \widetilde{\partial}_\L \left[G^\L(p)G^\L(p')\right] = G^\L(p)S^\L(p') + S^\L(p)G^\L(p'). 
    \end{equation}
    An interesting fact of the decomposition in Eq.~\eqref{eq_methods: flow Eq V k} is that each of the three terms, $\mathcal{T}_{ph}^\L$, $\mathcal{T}_{\phx}^\L$, and $\mathcal{T}_{pp}^\L$, depend on a "bosonic" variable appearing as a sum or as a difference of two fermionic variables. One can therefore write the vertex function $V^\L$ as the sum of three terms, each of which depends on one of the above mentioned bosonic momenta and two fermionic ones, and its flow equation is given by the $\mathcal{T}^\L$ depending on the corresponding combination of momenta~\cite{Karrasch2008}. In formulas, we have
    \begin{equation}
        V^\L(k_1',k_2',k_1) = \lambda(k_1',k_2',k_1) + \phi^{(ph)\L}_{k_{ph},k_{ph}'}(k_1-k_1') + \phi^{(\phx)\L}_{k_{\phx},k_{\phx}'}(k_2'-k_1) - \phi^{(pp)\L}_{k_{pp},k_{pp}'}(k_1'+k_2'),
        \label{eq_methods: V ph phx pp}
    \end{equation}
    where $\lambda$ represents the bare two-particle interaction, and the last sign is choice of convenience. Furthermore, we have defined
    \begin{subequations}
        \begin{align}
            &k_{ph} = \rndup{k_1+k_1'}, \hskip 2cm
            k_{ph}' = \rndup{k_2+k_2'},\\
            &k_{\phx} = \rndup{k_1+k_2'}, \hskip 2cm
            k_{\phx}' = \rndup{k_2+k_1'},\\
            &k_{pp} = \rndup{k_1'-k_2'}, \hskip 2cm
            k_{pp}' = \rndup{k_1-k_2},
        \end{align}
        \label{eq_methods: k k' pp ph phx}
    \end{subequations}
    where, at finite $T$, the symbol $\rndupnotwo{k}$ rounds up the frequency component of $k$ to the closest fermionic Matsubara frequency, while at $T=0$ it has no effect. This apparently complicated parametrization of momenta has the goal to completely disentangle the dependencies on fermionic and bosonic variables of the various terms in \eqref{eq_methods: V ph phx pp}. The flow equations of these terms read as
    \begin{subequations}
        \begin{align}
            &\deL \phi^{(ph)\L}_{k,k'}(q) = \mathcal{T}_{ph}^\L\left(k-\rndup{q},k'+\rnddo{q},k+\rndup{q}\right),\\
            &\deL \phi^{(\phx)\L}_{k,k'}(q) = \mathcal{T}_{\phx}^\L\left(k-\rndup{q},k'+\rnddo{q},k'-\rnddo{q}\right),\\
            &\deL \phi^{(pp)\L}_{k,k'}(q) = -\mathcal{T}_{pp}^\L\left(\rnddo{q}+k,\rndup{q}-k,\rnddo{q}-k'\right),
        \end{align}
    \end{subequations}
    where here (at $T>0$) $\rndup{q}$ ($\rnddo{q}$) rounds up (down) the frequency component of $\frac{q}{2}$ to the closest \emph{bosonic} Matsubara frequency. 
    \subsection{Instability analysis}
    \label{sec_methods: instability analysis}
    One of the main reasons of the success obtained by the application of the fRG to correlated fermions, and the Hubbard model in particular, is that it allows for an \emph{unbiased} analysis of the possible instabilities and competing orders of the system~\cite{Zanchi1998,Zanchi2000,Halboth2000,Halboth2000_PRL,Honerkamp2001}. Indeed, through the fRG flow one can detect the presence of an ordering tendency by looking at the evolution of the vertex as the scale $\L$ is lowered and the cutoff removed. In many cases $V^\L$ diverges at a finite scale $\L_\mathrm{cr}>\Lfin$, signaling the onset of some spontaneous symmetry breaking. Decomposition~\eqref{eq_methods: flow Eq V k}, though being very practical under a computational point of view, does not generally allow for understanding which kind of order is to be realized at scales $\L<\L_\mathrm{cr}$. In this perspective, instead of \eqref{eq_methods: flow Eq V k}, one can perform a \emph{physical} channel decomposition, first introduced in the context of the fRG by Husemann \emph{et al.}~\cite{Husemann2009,Husemann2012}:
    \begin{equation}
        \begin{split}
            V^\L(k_1',k_2',k_1) = &\lambda(k_1',k_2',k_1) \\
            &+ \frac{1}{2}\mathcal{M}^{\L}_{k_{ph},k_{ph}'}(k_1-k_1') 
            - \frac{1}{2}\mathcal{C}^{\L}_{k_{ph},k_{ph}'}(k_1-k_1') \\
            &+ \mathcal{M}^{\L}_{k_{\phx},k_{\phx}'}(k_2'-k_1) \\
            &- \mathcal{P}^{\L}_{k_{pp},k_{pp}'}(k_1'+k_2'),
        \end{split}
        \label{eq_methods: channel decomp physical}
    \end{equation}
    where $\mathcal{M}^\L=\phi^{(\phx)\L}$, $\mathcal{C}^\L=-2\phi^{(ph)\L}+\phi^{(\phx)\L}$, and $\mathcal{P}^\L=\phi^{(pp)\L}$ are referred to as magnetic, charge, and pairing channels. Thanks to this decomposition, when a vertex divergence occurs, one can understand whether the system is trying to realize some kind of magnetic, charge, or superconducting (or superfluid) order, depending on which among $\mathcal{M}^\L$, $\mathcal{C}^\L$, or $\mathcal{P}^\L$ diverges. Furthermore, more information on the ordering tendency can be inferred by analyzing the combination of fermionic and bosonic momenta for which the channel takes extremely large (formally infinite) values. If, for example, in a 2D lattice system, we would detect $\mathcal{M}^{\L\to\L_\mathrm{cr}}_{k,k'}\left(q=\left(\left(\frac{\pi}{a},\frac{\pi}{a}\right),0\right)\right)\to\infty$ ($a$ is the lattice spacing), this would signal an instability towards antiferromagnetism. Differently, $\mathcal{P}^{\L\to\L_\mathrm{cr}}_{((\kx,\ky),\nu),k'}(q=(\bzero,0))\to+\infty$, and $\mathcal{P}^{\L\to\L_\mathrm{cr}}_{((\ky,\kx),\nu),k'}(q=(\bzero,0))\to-\infty$ would imply the tendency to a superconducting state with $d$-wave symmetry. The flow equations for the physical channels read as:
    \begin{subequations}
        \begin{align}
            &\deL \mathcal{M}^\L_{k,k'}(q) = \mathcal{T}_{\phx}^\L\left(k-\rndup{q},k'-\rnddo{q},k'-\rnddo{q}\right),\\
            &\deL \mathcal{C}^\L_{k,k'}(q) = -2\mathcal{T}_{ph}^\L\left(k-\rndup{q},k'-\rnddo{q},k+\rndup{q}\right) \nonumber \\
            &\hskip2.25cm+\mathcal{T}_{\phx}^\L\left(k-\rndup{q},k'-\rnddo{q},k'-\rnddo{q}\right), \\
            &\deL \mathcal{P}^\L_{k,k'}(q) = -\mathcal{T}_{pp}^\L\left(\rnddo{q}+k,\rndup{q}-k,\rnddo{q}-k'\right),
        \end{align}
    \end{subequations}
    where \eqref{eq_methods: channel decomp physical} has to be inserted into \eqref{eq_methods: Tph Tphx Tpp}. In Appendix~\ref{app: symm V}, one can find the symmetry properties of the various channels. 
    %
    \section{Dynamical mean-field theory (DMFT)}
    While the fRG schemes are able to capture both long- and short-range correlation effects, their applicability is restricted to weakly interacting systems, as the unavoidable truncations can be justified only in this limit. In this section, we deal with a different approach, namely the dynamical mean-field theory (DMFT)~\cite{Georges1992,Georges1996}, which can be used to study even strongly interacting systems, but treats only \emph{local} (that is, extremely short-ranged) correlations. In this section we restrict our attention to a particular class of lattice models which exhibit a purely local interaction, that is, the Hubbard models:
    \begin{equation}
        \mathcal{H} = \sum_{jj',\sigma=\up,\down}t_{jj'} c^\dagger_{j,\sigma}c_{j',\sigma} + U \sum_j n_{j,\up} n_{j,\down}-\mu\sum_{j,\sigma} n_{j,\sigma} ,
        \label{eq_methods: Hubbard hamilt}
    \end{equation}
    where $c^\dagger_{j,\sigma}$ ($c_{j,\sigma}$) creates (annihilates) a spin-$\frac{1}{2}$ electron at site $j$ with spin projection $\sigma$, $t_{jj'}$ represents the probability amplitude for an electron to hop form site $j$ to site $j'$, $U$ is the strength of the onsite interaction, $n_{j,\sigma}=c^\dagger_{j,\sigma}c_{j,\sigma}$, and $\mu$ is the chemical potential. 
    
    In classical spin systems, such as the ferromagnetic Ising model, a \emph{mean-field} (MF) approximation consists in replacing all the spins surrounding a given site with a uniform background field, the Weiss field, whose value is obtained by a self-consistent equation. Similarly, in lattice quantum many-body systems, one can focus on a single site and replace the neighboring ones with a \emph{dynamical} field, which still fully embodies quantum fluctuation effects~\cite{Georges1992}. Similarly to MF for spin systems, DMFT is exact in the limit of large coordination number $z\to\infty$, or, equivalently, in the limit of infinite spatial dimensions~\cite{Metzner1989}.
    \subsection{Self-consistency relation}
    The key point of DMFT is to replace the action deriving from~\eqref{eq_methods: Hubbard hamilt}, $\mathcal{S}=\int_0^\beta\!d\tau[c^\dagger \partial_\tau c + \mathcal{H}]$, with a purely local one
    \begin{equation}
        \mathcal{S}_\mathrm{imp} = 
        -\int_0^\beta\!d\tau \int_0^\beta\!d\tau'
        \sum_\sigma c^\dagger_{0,\sigma}(\tau)\,\mathcal{G}_0^{-1}(\tau-\tau')\,
        c_{0,\sigma}(\tau')
        +U\int_0^\beta\!d\tau\,
        n_{0,\up}(\tau)n_{0,\down}(\tau),
        \label{eq_methods: local eff action}
    \end{equation}
    where the label 0 in the fermionic operators stands for a given fixed site of the lattice and $U$ takes the same value as in the original Hubbard model. This action is usually referred to as (quantum) impurity problem, as it describes a 0+1 dimensional system. Here, the function $\mathcal{G}_0^{-1}$ plays the role of the Weiss field and has to be determined self-consistently. Since~\eqref{eq_methods: local eff action} is a \emph{local} approximation of~\eqref{eq_methods: Hubbard hamilt}, we require the local Green's function of the Hubbard model, that is
    \begin{equation}
        G_{\mathrm{loc}}(\tau) = 
        -\Big\langle \mathcal{T}\left\{c_{j,\sigma}(\tau)c^\dagger_{j,\sigma}(0)\right\} \Big\rangle,
    \end{equation}
    with $\mathcal{T}\{\bullet\}$ the time ordering operator, to equal the one obtained from~\eqref{eq_methods: local eff action}, which, in imaginary frequency space, can be written as
    \begin{equation}
        \mathcal{G}(\nu) = \frac{1}{\mathcal{G}_0^{-1}(\nu)-\Sigma_\mathrm{imp}(\nu)},
    \end{equation}
    with $\Sigma_\mathrm{imp}(\nu)$ the self-energy of the local action. Furthermore, the self-energy of the Hubbard model, $\Sigma_{jj'}(\tau)$, is approximated to a purely local function, that is,
    \begin{equation}
        \Sigma_{jj'}(\tau) \simeq \Sigma_\mathrm{dmft}(\tau) \delta_{jj'},
    \end{equation}
    which becomes an exact statement in infinite dimensions $d\to\infty$, as shown in Ref.~\cite{Metzner1989} by means of diagrammatic arguments. In other words, we are requiring  the Luttinger-Ward functional (see Ref.~\cite{Abrikosov1965}) $\Phi[G_{jj'}(\tau)]$ to be a functional of the local Green's function $G_{jj}(\tau)$ only, so that
    \begin{equation}
        \Sigma_{jj'}(\tau) = \frac{\delta\Phi[G_{jj'}(\tau)]}{\delta G_{jj'}(\tau)}
        \simeq\frac{\delta\Phi[G_{jj}(\tau)]}{\delta G_{jj'}(\tau)} = \Sigma_\mathrm{dmft}(\tau)\delta_{jj'}.
    \end{equation}
    Essentially, we are claiming that if we neglect the nonlocal ($j\neq j'$) elements of the self-energy, this can be generated by the Luttinger-Ward functional of a purely local theory, which we choose to be the one defined by~\eqref{eq_methods: local eff action}. This leads us to conclude that $\Sigma_\mathrm{dmft}(\tau)=\Sigma_\mathrm{imp}(\tau)$.
    The self-consistency relation can be therefore expressed in the frequency domain as
    \begin{equation}
        G_{jj}(\nu) = \int_{k\in \mathrm{B.Z.}} \!\!\frac{d^2\bk}{(2\pi)^2}\,\frac{1}{i\nu-\xi_\bk-\Sigma_\mathrm{dmft}(\nu)} = 
        \frac{1}{\mathcal{G}_0^{-1}(\nu)-\Sigma_\mathrm{dmft}(\nu)}.
        \label{eq_methods: DMFT self consist}
    \end{equation}
    where $\xi_\bk=\epsilon_\bk-\mu$, with $\epsilon_\bk$ the Fourier transform of the hopping matrix $t_{jj'}$ and $\mu$ the chemical potential. 
    For a more detailed derivation of \eqref{eq_methods: DMFT self consist} and for a broader discussion, see Refs.~\cite{Georges1996,Georges1992}.
    
\begin{figure}[t]
        \centering
        \includegraphics[width=0.35\textwidth]{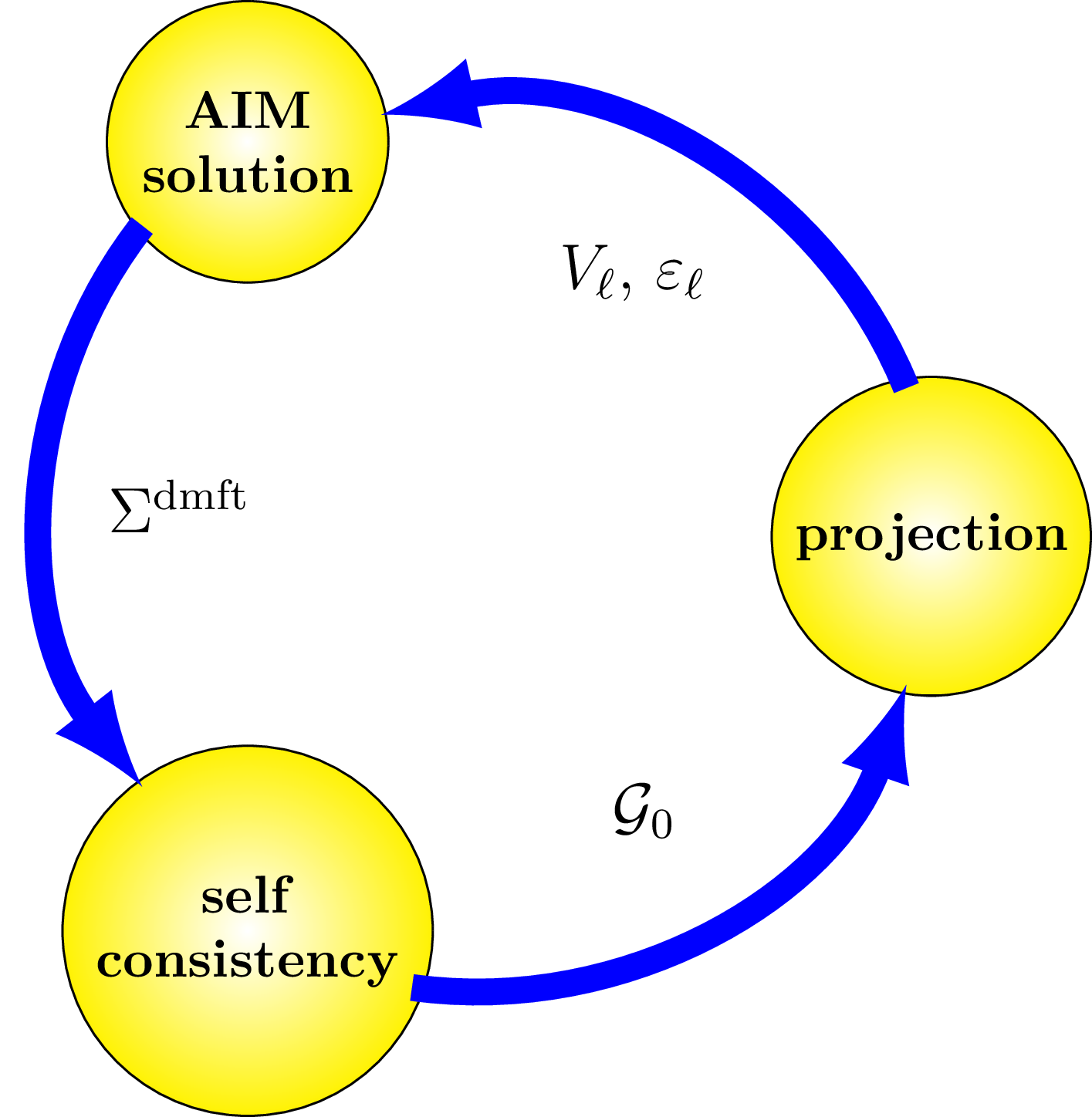}
        \caption{DMFT self-consistent loop with the AIM as impurity model.}
        \label{fig_methods: DMFT sc loop}
    \end{figure}

    Eq.~\eqref{eq_methods: DMFT self consist} closes the equations of the so-called DMFT loop. In essence, one starts with a guess for the Weiss field $\mathcal{G}_0^{-1}$, computes the self-energy of the action~\eqref{eq_methods: local eff action}, extracts a new $\mathcal{G}_0^{-1}$ from the self-consistency relation~\eqref{eq_methods: DMFT self consist}, and repeats this loop until convergence is reached, as shown in Fig.~\ref{fig_methods: DMFT sc loop}. 

    The main advantage of this computational scheme is that the action~\eqref{eq_methods: local eff action} is much easier to treat than the Hubbard model itself, and several reliable numerical methods (so-called impurity solvers) provide numerically exact solutions. Among those, we find quantum Monte Carlo (QMC) methods, originally adapted to quantum impurity problems by Hirsch and Fye~\cite{Hirsch1986}, exact diagonalization (ED)~\cite{Si1994,Caffarel1994,Rozenberg1994}, and the numerical renormalization group (NRG)~\cite{Wilson1975,Bulla2008}. 
    
    The ED and NRG methods require the impurity action~\eqref{eq_methods: local eff action} to descend from a Hamiltonian $\mathcal{H}$. This is provided by the Anderson impurity model (AIM)~\cite{Anderson1961}, which describes an impurity coupled to a bath of noninteracting electrons. Its Hamiltonian is given by
    \begin{equation}
        \mathcal{H}_\mathrm{AIM} = \sum_{\ell,\sigma} \varepsilon_\ell \,
        a^\dagger_{\ell,\sigma}a_{\ell,\sigma}
        +\sum_{\ell,\sigma}V_\ell\left[c^\dagger_{0,\sigma}a_{\ell,\sigma}+a^\dagger_{0,\sigma}c_{\ell,\sigma}\right]
        -\mu\sum_\sigma c^\dagger_{0,\sigma}c_{0,\sigma}
        +U n_{0,\up}n_{0,\down},
        \label{eq_methods: AIM Hamilt}
    \end{equation}
    where $a^\dagger_{\ell,\sigma}$ ($a_{\ell,\sigma}$) creates (annihilates) an electron on bath level $\ell$ with spin projection $\sigma$, $\varepsilon_\ell$ are the bath energy levels, $V_\ell$ represent the bath-impurity hybridization parameters, and $\mu$ is the impurity chemical potential. The set $\{\varepsilon_\ell,V_\ell\}$ is often referred to as Anderson parameters. Expressing $\mathcal{H}_\mathrm{AIM}$ as a functional integral, and integrating over the bath electrons, one obtains the impurity action~\eqref{eq_methods: local eff action}, with the Weiss field given by
    \begin{equation}
        \mathcal{G}_0^{-1}(\nu)=i\nu+\mu - \Delta(\nu), 
    \end{equation}
    where the hybridization function $\Delta(\nu)$ is related to $\varepsilon_\ell$ and $V_\ell$ by
    \begin{equation}
        \Delta(\nu) = \sum_\ell \frac{|V_\ell|^2}{i\nu-\varepsilon_\ell}.
    \end{equation}
    In the context of the AIM, the Weiss field is therefore expressed in terms of an optimally determined discrete set of Anderson parameters. 
    \subsection{DMFT two-particle vertex and susceptibilities}
    \label{subs_methods: DMFT susceptibilities}
    For many studies, the knowledge of the single-particle quantities such as the self-energy is not sufficient. The DMFT provides also a framework for the computation of two-particle quantities and response functions after the loop has converged and the optimal Weiss field (or Anderson parameters) has been found.
    
\begin{figure}
        \centering
        \includegraphics[width=1.0\textwidth]{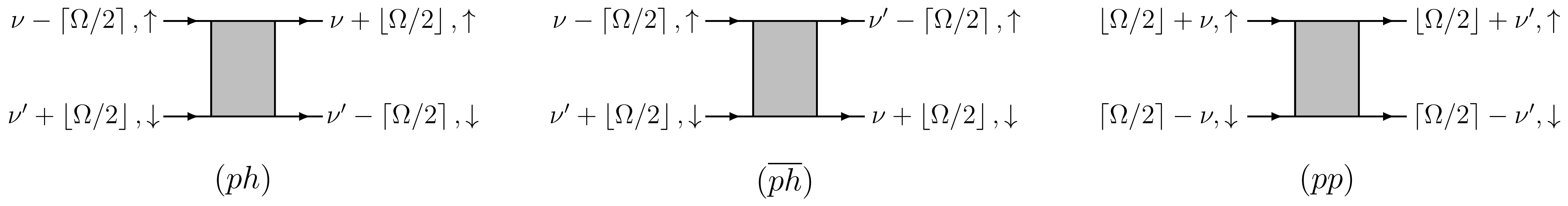}
        \caption{Schematic representation of the different notations for the two-particle vertex.}
        \label{fig_methods: notations}
    \end{figure}

    The impurity two-particle Green's function is defined as
    \begin{equation}
        G^{4,\mathrm{imp}}_{\sigma_1',\sigma_2',\sigma_1^{\phantom{'}},\sigma_2^{\phantom{'}}}(\tau_1',\tau_2',\tau_1,\tau_2)=
        \Big\langle \mathcal{T}\left\{
            c_{0,\sigma_1^{\phantom{'}}}(\tau_1)
            c_{0,\sigma_2^{\phantom{'}}}(\tau_2)
            c_{0,\sigma_1'}^\dagger(\tau_1')
            c_{0,\sigma_2'}^\dagger(\tau_2')
            \right\}
        \Big\rangle,
        \label{eq_methods: G4 DMFT}
    \end{equation}
    and it is by definition antisymmetric under the exchange of $(\tau_1',\sigma_1')$ with $(\tau_2',\sigma_2')$  or $(\tau_1,\sigma_1^{\phantom{'}})$ with $(\tau_2,\sigma_2^{\phantom{'}})$. Fourier transforming it with respect to the four imaginary time variables, one obtains
    \begin{equation}
        G^{4,\mathrm{imp}}_{\sigma_1',\sigma_2',\sigma_1^{\phantom{'}},\sigma_2^{\phantom{'}}}(\nu_1',\nu_2',\nu_1,\nu_2) = G^{4,\mathrm{imp}}_{\sigma_1',\sigma_2',\sigma_1^{\phantom{'}},\sigma_2^{\phantom{'}}}(\nu_1',\nu_2',\nu_1)\,\beta\delta_{\nu_1'+\nu_2'-\nu_1-\nu_2},
    \end{equation}
    where $\beta=1/T$ is the inverse temperature, and the delta function of the frequencies arises because of time translation invariance. Removing the disconnected terms, one obtains the connected two-particle Green's function
    \begin{equation}
        \begin{split}
            G^{4,c,\mathrm{imp}}_{\sigma_1',\sigma_2',\sigma_1^{\phantom{'}},\sigma_2^{\phantom{'}}}(\nu_1',\nu_2',\nu_1)=\,&G^{4,\mathrm{imp}}_{\sigma_1',\sigma_2',\sigma_1^{\phantom{'}},\sigma_2^{\phantom{'}}}(\nu_1',\nu_2',\nu_1)\\&-\beta\mathcal{G}(\nu_1')\mathcal{G}(\nu_2')\,
            \delta_{\nu_1',\nu_1}\,
            \delta_{\sigma_1',\sigma_1^{\phantom{'}}}\delta_{\sigma_2',\sigma_2^{\phantom{'}}}\\
            &+\beta \mathcal{G}(\nu_1')G(\nu_2')\delta_{\nu_1',\nu_2}\,\delta_{\sigma_1',\sigma_2^{\phantom{'}}}\delta_{\sigma_2',\sigma_1^{\phantom{'}}},
        \end{split}
        \label{eq_methods: G2 conn}
    \end{equation}
    with $\mathcal{G}(\nu)$ the single-particle Green's function of the impurity problem. The relation between the connected two-particle Green's function and the vertex is then given by~\cite{Rohringer2012}
    \begin{equation}
        \begin{split}
            G^{4,c,\mathrm{imp}}_{\sigma_1',\sigma_2',\sigma_1^{\phantom{'}},\sigma_2^{\phantom{'}}}(\nu_1',\nu_2',\nu_1)=
        -\mathcal{G}(\nu_1')\mathcal{G}(\nu_2')V^\mathrm{imp}_{\sigma_1',\sigma_2',\sigma_1^{\phantom{'}},\sigma_2^{\phantom{'}}}(\nu_1',\nu_2',\nu_1)\mathcal{G}(\nu_1)\mathcal{G}(\nu_2),
        \end{split}
        \label{eq_methods: G2 1PI}
    \end{equation}
    where $V^\mathrm{imp}$ is the impurity two-particle (1PI) vertex, and $\nu_2=\nu_1'+\nu_2'-\nu_1$ is fixed by energy conservation. Because of the spin-rotational invariance of the system, the spin dependence of the vertex can be simplified to
    \begin{equation}
        \begin{split}
            &V^\mathrm{imp}_{\sigma_1',\sigma_2',\sigma_1^{\phantom{'}},\sigma_2^{\phantom{'}}}(\nu_1',\nu_2',\nu_1)=
            V^\mathrm{imp}(\nu_1',\nu_2',\nu_1)\delta_{\sigma_1',\sigma_1^{\phantom{'}}}\delta_{\sigma_2',\sigma_2^{\phantom{'}}}
            -V^\mathrm{imp}(\nu_2',\nu_1',\nu_1)\delta_{\sigma_1',\sigma_2^{\phantom{'}}}\delta_{\sigma_2',\sigma_1^{\phantom{'}}}.
        \end{split}
    \end{equation}
    Furthermore, we can introduce three different notations for the two particle vertex, depending on the use one wants to make of it. We define the particle-hole ($ph$), particle-hole-crossed ($\overline{ph}$), and particle-particle ($pp$) notations as:
    \begin{subequations}
        \begin{align}
            &V^{\mathrm{imp},ph}_{\nu,\nu'}(\Omega)=V^\mathrm{imp}\left(\nu-\rndup{\Omega},\nu'+\rnddo{\Omega},\nu+\rnddo{\Omega}\right),\\
            &V^{\mathrm{imp},\overline{ph}}_{\nu,\nu'}(\Omega)=V^\mathrm{imp}\left(\nu-\rndup{\Omega},\nu'+\rnddo{\Omega},\nu'-\rndup{\Omega}\right),\\
            &V^{\mathrm{imp},pp}_{\nu,\nu'}(\Omega)=V^\mathrm{imp}\left(\rnddo{\Omega}+\nu,\rndup{\Omega}-\nu,\rnddo{\Omega}+\nu'\right),
        \end{align}
    \end{subequations}
    where, as explained previously, $\rndupnotwo{\bullet}$ ($\rnddonotwo{\bullet}$) rounds its argument up (down) to the closest \emph{bosonic} Matsubara frequency. In Fig.~\ref{fig_methods: notations}, we show a pictorial representation of the different notations for the vertex function. 
    
    Within the QMC methods, the two-particle Green's function can be directly sampled from the impurity action~\eqref{eq_methods: local eff action} with converged Weiss field $\mathcal{G}_0^{-1}$, while for an ED or a NRG solver, one has to employ the \emph{Lehmann representation} of $G^{4,\mathrm{imp}}$ ~\cite{Toschi2007}. Once the two-particle Green's function has been obtained, the vertex can be extract via~\eqref{eq_methods: G2 conn} and~\eqref{eq_methods: G2 1PI}.
    
    The computation of the susceptibilities or transport coefficients of the lattice system can be achieved, within DMFT, through the knowledge of the vertex function. For example, the charge/magnetic susceptibilities of a paramagnetic system can be expressed in terms of the \emph{generalized} susceptibility $\chi^{c|m}_{\nu\nu'}(\bq,\Omega)$ as
    \begin{equation}
        \chi^{c|m}(\bq,\Omega)=T^2\sum_{\nu\nu'}\chi^{c|m}_{\nu\nu'}(\bq,\Omega).
    \end{equation}
    The DMFT approximation for $\chi^{c|m}_{\nu\nu'}(\bq,\Omega)$ is obtained solving the integral equation
    \begin{equation}
        \begin{split}
            \chi^{c|m}_{\nu\nu'}(\bq,\Omega)=\beta\chi^0_\nu(\bq,\Omega)\delta_{\nu\nu'}
            -T\sum_{\nu''}\chi^0_\nu(\bq,\Omega)\,\widetilde{V}^{c|m}_{\nu\nu''}(\Omega)\,\chi^{c|m}_{\nu''\nu'}(\bq,\Omega),
        \end{split}
        \label{eq_methods: generalized chi DMFT}
    \end{equation}
    where $\chi^0_\nu(\bq,\Omega)$ is given by
    \begin{equation}
        \chi^0_\nu(\bq,\Omega)=-\int_\bk G\left(\bk+\frac{\bq}{2},\nu+\rnddo{\Omega}\right) G\left(\bk-\frac{\bq}{2},\nu-\rndup{\Omega}\right),
        \label{eq_methods: chi0 generalized}
    \end{equation}
    with $\int_\bk=\int_{\bk\in\mathrm{B.Z.}}\frac{d^d\bk}{(2\pi)^d}$, and $G(\bk,\nu)$ the lattice propagator evaluated with the local self-energy $\Sigma(\nu)$. Finally, in Eq.~\eqref{eq_methods: generalized chi DMFT}, $\widetilde{V}^{c|m}$ represents the two particle irreducible (2PI) vertex in the charge/magnetic channel at the DMFT level. It can be obtained inverting a Bethe-Salpeter equation, that is,
    \begin{equation}
        V^{c|m}_{\nu\nu'}(\Omega) = \widetilde{V}^{c|m}_{\nu\nu'}(\Omega) + T\sum_{\nu''} \widetilde{V}^{c|m}_{\nu\nu''}(\Omega)\,\chi^{0,\mathrm{imp}}_{\nu''}(\Omega)\, V^{c|m}_{\nu''\nu'}(\Omega),
    \end{equation}
    where $\chi^{0,\mathrm{imp}}$ must be evaluated similarly to~\eqref{eq_methods: chi0 generalized} with the \emph{local} (or impurity) Green's function, and 
    \begin{subequations}
        \begin{align}
            &V^c_{\nu\nu'}(\Omega)=V_{\up\up\up\up,\nu\nu'}^{\mathrm{imp},ph}(\Omega)+
            V_{\up\up\down\down,\nu\nu'}^{\mathrm{imp},ph}(\Omega)=2V^{\mathrm{imp},ph}_{\nu\nu'}(\Omega)-V^{\mathrm{imp},\overline{ph}}_{\nu\nu'}(\Omega),\\
            &V^m_{\nu\nu'}(\Omega)=V_{\up\up\up\up,\nu\nu'}^{\mathrm{imp},ph}(\Omega)-
            V_{\up\up\down\down,\nu\nu'}^{\mathrm{imp},ph}(\Omega)=-V^{\mathrm{imp},\overline{ph}}_{\nu\nu'}(\Omega).
        \end{align}
    \end{subequations}
    In $d\to\infty$, even though the two-particle vertex is generally momentum-dependent, Eq.~\eqref{eq_methods: generalized chi DMFT} with a purely local 2PI vertex, is exact, as it can be proven by means of diagrammatic arguments~\cite{Georges1996}.
    \subsection{Strong coupling effects: the Mott transition}
    One of the earliest successes of the DMFT was, unlike weak-coupling theories, its ability to correctly capture and the describe the occurrence of a metal-to-insulator (MIT) transition in the Hubbard model, the so called \emph{Mott transition}, named after Mott's early works~\cite{Mott1949} on this topic. 
    
\begin{figure}[t]
        \centering
        \includegraphics[width=0.75\textwidth]{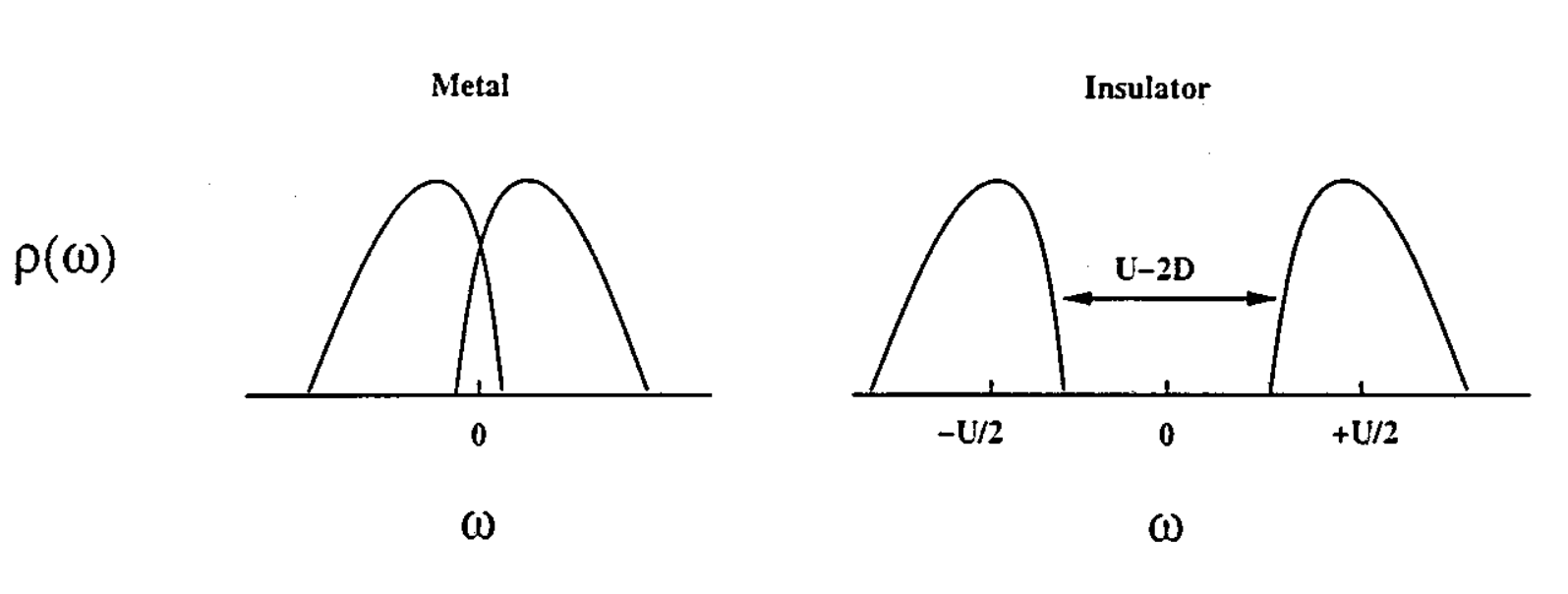}
        \caption{Schematic picture of Hubbard's attempt to describe the Mott MIT transition. Taken from Ref.~\cite{Georges1996}.}
        \label{fig_methods: Hubbard bands}
    \end{figure}

    In 1964, Hubbard~\cite{Hubbard1964} attempted to describe this transition within an \emph{effective band} picture. According to his view, the spectral function is composed of two "domes" which overlap in the metallic regime. As the interaction strength $U$ is increased, they move apart from each other, until, at the transition, they split into two separate bands, the so-called \emph{Hubbard bands} (see Fig.~\ref{fig_methods: Hubbard bands}). Despite this picture being \emph{qualitatively} correct in the insulating regime. it completely fails in reproducing the Fermi liquid properties of the metallic side. Differently, before the advent of the DMFT, other approaches could instead properly capture the transition approaching from the metallic regime~\cite{Brinkman1970}, but failed in describing the insulating phase. 

\begin{figure}[t]
        \centering
        \includegraphics[width=0.5 \textwidth]{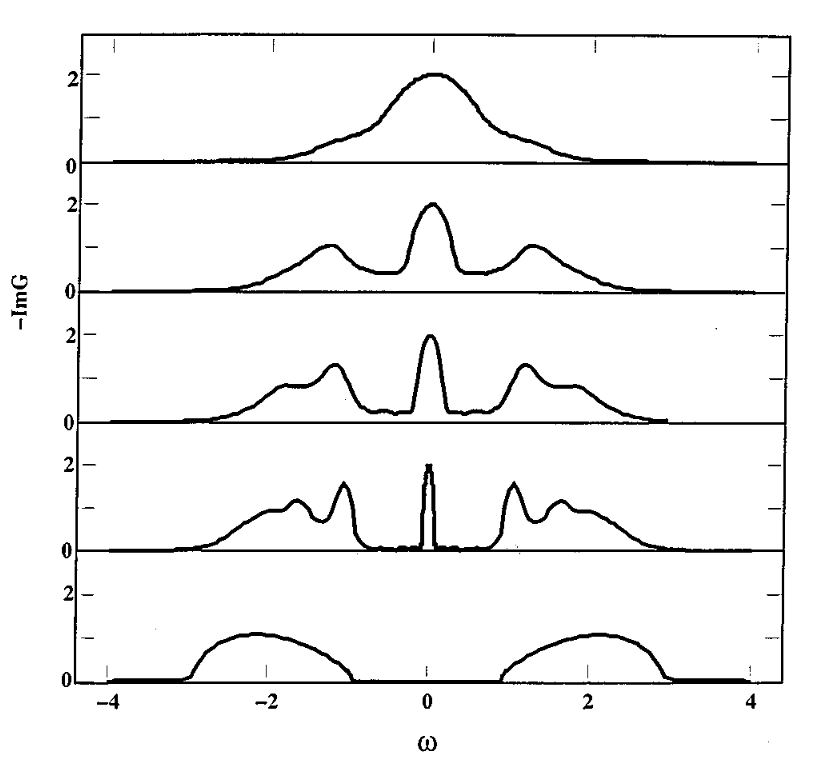}
        \caption{Evolution of the spectral function across the MIT, as predicted by DMFT. Between the metallic regime, marked by a large density of states at the Fermi level, and the insulating one, exhibiting a large charge gap, a Kondo peak is formed, signaling the onset of local moment screening. Taken from Ref.~\cite{Georges1996}.}
        \label{fig_methods: MIT DMFT}
    \end{figure}

    Within the DMFT, since no assumptions are made on the strength of the on-site repulsion $U$, both sides of the transitions can be studied qualitatively and quantitatively. In addition to the Fermi liquid regime and the insulating one, a new intermediate regime is predicted. In fact, for $U$ only \emph{slightly} smaller than the critical value (above which a gap in the excitation spectrum is generated) the spectral function already exhibits two evident precursors of the Hubbard bands in between of which, that is, at the Fermi level, a narrow peak appears (Fig.~\ref{fig_methods: MIT DMFT}). This feature, visible only at low temperatures, is a hallmark of the Kondo effect taking place. Indeed, in this regime, a local moment (spin) is already formed on the impurity site, as the charge excitations have been gapped out, and the (self-consistent) bath electrons screen it, leading to a singlet ground state. A broader discussion on the MIT as well as the antiferromagnetic properties of the (half-filled) Hubbard model as predicted by the DMFT can be found in Ref.~\cite{Georges1996}.  
    
    \subsection{Extensions of the DMFT}
    In this subsection we briefly list possible extensions of the DMFT to include the effects of \emph{nonlocal} correlations. For a more detailed overview, we refer to Ref.~\cite{Rohringer2018}. First of all, we find approximations that allow for the treatment of \emph{short-range} correlations, by replacing the single impurity atom with a cluster of few sites, either in real space, as in the cluster DMFT (CDMFT)~\cite{Kotliar2001}, or in reciprocal space, in the so-called dynamical cluster approximation (DCA)~\cite{Maier2005}. Even with few cluster sites, these approximations are sufficient to capture the interplay of antiferromagnetism and superconductivity in the Hubbard model~\cite{Foley2019}. Furthermore, we find the dual boson approach~\cite{Rubtsov2012}, that extends the applicability of the DMFT to systems with nonlocal interactions by adding to the impurity problem some local bosonic degrees of freedom. The dual fermion theory~\cite{Rubtsov2008}, instead allows for a \emph{perturbative} inclusion of nonlocal correlations on top of the DMFT, similarly to the \emph{vertex-based} approaches such as the dynamical vertex approximation (D$\Gamma$A)~\cite{Toschi2007} and the triply and quadruply irreducible local expansions (TRILEX and QUADRILEX)~\cite{Ayral2016}.
    %
    \section[The \texorpdfstring{DMF\textsuperscript2RG}{DMF2RG} approach]{Boosting the fRG to strong coupling: the \texorpdfstring{DMF\textsuperscript2RG}{DMF2RG} approach}
    In this section, we introduce another extension of the DMFT for the inclusion of nonlocal correlations, namely its fusion together with the fRG in the so-called DMF\textsuperscript{2}RG approach~\cite{Taranto2014,Metzner2014}. Alternatively, the DMF\textsuperscript{2}RG can be viewed as a development of the fRG that enlarges its domain of validity to strongly interacting systems. 
    
    We start by defining a scale-dependent action as
    \begin{equation}
        \mathcal{S}^\L\left[\psi,\overline{\psi}\right]=-\int_k\sum_\sigma\overline{\psi}_{k,\sigma}\left[G_0^\L(k)\right]^{-1}\psi_{k,\sigma} + U\int_0^\beta\!d\tau\,\sum_j n_{j,\up}(\tau)n_{j,\down}(\tau).
        \label{eq_methods: DMF2RG scale-dependent action}
    \end{equation}
    We notice that by choosing $G_0^\L(k)=\mathcal{G}_0(\nu)$, \eqref{eq_methods: DMF2RG scale-dependent action} becomes the action of $N_s$ (number of lattice sites) identical and uncoupled impurity problems. On the other hand, $G_0^\L(k)=(i\nu-\xi_\bk)^{-1}$, gives the Hubbard model action. The key idea of the DMF\textsuperscript{2}RG is therefore to set up a fRG flow, which interpolates between the self-consistent AIM and the Hubbard model. The boundary conditions for $G^\L_0(k)$ read therefore as
\begin{subequations}
    \begin{align}
        &G_0^{\Lini}(k)=\mathcal{G}_0(\nu),\\
        &G_0^{\Lfin}(k)=\frac{1}{i\nu-\xi_\bk}.
    \end{align}
\end{subequations}
    Furthermore, one requires the DMFT solution to be conserved at each fRG step~\cite{Vilardi2019,Vilardi_Thesis}, that is
    \begin{equation}
        \int_\bk G^\L(k)\big\rvert_{\Sigma^\L(k)=\Sigma_\mathrm{dmft}(\nu)}=
        \int_\bk \frac{1}{\left[G_0^\L(k)\right]^{-1}-\Sigma_\mathrm{dmft}(\nu)}=
        \frac{1}{\mathcal{G}_0^{-1}(\nu)-\Sigma_\mathrm{dmft}(\nu)}.
        \label{eq_methods: DMFT conservation}
    \end{equation}
    Possible cutoffs schemes satisfying the boundary conditions and the conservation of DMFT might be, for example,
    \begin{equation}
        \left[G_0^\L(k)\right]^{-1}=\Theta^\L(k)\left(i\nu-\xi_\bk\right)+\Xi^\L(k)\mathcal{G}_0^{-1}(\nu),
    \end{equation}
    or
    \begin{equation}
        G_0^\L(k)=\frac{\Theta^\L(k)}{i\nu-\xi_\bk}+\Xi^\L(k)\mathcal{G}_0(\nu),
    \end{equation}
    where $\Theta^\L(k)$ is an arbitrarily chosen cutoff satisfying $\Theta^{\Lini}(k)=0$, and $\Theta^{\Lfin}(k)=1$, and $\Xi^\L(k)$ is calculated at every step from~\eqref{eq_methods: DMFT conservation}. Obviously, at $\L=\Lini$ (when $\Theta^\L(k)=0$), one would get $\Xi^{\Lini}(k)=1$, while at $\L=\Lfin$, Eq.~\eqref{eq_methods: DMFT conservation} becomes the DMFT self-consistency condition, fulfilled by $G_0^\L(k)=(i\nu-\xi_\bk)^{-1}$, which returns $\Xi^{\Lfin}(k)=0$. 
    
    The choice $\mathcal{S}^{\Lini}[\psi,\overline{\psi}]=\sum_j\mathcal{S}_\mathrm{imp}[\psi_j,\overline{\psi}_j]$, imposes an initial condition for the fRG effective action, that is,
    \begin{equation}
        \Gamma^{\Lini}\left[\psi,\overline{\psi}\right] = 
        \sum_j \Gamma_\mathrm{imp}\left[\psi_j,\overline{\psi}_j\right],
    \end{equation}
    where $\Gamma_\mathrm{imp}$ is the effective action of the self-consistent impurity problem. Expanding it in power of the fields, we get
    \begin{equation}
        \begin{split}
            \Gamma_\mathrm{imp}\left[\psi,\overline{\psi}\right] = 
            &-\int_\nu\sum_\sigma               \overline{\psi}_{\nu,\sigma}\left[\mathcal{G}_0^{-1}(\nu)-\Sigma_\mathrm{dmft}(\nu)\right]\psi_{\nu,\sigma}\\
            &+\frac{1}{(2!)^2}\int_{\nu_1',\nu_2',\nu_1}\,\,\sum_{\substack{\sigma_1',\sigma_2',\\\sigma_1^{\phantom{'}},\sigma_2^{\phantom{'}}}}
            \overline{\psi}_{\nu_1',\sigma_1'}
            \overline{\psi}_{\nu_2',\sigma_2'}\,
            V^\mathrm{imp}_{\sigma_1',\sigma_2',\sigma_1^{\phantom{'}},\sigma_2^{\phantom{'}}}(\nu_1',\nu_2',\nu_1)\,
            \psi_{\nu_1'+\nu_2'-\nu_1,\sigma_2^{\phantom{'}}}\psi_{\nu_1,\sigma_1^{\phantom{'}}}\\
            &+\dots
        \end{split}
        \label{eq_methods: Gamma ini DMF2RG}
    \end{equation}

    Within the DMF\textsuperscript{2}RG, the flow equations for the 1PI vertices remain unchanged, while their initial conditions can be read from~\eqref{eq_methods: Gamma ini DMF2RG}:
    \begin{subequations}
        \begin{align}
            &\Sigma^{\Lini}(k) = \Sigma_\mathrm{dmft}(\nu),\\
            &V^\Lini(k_1,k_2,k_3)=V^\mathrm{imp}(\nu_1,\nu_2,\nu_3) = V^\mathrm{imp}_{\up\down\up\down}(\nu_1,\nu_2,\nu_3).
        \end{align}
    \end{subequations}

    The development of the DMF\textsuperscript{2}RG has enabled the study of the doped Hubbard model at strong coupling, with particular focus on the (generally incommensurate) antiferromagnetic and ($d$-wave) superconducting instabilities~\cite{Vilardi2019,Vilardi_Thesis}.
    %

\cleardoublepage
    \rhead[\fancyplain{}{\bfseries Charge carrier drop driven by spiral magnetism}]{\fancyplain{}{\bfseries\thepage}}
    \lhead[\fancyplain{}{\bfseries\thepage}]{\fancyplain{}{\bfseries Charge carrier drop driven by spiral magnetism}}
    \chapter{Charge carrier drop driven by spiral magnetism}
    \label{chap: spiral DMFT}
    In this chapter, we present a DMFT description of the so-called spiral magnetic state of the Hubbard model. This magnetically ordered phase is a candidate for the normal state of cuprate superconductors, emerging when superconductivity gets suppressed by strong magnetic fields, as realized in a series of recent experiments~\cite{Badoux2016,Laliberte2016,Collignon2017,Proust2019}. In particular, a sudden change in the charge carrier density, measured via the Hall number, is observed as the hole doping $p=1-n$ is varied across the value $p=p^*$, where the pseudogap phase is supposed to end. This observation is consistent with a drastic change in the Fermi surface topology, which can described, among others, by a transition from a spiral magnet to a paramagnet~\cite{Eberlein2016,Chatterjee2017,Verret2017}. Other possible candidates for the phase appearing for $p<p^*$ are N\'eel antiferromagnetism~\cite{Storey2016,Storey2017}, charge density waves~\cite{Caprara2017,Sharma2018}, or nematic order~\cite{Maharaj2017}. 
    
    The chapter is organized as it follows. First of all, we define the spiral magnetic state and provide a DMFT description of it. Secondly, we present results for the spiral order parameter as a function of doping at low temperatures, together with an analysis of the evolution of the Fermi surfaces. Finally, we compare our results with the experimental findings by computing the transport coefficients using the DMFT parameters as an input for the formulas derived in Ref.~\cite{Mitscherling2018}. This task has been carried out by J.~Mitscherling, who equally contributed to Ref.~\cite{Bonetti2020_I}, which contains the results presented in this chapter.

    \section{Spiral magnetism}
\begin{figure}[t]
        \centering
        \includegraphics[width=0.65\textwidth]{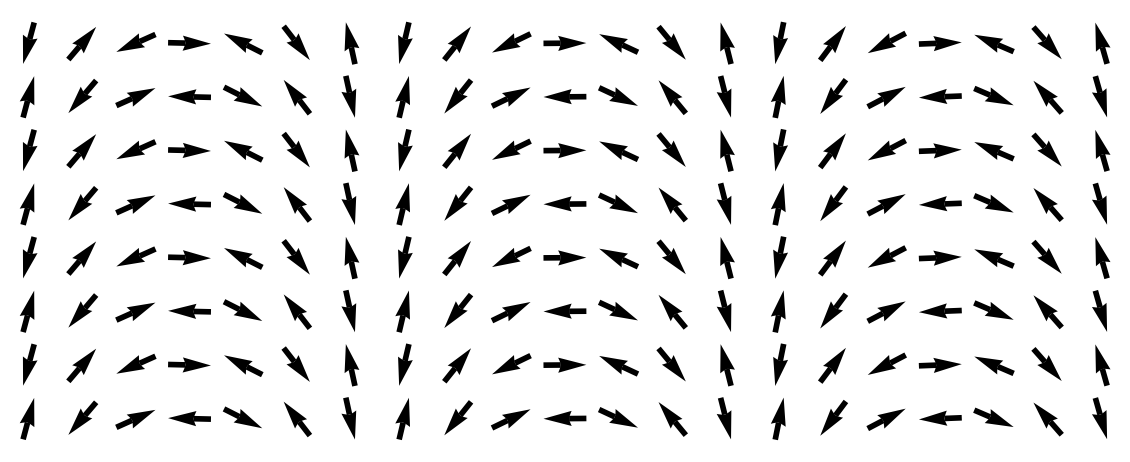}
        \caption{Magnetization pattern for a spiral magnetic state on a square lattice with lattice constant $a=1$, and $\bQ=(\pi-2\pi\eta,\pi)$, with $\eta\simeq0.07$.}
        \label{fig_spiral: spiral}
    \end{figure}

    Spiral magnetic order is defined by a finite expectation value of the spin operator of the form
    \begin{equation}
        \langle \vec{S}_j \rangle = m\hat{n}_j,
        \label{eq_spiral: <Sj>}
    \end{equation}
    where $m$ is the amplitude of the onsite magnetization, and $\hat{n}_j$ is a unitary vector indicating the magnetization direction on site $j$, which can be written as
    \begin{equation}
        \hat{n}_j = \cos(\bQ\cdot\mathbf{R}_j)\hat{v}_1
        +\sin(\bQ\cdot\mathbf{R}_j)\hat{v}_2,
        \label{eq_spiral: n_j}
    \end{equation}
    with $\hat{v}_1$ and $\hat{v}_2$ two constant mutually orthogonal unitary vectors. The magnetization lies therefore in the plane spanned by $\hat{v}_1$ and $\hat{v}_2$, and its direction on two neighboring sites $j$ and $j'$ differs by an angle $\bQ\cdot(\mathbf{R}_j-\mathbf{R}_{j'})$. The vector $\bQ$ is a parameter which must be determined microscopically. In the square lattice Hubbard model it often takes the form, in units of the inverse lattice constant $a^{-1}$, $\bQ=(\pi-2\pi\eta,\pi)$ or, in the case of a \emph{diagonal spiral}, $\bQ=(\pi-2\pi\eta,\pi-2\pi\eta)$, where the parameter $\eta$ is called incommensurability. If the system Hamiltonian exhibits SU(2) spin symmetry, as in case of the Hubbard model, the vectors $\hat{v}_1$ and $\hat{v}_2$ can be chosen arbitrarily, and we thus choose $\hat{v}_1=\hat{e}_1\equiv(1,0,0)$, and
    $\hat{v}_2=\hat{e}_2\equiv(0,1,0)$. The magnetization pattern resulting from this choice on a square lattice for a specific value of $\bQ$ is shown in Fig.~\ref{fig_spiral: spiral}.
    Within the Hubbard model, where the fundamental degrees of freedom are electrons rather than spins, the spin operator is expressed as 
    \begin{equation}
        \vec{S}_j = \frac{1}{2}\sum_{s,s'=\up,\down}c^\dagger_{j,s}\vec{\PauliMat}_{ss'}c_{j,s'},
        \label{eq_spiral: Sj Hubbard}
    \end{equation}
    with $\vec{\PauliMat}$ the Pauli matrices. Combining the above definition with \eqref{eq_spiral: <Sj>} and \eqref{eq_spiral: n_j}, one obtains the following expression for the onsite magnetization amplitude
    \begin{equation}
        m = \frac{1}{2}\int_\bk \Big\langle c^\dagger_{\bk,\up}c_{\bk+\bQ,\down}+c^\dagger_{\bk+\bQ,\down}c_{\bk,\up}\Big\rangle.
    \end{equation}
    From the equation above, it is evident that spiral magnetism couples the single particle states $(\bk,\up)$ and $(\bk+\bQ,\down)$, for each momentum $\bk$. It is thus convenient to use a Nambu-like basis $(c_{\bk,\up},c_{\bk+\bQ,\down})$, for which the inverse bare Green's function reads as
    \begin{equation}
        \mathbf{G}_0^{-1}(\bk,\nu)=\left(
        \begin{array}{cc}
            i\nu-\xi_{\bk} &  0\\
            0 & i\nu-\xi_{\bk+\bQ}
        \end{array}
        \right),
    \end{equation}
    with $\xi_\bk$ the single-particle dispersion relative to the chemical potential $\mu$. Within the above definitions, the 2D N\'eel state is recovered by setting $\bQ=\bQ_\mathrm{AF}=(\pi,\pi)$. In the Hubbard model a spiral magnetic (that is, with $\bQ$ close to $\bQ_\mathrm{AF}$) state has been found by several methods at finite doping: Hartree-Fock~\cite{Igoshev2010}, slave-boson mean-field~\cite{Fresard1991} calculations, as well as expansions in the hole density~\cite{Chubukov1995}, and moderate-coupling functional renormalization group~\cite{Yamase2016} calculations. Interestingly enough, normal state DMFT calculations have revealed that the ordering wave vector $\bQ$ is related to the shape of the Fermi surface geometry not only at weak but also at strong coupling~\cite{Vilardi2018}. Furthermore, spiral states are found to emerge upon doping also in the $t$-$J$ model~\cite{Shraiman1989,Kotov2004}.
    \section{DMFT for spiral states}
    The single impurity DMFT equations presented in Chap.~\ref{chap: methods} can be easily extended to magnetically ordered states~\cite{Georges1996}. The particular case of spiral magnetism has been treated in Refs.~\cite{Fleck1999,Goto2016} for the square- and triangular-lattice Hubbard model, respectively.
    
    In the Nambu-like basis introduced previously, the self-consistency equation takes the form 
    \begin{equation}
        \int_\bk \left[\mathbf{G}_0^{-1}(\bk,\nu)-\boldsymbol{\Sigma}_\mathrm{dmft}(\nu)\right]^{-1}=
        \left[\boldsymbol{\mathcal{G}}_0^{-1}(\nu)-\boldsymbol{\Sigma}_\mathrm{dmft}(\nu)\right]^{-1},
    \end{equation}
    where $\boldsymbol{\Sigma}_\mathrm{dmft}(\nu)$ is the local self-energy, and $\boldsymbol{\mathcal{G}}_0(\nu)$ the bare propagator of the self-consistent AIM. The self-energy is a $2\times2$ matrix of the form
    \begin{equation}
        \boldsymbol{\Sigma}_\mathrm{dmft}(\nu) = \left(
        \begin{array}{cc}
            \Sigma(\nu) & \Delta(\nu) \\
            \Delta^*(-\nu) & \Sigma(\nu)
        \end{array}
        \right),
    \end{equation}
    with $\Sigma(\nu)$ the normal self-energy, and $\Delta(\nu)$ the gap function. Since the impurity model lives in 0+1 dimensions, there can be no spontaneous symmetry breaking, leading to off diagonal elements in the self-energy with a diagonal Weiss field $\boldsymbol{\mathcal{G}}_0$. We therefore explicitly break the SU(2) symmetry in the impurity model, allowing for a non-diagonal bare propagator. The corresponding AIM can be then written as (cf.~Eq.~\eqref{eq_methods: AIM Hamilt})
    \begin{equation}
        \begin{split}
            \mathcal{H}_\mathrm{AIM}=\sum_{\ell,\sigma} \varepsilon_\ell\, a^\dagger_{\ell,\sigma}a_{\ell,\sigma}+
            \sum_{\ell,\sigma,\sigma'}\left[V_\ell^{\sigma\sigma'}c^\dagger_{0,\sigma}a_{\ell,\sigma'}+\mathrm{h.c.}\right]
            -\mu\sum_\sigma c^\dagger_{0,\sigma}c_{0,\sigma}
            +Un_{0,\up}n_{0,\down},
        \end{split}
        \label{eq_spiral: spiral AIM}
    \end{equation}
    with $V_\ell^{\sigma\sigma'}$ a hermitian matrix describing spin-dependent hoppings. By means of a suitable global spin rotation around the axis perpendicular to the magnetization plane, one can impose $\Delta(\nu)=\Delta^*(-\nu)$, and therefore require the $V_\ell^{\sigma\sigma'}$ to be real symmetric matrices. The self-consistent loop will then return nonzero off-diagonal hoppings ($V_\ell^{\up\down}$ and $V_\ell^{\down\up}$), and therefore a finite gap function $\Delta(\nu)$, only if symmetry breaking occurs in the original lattice system. Integrating out the bath fermions in Eq.~\eqref{eq_spiral: spiral AIM}, one obtains the Weiss field
    \begin{equation}
        \boldsymbol{\mathcal{G}}_0(\nu) = 
        (i\nu+\mu)\mathbb{1}-\sum_\ell \frac{\boldsymbol{V}^\dagger_\ell \boldsymbol{V}^{\phantom{\dagger}}_\ell}{i\nu-\varepsilon_\ell},
    \end{equation}
    which, in general, exhibits off-diagonal elements. 
    
\begin{figure}[t]
        \centering
        \includegraphics[width=0.7\textwidth]{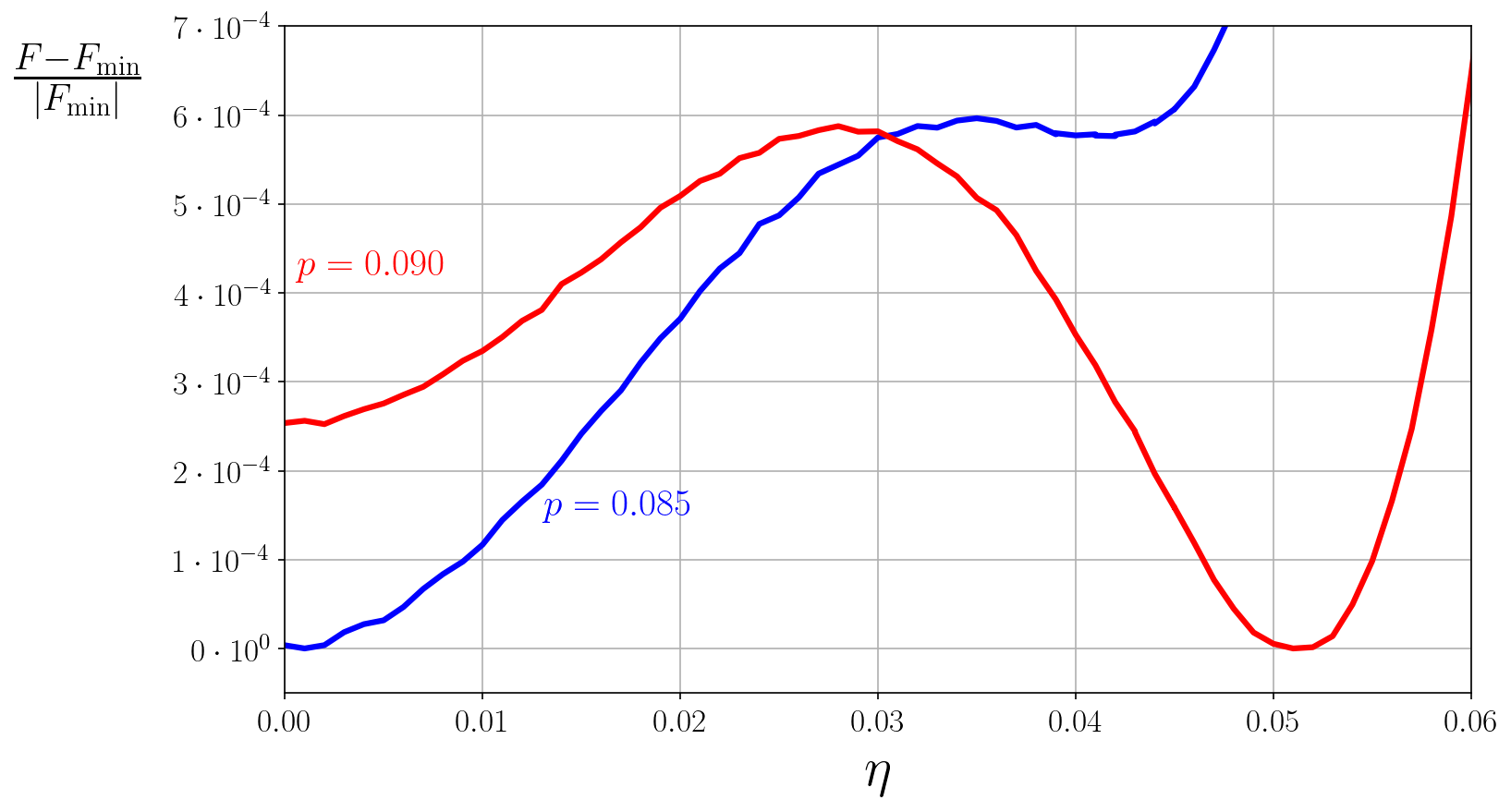}
        \caption{Free energy potential relative to its minimum value $F_\mathrm{min}$ at two different doping values, as a function of the incommensurability $\eta$ for a $\bQ=(\pi-2\pi\eta,\pi)$ spiral.}
        \label{fig_spiral: Free en vs eta}
    \end{figure}

    Using ED with four bath sites as impurity solver, we converge several loops for various values of $\bQ$, and we retain the one that minimizes the grand-canonical potential. For its computation we use the formula~\cite{Georges1996}
    \begin{equation}
        \frac{\Omega}{V} = \Omega_\mathrm{imp}-T\sum_\nu\int_\bk\Tr\log\left[\mathbf{G}_0^{-1}(\bk,\nu)-\boldsymbol{\Sigma}_\mathrm{dmft}(\nu)\right]
        +T\sum_\nu\Tr\log\left[\boldsymbol{\mathcal{G}}_0^{-1}(\nu)-\boldsymbol{\Sigma}_\mathrm{dmft}(\nu)\right],
    \end{equation}
    with $V$ the system volume, and $\Omega_\mathrm{imp}$ the impurity grand-canonical potential per unit volume, which can be computed within the ED solver as
    \begin{equation}
        \Omega_\mathrm{imp} = -2T\sum_n \log \left(1+e^{-\beta \epsilon_n}\right),
    \end{equation}
    where the factor 2 comes from the spin degeneracy, and $\epsilon_n$ are the eigenenergies of the AIM Hamiltonian. In the case of calculations performed at finite density $n$ rather than at fixed chemical potential $\mu$, the function to be minimized is the free energy per unit volume $F/V=\Omega/V+\mu n$. In Fig.~\ref{fig_spiral: Free en vs eta} we show a typical behavior of $F/V$ as a function of the incommensurability $\eta$ for a $\bQ=(\pi-2\pi\eta,\pi)$ spiral. We notice that the variation of microscopic parameters such as the hole doping $p=1-n$ can drive the system from a N\'eel state ($\eta=0$) to a spiral one ($\eta\neq0$). 
    \section{Hubbard model parameters}
    In order to mimic the behavior of real materials, namely YBa\textsubscript 2Cu\textsubscript 3O\textsubscript y (YBCO), and La\textsubscript{2-x}Sr\textsubscript xCuO\textsubscript4 (LSCO), we use hopping parameters ($t'$ and $t''$) calculated by downfolding \emph{ab initio} band structures on the single-band Hubbard model~\cite{Andersen1995,Pavarini2001}. For LSCO we choose $t'=-0.17t$, $t''=0.05t$, and $U=8t$, while for YBCO we have $t'=-0.3t$, $t''=0.15t$, and $U=10t$. Furthermore, since YBCO is a bilayer compound, its band structure must be extended to 
    \begin{equation}
        \xi_{\bk,k_z}=\xi_\bk-t^\perp_\bk\cos k_z,
    \end{equation}
    where $k_z\in\{0,\pi\}$ is the $z$-axis component of the momentum, and $t^\perp_\bk$ is an interlayer hopping amplitude taking the form 
    \begin{equation}
        t^\perp_\bk=t^\perp(\cos k_x - \cos k_y )^2, 
    \end{equation}
    with $t^\perp=0.15t$. The dispersion obtained with $k_z=0$ is often referred to as \emph{bonding band}, and the one with $k_z=\pi$ as \emph{antibonding band}. The self-consistency equation must be then modified to 
    \begin{equation}
        \frac{1}{2}\sum_{k_z=0,\pi}\int_\bk \left[\mathbf{G}_0^{-1}(\bk,k_z,\nu)-\boldsymbol{\Sigma}_\mathrm{dmft}(\nu)\right]^{-1}=
        \left[\boldsymbol{\mathcal{G}}_0^{-1}(\nu)-\boldsymbol{\Sigma}_\mathrm{dmft}(\nu)\right]^{-1},
    \end{equation}
    where the bare lattice Green's function is now given by
    \begin{equation}
        \mathbf{G}_0^{-1}(\bk,k_z,\nu)=\left(
        \begin{array}{cc}
            i\nu-\xi_{\bk,k_z} &  0\\
            0 & i\nu-\xi_{\bk+\bQ,k_z+Q_z}
        \end{array}
        \right),
    \end{equation}
    with $Q_z=\pi$, that is, we require the interlayer dimers to be antiferromagnetically ordered. In the rest of this chapter, all the quantities with energy dimensions will be given in units of the hopping $t$ when not explicitly stated otherwise.
    \section{Order parameter and incommensurability}
    In this section, we show results obtained from calculations at the lowest temperatures reachable by the ED algorithm with $N_s=4$ bath sites, namely $T=0.027t$ for LSCO, and $T=0.04t$ for YBCO. Notice that decreasing $T$ below these two values leads, at least for some dopings, to an unphysical decrease and eventual vanishing of the order parameter $m$. Lower temperatures could be reached increasing $N_s$. However, the exponential scaling of the ED algorithm makes low-$T$ calculations computationally involved. We obtain homogeneous solutions for any doping, that is, for all values of $p$ shown, we have $\frac{\partial\mu}{\partial n}>0$. By contrast, in Hartree-Fock studies~\cite{Igoshev2010} phases with two different densities have been found over broad doping regions.

    \begin{figure}[t]
        \centering
        \includegraphics[width=0.6\textwidth]{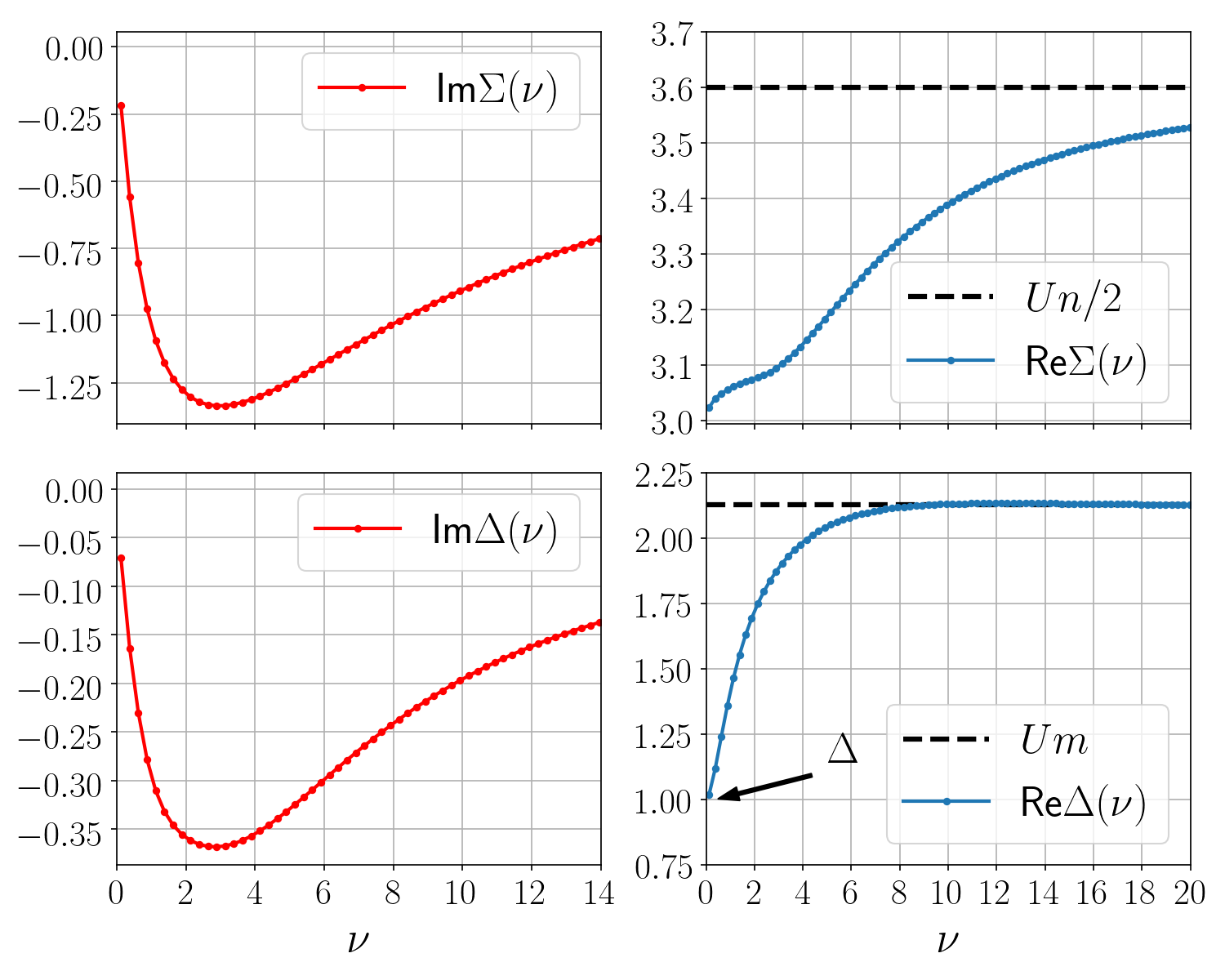}
        \caption{Diagonal (top) and off-diagonal (bottom) component of the DMFT self-energy as a function of the Matsubara frequency for LSCO parameters at $p=0.10$, and $T=0.04t$.}
        \label{fig_spiral: sigma & delta vs nu}
    \end{figure}

    Differently than static mean-field theory, where the off-diagonal self-energy is given by a simple number which can be chosen as purely real, within DMFT it acquires a frequency dependency, and, in general, an imaginary part. A particular case when $\Delta(\nu)$ can be chosen as a purely real function of the frequency is the half-filled Hubbard model with only nearest neighbor hoppings ($t'=t''=0$), where a particle-hole transformation can map the antiferromagnetic state onto a superconducting one, for which it is always possible to choose a real gap function. In Fig.~\ref{fig_spiral: sigma & delta vs nu}, we plot the normal and anomalous self-energies as functions of the Matsubara frequency $\nu$. $\Sigma(\nu)$ displays a behavior qualitatively similar to the one of the paramagnetic state, with a negative imaginary part, and a real one approaching the Hartree-Fock expression $Un/2$ for $\nu\to\infty$. The anomalous self-energy $\Delta(\nu)$ exhibits a sizable frequency dependence with its real part interpolating between its value at the Fermi level $\Delta\equiv\Delta(\nu\to0)$, and an Hartree-Fock-like expression $Um$, with $m$ the onsite magnetization. We notice that within the DMFT (local) charge and pairing fluctuations are taken into account, leading to an overall suppression of $\Delta$ compared to the Hartree-Fock result. This is the magnetic equivalent of the Gor'kov-Melik-Barkhudarov effect found in superconductors~\cite{Gorkov1961}. The observation that $\Delta<Um$ is another manifestation of these fluctuations.

    \begin{figure}[t]
        \centering
        \includegraphics[width=0.9\textwidth]{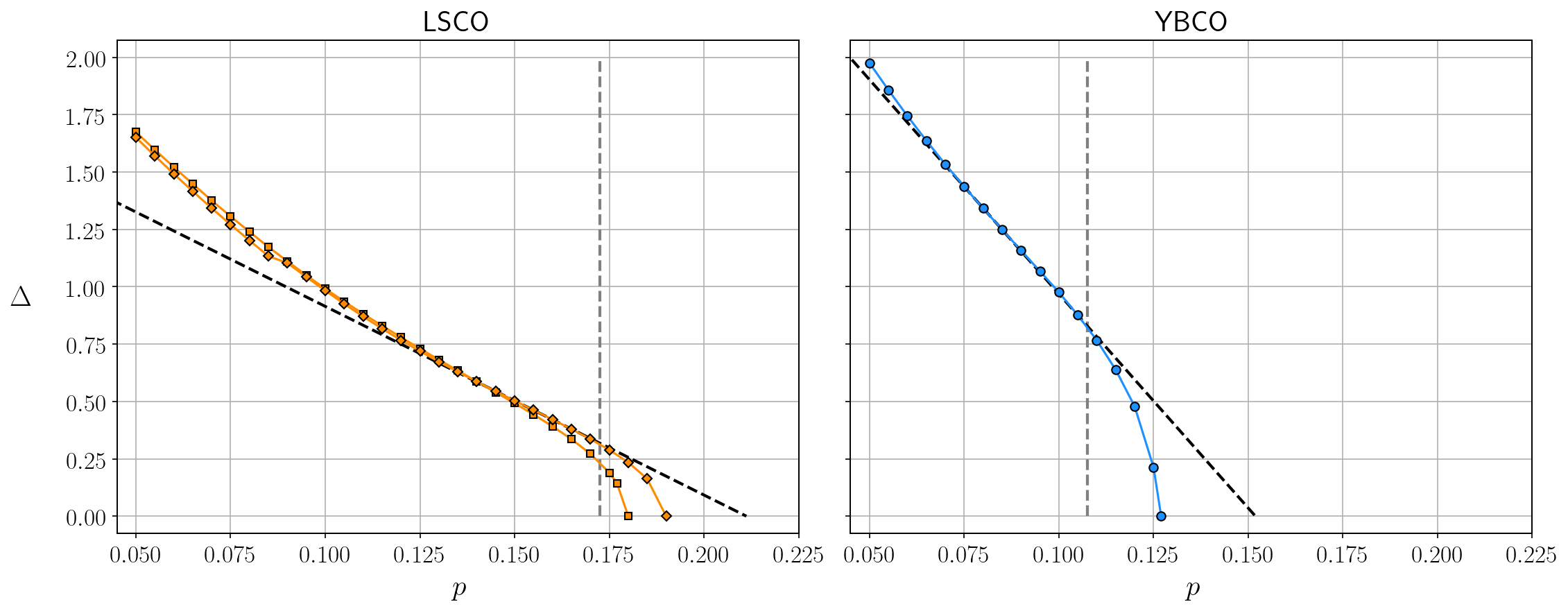}
        \caption{Magnetic gaps for LSCO (left panel) and YBCO (right panel) as functions of doping. For LSCO, we show results at $T=0.04t$ (squares), and $T=0.027t$ (diamonds). The dashed black lines represent estimations of the gaps at $T=0$ via a linear extrapolation, while the dashed gray lines indicate the doping above which electron pockets are present, together with hole pockets, in the Fermi surface}
        \label{fig_spiral: delta vs p}
    \end{figure}
    In Fig.~\ref{fig_spiral: delta vs p}, we show the extrapolated zero frequency gap $\Delta$ as a function of the doping for the two materials under study. As expected, the gap is maximal at half filling, and decreases monotonically upon doping, until it vanishes continuously at $p=p^*$. Due to the mean-field character of the DMFT, the magnetic gap is expected to behave proportionally to $(p^*-p)^{1/2}$ for $p$ slightly below $p^*$ at finite temperature. Examining the temperature trend for LSCO (left panel of Fig.~\ref{fig_spiral: delta vs p}), lowering the temperature, we expect $p^*$ to increase, and the approximately linear behavior of $\Delta$ to extend up to the critical doping, as indicated by the extrapolation in the figure. In principle, a weak first-order transition is also possible at $T=0$.

    \begin{figure}[t]
        \centering
        \includegraphics[width=0.5\textwidth]{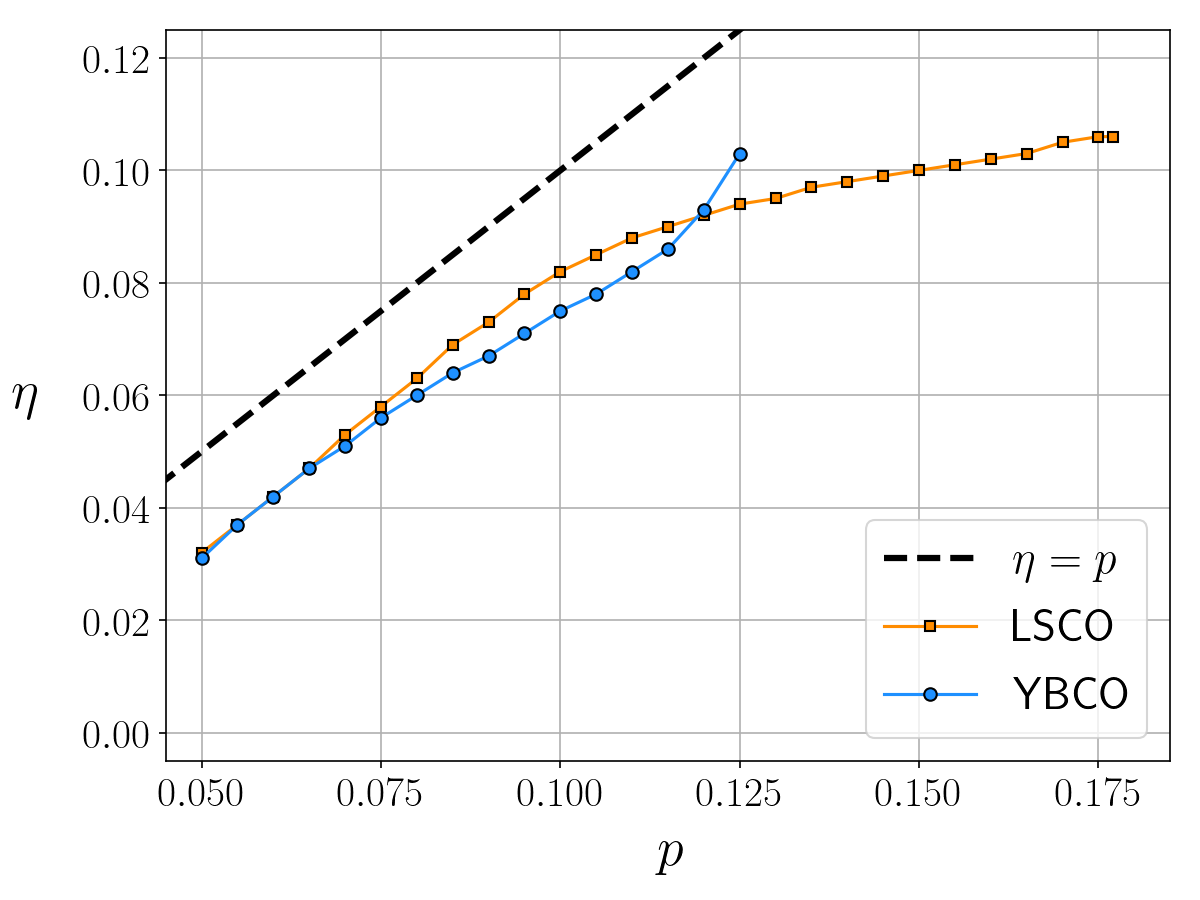}
        \caption{Incommensurability $\eta$ as a function of doping for LSCO and YBCO parameters at $T=0.04t$.}
        \label{fig_spiral: eta vs p}
    \end{figure}
    Within the parameter ranges under study, the ordering wave vector always takes the form $\bQ=(\pi-2\pi\eta,\pi)$ (or symmetry related), with the incommensurability $\eta$ varying with doping, as shown in Fig.~\ref{fig_spiral: eta vs p}. For both compounds we find that $\eta$ is lower than $p$. Experimentally, the relation $\eta(p)\simeq p$ has been found to hold for LSCO for $0.06<p<0.12$, saturating to $\eta\simeq1/8$ for larger dopings~\cite{Yamada1998}. Differently, experiments on YBCO have found $\eta(p)$ being significantly smaller than $p$~\cite{Haug2010}.
    \section{Fermi surfaces}
    The onset of spiral magnetic order leads to a band splitting and therefore to a fractionalization of the Fermi surface. In the vicinity of the Fermi level, we can approximate the anomalous and normal self-energies as constants, $\Delta$ and $\Sigma_0\equiv\mathrm{Re}\Sigma(\nu\to0)$, which leads to a mean-field expression~\cite{Igoshev2010} for the quasiparticle bands reading as
    \begin{equation}
        E^\pm_\bk=\frac{\epsilon_\bk+\epsilon_{\bk+\bQ}}{2}\pm\sqrt{\left(\frac{\epsilon_\bk-\epsilon_{\bk+\bQ}}{2}\right)^2+\Delta^2}-\widetilde{\mu},
    \end{equation}
    with $\widetilde{\mu}=\mu-\Sigma_0$. The quasiparticle Fermi surfaces are then given by $E_\bk^\pm=0$. In the case of the bilayer compound YBCO, there are two sets of Fermi surfaces corresponding to the bonding and antibonding bands. We remark that the above expression for the quasiparticle dispersions holds only in the vicinity of the Fermi level, where the expansion of the DMFT self-energies is justifiable.

    \begin{figure}[t]
        \centering
        \includegraphics[width=0.8\textwidth]{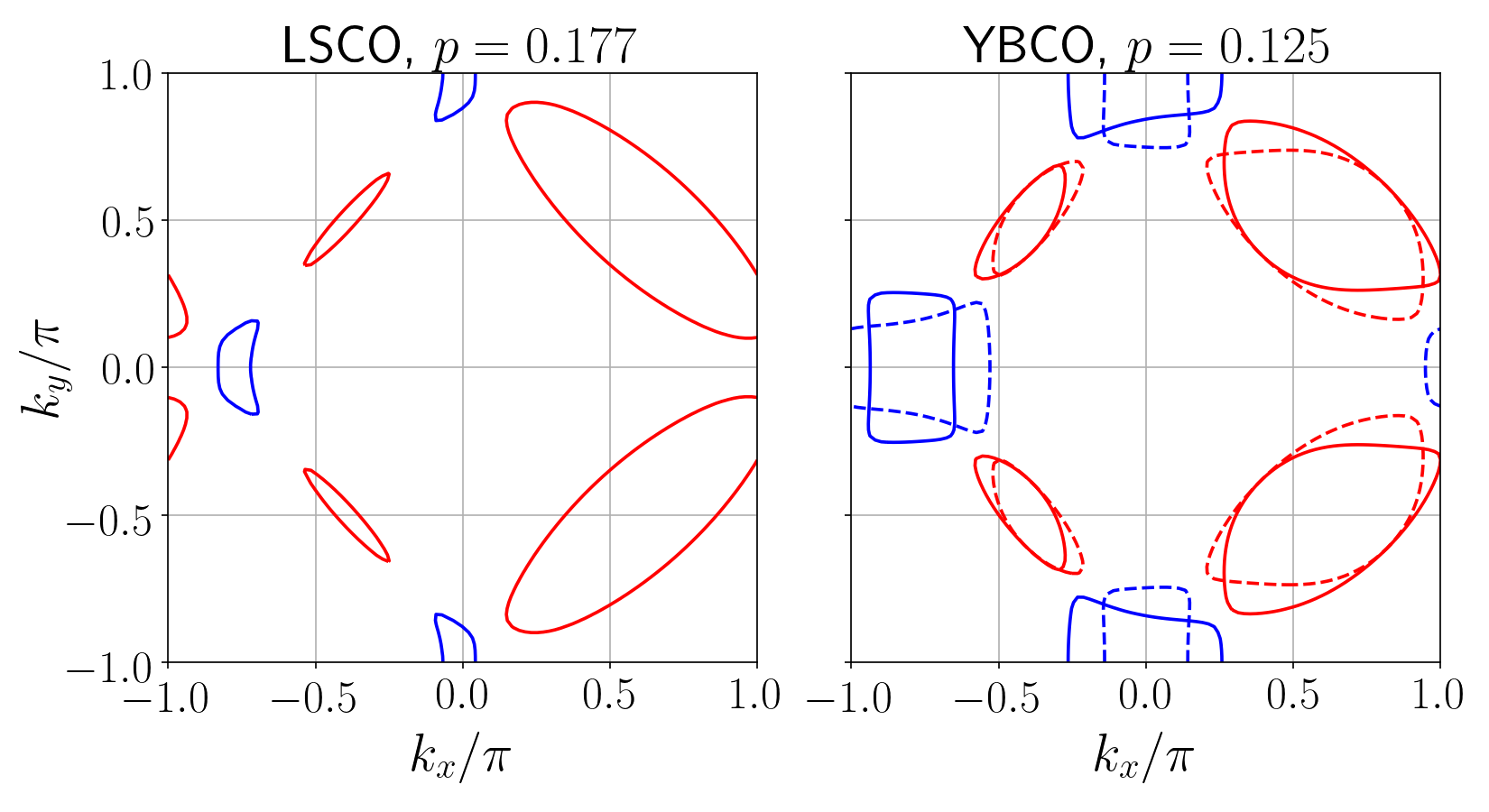}
        \caption{Fermi surfaces for LSCO and YBCO slightly below their critical doping at $T=0.04t$. The read lines indicate hole pockets, while the blue ones electron pockets. For YBCO solid and dashed lines denote the bonding and antibonding bands, respectively.}
        \label{fig_spiral: FS p*}
    \end{figure}
    In Fig.~\ref{fig_spiral: FS p*} the quasiparticle Fermi surfaces for LSCO and YBCO band parameters are shown for doping values slightly smaller than their respective critical doping $p^*$. In all cases, due to the small value of $\Delta$ in the vicinity of $p^*$, both electron and hole pockets are present.

    \begin{figure}[t]
        \centering
        \includegraphics[width=0.5\textwidth]{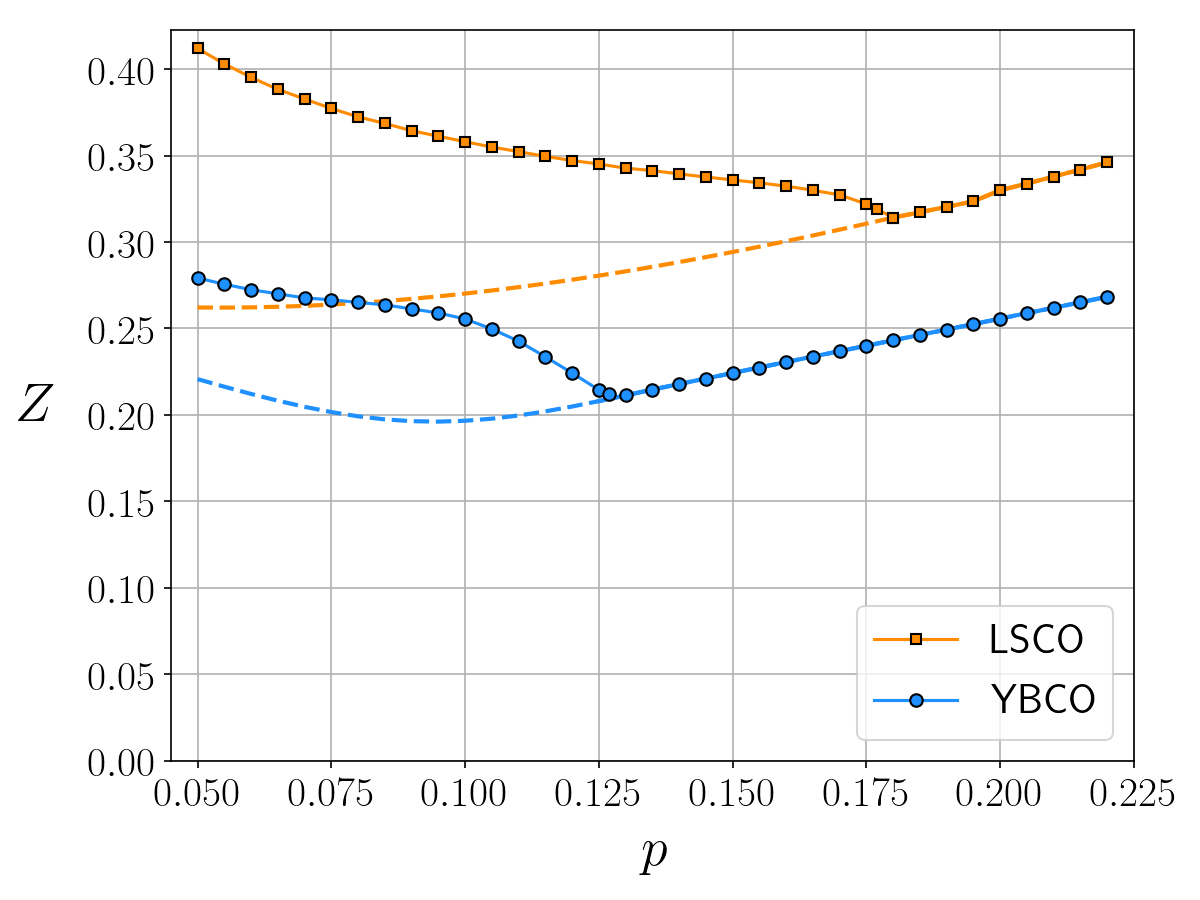}
        \caption{$Z$ factors as functions of the doping for LSCO and YBCO band parameters at $T=0.04t$. The dashed lines represent the $Z$ factors in the unstable paramagnetic phase.}
        \label{fig_spiral: Z factors}
    \end{figure}
    The quasiparticle Fermi surface differs from the Fermi surface observed in photoemission experiments. The latter is determined by poles of the diagonal elements of the Green's function, corresponding to peaks in the spectral function at zero frequency $A(\bk,0)$. Discarding the frequency dependence of the self-energies, the spectral functions in the vicinity of the Fermi level can be expressed as~\cite{Eberlein2016}
    \begin{subequations}
        \begin{align}
            &A_\up(\bk,\omega)=\sum_{\eta=\pm}\frac{\Delta^2}{\Delta^2+\left(\xi_{\bk-\bQ}-E^{-\eta}_{\bk-\bQ}\right)^2}\,\delta\!\left(\omega-E^{\eta}_{\bk-\bQ}\right),\\
            &A_\down(\bk,\omega)=\sum_{\eta=\pm}\frac{\Delta^2}{\Delta^2+\left(\xi_{\bk}-E^{\eta}_{\bk}\right)^2}\,\delta\!\left(\omega-E^{\eta}_{\bk}\right),
        \end{align}
    \end{subequations}
    where $\omega/t\ll1$, $\xi_\bk=\epsilon_\bk-\widetilde{\mu}$, and $\delta(x)$ denotes the Dirac delta function. The total spectral function, $A(\bk,\omega)=A_\up(\bk,\omega)+A_\down(\bk,\omega)$, is inversion symmetric ($A(-\bk,\omega)=A(\bk,\omega)$) for band dispersions obeying $\epsilon_{-\bk}=\epsilon_\bk$, while the quasiparticle bands are not~\cite{Bonetti2020_I}. Furthermore, the spectral weight on the Fermi surface is given by $\frac{\Delta^2}{\Delta^2+\xi_\bk^2}$, which is maximal for momenta close to the "bare" Fermi surface, $\xi_\bk=0$. 
    
    At low temperatures and in the vicinity of the Fermi level, the main effect of the normal self-energy $\Sigma(\nu)$ is a renormalization of the quasiparticle weight by the $Z$ factor
    \begin{equation}
        Z=\left[1-\frac{\partial\,\mathrm{Im}\Sigma(\nu)}{\partial\nu}\bigg\rvert_{\nu=0}\right]^{-1}\leq1,
    \end{equation}
    where the derivative can be approximated by $\mathrm{Im}\Sigma(\pi T)/(\pi T)$ at finite temperatures. The $Z$ factor reduces the bare dispersion to $\bar{\xi}_\bk=Z\xi_\bk$, the magnetic gap to $\bar{\Delta}=Z\Delta$, and the quasiparticle energies to $\bar{E}^\pm_\bk=ZE^\pm_\bk$. Moreover, the quasiparticle contributions to the spectral function get suppressed by a global factor $Z$. The missing spectral weight is then shifted to incoherent contributions at higher energies. The resulting spectral function will then read as 
    \begin{equation}
        \begin{split}
            A(\bk,\omega)=&Z\sum_{\eta=\pm}\left[
                \frac{\bar{\Delta}^2}{\bar{\Delta}^2+\left(\bar{\xi}_{\bk-\bQ}-\bar{E}^{-\eta}_{\bk-\bQ}\right)^2}\,\delta\!\left(\omega-\bar{E}^{\eta}_{\bk-\bQ}\right)
                +\frac{\bar{\Delta}^2}{\bar{\Delta}^2+\left(\bar{\xi}_{\bk}-\bar{E}^{\eta}_{\bk}\right)^2}\,\delta\!\left(\omega-\bar{E}^{\eta}_{\bk}\right)
            \right]\\
            &+A^\mathrm{inc}(\bk,\omega).
        \end{split}
    \end{equation}
    In Fig.~\ref{fig_spiral: Z factors}, we plot the $Z$ factors for LSCO and YBCO parameters computed at $T=0.04t$ as functions of the doping. The values computed for the (enforced) unstable paramagnetic solution are also shown for comparison (dashed lines). We notice that the $Z$ factors exhibit a quite weak doping dependence, and, depending on the material, take values between 0.2 and 0.4, with the strongest renormalization occurring for YBCO. We remark that for $p\to0$ the paramagnetic $Z$ factors are not expected to vanish as the choice of parameters for both materials makes them lie on the metallic side of the Mott transition at half filling.

    \begin{figure}[t]
        \centering
        \includegraphics[width=0.9\textwidth]{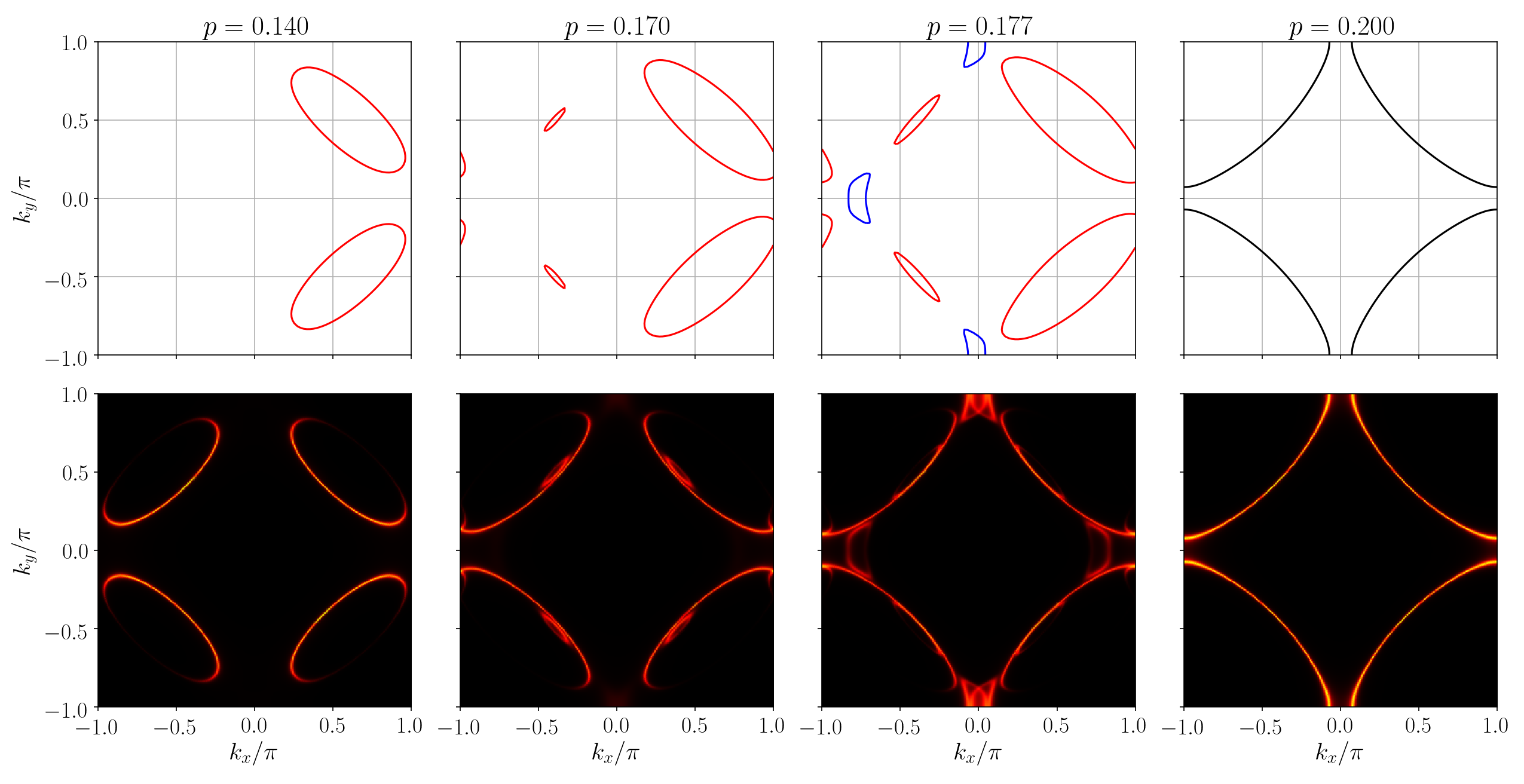}
        \caption{Quasiparticle Fermi surfaces (top) and spectral functions $A(\bk,0)$ (bottom) for LSCO parameters and for different doping values. The spectral functions have been broadened by a constant scattering rate $\Gamma=0.025t$.}
        \label{fig_spiral: FS sp fun LSCO}
    \end{figure}
    In Fig.~\ref{fig_spiral: FS sp fun LSCO}, we show the quasiparticle Fermi surfaces and spectral functions for various doping values across the spiral-to-paramagnetic transition. Electron pockets are present in the Fermi surface only in a narrow doping region below $p^*$ (see also Fig.~\ref{fig_spiral: delta vs p}). The spectral function exhibits visible peaks only on the inner sides of the pockets, as the outer sides are strongly suppressed by the spectral weight. Therefore, the Fermi surface observed in photoemission experiments smoothly evolves from Fermi arcs, characteristic of the pseudogap phase, to a large Fermi surface upon increasing doping. 
    \section{Application to transport experiments in Cuprates}
    \begin{figure}[t]
        \centering
        \includegraphics[width=0.9\textwidth]{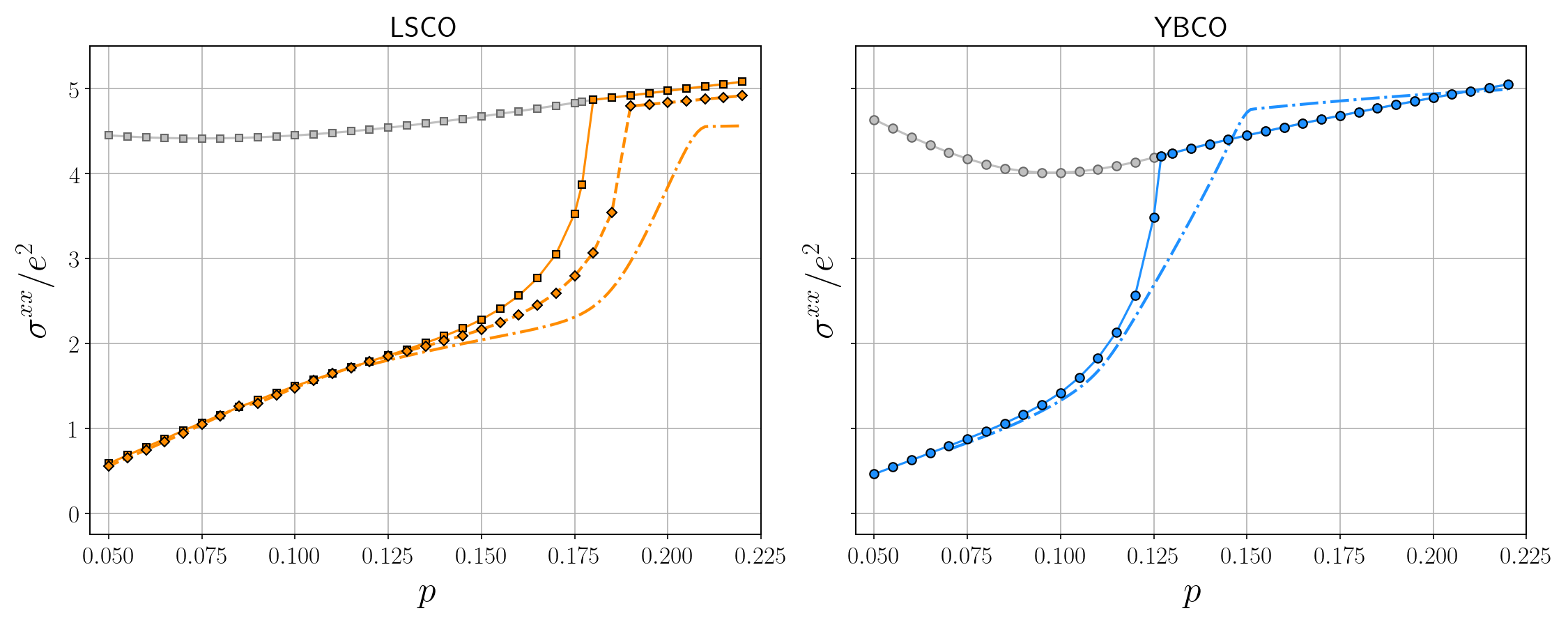}
        \caption{Longitudinal conductivity as a function of doping for LSCO (left panel) at $T=0.04t$ (solid line, squares), and $T=0.027t$ (dashed line, diamonds) and YBCO (right panel) at $T=0.04t$. The dashed-dotted lines represent extrapolations at $T=0$. The conductivity in the unstable paramagnetic phase (gray symbols) is shown for comparison.}
        \label{fig_spiral: sigma_xx LSCO YBCO}
    \end{figure}
    Transport coefficients can in principle be computed within the DMFT. However, this involves a delicate analytic continuation from Matsubara to real frequencies. Furthermore, the quasiparticle lifetimes calculated within this approach are due to electron-electron scattering processes, while in the real systems important contributions also come from phonons and impurities.
    We therefore compute the magnetic gap $\Delta$, the incommensurability $\eta$, and the $Z$ factor as functions of the doping $p$ within the DMFT, and plug them in a mean-field Hamiltonian, while taking estimates for the scattering rates from experiments. 
    The mean-field Hamiltonian reads as
    \begin{equation}
        \mathcal{H}_\mathrm{MF}=\int_\bk\sum_\sigma\left(\bar{\epsilon}_\bk-\mu\right) c^\dagger_{\bk,\sigma}c_{\bk,\sigma}+\int_\bk \bar{\Delta}\left(c^\dagger_{\bk,\up}c_{\bk+\bQ,\down}+c^\dagger_{\bk+\bQ,\down}c_{\bk,\up}\right),
        \label{eq_spiral: H MF}
    \end{equation}
    with $\bar{\epsilon}_\bk=Z\epsilon_\bk$. The chemical potential $\mu$ is then adapted such that the doping calculated from~\eqref{eq_spiral: H MF} coincides with the one computed within the DMFT. The scattering rate is then implemented by adding a constant imaginary part $i\Gamma$ to the inverse retarded bare propagator, with $\Gamma$ fixed to $0.025t$.
    
    The transport coefficients are obtained by coupling the system to the U(1) electromagnetic gauge potential $\boldsymbol{A}(\mathbf{r},t)$ through the \emph{Peierls substitution}, that is
    \begin{equation}
        t_{jj'}\to t_{jj'}\exp\left[ie\int_{\mathbf{R}_j}^{\mathbf{R}_{j'}}\boldsymbol{A}(\mathbf{r},t)\cdot d\mathbf{r}\right],
        \label{eq_spiral: Peierls subst}
    \end{equation}
    with $t_{jj'}$ the hopping matrix, that is the Fourier transform of $\bar{\epsilon}_\bk$, and $e<0$ the electron charge. The ordinary and Hall conductivities are defined as
    \begin{equation}
        j^\alpha = \left[\sigma^{\alpha\beta}+\sigma_H^{\alpha\beta\gamma}B^\gamma\right]E^\beta,
    \end{equation}
    with $j^\alpha$ the electrical current, and $E^\beta$ and $B^\gamma$ the electric and magnetic field, respectively. The Hall coefficient is then given by
    \begin{equation}
        R_H=\frac{\sigma_H^{xyz}}{\sigma^{xx}\,\sigma^{yy}},
    \end{equation}
    and the Hall number as $n_H=1/(|e|R_H)$. Exact expressions for the conductivities of the Hamiltonian~\eqref{eq_spiral: H MF} can be obtained, and we refer to Ref.~\cite{Mitscherling2018} for a derivation and more details. These formulas go well beyond the independent band picture often used in the calculation of transport properties, as they include \emph{interband} and \emph{intraband} contributions on equal footing. For a broad discussion on these different terms in \emph{general} two-band models, we refer to Refs.~\cite{Mitscherling2020,Mitscherling_Thesis}. 
    
    In Fig.~\ref{fig_spiral: sigma_xx LSCO YBCO}, we show the longitudinal conductivity as a function of doping for the two materials under study and for different temperatures, together with an extrapolation at zero temperature, obtained by inserting the guess for the doping dependence of $\Delta$ at $T=0$ sketched in Fig.~\ref{fig_spiral: delta vs p}. The expected drop at $p=p^*$ is  particularly steep at $T>0$ due to the square root onset of $\Delta(p)$, while it is smoother at $T=0$. Since in the present calculation the scattering rate does not depend on doping, the drop in $\sigma^{xx}$ is exclusively due to the Fermi surface reconstruction.

    \begin{figure}[t]
        \centering
        \includegraphics[width=0.9\textwidth]{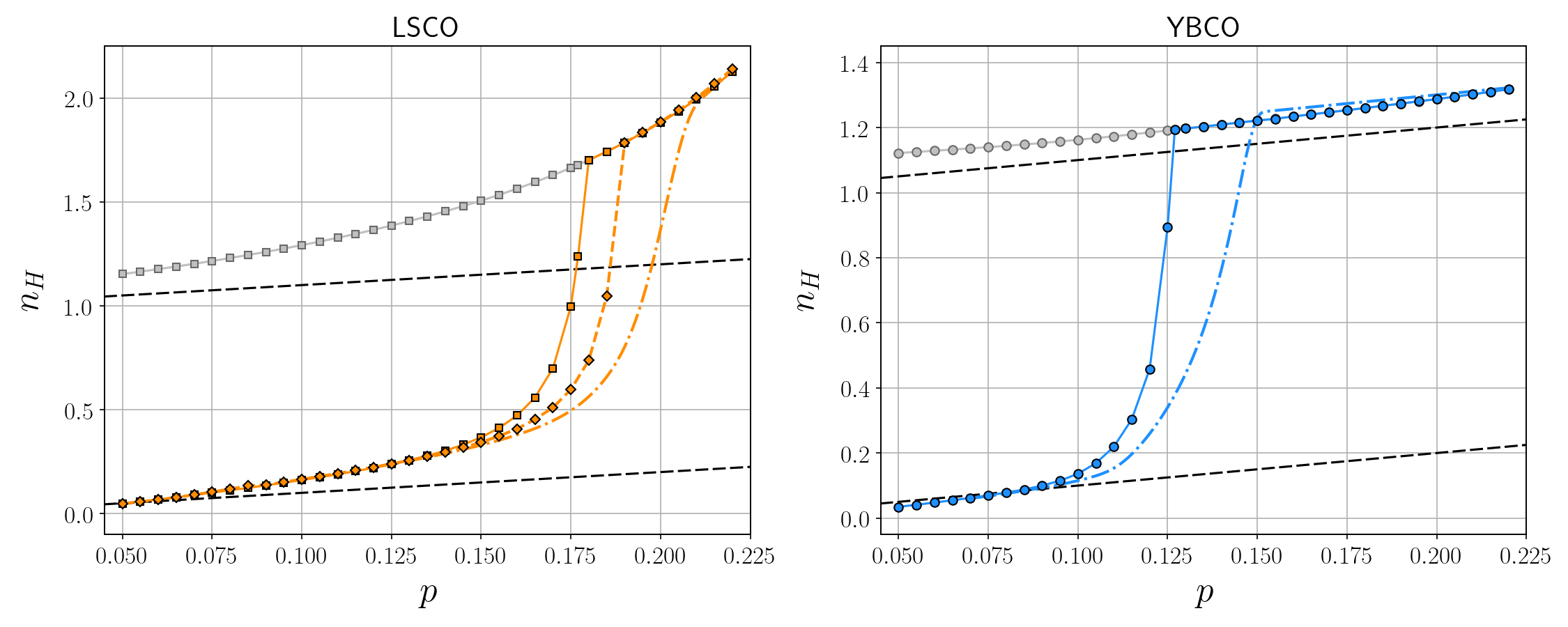}
        \caption{Hall number $n_H$ as a function of doping for LSCO (left panel) and YBCO (right panel). The symbol code is the same as in Fig.~\ref{fig_spiral: sigma_xx LSCO YBCO}. The black dashed lines correspond to the na\"ive expectations $n_H=p$ for the hole pockets and $n_H=1+p$ for a large Fermi surface.}
        \label{fig_spiral: nH LSCO YBCO}
    \end{figure}
    The Hall number as a function of doping is plotted in Fig.~\ref{fig_spiral: nH LSCO YBCO} for different temperatures, together with an extrapolation at $T=0$. A pronounced drop is found for $p<p^*$, indicating once again a drop in the charge carrier concentration. In the high-field limit $\omega_c\tau\gg1$, with $\omega_c$ the cyclotron frequency and $\tau\propto1/\Gamma$ the quasiparticle lifetime, the Hall number exactly equals the charge carrier density enclosed by the Fermi pockets. However, the experiments are performed in the \emph{low-field} limit $\omega_c\tau\ll 1$. In this limit, $n_H$ equals the charge carrier density \emph{only} for parabolic dispersions. For low doping, the Hall number approaches the value $p$, indicating that for small $p$ the hole pockets are well approximated by ellipses. In the paramagnetic phase emerging at $p>p^*$, $n_H$ is slightly above the na\"ive expectation $1+p$ for YBCO, while for LSCO it is completely off, a sign that in this regime the dispersion is far from being parabolic. In fact, the large values of $n_H$ are a precursor of a divergence occurring for $p=0.33$, well above the van Hove doping at $p=0.23$. 

    \begin{figure}[b!]
        \centering
        \includegraphics[width=0.5\textwidth]{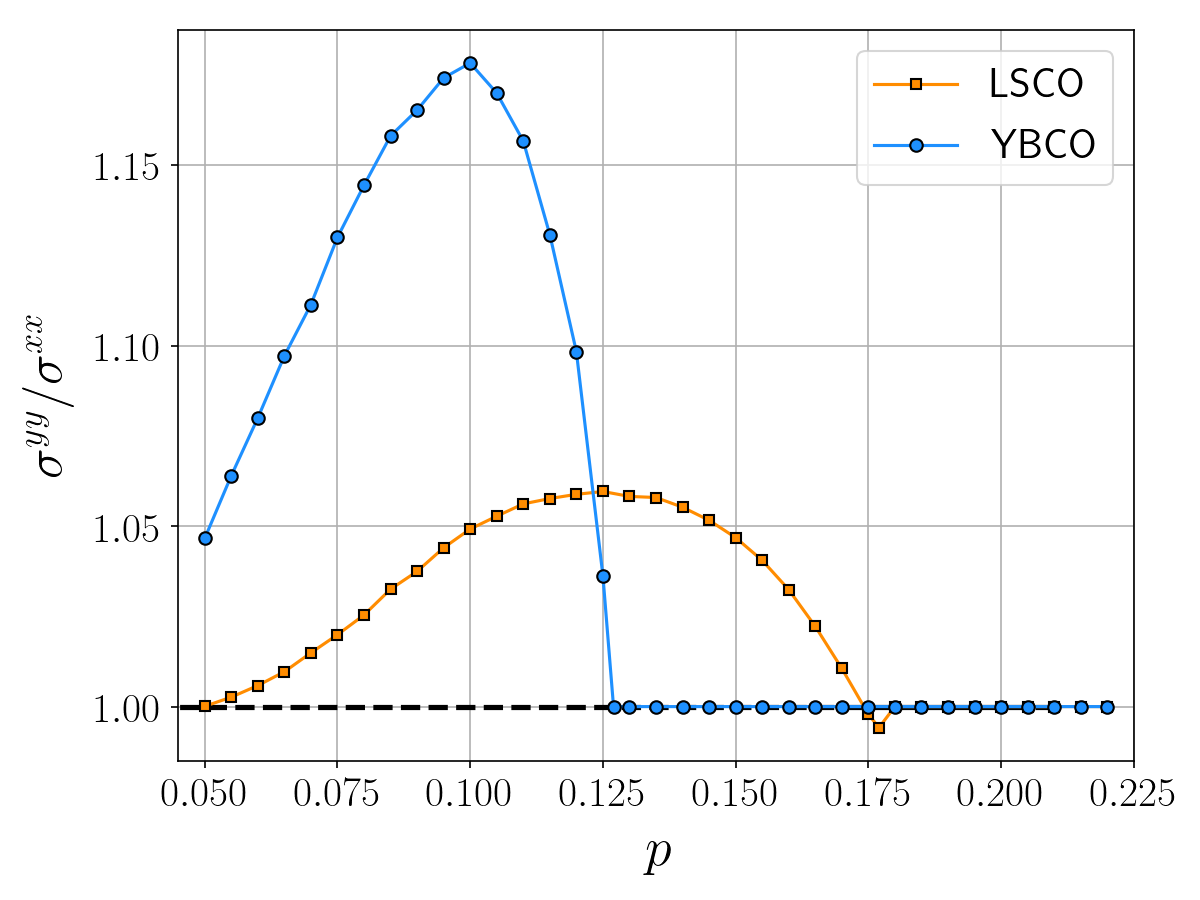}
        \caption{Ratio $\sigma^{yy}/\sigma^{xx}$ as a function of doping for LSCO (orange symbols) and YBCO (blue symbols) at $T=0.04t$.}
        \label{fig_spiral: sigma_yy/sigma_xx}
    \end{figure}
    In Fig.~\ref{fig_spiral: sigma_yy/sigma_xx}, we show the ratio $\sigma^{yy}/\sigma^{xx}$ as a function of doping for LSCO and YBCO at $T=0.04t$. The breaking of the square lattice symmetry due to the onset of spiral order leads to an anisotropy, or nematicity, in the longitudinal conductivity. This behavior has also been experimentally observed in Ref.~\cite{Ando2002}, where the values for the ratio were however much larger than the ones of the present calculation. For a wavevector of the form $\bQ=(\pi-2\pi\eta,\pi)$, the longitudinal conductivity in the $y$ direction is larger than the one in the $x$ direction. Lowering $p$ below $p^*$, the decrease in $\eta$ is compensated by an increase in $\Delta$, leading to an overall increase in the ratio $\sigma^{yy}/\sigma^{xx}$, until a point where the incommensurability becomes too small and the ratio decreases again, saturating to 1 for small values of the doping, where $\eta=0$.
    

\cleardoublepage
    \rhead[\fancyplain{}{\bfseries fRG+MF approach to the Hubbard model}]{\fancyplain{}{\bfseries\thepage}}
    \lhead[\fancyplain{}{\bfseries\thepage}]{\fancyplain{}{\bfseries fRG+MF approach to the Hubbard model}}
    \chapter{fRG+MF approach to the Hubbard model}
    \label{chap: fRG+MF}
    Aim of this chapter is to present a framework that allows to continue the fRG flow into phases exhibiting spontaneous symmetry breaking (SSB). This can be achieved by means of a simplified truncation that neglects the interplay of the different channels below the critical scale $\Lamc$, at which symmetry breaking occurs. This set of flow equations can be shown to be equivalent to a mean-field (MF) approach with renormalized couplings computed for RG scales ($\L$) larger than $\Lamc$~\cite{Wang2014,Yamase2016}. Therefore, we call this approach fRG+MF. Neglecting the channel competition also for $\L>\Lamc$ leads to the Hartree-Fock approximation for the order parameter~\cite{Salmhofer2004}.
    
    This chapter is divided into three parts. In the first one, we introduce the fRG+MF equations. In the second one, we perform a fRG+MF calculation of the phase diagram of the Hubbard model. We treat all the frequency dependencies of the two-particle vertex, therefore extending the static results of Ref.~\cite{Yamase2016}. The results of this part have been published in Ref.~\cite{Vilardi2020}. In the third part, we tackle the problem of reformulating the fRG+MF approach in a mixed boson-fermion representation, where the explicit presence of a bosonic field allows for a systematic inclusion of the collective fluctuations on top of the MF. Ref.~\cite{Bonetti2020_II} contains the results presented in this part. %
    \section{fRG+MF formalism}
    \label{sec_fRG_MF: fRG+MF equations}
    In this section, we derive the fRG+MF equations assuming that at $\L=\Lamc$ the particle-particle channel $\phi^{(pp)\L}$ diverges, signaling the onset of superconductivity for $\L<\Lamc$, arising from the breaking of the global U(1) charge symmetry. Generalizations to other orders and symmetries are straightforward, as shown for example in Sec.~\ref{sec_fRG_MF: Phase diag} for the case of N\'eel and spiral antiferromagnetism. 
    
    We now derive the equations for the fRG+MF approach that neglects any kind of order parameter (thermal or quantum) fluctuations. In order to deal with the breaking of the global U(1) symmetry, we introduce the Nambu spinors
    \begin{equation}
        \Psi_k=\left(
        \begin{array}{c}
            \psi_{k,\uparrow}  \\
            \overline{\psi}_{-k,\downarrow}  
        \end{array}
        \right) \hskip 1cm 
        \overline{\Psi}_k=\left(
        \begin{array}{c}
            \overline{\psi}_{k,\uparrow}  \\
            \psi_{-k,\downarrow}  
        \end{array}
        \right),
        \label{eq_fRG+MF: Nambu spinors}
    \end{equation}
    where $\psi_{k,\sigma}$ ($\overline{\psi}_{k,\sigma}$) is a Grassmanian field corresponding to the annihilation (creation) of an electron, $k=(\bk,\nu)$ a collective variable comprising the lattice momentum and a fermionic Matsubara frequency, and $\sigma=\up,\down$ the spin quantum number. 
    
    \subsection{Flow equations and integration}
    In the SSB phase, the vertex function $V$ acquires anomalous components due to the violation of particle number conservation. In particular, besides the normal vertex describing scattering processes with two incoming and two outgoing particles ($V_{2+2}$), in the superfluid phase also components with three ($V_{3+1}$) or four ($V_{4+0}$) incoming or outgoing particles can arise. We avoid to treat the 3+1 components, since they are related to the coupling of the order parameter to charge fluctuations~\cite{Eberlein2013}, which do not play any role in a MF-like approximation for the superfluid state. It turns out to be useful to work with linear combinations 
    \begin{equation}
        \begin{split}
            &V_\mathcal{A}=\Re\left\{V_{2+2}+V_{4+0}\right\},\\
            &V_\Phi=\Re\left\{V_{2+2}-V_{4+0}\right\},
            \label{eq_fRG+MF:  A and Phi vertex combination}
        \end{split}
    \end{equation}
    that represent two fermion interactions in the longitudinal and transverse order parameter channels, respectively. They are related to the amplitude and phase fluctuations of the superfluid order parameter, respectively. In principle, a longitudinal-transverse mixed interaction can also appear, from the imaginary parts of the vertices in Eq.~\eqref{eq_fRG+MF:  A and Phi vertex combination}, but it has no effect in the present MF approximation because it vanishes at zero center of mass frequency~\cite{Eberlein_Thesis}.
    
    Below the critical scale, $\Lambda <\Lamc$, we consider a truncation of the effective action of the form 
    \begin{equation}
        \begin{split}
            \Gamma^{\Lambda}_{\text{SSB}}[\Psi,\overline{\Psi}]=-\int_{k} \overline{\Psi}_{k} \, \left[\mathbf{G}^{\Lambda}(k)\right]^{-1} \Psi_{k}\,\,
            +&\int_{k,k',q}V^{\Lambda}_{\mathcal{A}}(k,k';q)\, S^1_{k,q}\,S^1_{k',-q}\\
            +&\int_{k,k',q}V^{\Lambda}_{\Phi}(k,k';q)\, S^2_{k,q}\,S^2_{k',-q} ,
        \end{split}
        \label{eq_fRG+MF: fermionic SSB truncation}
    \end{equation}
    with the Nambu bilinears defined as
    \begin{equation}
        S^\alpha_{k,q}=\overline{\Psi}_{k+\rnddo{q}}\, \tau^\alpha \,\Psi_{k-\rndup{q}},
        \label{eq_fRG+MF: fermion bilinear}
    \end{equation}
    where the Pauli matrices $\tau^\alpha$ are contracted with Nambu spinor indexes. The fermionic propagator $\mathbf{G}^\Lambda(k)$ is given by the matrix
    \begin{equation}
        \left(
        \begin{array}{cc}
            Q_{0}^\Lambda(k)-\Sigma^\Lambda(k) &  \Delta^\Lambda(k)\\
            \Delta^\Lambda(k) & -Q_0^\Lambda(-k)+\Sigma^\Lambda(-k)
        \end{array}
        \right)^{-1},
    \end{equation}
    where $Q_{0}^\Lambda(k)=i\nu-\xi_\mathbf{k}+R^\Lambda(k)$, $\xi_\bk$ is the single particle dispersion relative to the chemical potential, $R^\Lambda(k)$ the fRG regulator, $\Sigma^\Lambda(k)$ the normal self energy, and $\Delta^\Lambda(k)$ the superfluid gap. The initial conditions at the scale $\Lambda=\Lamc$ require $\Delta^{\Lamc}$ to be zero and both $V^{\Lamc}_\mathcal{A}$ and $V^{\Lamc}_\Phi$ to equal the vertex $V^{\Lamc}$ in the symmetric phase.
    
    We are now going to introduce the MF approximation to the symmetry broken state, that means that we focus on the $q=0$ component of $V_\mathcal{A}$ and $V_\Phi$ and neglect all the rest. So, from now on we keep only the $q=0$ terms. We also neglect the flow of the normal self-energy below $\Lamc$. In order to simplify the presentation, we introduce a matrix-vector notation for the gaps and vertices. In particular, the functions $V_\mathcal{A}$ and $V_\Phi$ are matrices in the indices $k$ and $k'$, while the gap and the fermionic propagator behave as vectors. For example, in this notation an object of the type $\int_{k'}V_\mathcal{A}^\Lambda(k,k')\Delta^\Lambda(k')$ can be viewed as a matrix-vector product, $V_\mathcal{A}^\Lambda \Delta^\Lambda$. 
    
    Within our MF approximation, we consider in the set of flow equations only the terms that involve only the $q=0$ components of the functions $V_\mathcal{A}$ and $V_\Phi$. This means that in a generalization of Eq.~\eqref{eq_methods: flow eq vertex xx'} to the SSB phase, we consider only the particle-particle contributions. In formulas we have:
    \begin{align}
        &\partial_\Lambda V_\mathcal{A}^\Lambda=V_\mathcal{A}^\Lambda\left[\widetilde{\partial}_\Lambda\Pi^\Lambda_{11}\right] V_\mathcal{A}^\Lambda+\Gamma^{(6)\Lambda} \circ \widetilde{\partial}_\Lambda G^\Lambda,
        \label{eq_fRG+MF: flow eq Va fermionic}\\
        &\partial_\Lambda V_\Phi^\Lambda=V_\Phi^\Lambda \left[\widetilde{\partial}_\Lambda\Pi^\Lambda_{22}\right] V_\Phi^\Lambda+\Gamma^{(6)\Lambda} \circ \widetilde{\partial}_\Lambda G^\Lambda,
        \label{eq_fRG+MF: flow eq Vphi fermionic}
    \end{align}
    where we have defined the bubbles 
    \begin{equation}
        \Pi^\Lambda_{\alpha\beta}(k,k')=-\frac{1}{2}\Tr\left[\tau^\alpha\,\mathbf{G}^\Lambda(k)\,\tau^\beta\,\mathbf{G}^\Lambda(k)\right]\delta_{k,k'},
    \end{equation}
    with $\delta_{k,k'}=(2\pi)^2/T \,\delta(\mathbf{k}-\mathbf{k}')\delta_{\nu\nu'}$, and the trace runs over Nambu spin indexes.
    The last terms of Eqs.~\eqref{eq_fRG+MF: flow eq Va fermionic} and~\eqref{eq_fRG+MF: flow eq Vphi fermionic} involve the 6-particle interaction, which we treat here in the Katanin approximation, that allows us to replace the derivative acting on the regulator $\widetilde{\partial}_\Lambda$ of the bubbles with the full scale derivative $\partial_\Lambda$~\cite{Katanin2004}. This approach is useful for it provides the exact solution of mean-field models, such as the reduced BCS model, in which the bare interaction is restricted to the zero center of mass momentum channel~\cite{Salmhofer2004}. 
    In this way, the flow equation~\eqref{eq_fRG+MF: flow eq Va fermionic} for the vertex $V_\mathcal{A}$, together with the initial condition $V_\mathcal{A}^{\Lamc}=V^{\Lamc}$ can be integrated analytically, yielding
    \begin{equation}
        \begin{split}
            V_\mathcal{A}^\Lambda = &\left[1+V^{\Lamc}(\Pi^{\Lamc}-\Pi_{11}^\Lambda)\right]^{-1}V^{\Lamc} =\left[1-\widetilde{V}^{\Lamc}\Pi_{11}^\Lambda\right]^{-1}\widetilde{V}^{\Lamc},
        \end{split}
        \label{eq_fRG+MF: Va solution fermionic}
    \end{equation}
    where  
    \begin{equation}
        \Pi^{\Lamc}(k,k')=G^{\Lamc}(k)G^{\Lamc}(-k)\delta_{k,k'},
        \label{eq_fRG+MF: bubble at Lambda_s}
    \end{equation}
    is the (normal) particle-particle bubble at zero center of mass momentum, 
    \begin{equation}
        G^{\Lambda}(k)=\frac{1}{Q_0^{\Lambda}(k)-\Sigma^{\Lamc}(k)},
        \label{eq_fRG+MF: G at Lambda_s}
    \end{equation}
    
    is the fermionic normal propagator, and
    \begin{equation}
        \widetilde{V}^{\Lamc}=\left[1+V^{\Lamc}\Pi^{\Lamc}\right]^{-1}V^{\Lamc}
        \label{eq_fRG+MF: irr vertex fermionic}
    \end{equation}
    is the irreducible (normal) vertex in the particle-particle channel at the critical scale. The flow equation for the transverse vertex $V_\Phi$ exhibits a formal solution similar to the one in Eq.~\eqref{eq_fRG+MF: Va solution fermionic}, but the matrix $[1-\widetilde{V}^{\Lamc}\Pi_{22}^\Lambda]$ is not invertible. We will come to this aspect later. 
    \subsection{Gap equation}
    Similarly to the flow equations for vertices, in the flow equation of the superfluid gap we neglect the contributions involving the vertices at $q\neq 0$. We are then left with
    \begin{equation}
        \partial_\Lambda\Delta^\Lambda(k)=\int_{k'}V_\mathcal{A}^\Lambda(k,k')\,\widetilde{\partial}_\Lambda F^\Lambda(k'),
        \label{eq_fRG+MF: gap flow equation}
    \end{equation}
    where  
    \begin{equation}
        F^\Lambda(k)=\frac{\Delta^\Lambda(k)}{[G^\Lambda(k)\,G^\Lambda(-k)]^{-1}+\left[\Delta^\Lambda(k)\right]^2}
        \label{eq_fRG+MF: F definition}
    \end{equation}
    is the anomalous fermionic propagator, with $G^\L$ defined as in Eq.~\eqref{eq_fRG+MF: G at Lambda_s}, and with the normal self-energy kept fixed at its value at the critical scale. By inserting Eq.~\eqref{eq_fRG+MF: Va solution fermionic} into Eq.~\eqref{eq_fRG+MF: gap flow equation} and using the initial condition $\Delta^{\Lamc}=0$, we can analytically integrate the flow equation, obtaining the gap equation~\cite{Wang2014}
    \begin{equation}
        \Delta^\Lambda(k)=\int_{k'}\widetilde{V}^{\Lamc}(k,k')\, F^\Lambda(k').
        \label{eq_fRG+MF: gap equation fermionic}
    \end{equation}
    In the special case in which the contributions to the vertex flow equation from other channels (different from the particle-particle) are neglected also above the critical scale, the irreducible vertex is nothing but the bare interaction, and Eq.~\eqref{eq_fRG+MF: gap equation fermionic} reduces to the standard Hartree-Fock approximation to the SSB state. 
    \subsection{Goldstone Theorem}
    \label{sec_fRG+MF: Goldstone fermi}
    In this subsection we prove that the present truncation of flow equations fulfills the Goldstone theorem. We revert our attention on the transverse vertex $V_\Phi$. Its flow equation in Eq.~\eqref{eq_fRG+MF: flow eq Vphi fermionic} can be (formally) integrated, too, together with the initial condition $V_\Phi^{\Lamc}=V^{\Lamc}$, giving 
    \begin{equation}
        \begin{split}
            V_\Phi^\Lambda = \left[1+V^{\Lamc}(\Pi^{\Lamc}-\Pi_{22}^\Lambda)\right]^{-1}V^{\Lamc} =\left[1-\widetilde{V}^{\Lamc}\Pi_{22}^\Lambda\right]^{-1}\widetilde{V}^{\Lamc}.
        \end{split}
        \label{eq_fRG+MF: Vphi solution fermionic}
    \end{equation}
    However, by using the relation 
    \begin{equation}
        \Pi_{22}^\Lambda(k,k')=\frac{F^\Lambda(k)}{\Delta^\Lambda(k)}\,\delta_{k,k'},
        \label{eq_fRG+MF: Pi22=F/delta}
    \end{equation}
    one can rewrite the matrix in square brackets on the right hand side of Eq.~\eqref{eq_fRG+MF: Vphi solution fermionic} as
    \begin{equation}
        \delta_{k,k'}-\widetilde{V}^{\Lamc}(k,k')\,\frac{F^\Lambda(k')}{\Delta^\Lambda(k')}.
    \end{equation}
    Multiplying this expression by $\Delta^\Lambda(k')$ and integrating over $k'$, we see that it vanishes if the gap equation~\eqref{eq_fRG+MF: gap equation fermionic} is obeyed. Thus, the matrix in square brackets in Eq.~\eqref{eq_fRG+MF: Vphi solution fermionic} has a zero eigenvalue with the superfluid gap as eigenvector. In matrix notation this property can be expressed as
    \begin{equation}
        \left[ 1 - \widetilde{V}^{\Lamc}\Pi^\Lambda_{22}\right]\Delta^\Lambda=0.
    \end{equation}
    Due to the presence of this zero eigenvalue, the matrix $[1-\widetilde{V}^{\Lamc}\Pi_{22}^\Lambda]$ is not invertible. This is nothing but a manifestation of the Goldstone theorem. Indeed, due to the breaking of the global U(1) symmetry, transverse fluctuations of the order parameter become massless at $q=0$, leading to the divergence of the transverse two fermion interaction $V_\Phi$.
    \section{Interplay of antiferromagnetism and superconductivity}
    \label{sec_fRG_MF: Phase diag}
    In this section, we present an application of the fRG+MF approach to the phase diagram of the two-dimensional Hubbard model. We parametrize the vertex function by \emph{fully} taking into account its frequency dependence. In Refs.~\cite{Husemann2012,Vilardi2017} the frequency dependence of the vertex function has been shown to be important, as a static approximation underestimates the size of magnetic fluctuations, while overestimating the $d$-wave pairing scale. The present dynamic computation therefore extends and improves the static results obtained in Ref.~\cite{Yamase2016}.
    \subsection{Symmetric regime}
    \label{sec_fRG_MF: symmetric regime}
    In the symmetric regime, that is, for $\L>\Lamc$, we perform a weak-coupling fRG calculation within a 1-loop truncation. For the vertex function, we start from the parametrization in Eq.~\eqref{eq_methods: channel decomp physical}, and simplify the dependencies of the three channels on $\bk$, $\bk'$. We perform a form factor expansion in these dependencies and retain only the $s$-wave terms for the magnetic and charge channels, and $s$-wave and $d$-wave terms for the pairing one. In formulas, we approximate 
    \begin{subequations}
        \begin{align}
            &\mathcal{M}^\L_{k,k'}(q)=\mathcal{M}^\L_{\nu\nu'}(q),\\
            &\mathcal{C}^\L_{k,k'}(q)=\mathcal{C}^\L_{\nu\nu'}(q),\\
            &\mathcal{S}^\L_{k,k'}(q)=\mathcal{S}^\L_{\nu\nu'}(q) + d_\bk d_{\bk'}\mathcal{D}^\L_{\nu\nu'}(q),
        \end{align}
    \end{subequations}
    where the $d$-wave form factor reads as $d_\bk=\cos k_x-\cos k_y$, and $q=(\bq,\Omega)$ is a collective variable comprising a momentum and a bosonic Matsubara frequency. Furthermore, we set the initial two-particle vertex equal to the bare interaction $U$, that is, in Eq.~\eqref{eq_methods: channel decomp physical} we set $\lambda(k_1',k_2',k_1)=U$. The parametrization of the vertex function described above has been used in Ref.~\cite{Vilardi2017}, with a slightly different notation, and we refer to this publication and to Appendix~\ref{app: symm V} for the flow equations for $\mathcal{M}^\L$, $\mathcal{C}^\L$, $\mathcal{S}^\L$, and $\mathcal{D}^\L$.
    \subsection{Symmetry broken regime}
    In the $\L<\Lamc$ regime, at least one of the symmetries of the Hubbard Hamiltonian is spontaneously broken. The flow in the symmetric phase~\cite{Metzner2012}, and other methods~\cite{Scalapino2012} indicate antiferromagnetism, of N\'eel type or incommensurate, and $d$-wave pairing as the leading instabilities. Among all possible incommensurate antiferromagnetic orderings, we restrict ourselves to spiral  order, exhaustively described in Chap.~\ref{chap: spiral DMFT}, characterized by the order parameter $\langle \overline{\psi}_{k,\up}\psi_{k+Q,\down} \rangle$, with $Q=(\bQ,0)$, and $\bQ$ the ordering wave vector. N\'eel antiferromagnetism is then recovered by setting $\bQ=(\pi,\pi)$. 
    
    Allowing for the formation of spiral order and $d$-wave pairing, the quadratic part of the effective action takes the form
    \begin{equation}
        \begin{split}
            \Gamma^{(2)\L}\left[\psi,\overline{\psi}\right]=&-\int_k\sum_\sigma \overline{\psi}_{k,\sigma}\left[\left(G_0^\L(k)\right)^{-1}-\Sigma^\L(k)\right]\psi_{k,\sigma}\\
            &-\int_k \left[\left(\Delta_m^\L(k^*)\right)^*m(k)+\Delta_m^\L(k)m^*(k)\right]\\
            &-\int_k \left[\left(\Delta_p^\L(k^*)\right)^*p(k)+\Delta_p^\L(k)p^*(k)\right],
        \end{split}
    \end{equation}
    where $\Delta_m^\L(k)$ and $\Delta_p^\L(k)$ are the spiral and pairing anomalous self-energies, respectively, and $k^*=(\bk,-\nu)$. We have defined the bilinears $m(k)$ and $p(k)$ as
    \begin{subequations}
        \begin{align}
            &m(k) = \overline{\psi}_{k,\up}\psi_{k+Q,\down}, \hskip 1,05cm m^*(k)= \overline{\psi}_{k+Q,\down}\psi_{k,\up},\\
            &p(k) = \psi_{k,\up}\psi_{-k,\down}, \hskip 1.5cm p^*(k)=\overline{\psi}_{-k,\down}\overline{\psi}_{k,\up}.
        \end{align}
    \end{subequations}
    From now on, we neglect the normal self-energy $\Sigma^\L(k)$ both above and below $\Lamc$. It is more convenient to employ a 4-component Nambu-like basis, reading as 
    \begin{equation}
        \Psi_k = \left(
        \begin{array}{c}
            \psi_{k,\up} \\
            \overline{\psi}_{-k,\down}  \\
            \psi_{k+Q,\down}\\
            \overline{\psi}_{-k-Q,\up}  
        \end{array}
        \right)\hskip2cm
        \overline{\Psi}_k = \left(
        \begin{array}{c}
            \overline{\psi}_{k,\up} \\
            \psi_{-k,\down}  \\
            \overline{\psi}_{k+Q,\down}\\
            \psi_{-k-Q,\up}  
        \end{array}
        \right).
    \end{equation}
    In this way, the quadratic part of the action can be expressed as 
    \begin{equation}
        \Gamma^{(2)\L}\left[\Psi,\overline{\Psi}\right] = 
        -\int_k \overline{\Psi}_k \left[\mathbf{G}^\L(k)\right]^{-1}\Psi_k,
    \end{equation}
    with 
    \begin{equation}
        \left[\mathbf{G}^\L(k)\right]^{-1} = {\footnotesize\left(
        \begin{array}{cccc}
            \left[G_0^\L(k)\right]^{-1} &  \Delta_p^\L(k) & \Delta_m^\L(k) & 0\\
            \left[\Delta_p^\L(k^*)\right]^* & -\left[G_0^\L(-k)\right]^{-1} & 0 & -\Delta_m^\L(-k-Q)\\
            \left[\Delta_m^\L(k^*)\right]^* & 0 &  \left[G_0^\L(k+Q)\right]^{-1} & -\Delta_p^\L(-k-Q) \\
            0 & -\left[\Delta_m^\L(-k^*-Q^*)\right]^* & -\left[\Delta_m^\L(-k^*-Q^*)\right]^* &  -\left[G_0^\L(-k-Q)\right]^{-1}
        \end{array}
        \right)}.
        \label{eq_fRG+MF: 4x4 G}
    \end{equation}
    The fRG+MF approach introduced in Sec.~\ref{sec_fRG_MF: fRG+MF equations} can be easily generalized to the present case. Neglecting the $q\neq0$ ($q\neq Q$) contributions to the pairing (spiral) channel, the quartic part of the effective action reads as
    \begin{equation}
        \begin{split}
            \Gamma^{(4)\L}\left[\psi,\overline{\psi}\right]=
            &+\frac{1}{2}\int_k V_m^\L(k,k')\left[m^*(k)m(k')+m^*(k')m(k)\right]\\
            &+\frac{1}{2}\int_k W_m^\L(k,k')\left[m^*(k)m^*(k')+m(k')m(k)\right]\\
            &+\frac{1}{2}\int_k V_p^\L(k,k')\left[p^*(k)p(k')+p^*(k')p(k)\right]\\
            &+\frac{1}{2}\int_k W_p^\L(k,k')\left[p^*(k)p^*(k')+p(k')p(k)\right],
        \end{split}
    \end{equation}
    where $W_m^\L$ and $W^\L_p$ represent anomalous interaction terms in the SSB phase. At the critical scale, the normal interactions are given by
    \begin{subequations}
        \begin{align}
            &V_m^\Lamc(k,k')=V^\Lamc_{\up\down\down\up}(k+Q,k',k),\\
            &V_p^\Lamc(k,k')=\frac{1}{2}\left[V^\Lamc_{\up\down\up\down}(k,-k,k')-V^\Lamc_{\up\down\down\up}(k,-k,k')\right],
        \end{align}
    \end{subequations}
    where the pairing vertex has been projected onto the singlet component. One can then define the longitudinal and transverse interactions as
    \begin{subequations}
        \begin{align}
            &\mathcal{A}_m^\L(k,k') = V_m^\L(k,k') + W_m^\L(k,k'),\\
            &\Phi_m^\L(k,k') = V_m^\L(k,k') - W_m^\L(k,k'),\\
            &\mathcal{A}_p^\L(k,k') = V_p^\L(k,k') + W_p^\L(k,k'),\\
            &\Phi_p^\L(k,k') = V_p^\L(k,k') - W_p^\L(k,k').
        \end{align}
        \label{eq_fRG+MF: A e Phi m e p}
    \end{subequations}
    While $\Phi_m^\L$ and $\Phi_p^\L$ are decoupled from the flow equations for the gap functions within the fRG+MF approach, they are crucial for the fulfillment of the Goldstone theorem, as shown in Sec.~\ref{sec_fRG+MF: Goldstone fermi}.
    
    In line with the parametrization performed in the symmetric regime, we approximate
    \begin{subequations}
        \begin{align}
            &\Delta_m^\L(k) = \Delta_m^\L(\nu),\\
            &\Delta_p^\L(k) = \Delta_p^\L(\nu)d_\bk,
        \end{align}
    \end{subequations}
    and 
    \begin{subequations}
        \begin{align}
            &\mathcal{A}_m^\L(k,k')=\mathcal{A}_m^\L(\nu,\nu'),\\
            &\mathcal{A}_p^\L(k,k')=\mathcal{A}_p^\L(\nu,\nu')d_\bk d_{\bk'}.
        \end{align}
    \end{subequations}
    Notice that for the pairing gap we do not have considered an $s$-wave term because in the repulsive Hubbard model the interaction in the $s$-wave particle-particle channel is always repulsive.
    
    Within the matrix notation previously introduced, the flow equations for the longitudinal interactions read as
    \begin{equation}
        \deL \mathcal{A}_X^\L = \mathcal{A}_X^\L \left[\partial_\L\Pi_X^\L\right] \mathcal{A}_X^\L,
    \end{equation}
    with $X=m,p$. The longitudinal bubbles are defined as
    \begin{subequations}
        \begin{align}
            & \Pi_m^\L(\nu,\nu') = T\delta_{\nu\nu'}\int_\bk \left\{G^\L(k)G^\L(k+Q)+\left[F_m(k)\right]^2\right\},\\
            & \Pi_p^\L(\nu,\nu') = T\delta_{\nu\nu'}\int_\bk d_\bk^2\left\{-G^\L(k)G^\L(-k)+\left[F_p(k)\right]^2\right\},
        \end{align}
    \end{subequations}
    with 
    \begin{subequations}
        \begin{align}
            &G^\L(k) = \left[\mathbf{G}^\L(k)\right]_{11},\\
            &F_m^\L(k) = \left[\mathbf{G}^\L(k)\right]_{13},\\
            &F_p^\L(k) = \left[\mathbf{G}^\L(k)\right]_{12},
        \end{align}
    \end{subequations}
    where $\mathbf{G}^\L(k)$ is obtained inverting Eq.~\eqref{eq_fRG+MF: 4x4 G}. As shown in Sec.~\ref{sec_fRG_MF: fRG+MF equations}, the flow equations for $\mathcal{A}_X^\L$ can be analytically solved, giving
    \begin{equation}
        \mathcal{A}_X^\L = \left[1-\widetilde{V}^\Lamc_X\Pi_X^\L\right]^{-1}\widetilde{V}^\Lamc_X,
        \label{eq_fRG+MF: integrated Ax}
    \end{equation}
    with the irreducible vertex at the critical scale reading as
    \begin{equation}
        \widetilde{V}^\Lamc_X = \left[1+V^\Lamc_X\Pi_X^\Lamc\right]^{-1}V^\Lamc_X.
        \label{eq_fRG+MF: irr V fermionic}
    \end{equation}
    The flow equations for the gap functions are given by
    \begin{subequations}
        \begin{align}
            &\deL\Delta_m^\L(\nu)=-T\sum_{\nu'}\int_\bk A_m^\L(\nu,\nu')\widetilde{\partial}_\L F_m(\bk,\nu'),\\
            &\deL\Delta_p^\L(\nu)=-T\sum_{\nu'}\int_\bk A_p^\L(\nu,\nu')\widetilde{\partial}_\L F_p(\bk,\nu')d_\bk, 
        \end{align}
        \label{eq_fRG+MF: gap flow eqs}
    \end{subequations}
    with $\widetilde{\partial}_\L$ the single-scale derivative. The integration of the above equations returns
    \begin{subequations}
        \begin{align}
            &\Delta_m^\L(\nu) = -T\sum_{\nu'}\int_\bk \widetilde{V}_m^\Lamc(\nu,\nu')F_m(\bk,\nu'),\\
            &\Delta_p^\L(\nu) = -T\sum_{\nu'}\int_\bk \widetilde{V}_p^\Lamc(\nu,\nu')F_p(\bk,\nu')d_\bk.
        \end{align}
    \end{subequations}
    Since a solution of the above nonlinear integral equations is hard to converge when both order parameters are finite, we compute the gap functions from their flow equations~\eqref{eq_fRG+MF: gap flow eqs}, plugging in the integrated form of the function $\mathcal{A}_X^\L$, in Eq.~\eqref{eq_fRG+MF: integrated Ax}. By means of global transformations, one can impose $[\Delta_m(-\nu)]^*=\Delta_m(\nu)$, and $[\Delta_p^\L(-\nu)]^*=\Delta_p(\nu)$. Since the relation $\Delta_p(-\nu)=\Delta_p(\nu)$, descending from the singlet nature of the pairing, is general, one can always remove the imaginary part of the pairing gap function. Notice that for the magnetic one this is in general not possible. Concerning the computation of the spiral wave vector $\bQ$, we fix it to the momentum at which the magnetic channel $\mathcal{M}^\L$ peaks (or even diverges) at $\L=\Lamc$. 

    \subsection{Order parameters}
    \begin{figure}[t]
        \centering
        \includegraphics[width=0.75\textwidth]{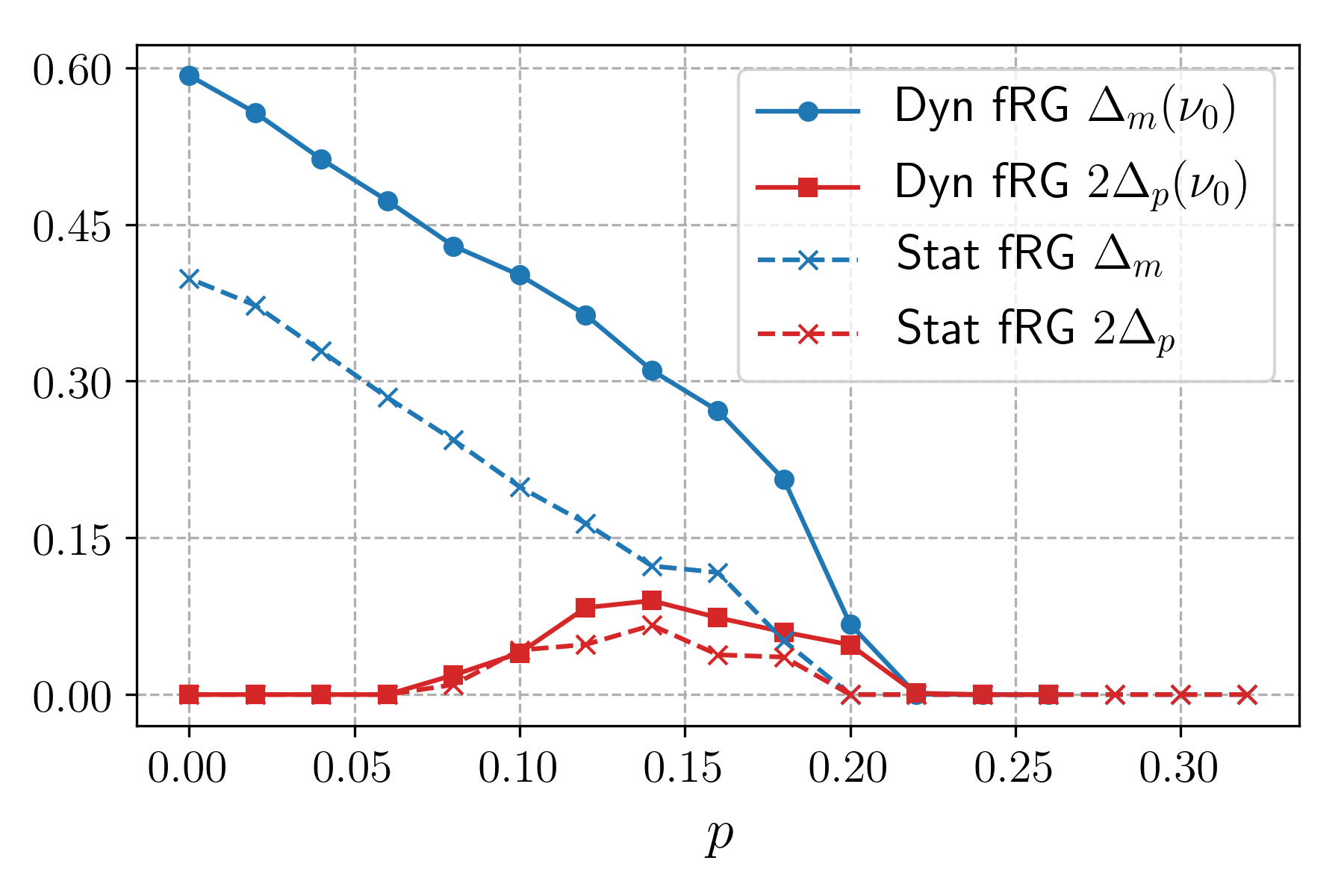}
        \caption{Gap function amplitudes at the lowest fermionic Matsubara frequency $\nu_0=\pi T$ as functions of the hole doping $p$ at $T=0.027t$. Static results are shown for comparison. The factor 2 in the pairing gap is due to the fact that $\mathrm{max}_\bk(d_\bk)=2$.}
        \label{fig_fRG+MF: gaps vs p}
    \end{figure}
    We run a fRG flow in the symmetric phase keeping, for each of the channel functions $\mathcal{M}^\L$, $\mathcal{C}^\L$, $\mathcal{S}^\L$, and $\mathcal{D}^\L$, about 90 values for each of the three frequency arguments and 320 patches in the Brillouin zone for the momentum dependence. The critical scale $\Lamc$ has been determined as the scale where the largest of the channels exceeds the value of $400t$ ($t$ is the nearest neighbor hopping). The electron density $n$ has been calculated along the flow from the first diagonal entry of the matrix propagator~\eqref{eq_fRG+MF: 4x4 G} and kept fixed at each scale $\L$ by tuning the chemical potential $\mu$. The chosen Hubbard model parameters are $t'=-0,16t$, $t''=0$, and $U=3t$. The lowest temperature reached by dynamical calculations has been set to $T=0.027t$. When not explicitly stated, we use $t$ as our energy unit all along this section. All quantities without the superscript $\L$ have to be understood as computed at the final scale $\L=0$.
    %
    %
    In Fig.~\ref{fig_fRG+MF: gaps vs p}, we show the order parameters computed at the lowest Matsubara frequency $\nu_0=\pi T$, as functions of the hole doping $p=1-n$, and at fixed temperature $T=0.027t$. For the pairing gap, we consider its maximum in momentum space, that is, $\Delta_p(\nu)$ multiplied by a factor 2, coming from $\mathrm{max}_\bk(d_\bk)=2$, occurring at $\bk=(0,\pi)$, or symmetry related. While $\Delta_p(\nu)$ is purely real, $\Delta_m(\nu)$ has an imaginary part, whose continuation to real frequencies vanishes for $\nu\to 0$. $\mathrm{Im}\Delta_m(\nu_0)$ is therefore always small for low $T$. 
    
    Magnetic order is found from half filling to about $p=0.20$, with the size of the gap monotonically decreasing upon doping, and with spiral replacing N\'eel order at about $p=0.14$. The ordering wave vector is always of the form $\bQ=(\pi-2\pi\eta,\pi)$, or symmetry related, with the incommensurability $\eta$ exhibiting a sudden jump at the N\'eel-to-spiral transition. A sizable $d$-wave pairing state is found for dopings between $0.08$ and $0.20$ coexisting with antiferromagnetic ordering, therefore confirming the previous static results obtained in Refs.~\cite{Reiss2007,Wang2014,Yamase2016}. 
    
    From Fig.~\ref{fig_fRG+MF: gaps vs p} we deduce that the inclusion of dynamic effects enhances the order parameter magnitudes. This is due to multiple effects. First of all the function $\mathcal{M}^\L_{\nu\nu'}(\bQ,0)$ has a minimum at $\nu=\nu'=\pm\pi T$, which in the static approximation is extended to the whole frequency range, leading to reduced magnetic correlations. Secondly, the static approximation largely overestimates the screening of the magnetic channel by the other channels for $\L>\Lamc$~\cite{Vilardi2017,Tagliavini2019}. On the other hand, the function $\mathcal{D}^\L_{\nu\nu'}(0)$ is maximal at $\nu=\nu'=\pm\pi T$ and rapidly decays to zero for large $\nu$, $\nu'$. This implies that, conversely to the magnetic channel, $d$-wave correlations are enhanced in the static limit~\cite{Husemann2012}. In this approximation, however, as previously discussed, the magnetic fluctuations providing the seed for the pairing are weaker, leading to a mild overall enhancement of the $d$-wave pairing gap when dynamical effects are included. 

    \begin{figure}[t]
        \centering
        \includegraphics[width=0.75\textwidth]{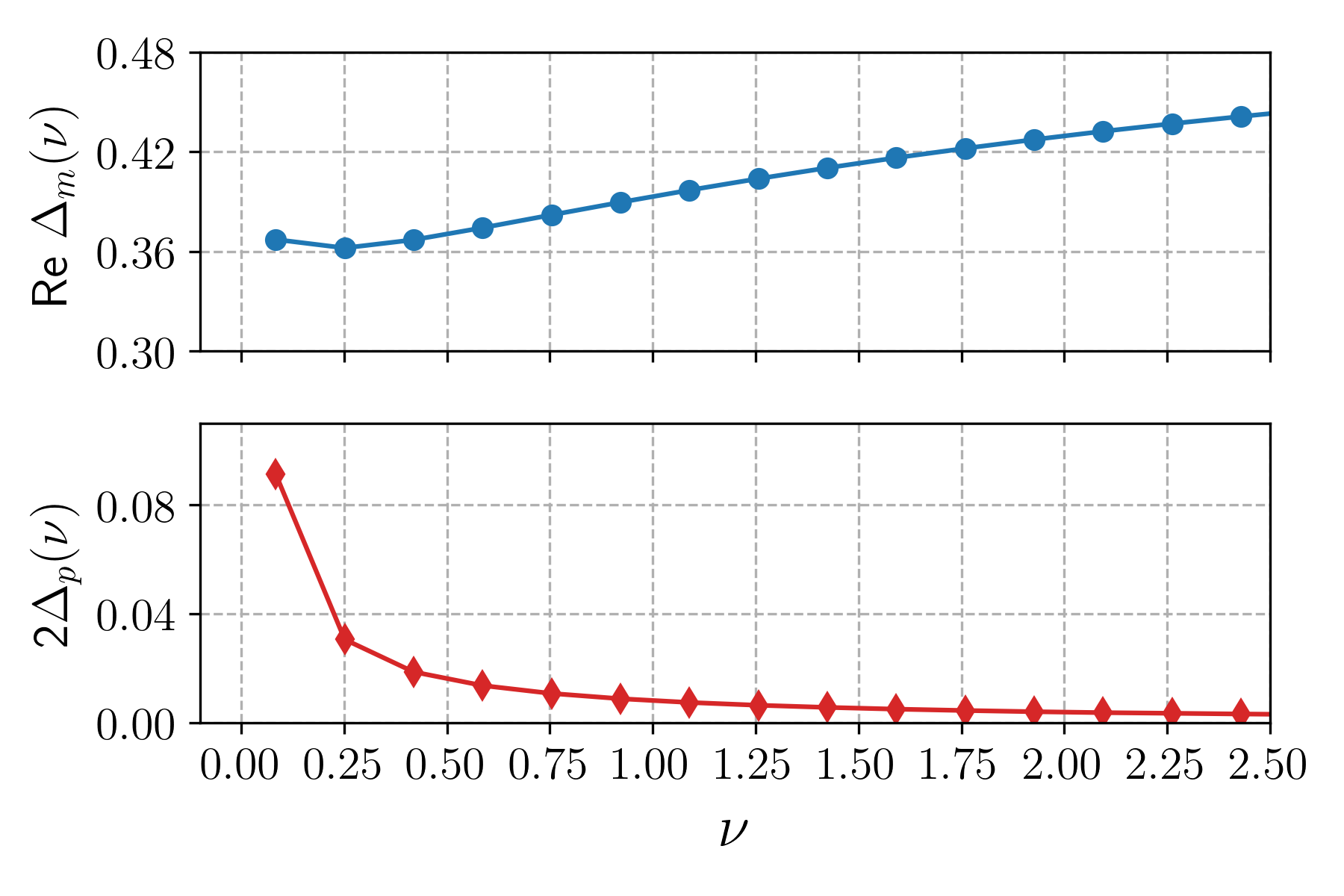}
        \caption{Frequency dependence of the real parts of the gap functions at $p=0.12$ and $T=0.027$.}
        \label{fig_fRG+MF: gaps vs nu}
    \end{figure}
    A typical behavior of the gap functions as a function of the Matsubara frequency $\nu$ is shown in Fig.~\ref{fig_fRG+MF: gaps vs nu}. Similarly to what has been discussed in Chap.~\ref{chap: spiral DMFT}, the magnetic gap interpolates between its value at the Fermi level and the constant Hartree-Fock-like expression $Um$, with $m$ the onsite magnetization, at $\nu\to\infty$. By contrast, the $d$-wave pairing gap is maximal for $\nu\to0$ and rapidly decays to zero for large frequencies, related to the fact that the Hartree-Fock approximation would yield no $d$-wave pairing at all in the Hubbard model. Finally, the magnetic gap function shows a generally small imaginary part (not shown) obeying $\mathrm{Im}\Delta_m(-\nu)=-\mathrm{Im}\Delta_m(\nu)$ therefore extrapolating to zero for $\nu\to 0$. 

    \begin{figure}[t]
        \centering
        \includegraphics[width=0.75 \textwidth]{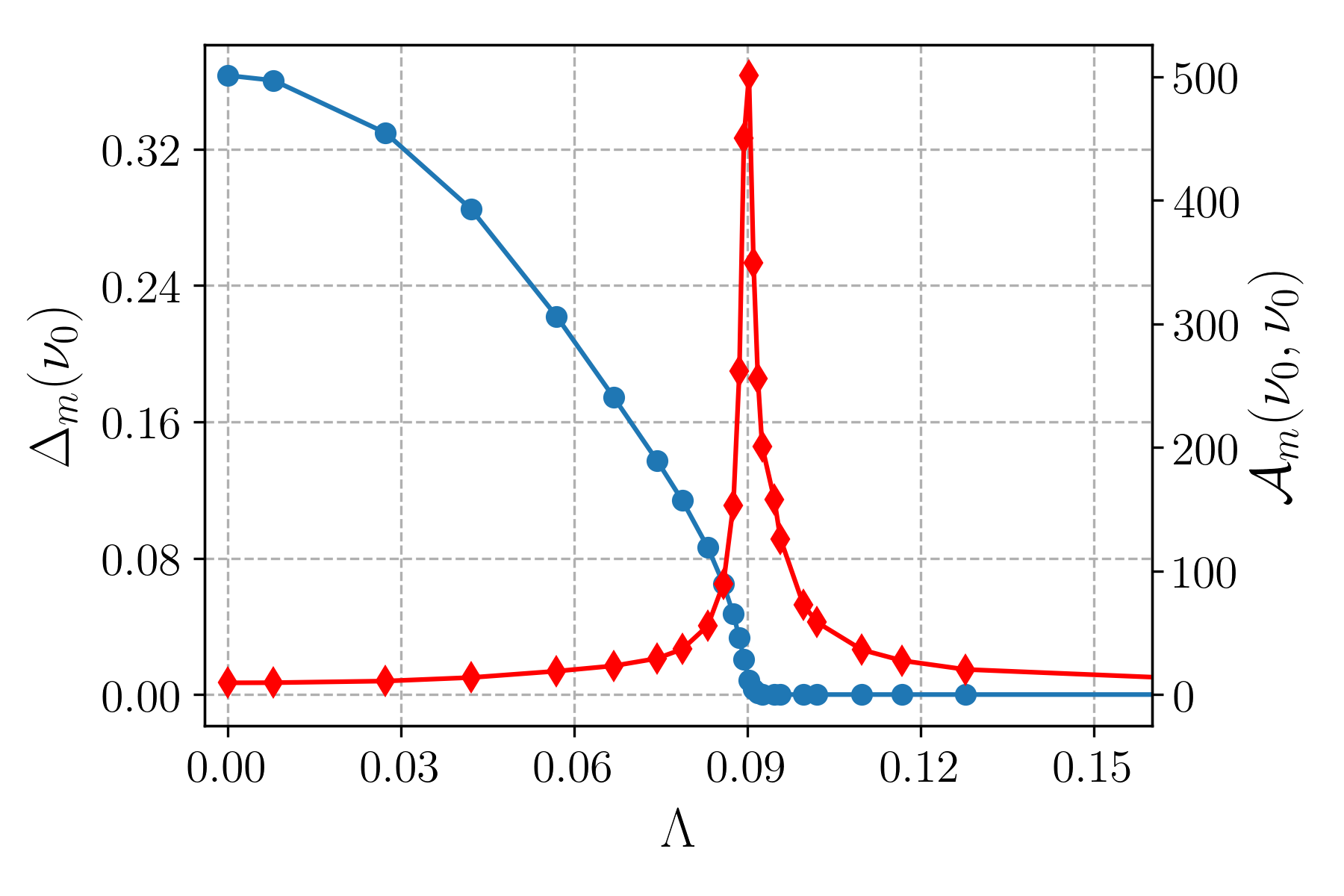}
        \caption{Scale dependence of the real part of the magnetic gap function $\mathrm{Re}\Delta_m(\nu)$ (blue dots) and longitudinal two-particle interaction $\mathrm{Re}\mathcal{A}_m(\nu,\nu')$ (red diamonds) at the lowest Matsubara frequency $\nu_0=\pi T$, at doping $p=0.12$, and temperature $T=0.027t$.}
        \label{fig_fRG+MF: gap vs Lambda}
    \end{figure}
    In Fig.~\ref{fig_fRG+MF: gap vs Lambda} we show the behavior of the magnetic gap function and of the longitudinal magnetic interaction computed at $\nu_0=\pi T$ as functions of the fRG scale. In the symmetric phase, the effective interaction grows until it diverges at the critical scale $\L=\Lamc$. In the SSB regime, a magnetic order parameters forms, leading to a quick decrease of the longitudinal interaction, that saturates to a finite value at the end of the flow. By contrast, the transverse interaction (not shown) remains infinite for all $\L<\Lamc$, in agreement with the Goldstone theorem. The flow of the analogous quantities in the pairing channel looks similar, but the divergence occurs at a scale smaller than $\Lamc$, as the leading instability in the present parameter regime is always a magnetic one. 
    \subsection{Berezinskii-Kosterlitz-Thouless transition and phase diagram}
    In this section, we compute the superfluid phase stiffness, which enables us to estimate the Berezinskii-Kosterlitz-Thouless, or simply Kosterlitz-Thouless, (KT) transition temperature ($T_\mathrm{KT}$) for the onset of superconductivity~\cite{Berezinskii1971,Kosterlitz1973}. $T_\mathrm{KT}$, together with the temperature for the onset of magnetism, $T^*$, allows us to draw a phase diagram for the Hubbard model at intermediate coupling. 
    
    Coupling the system to an external U(1) electromagnetic gauge field $\boldsymbol{A}(\mathbf{r},t)$, via, for example, the Peierls substitution (see Eq.~\eqref{eq_spiral: Peierls subst}), one is able to compute the electromagnetic response kernel $K_{\alpha\alpha'}(\bq,\omega)$, defined via
    \begin{equation}
        j_\alpha(\bq,\omega)=-\sum_{\alpha'}K_{\alpha\alpha'}(\bq,\omega)A_{\alpha'}(\bq,\omega),
    \end{equation}
    with $j_\alpha$ the electromagnetic current. The superfluid stiffness is then given by
    \begin{equation}
        J_{\alpha\alpha'} = \frac{1}{(2e)^2}\lim_{\bq\to\bzero}K_{\alpha\alpha'}(\bq,0),
    \end{equation}
    with $e$ the electron charge. 
    If the \emph{global} U(1) charge symmetry is broken via the formation of a pairing gap, the limit in the equation above is finite, and one finds a finite stiffness. Writing the superconducting order parameter as $\Phi(x)=\sqrt{\alpha^2+\varrho(x)}e^{2ie\theta(x)}$, with $\alpha=\langle\Phi(x)\rangle\in \mathbb{R}$, and neglecting the amplitude fluctuations described by $\varrho(x)$, one can derive a long-wavelength classical effective action for phase fluctuations only
    \begin{equation}
        \mathcal{S}_\mathrm{eff}[\theta]=
        \frac{1}{2}\sum_{\alpha\alpha'}J_{\alpha\alpha'}
        \int \!d^2 \mathbf{x}\,[\nabla_\alpha \theta(\mathbf{x})] [\nabla_{\alpha'} \theta(\mathbf{x})],
        \label{eq_fRG+MF: S BKT}
    \end{equation}
    where $\theta(\mathbf{x})\in[0,2\pi]$, and the superfluid stiffness plays the role of a coupling constant. This action is well known to display a \emph{topological phase transition} at finite temperature $T_\mathrm{KT}$, above which topological vortex configurations proliferate and reduce $J_{\alpha\alpha'}$ to zero, causing an exponential decay in the correlation function. Differently, for $0<T<T_\mathrm{KT}$, the vortices get bound in pairs and form a \emph{quasi-long-range ordered} phase, marked by a power law decay of the order parameter correlation function. The power law exponent is found to scale linearly with temperature, eventually vanishing at $T=0$, marking the onset of a true off-diagonal long-range order (ODLRO). The $0<T<T_\mathrm{KT}$ phase does not exhibit ODLRO, according to the Mermin-Wagner theorem~\cite{Mermin1966}, but an infinite correlation length, due to the slower than exponential decay of the correlation function. For isotropic systems, that is,  when $J_{\alpha\alpha'}=J\delta_{\alpha\alpha'}$, the transition temperature can be computed by the universal formula~\cite{ChaikinLubensky}
    \begin{equation}
        T_\mathrm{KT}=\frac{\pi}{2}J(T_\mathrm{KT}).
        \label{eq_fRG+MF: T KT}
    \end{equation}

    If the system is non-isotropic, one can introduce some rotated spatial coordinates as
    \begin{equation}
        x'_{\alpha}=\sum_{\alpha'=1}^2\left[\boldsymbol J^{\frac{1}{2}}\right]_{\alpha\alpha'}x_{\alpha'},
    \end{equation}
    with $\boldsymbol J$ the stiffness tensor. Action~\eqref{eq_fRG+MF: S BKT} is thus rotationally invariant in the new basis, with a new stiffness given by $J_\mathrm{eff}=\sqrt{\mathrm{det}[\boldsymbol J]}$, coming from the Jacobian of the coordinate change. The BKT temperature for a non-isotropic system therefore reads as
    \begin{equation}
        T_\mathrm{KT}=\frac{\pi}{2}\sqrt{\mathrm{det}[\boldsymbol{J}(T_\mathrm{KT})]}.
        \label{eq_fRG+MF: T KT anisotr}
    \end{equation}

    The authors of Ref.~\cite{Metzner2019} have derived formulas for the superfluid stiffness in a \emph{mean-field} state in which antiferromagnetism and superconductivity coexist. Since these equations have been derived in the static limit, we compute the superfluid phase stiffness by plugging into them the (real parts of) the gap functions calculated at the lowest Matsubara frequency. For a spiral state with $\bQ=(\pi-2\pi\eta,\pi)$, we have $J_{xx}\neq J_{yy}$, and $J_{xy}=J_{yx}=0$, while for a N\'eel state, $J_{xx}=J_{yy}$, and $J_{xy}=J_{yx}=0$. 

    \begin{figure}[t]
        \centering
        \includegraphics[width=0.75\textwidth]{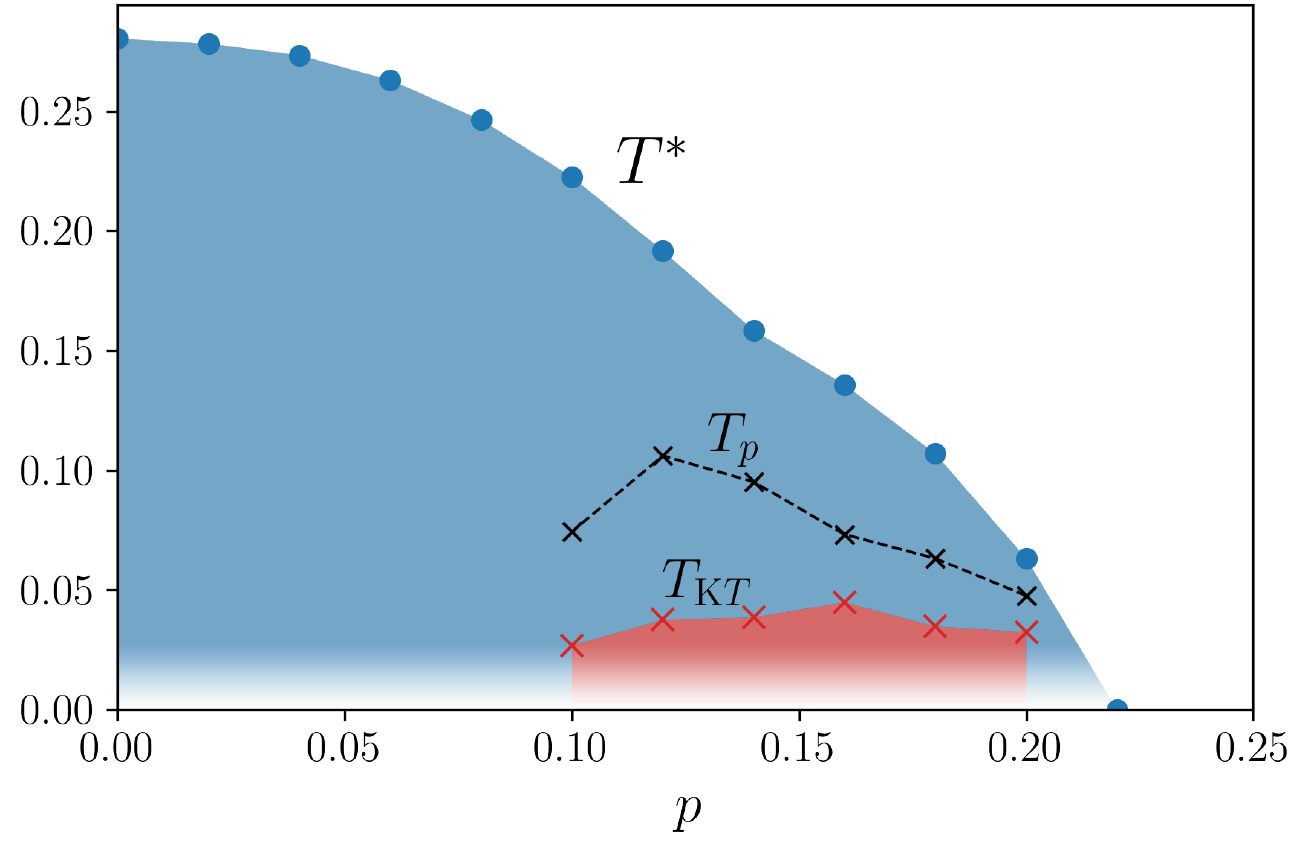}
        \caption{Doping-temperature phase diagram with the Kosterlitz-Thouless temperature, $T_\mathrm{KT}$, and the antiferromagnetic ($T^*$) and pairing ($T_p$) onset temperatures. The fading colors at low temperatures indicate that our dynamical fRG code is not able to access temperatures lower than $T=0.027t$.}
        \label{fig_fRG+MF: phase diag}
    \end{figure}
    In Fig.~\ref{fig_fRG+MF: phase diag}, we plot the computed $T_\mathrm{KT}$, together with $T^*$, that is, the lowest temperature at which the flow does not encounter a divergence in the magnetic channel down to $\L=0$, and $T_p$, at which the pairing gap vanishes, as functions of the hole doping $p$. We notice a large difference between $T_\mathrm{KT}$ and $T_p$, marking a sizable window ($T_\mathrm{KT}<T<T_p$) of the phase diagram where strong superconducting fluctuations open a gap in the spectrum, but the order parameter correlation function remains exponentially decaying, resulting in the lack of superconducting properties, which, instead, are present for $T<T_\mathrm{KT}$.
    
    The Mermin-Wagner theorem prevents the formation of long-range order at finite temperature. We therefore expect that, upon including order parameter fluctuations, the antiferromagnetic state found for $T<T^*$ will be transformed into a \emph{pseudogap state} with short-range correlations. We can thus interpret $T^*$ as the temperature for the onset of the pseudogap behavior. This topic is the subject of Chap.~\ref{chap: pseudogap}
    \section{Broken symmetry phase: bosonic formalism}
    \label{sec_fRG+MF: bosonic formalism}
    The SSB phase can be accessed also via the introduction of a bosonic field, describing the fluctuations of the order parameter~\cite{Baier2004,Strack2008,Friederich2011,Obert2013}, and whose finite expectation value is related to the formation of anomalous components in the fermionic propagator. Similarly to Sec.~\ref{sec_fRG_MF: fRG+MF equations}, we focus here on superconducting order, while generalizations to other order parameters are straightforward.
    
    In order to introduce this bosonic field, we express the vertex at the critical scale in the following form:
    \begin{equation}
        V^{\Lamc}(k,k';q)=\frac{h^{\Lamc}(k;q)\,h^{\Lamc}(k';q)}{m^{\Lamc}(q)}+\mathcal{Q}^{\Lamc}(k,k';q).
        \label{eq_fRG+MF: vertex at Lambda crit}
    \end{equation}
    We assume from now on that the divergence of the vertex, related to the appearance of a massless mode, is absorbed into the first term, while the second one remains finite.  In other words, we assume that at the critical scale $\Lamc$, at which the vertex is formally divergent, the (inverse) bosonic propagator $m^{\Lamc}(q)$ vanishes at zero frequency and momentum, while the Yukawa coupling $h^{\Lamc}(k;q)$ and the residual two fermion interaction $\mathcal{Q}^{\Lamc}(k,k';q)$ remain finite. 
    
    In Sec.~\ref{sec_fRG+MF: vertex bosonization} we introduce a systematic scheme to extract the decomposition~\eqref{eq_fRG+MF: vertex at Lambda crit} from a given vertex at the critical scale.
    \subsection{Hubbard-Stratonovich transformation and MF-truncation}
    Since the effective action at a given scale $\Lambda$ can be viewed as a bare action with bare propagator $G_0-G_0^\Lambda$ (with $G_0^\Lambda$ the regularized bare propagator)\footnote{One can prove it by considering the effective interaction functional $\mathcal{V}$, as shown in Ref.~\cite{Metzner2012}.}, one can decouple the factorized (and singular) part of the vertex at $\Lamc$ via a Gaussian integration, thus introducing a bosonic field. By adding source terms which couple linearly to this field and to the fermionic ones, one obtains the generating functional of connected Green's functions, whose Legendre transform at the critical scale reads as
    \begin{equation}
        \begin{split}
            \Gamma^{\Lamc}[\psi,\overline{\psi},\phi]=
            &-\int_{k,\sigma} \overline{\psi}_{k,\sigma} \left[G^{\Lamc}(k)\right]^{-1} \psi_{k,\sigma}
            -\int_{q} \phi^*_q \, m^{\Lamc}(q)\, \phi_q\\
            &+\int_{k,k',q}\mathcal{Q}^{\Lamc}(k,k';q)\,\overline{\psi}_{k,\uparrow}  \overline{\psi}_{q-k,\downarrow} \psi_{q-k',\downarrow} \psi_{k',\uparrow}\\
            &+\int_{k,q}h^{\Lamc}(k;q)\left[ \overline{\psi}_{k,\uparrow} \overline{\psi}_{q-k,\downarrow} \phi_q + \text{h.c.}\right],
        \end{split}
        \label{eq_fRG+MF: gamma lambda crit bos}
    \end{equation}
    where $\phi$ represents the expectation value (in presence of sources) of the Hubbard-Stratonovich field.
    Note that we have avoided to introduce an interaction between equal spin fermions. Indeed, since we are focusing on a spin singlet superconducting order parameter, within the MF approximation this interaction does not contribute to the flow equations. 
    
    The Hubbard-Stratonovich transformation introduced in Eq.~\eqref{eq_fRG+MF: gamma lambda crit bos} is free of the so-called Fierz ambiguity, according to which different ways of decoupling of the bare interaction can lead to different mean-field results for the gap (see, for example, Ref.~\cite{Baier2004}).  Indeed, through the inclusion of the residual two fermion interaction, we are able to recover the same equations that one would get without bosonizing the interactions, as proven in Sec.~\ref{subsec_fRG+MF: equivalence bos and fer}. In essence, the only ambiguity lies in selecting what to assign to the bosonized part of the vertex and what to $\mathcal{Q}$, but by keeping both of them all along the flow, the results will not depend on this choice.
    
    We introduce Nambu spinors as in Eq.~\eqref{eq_fRG+MF: Nambu spinors} and we decompose the bosonic field into its (flowing) expectation value plus longitudinal ($\sigma$) and transverse ($\pi$) fluctuations~\cite{Strack2008,Obert2013}:
    \begin{subequations}
        \begin{align}
            &\phi_q=\alpha^\Lambda\,\delta_{q,0} + \sigma_q + i\, \pi_q, \\
            &\phi^*_q=\alpha^\Lambda\,\delta_{q,0} + \sigma_{-q} - i\, \pi_{-q},
        \end{align}
    \end{subequations}
    where we have chosen $\alpha^\Lambda$ to be real. For the effective action at $\Lambda<\Lamc$ in the SSB phase, we use the following \textit{ansatz}
    \begin{equation}
        \begin{split}
            \Gamma^{\Lambda}_\text{SSB}[\Psi,\overline{\Psi},\sigma,\pi]=\Gamma^\Lambda_{\Psi^2}+\Gamma^\Lambda_{\sigma^2}+\Gamma^\Lambda_{\pi^2}
            +\Gamma^\Lambda_{\Psi^2\sigma} + \Gamma^\Lambda_{\Psi^2\pi}
            +\Gamma^\Lambda_{\Psi^4},
        \end{split}
        \label{eq_fRG+MF: bosonic eff action}
    \end{equation}
    where the first three quadratic terms are given by
    \begin{subequations}
        \begin{align}
            &\Gamma^\Lambda_{\Psi^2}=-\int_{k} \overline{\Psi}_{k} \left[\mathbf{G}^{\Lambda}(k)\right]^{-1} \Psi_{k},\\
            &\Gamma^\Lambda_{\sigma^2}=-\frac{1}{2}\int_q \sigma_{-q}\,m_\sigma^{\Lambda}(q)\, \sigma_q,\\
            &\Gamma^\Lambda_{\pi^2}=-\frac{1}{2}\int_q \pi_{-q}\,m_\pi^{\Lambda}(q)\, \pi_q,    
        \end{align}
    \end{subequations}
    and the fermion-boson interactions are 
    \begin{subequations}
        \begin{align}
            &\Gamma^\Lambda_{\Psi^2\sigma}=\int_{k,q}h^{\Lambda}_\sigma(k;q)\left\{S^1_{k,-q}\,\sigma_q+ \text{h.c.} \right\},\\
            &\Gamma^\Lambda_{\Psi^2\pi}=\int_{k,q}h^{\Lambda}_\pi(k;q)\left\{S^2_{k,-q}\,\pi_q+ \text{h.c.} \right\},
        \end{align}
    \end{subequations}
    with $S^\alpha_{k,q}$ as in Eq.~\eqref{eq_fRG+MF: fermion bilinear}.
    The residual two fermion interaction term is written as
    \begin{equation}
        \begin{split}
            \Gamma^\Lambda_{\Psi^{4}}=
            \int_{k,k',q}\mathcal{A}^{\Lambda}(k,k';q)\,S^1_{k,q}\,S^1_{k',-q}
            +\int_{k,k',q}\hskip - 5mm\Phi^{\Lambda}(k,k';q) \,S^2_{k,q}\,S^2_{k',-q}.
        \end{split}    
    \end{equation}
    Notice that in the above equation the terms $\mathcal{A}^\L$ and $\Phi^\L$ have a different physical meaning than those in Eq.~\eqref{eq_fRG+MF: A e Phi m e p}. While the former represent only a residual interaction term, the latter embody \emph{all} the interaction processes in the longitudinal and transverse channels. 
    
    As in the fermionic formalism, in the truncation in Eq.~\eqref{eq_fRG+MF: bosonic eff action} we have neglected any type of longitudinal-transverse fluctuation mixing in the Yukawa couplings, bosonic propagators and two fermion interactions because at $q=0$ they are identically zero. In the bosonic formulation, as well as for the fermionic one, the MF approximation selects only the $q=0$ components of the various terms appearing in the effective action and neglects all the rest. So, from now on we keep only the $q=0$ terms. We will make use of the matrix notation introduced in Sec.~\ref{sec_fRG_MF: fRG+MF equations}, where the newly introduced Yukawa couplings behave as vectors and bosonic inverse propagators as scalars. 
    \subsection{Flow equations and integration}
    \label{sec_fRG+MF: bosonic flow and integration}
    Here we focus on the flow equations for two fermion interactions, Yukawa couplings and bosonic propagators in the longitudinal and transverse channels within a MF approximation, that is, we focus only on the Cooper channel ($q=0$) and neglect all the diagrams containing internal bosonic lines or couplings $\mathcal{A}^\L$, $\Phi^\L$ at $q\neq 0$. Furthermore, we introduce a generalized Katanin approximation to account for higher order couplings in the flow equations. This approximation allows to replace the single-scale derivatives in the bubbles with full scale derivatives. We refer to Appendix~\ref{app: fRG+MF app} for more details and a derivation of the latter. We now show that our reduced set of flow equations for the various couplings can be integrated. We first focus on the longitudinal channel, while in the transverse one the flow equations possess the same structure. 
    
\begin{figure}[t]
        \centering
        \includegraphics[width=0.35\textwidth]{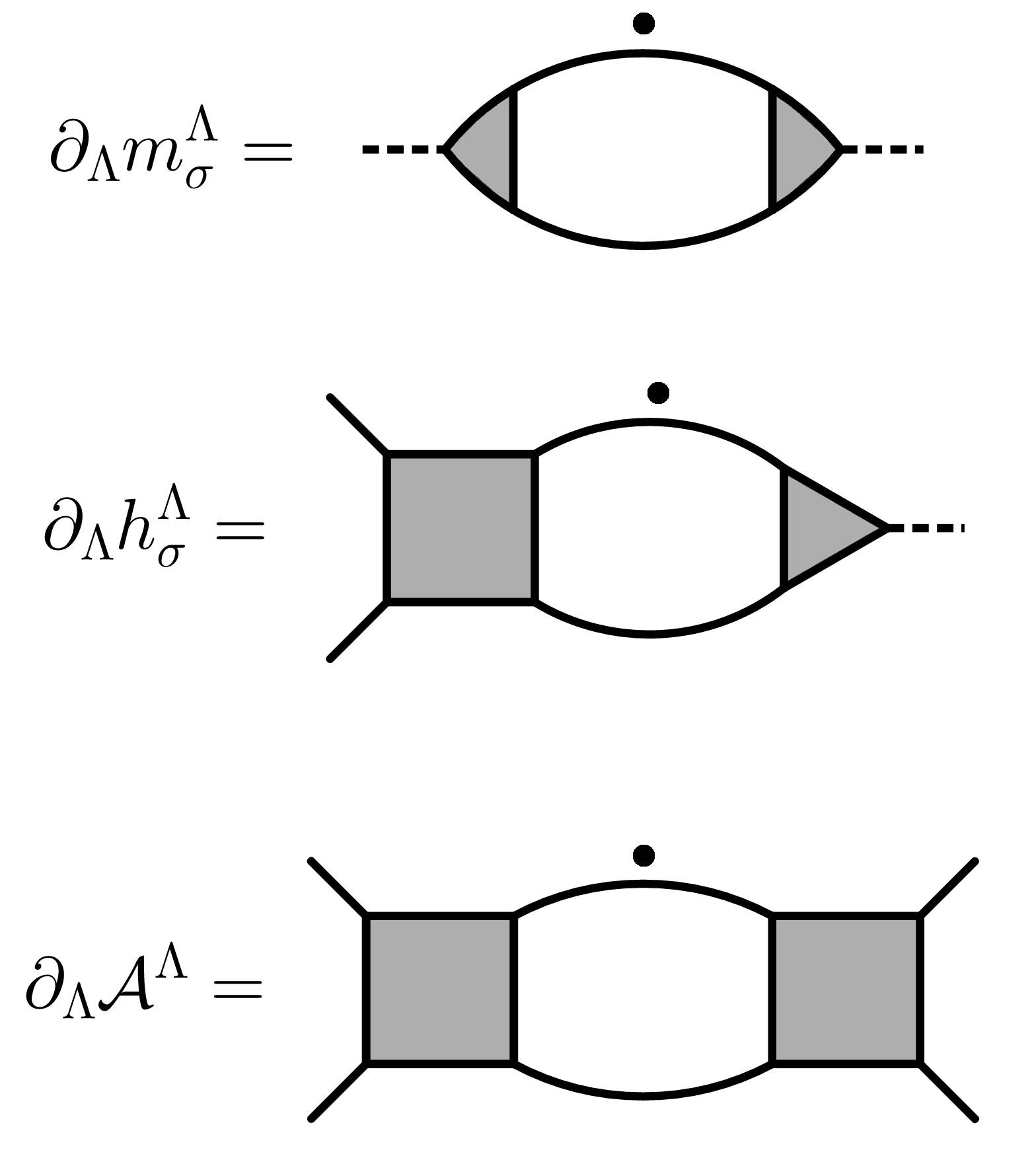}
        \caption{Schematic representation of flow equations for the mass and the couplings in the longitudinal channel. Full lines represent Nambu matrix propagators, triangles the Yukawa coupling $h_\sigma$ and squares the residual interaction $\mathcal{A}$. The black dots over fermionic legs represent full derivatives with respect to the scale $\Lambda$.}
        \label{fig_fRG+MF: flow eqs}
    \end{figure}

    The flow equation for the longitudinal bosonic mass (inverse propagator at $q=0$) reads as
    \begin{equation}
        \begin{split}
            \partial_\Lambda m_\sigma^\Lambda=\int_{k,k'} h^\Lambda_\sigma(k) \left[\partial_\Lambda\Pi^\Lambda_{11}(k,k')\right] h^\Lambda_\sigma(k')
            \equiv \left[h^\Lambda_\sigma\right]^T\left[\partial_\Lambda\Pi^\Lambda_{11}\right] h^\Lambda_\sigma.
        \end{split}
        \label{eq_fRG+MF: flow P sigma}
    \end{equation}
    Similarly, the equation for the longitudinal Yukawa coupling is
    \begin{equation}
        \partial_\Lambda h^\Lambda_\sigma=\mathcal{A}^\Lambda\left[\partial_\Lambda\Pi^\Lambda_{11}\right]h^\Lambda_\sigma,
        \label{eq_fRG+MF: flow h sigma}
    \end{equation}
    and the one for the residual two fermion longitudinal interaction is given by
    \begin{equation}
        \partial_\Lambda\mathcal{A}^\Lambda=\mathcal{A}^\Lambda\left[\partial_\Lambda\Pi^\Lambda_{11}\right]\mathcal{A}^\Lambda.
        \label{eq_fRG+MF: A flow eq}
    \end{equation}
    The above flow equations are pictorially shown in Fig.~\ref{fig_fRG+MF: flow eqs}. The initial conditions at $\Lambda=\Lamc$ read, for both channels,
    \begin{subequations}
        \begin{align}
            &m_\sigma^{\Lamc}=m_\pi^{\Lamc}=m^{\Lamc},\\
            &h_\sigma^{\Lamc}=h_\pi^{\Lamc}=h^{\Lamc},\\
            &\mathcal{A}^{\Lamc}=\Phi^{\Lamc}=\mathcal{Q}^{\Lamc}.
        \end{align}
    \end{subequations}
    We start by integrating the equation for the residual two fermion longitudinal interaction $\mathcal{A}^\L$. Eq.~\eqref{eq_fRG+MF: A flow eq} can be solved exactly as we have done in the fermionic formalism, obtaining for $\mathcal{A}^\L$
    \begin{equation}
        \mathcal{A}^\Lambda = \left[1-\widetilde{\mathcal{Q}}^{\Lamc}\Pi_{11}^\Lambda\right]^{-1}\widetilde{\mathcal{Q}}^{\Lamc},
        \label{eq_fRG+MF: A}
    \end{equation}
    where we have introduced a reduced residual two fermion interaction $\widetilde{\mathcal{Q}}$
    \begin{equation}
        \widetilde{\mathcal{Q}}^{\Lamc}=\left[1+\mathcal{Q}^{\Lamc}\Pi^{\Lamc}\right]^{-1}\mathcal{Q}^{\Lamc}.
        \label{eq_fRG+MF: reduced C tilde}
    \end{equation}
    We are now in the position to employ this result and plug it in Eq.~\eqref{eq_fRG+MF: flow h sigma} for the Yukawa coupling. The latter can be integrated as well. Its solution reads as
    \begin{equation}
        h_\sigma^\Lambda= \left[1-\widetilde{\mathcal{Q}}^{\Lamc}\Pi_{11}^\Lambda\right]^{-1}\widetilde{h}^{\Lamc},
        \label{eq_fRG+MF: h_sigma}
    \end{equation}
    where the introduction of a "reduced" Yukawa coupling
    \begin{equation}
        \widetilde{h}^{\Lamc}=\left[1+\mathcal{Q}^{\Lamc}\Pi^{\Lamc}\right]^{-1}h^{\Lamc}
        \label{eq_fRG+MF: reduced yukawa}
    \end{equation}
    is necessary. This Bethe-Salpeter-like equation for the Yukawa coupling is similar in structure to the parquetlike equations for the three-leg vertex derived in Ref.~\cite{Krien2019_II}.
    Finally, we can use the two results of Eqs.~\eqref{eq_fRG+MF: A} and~\eqref{eq_fRG+MF: h_sigma} and plug them in the equation for the bosonic mass, whose integration provides
    \begin{equation}
        m_\sigma^\Lambda=\widetilde{m}^{\Lamc}-\left[\widetilde{h}^{\Lamc}\right]^T\Pi_{11}^\Lambda\,h_\sigma^\Lambda,
        \label{eq_fRG+MF: P_sigma}
    \end{equation}
    where, by following definitions introduced above, the "reduced" bosonic mass is given by
    \begin{equation}
        \widetilde{m}^{\Lamc}=m^{\Lamc}+\left[\widetilde{h}^{\Lamc}\right]^T\Pi^{\Lamc}\,h^{\Lamc}.
        \label{eq_fRG+MF: reduced mass P tilde}
    \end{equation}

    In the transverse channel, the equations have the same structure and can be integrated in the same way. Their solutions read as
    \begin{subequations}
        \begin{align}
            &\Phi^\Lambda = \left[1-\widetilde{\mathcal{Q}}^{\Lamc}\Pi_{22}^\Lambda\right]^{-1}\widetilde{\mathcal{Q}}^{\Lamc},
            \label{eq_fRG+MF: Phi}\\
            &h_\pi^\Lambda= \left[1-\widetilde{\mathcal{Q}}^{\Lamc}\Pi_{22}^\Lambda\right]^{-1}\widetilde{h}^{\Lamc},
            \label{eq_fRG+MF: h_pi}\\
            &m_\pi^\Lambda=\widetilde{m}^{\Lamc}-\left[\widetilde{h}^{\Lamc}\right]^T\Pi_{22}^\Lambda\,h_\pi^\Lambda.
            \label{eq_fRG+MF: goldstone mass}
        \end{align}
    \end{subequations}
    Eq.~\eqref{eq_fRG+MF: goldstone mass} provides the mass of the transverse mode, which, according to the Goldstone theorem, must be zero. We will show later that this is indeed fulfilled. 
    
    The combinations
    \begin{subequations}
        \begin{align}
            &\frac{h_\sigma^\Lambda \left[h_\sigma^\Lambda\right]^T}{m_\sigma^\Lambda}+\mathcal{A}^\Lambda,\\
            &\frac{h_\pi^\Lambda \left[h_\pi^\Lambda\right]^T}{m_\pi^\Lambda}+\Phi^\Lambda,
        \end{align}
        \label{eq_fRG+MF: eff fer interactions}
    \end{subequations}
    obey the same flow equations, Eqs.~\eqref{eq_fRG+MF: flow eq Va fermionic} and~\eqref{eq_fRG+MF: flow eq Vphi fermionic}, as the vertices in the fermionic formalism and share the same initial conditions. Therefore the solutions for these quantities coincide with expressions~\eqref{eq_fRG+MF: Va solution fermionic} and~\eqref{eq_fRG+MF: Vphi solution fermionic}, respectively. Within this equivalence, it is interesting to express the irreducible vertex $\widetilde{V}^{\Lamc}$ of Eq.~\eqref{eq_fRG+MF: irr vertex fermionic} in terms of the quantities, $\mathcal{Q}^{\Lamc}$, $h^{\Lamc}$ and $m^{\Lamc}$, introduced in the factorization in Eq.~\eqref{eq_fRG+MF: vertex at Lambda crit}:
    \begin{equation}
        \widetilde{V}^{\Lamc}=\frac{\widetilde{h}^{\Lamc}\left[\widetilde{h}^{\Lamc}\right]^T}{\widetilde{m}^{\Lamc}}+\widetilde{\mathcal{Q}}^{\Lamc},
        \label{eq_fRG+MF: irr V bosonic formalism}
    \end{equation}
    where $\widetilde{\mathcal{Q}}^{\Lamc}$, $\widetilde{h}^{\Lamc}$ and $\widetilde{m}^{\Lamc}$ were defined in Eqs.~\eqref{eq_fRG+MF: reduced C tilde},~\eqref{eq_fRG+MF: reduced yukawa} and~\eqref{eq_fRG+MF: reduced mass P tilde}. For a proof see Appendix~\ref{app: fRG+MF app}. Relation~\eqref{eq_fRG+MF: irr V bosonic formalism} is of particular interest because it states that when the full vertex is expressed as in Eq.~\eqref{eq_fRG+MF: vertex at Lambda crit}, then the irreducible one will obey a similar decomposition, where the bosonic propagator, Yukawa coupling and residual two fermion interaction are replaced by their "reduced" counterparts. This relation holds even for $q\neq 0$.
    \subsection{Ward identity for the gap and Goldstone theorem}
\begin{figure}[t]
        \centering
        \includegraphics[width=0.35\textwidth]{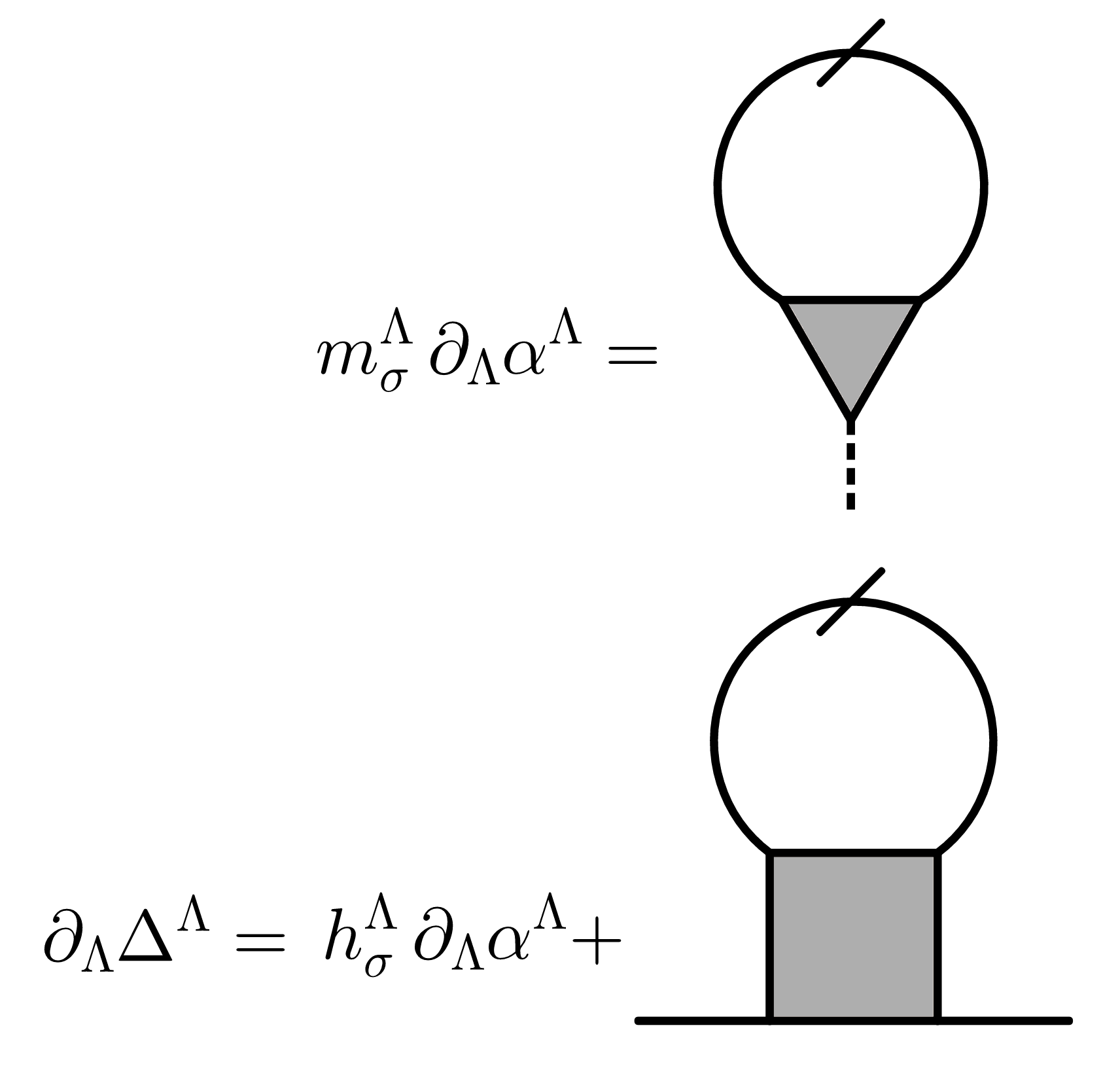}
        \caption{Schematic representation of flow equations for the bosonic expectation value $\alpha^\Lambda$ and fermionic gap $\Delta^\Lambda$. Aside from the slashed lines, representing Nambu matrix propagators with a scale derivative acting only on the regulator, the conventions for the symbols are the same as in Fig.~\ref{fig_fRG+MF: flow eqs}.}
        \label{fig_fRG+MF: flow eqs gaps}
    \end{figure}

    We now focus on the flow of the fermionic gap and the bosonic expectation value and express a relation that connects them. Their flow equations are given by (see Appendix~\ref{app: fRG+MF app}) 
    \begin{equation}
        \partial_\Lambda \alpha^\Lambda=\frac{1}{m_\sigma^\Lambda}\left[h_\sigma^\Lambda\right]^T\widetilde{\partial}_\Lambda F^\Lambda,
        \label{eq_fRG+MF: dalpha dLambda main text}
    \end{equation}
    and
    \begin{equation}
        \begin{split}
            \partial_\Lambda \Delta^\Lambda = \partial_\Lambda \alpha^\Lambda\, h_\sigma^\Lambda+\mathcal{A}^\Lambda\widetilde{\partial}_\Lambda F^\Lambda
            = \left[\frac{h_\sigma^\Lambda \left[h_\sigma^\Lambda\right]^T}{m_\sigma^\Lambda}+\mathcal{A}^\Lambda\right]\widetilde{\partial}_\Lambda F^\Lambda,
            \label{eq_fRG+MF: gap eq main text}
        \end{split}
    \end{equation}
    with $F^\Lambda$ given by Eq.~\eqref{eq_fRG+MF: F definition}. In Fig.~\ref{fig_fRG+MF: flow eqs gaps} we show a pictorial representation.
    Eq.~\eqref{eq_fRG+MF: dalpha dLambda main text} can be integrated, with the help of the previously obtained results for $\mathcal{A}$, $h_\sigma$ and $m_\sigma$, yielding
     \begin{equation}
        \alpha^\Lambda=\frac{1}{\widetilde{m}^{\Lamc}}\left[\widetilde{h}^{\Lamc}\right]^T F^\Lambda.
        \label{eq_fRG+MF: alpha solution}
    \end{equation}
    In the last line of Eq.~\eqref{eq_fRG+MF: gap eq main text}, as previously discussed, the object in square brackets equals the full vertex $V_\mathcal{A}$ of the fermionic formalism. Thus, integration of the gap equation is possible and the result is simply Eq.~\eqref{eq_fRG+MF: gap equation fermionic} of the fermionic formalism. However, if we now insert the expression in Eq.~\eqref{eq_fRG+MF: irr V bosonic formalism} for the irreducible vertex within the "fermionic" form (Eq.~\eqref{eq_fRG+MF: gap equation fermionic}) of the gap equation, and use relation~\eqref{eq_fRG+MF: Pi22=F/delta}, we get:
    \begin{equation}
        \Delta^\Lambda(k)=\alpha^\Lambda h_\pi^\Lambda(k).
        \label{eq_fRG+MF: Ward Identity}
    \end{equation}
    This equation is the Ward identity for the mixed boson-fermion system related to the global U(1) symmetry~\cite{Obert2013}. In Appendix~\ref{app: fRG+MF app} we propose a self consistent loop for the calculation of $\alpha$, $h_{\pi}$, through Eqs.~\ref{eq_fRG+MF: alpha solution} and~\ref{eq_fRG+MF: h_pi}, and subsequently the superfluid gap $\Delta$.
    Let us now come back to the question of the Goldstone theorem. For the mass of the Goldstone boson to be zero, it is necessary for Eq.~\eqref{eq_fRG+MF: goldstone mass} to vanish. We show that this is indeed the case. With the help of Eq.~\eqref{eq_fRG+MF: Pi22=F/delta}, we can reformulate the equation for the transverse mass in the form 
    \begin{equation}
        \begin{split}
            m^\Lambda_\pi = \widetilde{m}^{\Lamc}-\int_k \widetilde{h}^{\Lamc}(k)F^\Lambda(k)\frac{h^\Lambda_\pi(k)}{\Delta^\Lambda(k)}
            =\widetilde{m}^{\Lamc}-\frac{1}{\alpha^{\Lambda}}\int_k \widetilde{h}^{\Lamc}(k)F^\Lambda(k),
        \end{split}
    \end{equation}
    where the Ward Identity $\Delta=\alpha h_\pi$ was applied in the last line. We see that the expression for the Goldstone boson mass vanishes when $\alpha$ obeys its self consistent equation, Eq.~\eqref{eq_fRG+MF: alpha solution}. This proves that our truncation of flow equations fulfills the Goldstone theorem. \\

    Constructing a truncation of the fRG flow equations which fulfills the Ward identities and the Goldstone theorem is, in general, a nontrivial task. In Ref.~\cite{Bartosch2009}, in which the order parameter fluctuations have been included on top of the Hartree-Fock solution, no distinction has been made between the longitudinal and transverse Yukawa couplings and the Ward identity~\eqref{eq_fRG+MF: Ward Identity} as well as the Goldstone theorem have been enforced, by calculating the gap and the bosonic expectation values from them rather than from their flow equations. Similarly, in Ref.~\cite{Obert2013}, in order for the flow equations to fulfill the Goldstone theorem, it was necessary to impose $h_\sigma=h_\pi$ and use only the flow equation of $h_\pi$ for both Yukawa couplings. Within the present approximation, due to the mean-field-like nature of the truncation, the Ward identity~\eqref{eq_fRG+MF: Ward Identity} and the Goldstone theorem are automatically fulfilled by the flow equations.
    \subsection{Equivalence of bosonic and fermionic formalisms}
    \label{subsec_fRG+MF: equivalence bos and fer}
    As we have proven in the previous sections, within the MF approximation the fully fermionic formalism of Sec.~\ref{sec_fRG_MF: fRG+MF equations} and the bosonized approach introduced in the present section provide the same results for the superfluid gap and for the effective two fermion interactions. 
    
    Notwithstanding the formal equivalence, the bosonic formulation relies on a further requirement. In Eqs.~\eqref{eq_fRG+MF: Phi} and~\eqref{eq_fRG+MF: h_pi} we assumed the matrix $\left[1-\widetilde{\mathcal{Q}}^{\Lamc}\Pi_{22}^\Lambda\right]$ to be invertible. This statement is exactly equivalent to assert that the two fermion residual interaction $\Phi$ remains finite. Otherwise the Goldstone mode would lie in this coupling and not (only) in the Hubbard-Stratonovich boson. This cannot happen if the flow is stopped at a scale $\Lamc$ coinciding with the critical scale $\Lambda_c$ at which the (normal) bosonic mass $m^\Lambda$ turns zero, but it could take place if one considers symmetry breaking in more than one channel, as we have done in Sec.~\ref{sec_fRG_MF: Phase diag}. In particular, if one allows the system to develop two different orders and stops the flow when the mass of one of the two associated bosons becomes zero, it could happen that, within a MF approximation for both order types, the appearance of a finite gap in the first channel makes the two fermion transverse residual interaction in the other channel diverging. In that case one can apply the technique of the \textit{flowing bosonization}~\cite{Friederich2010,Friederich2011}, by reassigning to the bosonic sector the (most singular part of the) two fermion interactions that are generated during the flow. It can be proven that also this approach gives the same results for the gap and the effective fermionic interactions in Eq.~\eqref{eq_fRG+MF: eff fer interactions} as the fully fermionic formalism. 
    \subsection{Vertex bosonization}
    \label{sec_fRG+MF: vertex bosonization}
    In this section we present a systematic procedure to extract the quantities in Eq.~\eqref{eq_fRG+MF: vertex at Lambda crit} from a given vertex, within an approximate framework. 
    
    Starting from the channel decomposition in Eq.~\eqref{eq_methods: channel decomp physical}, we simplify the treatment of the dependence on fermionic spatial momenta of the various channels expanding them in a complete basis of Brillouin zone form factors $\{f^\ell_\mathbf{k}\}$~\cite{Lichtenstein2017}
    \begin{equation}
        \begin{split}
            \phi^\Lambda_X(k,k';q)=\sum_{\ell\ell'} \phi^{\Lambda}_{X,\ell\ell'}(\nu,\nu';q)f^\ell_{\mathbf{k}}\,f^{\ell'}_{\mathbf{k'}},
        \end{split}
        \label{eq_fRG+MF: form factor expansion}
    \end{equation}
    with $X=p$, $m$ or $c$, corresponding to pairing, magnetic, and charge channels. For practical calculations the above sum is truncated to a finite number of form factors and often only diagonal terms, $\ell=\ell'$, are considered. Within the form factor truncated expansion, one is left with the calculation of a finite number of channels that depend on a bosonic collective variable $q=(\mathbf{q},\Omega)$ and two fermionic Matsubara frequencies $\nu$ and $\nu'$.
    
    We will now show how to obtain the decomposition introduced in Eq.~\eqref{eq_fRG+MF: vertex at Lambda crit} within the form factor expansion.
    We focus on only one of the three channels, depending on the type of order we are interested in, and factorize its dependence on the two fermionic Matsubara frequencies. We introduce the so called channel asymptotics, that is, the functions that describe the channels for large $\nu$, $\nu'$. From now on, we adopt the shorthand $\lim_{\nu\rightarrow\infty}g(\nu)=g(\infty)$ for whatever $g$, function of $\nu$. By considering only diagonal terms in the form factor expansion in Eq.~\eqref{eq_fRG+MF: form factor expansion}, we can write the $\ell=\ell'$ components of the channels as~\cite{Wentzell2016}:
    \begin{equation}
        \begin{split}
            \phi_{X,\ell}^\Lambda(\nu,\nu';q)=\mathcal{K}_{X,\ell}^{(1)\Lambda}(q)+\mathcal{K}_{X,\ell}^{(2)\Lambda}(\nu;q)
            +\overline{\mathcal{K}}_{X,\ell}^{(2)\Lambda}(\nu';q)
            +\delta\phi^\Lambda_{X,\ell}(\nu,\nu';q),    
        \end{split}
        \label{eq_fRG+MF: vertex asymptotics}
    \end{equation}
    with
    \begin{subequations}
        \begin{align}
            &\mathcal{K}_{X,\ell}^{(1)\Lambda}(q)=\phi_{X,\ell}^\Lambda(\infty,\infty;q)\\
            &\mathcal{K}_{X,\ell}^{(2)\Lambda}(\nu;q)=\phi_{X,\ell}^\Lambda(\nu,\infty;q)-\mathcal{K}_{X,\ell}^{(1)\Lambda}(q)\\
            &\overline{\mathcal{K}}_{X,\ell}^{(2)\Lambda}(\nu';q)=\phi_{X,\ell}^\Lambda(\infty,\nu';q)-\mathcal{K}_{X,\ell}^{(1)\Lambda}(q)\\
            &\delta\phi^\Lambda_{X,\ell}(\nu,\infty;q)=\delta\phi^\Lambda_{X,\ell}(\infty,\nu';q)=0.
        \end{align}
        \label{eq_fRG+MF: asymptotics properties}
    \end{subequations}
    According to Ref.~\cite{Wentzell2016}, these functions are related to physical quantities. $\mathcal{K}_{X,\ell}^{(1)}$ turns out to be proportional to the relative susceptibility and the combination $\mathcal{K}_{X,\ell}^{(1)}+\mathcal{K}_{X,\ell}^{(2)}$ (or $\mathcal{K}_{X,\ell}^{(1)}+\overline{\mathcal{K}}_{X,\ell}^{(2)}$) to the boson-fermion vertex, that describes both the response of the Green's function to an external field~\cite{VanLoon2018} and the coupling between a fermion and an effective boson. In principle one should be able to calculate the above quantities directly from the vertex (see Ref.~\cite{Wentzell2016} for the details) without performing any limit. However, it is well known how fRG truncations, in particular the 1-loop approximation, do not properly weigh all the Feynman diagrams contributing to the vertex, so that the diagrammatic calculation and the high frequency limit give two different results. To keep the property in the last line of Eq.~\eqref{eq_fRG+MF: asymptotics properties}, we choose to perform the limits. We rewrite Eq.~\eqref{eq_fRG+MF: vertex asymptotics} in the following way:
    \begin{equation}
        \begin{split}
            \phi_{X,\ell}^\Lambda(\nu,\nu';q)=
            &\frac{\left[\mathcal{K}_{X,\ell}^{(1)\Lambda}+\mathcal{K}_{X,\ell}^{(2)\Lambda}\right]\left[\mathcal{K}_{X,\ell}^{(1)\Lambda}+\overline{\mathcal{K}}_{X,\ell}^{(2)\Lambda}\right]}{\mathcal{K}_{X,\ell}^{(1)\Lambda}}+\mathcal{R}_{X,\ell}^\Lambda\\
            =&\frac{\phi_{X,\ell}^\Lambda(\nu,\infty;q)\phi_{X,\ell}^\Lambda(\infty,\nu';q)}{\phi_{X,\ell}^\Lambda(\infty,\infty;q)}+\mathcal{R}_{X,\ell}^\Lambda(\nu,\nu';q),
        \end{split}
        \label{eq_fRG+MF: vertex separation}
    \end{equation}
    where we have made the frequency and momentum dependencies explicit only in the second line, and we have defined
    \begin{equation}
        \mathcal{R}_{X,\ell}^\Lambda(\nu,\nu';q)=\delta\phi^\Lambda_{X,\ell}(\nu,\nu';q)-\frac{\mathcal{K}_{X,\ell}^{(2)\Lambda}(\nu;q)\overline{\mathcal{K}}_{X,\ell}^{(2)\Lambda}(\nu';q)}{\mathcal{K}_{X,\ell}^{(1)\Lambda}(q)}.
    \end{equation}
    From the definitions given above, it is obvious that the rest function $\mathcal{R}_{X,\ell}$ decays to zero in all frequency directions.
    
    Since the first term of Eq.~\eqref{eq_fRG+MF: vertex separation} is separable by construction, we choose to identify this term with the first one of Eq.~\eqref{eq_fRG+MF: vertex at Lambda crit}. Indeed, in many cases the vertex divergence is manifest already in the asymptotic $\mathcal{K}_{X,\ell}^{(1)\Lambda}$, that we recall to be proportional to the susceptibility of the channel. There are however situations in which the functions $\mathcal{K}^{(1)}$ and $\mathcal{K}^{(2)}$ are zero even close to an instability in the channel, an important example being the $d$-wave superconducting instability in the repulsive Hubbard model. In general, this occurs for those channels that, within a Feynman diagram expansion, cannot be constructed with a ladder resummation with the bare vertex. In the Hubbard model, due to the locality of the bare interaction, this happens for every $\ell\neq 0$, that is, for every term in the form factor expansion different than the $s$-wave contribution. In this case one should adopt a different approach and, for example, replace the limits to infinity in Eq.~\eqref{eq_fRG+MF: vertex separation} by some given values of the Matsubara frequencies, $\pm \pi T$ for example. In Chap.~\ref{chap: Bos Fluct Norm}, we will present an alternative approach to the vertex factorization, by means of a diagrammatic decomposition called \emph{single boson exchange} (SBE) decomposition~\cite{Krien2019_I}.
    \subsection{Results for the attractive Hubbard model at half filling}
    \label{sec_fRG+MF: results}
    In this section we report some exemplary results of the equations derived within the bosonic formalism, for the two-dimensional attractive Hubbard model. We focus on the half-filled case. For pure nearest neighbor hopping with amplitude $-t$, the band dispersion $\xi_\mathbf{k}$ is given by
    \begin{equation}
        \xi_\mathbf{k} = - 2 t \left( \cos k_x + \cos k_y \right) -\mu,
        \label{eq_fRG+MF: dispersion band}
    \end{equation}
    with $\mu=0$ at half filling. We choose the onsite attraction and the temperature to be $U=-4t$ and $T=0.1t$, respectively. All results are presented in units of the hopping parameter $t$. 
    \subsubsection{Symmetric phase}
    In the symmetric phase, in order to run a fRG flow, we introduce the $\Omega$-regulator~\cite{Husemann2009}
    \begin{equation}
        R^\Lambda(k) = \left(i\nu-\xi_\mathbf{k}\right) \frac{\Lambda^2}{\nu^2},
    \end{equation}
    so that the initial scale is $\Lini=+\infty$ (fixed to a large number in the numerical calculation) and the final one $\Lfin=0$. We choose a 1-loop truncation, and use the physical channel decomposition in Eq.~\eqref{eq_methods: channel decomp physical}, with a form factor expansion. We truncate Eq.~\eqref{eq_fRG+MF: form factor expansion} only to the first term, that is, we use only s-wave, $f^{(0)}_\mathbf{k}\equiv 1$, form factors. Within these approximations, the vertex reads as
    \begin{equation}
        \begin{split}
            V^\Lambda(k_1,k_2,k_3) = &- U - \mathcal{P}^{\Lambda}_{\nu_1\nu_3}(k_1+k_2) \\
            &+ \mathcal{M}^{\Lambda}_{\nu_1\nu_2}(k_2-k_3)\\
            &+\frac{1}{2} \mathcal{M}^{\Lambda}_{\nu_1\nu_2}(k_3-k_1)
            -\frac{1}{2} \mathcal{C}^{\Lambda}_{\nu_1\nu_2}(k_3-k_1),
        \end{split}
        \label{eq_fRG+MF: channel decomposition attractive model}
    \end{equation}
    where $\mathcal{P}$, $\mathcal{M}$, $\mathcal{C}$, are referred as pairing, magnetic and charge channel, respectively.
    Furthermore, we focus only on the spin-singlet component of the pairing (the triplet one is very small in the present parameter region), so that we require the pairing channel to obey~\cite{Rohringer2012}
    \begin{equation}
        \mathcal{P}^{\Lambda}_{\nu\nu'}(q) = \mathcal{P}^{\Lambda}_{-\nu+\Omega\,\mathrm{m}\,2,\nu'}(q) = \mathcal{P}^{\Lambda}_{\nu,-\nu'+\Omega\,\mathrm{m}\,2}(q),
    \end{equation}
    where $q=(\mathbf{q},\Omega)$, and $\Omega\,\mathrm{m}\,2=2(n\,\mathrm{mod}\,2)\pi T$, and $n\in\mathbb{Z}$ is the Matusbara frequency index. The initial condition for the vertex reads as
    \begin{equation}
        V^{\Lini}(k_1,k_2,k_3) = - U,
    \end{equation}
    so that $\mathcal{P}^{\Lini}=\mathcal{M}^{\Lini}=\mathcal{C}^{\Lini}=0$.
    Neglecting the fermionic self-energy, $\Sigma^\Lambda(k)\equiv0$, we run a flow for these three quantities until one (ore more) of them diverges. Each channel is computed by keeping 50 positive and 50 negative values for each of the three Matsubara frequencies (two fermionic, one bosonic) on which it depends. Frequency asymptotics are also taken into account by following Ref.~\cite{Wentzell2016}. The momentum dependence of the channel is treated by discretizing with 38 patches the region $\mathcal{B}=\{(k_x,k_y): 0\leq k_y\leq k_x\leq\pi\}$ in the Brillouin zone and extending to the other regions by using lattice symmetries. 
    
    Due to particle-hole symmetry occurring at half filling, pairing fluctuations at $\mathbf{q}=0$ combine with charge fluctuations at $\mathbf{q}=(\pi,\pi)$ to form an order parameter with SO(3) symmetry~\cite{Micnas1990}. Indeed, with the help of a canonical particle-hole transformation, one can map the attractive half-filled Hubbard model onto the repulsive one. Within this duality, the SO(3)-symmetric magnetic order parameter is mapped onto the above mentioned combined charge-pairing order parameter and vice versa. This is the reason why we find a critical scale, $\Lambda_c$, at which \emph{both} $\mathcal{C}((\pi,\pi),0)$ and $\mathcal{P}(\mathbf{0},0)$ diverge.
    On a practical level, we define the critical scale $\Lamc$ as the scale at which one (or more, in this case) channel exceeds $10^3t$. With our choice of parameters, we find that at $\Lamc \simeq 0.378t$ both $\mathcal{C}$ and $\mathcal{P}$ cross our threshold.
    \ifx
    \begin{figure}[t]
        \centering
        \includegraphics[width=0.6\textwidth]{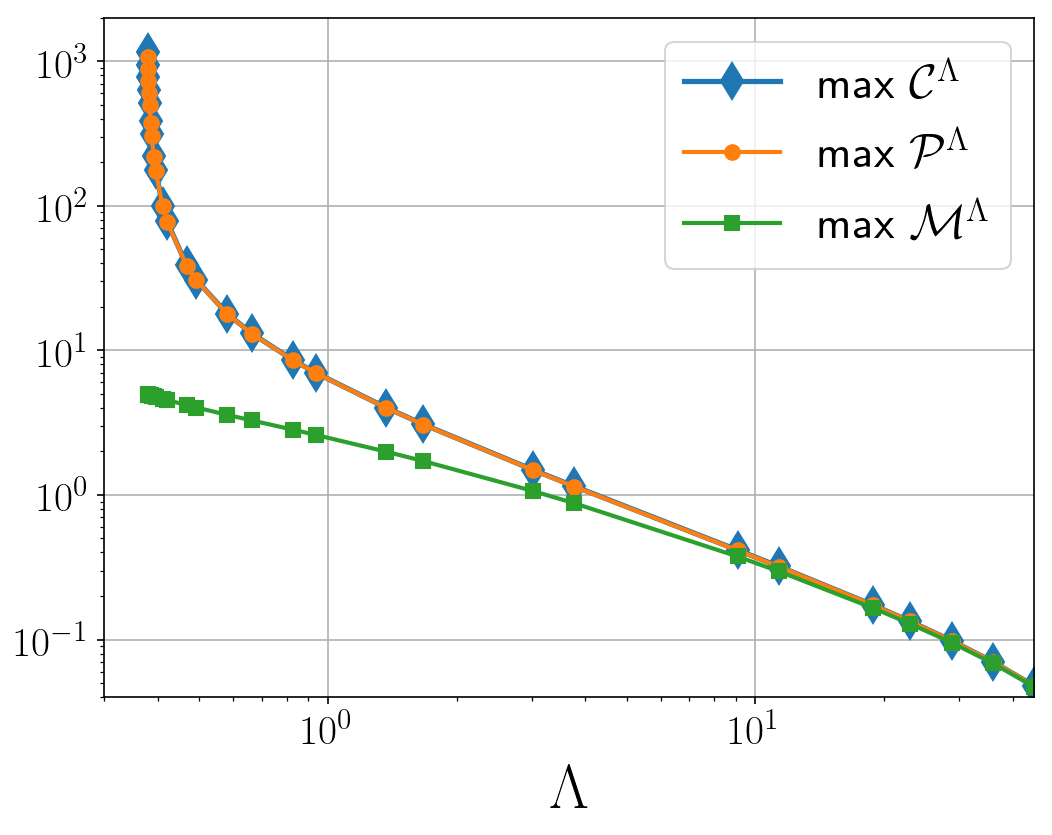}
        \caption{Flow of the maximum values of the pairing, magnetic and charge channel. The maximum value of the charge channel at zero frequency and momentum $(\pi,\pi)$ and the one for the pairing channel at $q=0$ coincide, within the numerical accuracy, and both exceed the threshold $10^3t$ at the critical scale, signaling an instability to the formation of an order parameter given by any linear combination of the superfluid and the charge density wave ones.}
        \label{fig_fRG+MF: flow channels}
    \end{figure}
    \fi
    In the SSB phase, we choose to restrict the ordering to the pairing channel, thus excluding the formation of charge density waves. This choice is always possible because we have the freedom to choose the "direction" in which our order parameter points. In the particle-hole dual repulsive model, our choice would be equivalent to choose the (antiferro-) magnetic order parameter to lie in the $xy$ plane. This choice is implemented by selecting the particle-particle channel as the only one contributing to the flow in the SSB phase, as exposed in Secs.~\ref{sec_fRG_MF: fRG+MF equations} and~\ref{sec_fRG+MF: bosonic flow and integration}.
    
    In order to access the SSB phase with our bosonic formalism, we need to perform the decomposition in Eq.~\eqref{eq_fRG+MF: vertex at Lambda crit} for our vertex at $\Lamc$. Before proceeding, in order to be consistent with our form factor expansion in the SSB phase, we need to project $V$ in Eq.~\eqref{eq_fRG+MF: channel decomposition attractive model} onto the $s$-wave form factors, because we want the quantities in the ordered phase to be functions of Matsubara frequencies only. Therefore we define the total vertex projected onto $s$-wave form factors 
    \begin{equation}
        \overline{V}^{\Lamc}_{\nu\nu'}(q)=\int_{\mathbf{k},\mathbf{k}'}V^{\Lamc}\hskip -1mm\left(\rnddo{q}+k,\rndup{q}-k,k'\right).
    \end{equation}
    Furthermore, since we are interested only in spin singlet pairing, we symmetrize it with respect to one of the two fermionic frequencies, so that in the end we are dealing with
    \begin{equation}
        V^{\Lamc}_{\nu\nu'}(q)=\frac{\overline{V}^{\Lamc}_{\nu\nu'}(q)+\overline{V}^{\Lamc}_{\nu,-\nu'+\Omega\,\mathrm{m}\,2}(q)}{2}.
    \end{equation}
    In order to extract the Yukawa coupling $h^{\Lamc}$ and bosonic propagator $m^{\Lamc}$, we employ the strategy described in Sec.~\ref{sec_fRG+MF: vertex bosonization}. Here, however, instead of factorizing the pairing channel $\mathcal{P}^{\Lamc}$ alone, we subtract from it the bare interaction $U$. In principle, $U$ can be assigned both to the pairing channel, to be factorized, or to the residual two fermion interaction, giving rise to the same results in the SSB phase. However, when in a real calculation the vertices are calculated on a finite frequency box, it is more convenient to have the residual two fermion interaction $\mathcal{Q}^{\Lamc}$ as small as possible, in order to reduce finite size effects in the matrix inversions needed to extract the reduced couplings in Eqs.~\eqref{eq_fRG+MF: reduced C tilde},~\eqref{eq_fRG+MF: reduced yukawa} and~\eqref{eq_fRG+MF: reduced mass P tilde}, and in the calculation of $h_\pi$, in Eq.~\eqref{eq_fRG+MF: h_pi}.  
    Furthermore, since it is always possible to rescale the bosonic propagators and Yukawa couplings by a constant such that the vertex constructed with them (Eq.~\eqref{eq_fRG+MF: vertex separation}) is invariant, we impose the normalization condition $h^{\Lamc}(\nu\rightarrow\infty;q)=1$.
    In formulas, we thus have
    \begin{equation}
        m^{\Lamc}(q)=\frac{1}{\mathcal{K}_{p,\ell=0}^{(1)\Lamc}(q)-U}=\frac{1}{\mathcal{P}^{\Lamc}_{\infty,\infty}(q)-U},
    \end{equation}
    and 
    \begin{equation}
        \begin{split}
            h^{\Lamc}(\nu;q)=\frac{\mathcal{K}_{p,\ell=0}^{(2)\Lamc}(\nu;q)+\mathcal{K}_{p,\ell=0}^{(1)\Lamc}(q)-U}{\mathcal{K}_{p,\ell=0}^{(1)\Lamc}(q)-U}
            =\frac{\mathcal{P}^{\Lamc}_{\nu,\infty}(q)-U}{\mathcal{P}^{\Lamc}_{\infty,\infty}(q)-U}.
        \end{split}
    \end{equation}
    The limits are numerically performed by evaluating the pairing channel at large values of the fermionic frequencies.
    The extraction of the factorizable part from the pairing channel minus the bare interaction defines the rest function
    \begin{equation}
        \mathcal{R}^{\Lamc}_{\nu\nu'}(q)=\mathcal{P}^{\Lamc}_{\nu\nu'}(q)-U-\frac{h^{\Lamc}(\nu;q)h^{\Lamc}(\nu';q)}{m^{\Lamc}(q)},
    \end{equation}
    and the residual two fermion interaction $\mathcal{Q}$
    \begin{equation}
        \begin{split}
            \mathcal{Q}^{\Lamc}_{\nu\nu'}(q)=&\left[V^{\Lamc}_{\nu\nu'}(q)-\mathcal{P}^{\Lamc}_{\nu\nu'}(q)+U\right]+\mathcal{R}^{\Lamc}_{\nu\nu'}(q)
            =V^{\Lamc}_{\nu\nu'}(q)-\frac{h^{\Lamc}(\nu;q)h^{\Lamc}(\nu';q)}{m^{\Lamc}(q)}.
        \end{split}
    \end{equation}
    We are now in the position to extract the reduced couplings, $\widetilde{\mathcal{Q}}^{\Lamc}$, $\widetilde{h}^{\Lamc}$ and $\widetilde{m}^{\Lamc}$, defined in Eqs.~\eqref{eq_fRG+MF: reduced C tilde},~\eqref{eq_fRG+MF: reduced yukawa},~\eqref{eq_fRG+MF: reduced mass P tilde}. This is achieved by numerically inverting the matrix (we drop the $q$-dependence from now on, assuming always $q=0$)
    \begin{equation}
        \delta_{\nu\nu'} + \mathcal{Q}^{\Lamc}_{\nu\nu'}\, \chi^{\Lamc}_{\nu'}, 
    \end{equation}
    with
    \begin{equation}
        \chi^{\Lamc}_{\nu} = T\int_{\mathbf{k}}G_0^{\Lamc}(k)G_0^{\Lamc}(-k), 
    \end{equation}
    and
    \begin{equation}
        G_0^{\Lamc}(k) = \frac{1}{i\nu-\xi_\mathbf{k}+R^{\Lamc}(k)} =\frac{\nu^2}{\nu^2+\Lamc^2}\frac{1}{i\nu-\xi_\mathbf{k}}.
    \end{equation}
    In Fig.~\ref{fig_fRG+MF: vertices Lambda s} we show the results for the pairing channel minus the bare interaction, the rest function, the residual two fermion interaction $\mathcal{Q}$ and the reduced one $\widetilde{\mathcal{Q}}$ at the critical scale. One can see that in the present parameter region the pairing channel (minus $U$) is highly factorizable. Indeed, despite the latter being very large because of the vicinity to the instability, the rest function $\mathcal{R}$ remains very small, a sign that the pairing channel is well described by the exchange of a single boson. Furthermore, thanks to our choice of assigning the bare interaction to the factorized part, as we see in Fig.~\ref{fig_fRG+MF: vertices Lambda s}, both $\mathcal{Q}$ and $\widetilde{\mathcal{Q}}$ possess frequency structures that arise from a background that is zero. 
    
    Finally, the full bosonic mass at the critical scale is close to zero, $m^{\Lamc}\simeq10^{-3} $, due to the vicinity to the instability, while the reduced one is finite, $\widetilde{m}^{\Lamc}\simeq 0.237$.
    \subsubsection{SSB Phase}
    \begin{figure}[t!]
        \centering
        \includegraphics[width=0.6\textwidth]{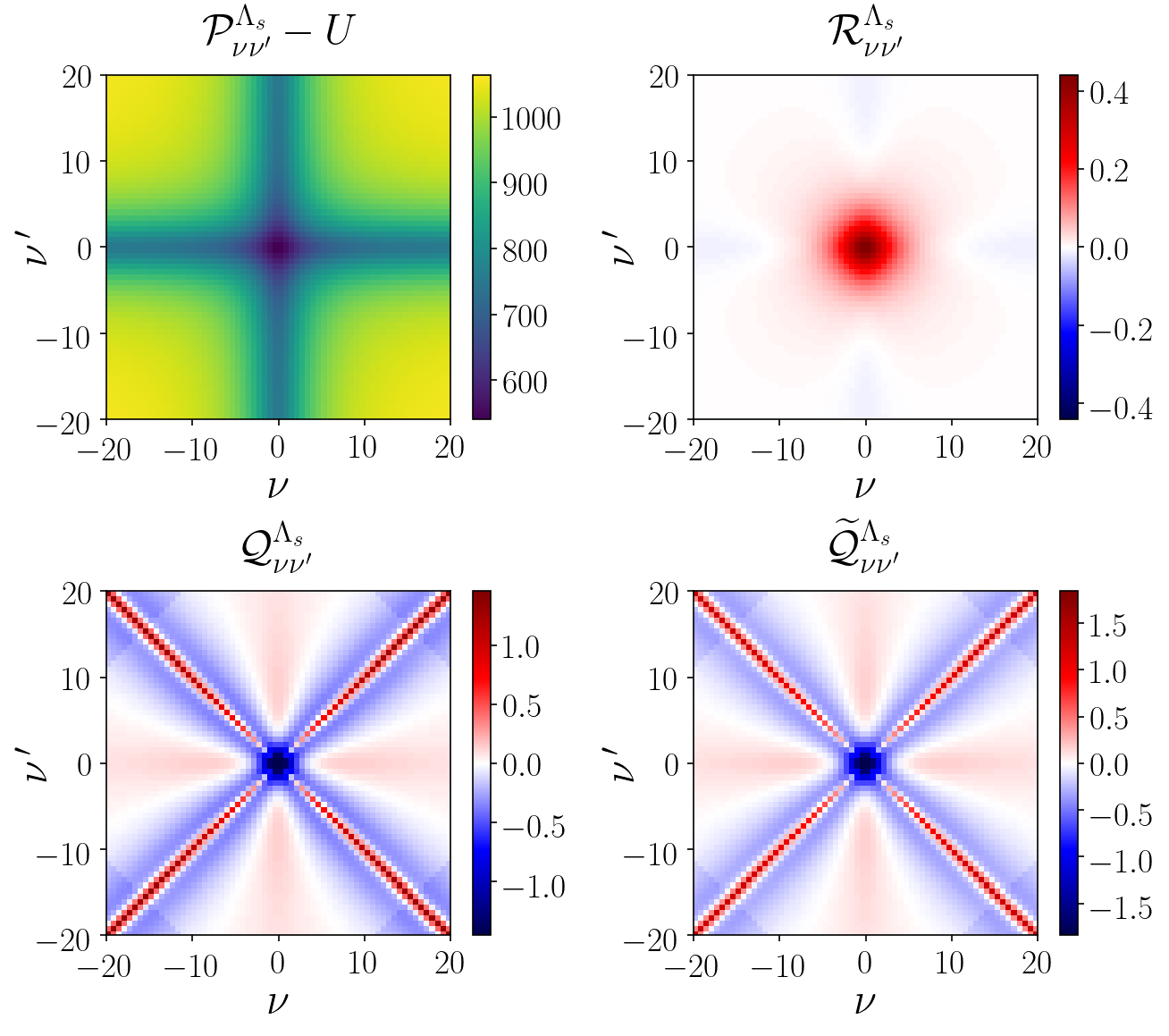}
        \caption{Couplings contributing to the total vertex at the critical scale. \\
        \textit{Upper left}: pairing channel minus the bare interaction. At the critical scale this quantity acquires very large values due to the vicinity to the pairing instability. \\
        \textit{Upper right}: rest function of the pairing channel minus the bare interaction. In the present regime the pairing channel is very well factorizable, giving rise to a small rest function.\\ 
        \textit{Lower left}: residual two fermion interaction. The choice of factorizing $\mathcal{P}^{\Lamc}-U$ instead of $\mathcal{P}^{\Lamc}$ alone makes the background of this quantity zero.\\
        \textit{Lower right}: reduced residual two fermion interaction. As well as the full one, this coupling has a zero background value, making calculations of couplings in the SSB phase more precise by reducing finite number of Matsubara frequencies effects in the matrix inversions.}
        
        \label{fig_fRG+MF: vertices Lambda s}
    \end{figure}
    In the SSB phase, instead of running the fRG flow, we employ the analytical integration of the flow equations described in Sec.~\ref{sec_fRG+MF: bosonic flow and integration}. On a practical level, we implement a solution of the loop described in Appendix~\ref{app: fRG+MF app}, that allows for the calculation of the bosonic expectation value $\alpha$, the transverse Yukawa coupling $h_\pi$ and subsequently the fermionic gap $\Delta$ through the Ward identity $\Delta=\alpha h_\pi$. In this section we drop the dependence on the scale, since we have reached the final scale $\Lfin=0$. Note that, as exposed previously, in the half-filled attractive Hubbard model the superfluid phase sets in by breaking a SO(3) rather than a U(1) symmetry. This means that one should expect the appearance of two massless Goldstone modes. Indeed, besides the Goldstone boson present in the (transverse) particle-particle channel, another one appears in the particle-hole channel and it is related to the divergence of the charge channel at momentum $(\pi,\pi)$. However, within our choice of considering only superfluid order and within the MF approximation, this mode is decoupled from our equations.

    \begin{figure}[t]
        \centering
        \includegraphics[width=0.65\textwidth]{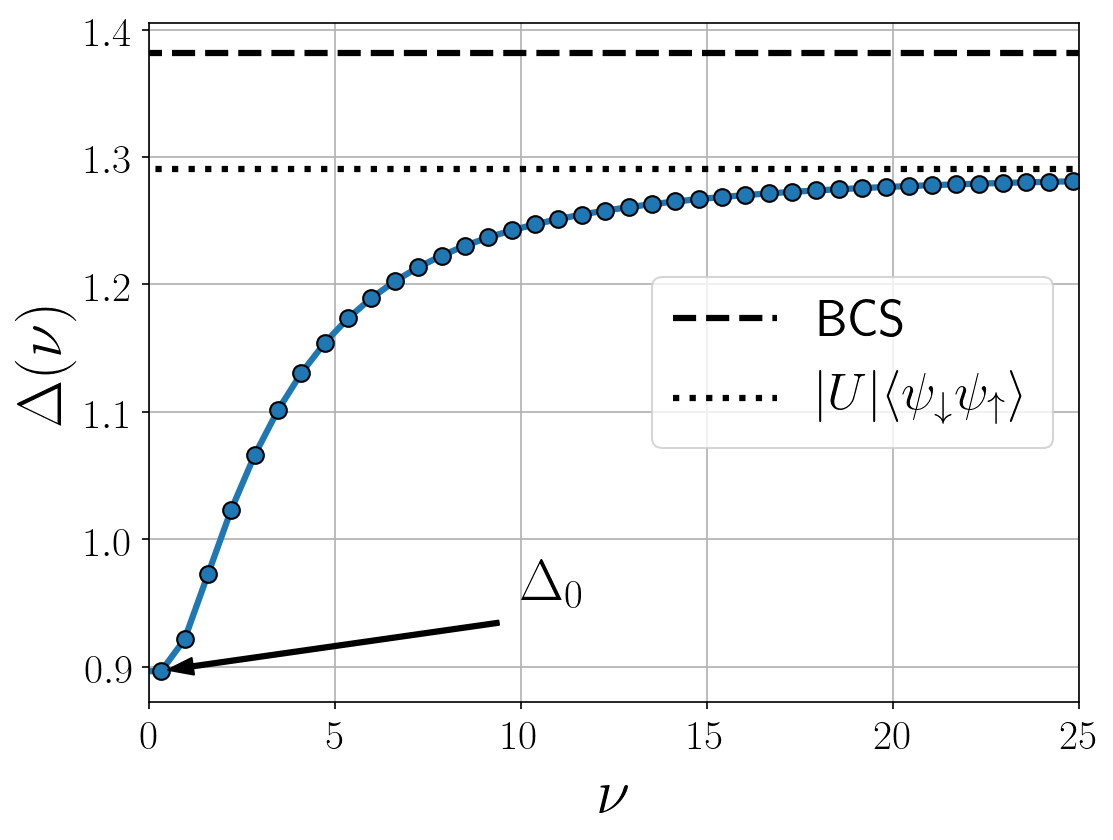}
        \caption{Frequency dependence of the superfluid gap. It interpolates between its value at the Fermi level, $\Delta_0$, and its asymptotic one. The dashed line marks the BCS value, while the dotted one  $|U|$ times the condensate fraction.}
        \label{fig_fRG+MF: gap}
    \end{figure}
    Within our previously discussed choice of bosonizing $\mathcal{P}^{\Lamc}-U$ instead of $\mathcal{P}^{\Lamc}$ alone, the self consistent loop introduced in Appendix~\ref{app: fRG+MF app} converges extremely fast, 15 iterations for example are sufficient to reach a precision of $10^{-7}$ in $\alpha$. 
    Once convergence is reached and the gap $\Delta(\nu)$ obtained, we are in the position to evaluate the remaining couplings introduced in Sec.~\ref{sec_fRG+MF: bosonic flow and integration} through their integrated flow equations. In Fig.~\ref{fig_fRG+MF: gap} we show the computed frequency dependence of the gap. It interpolates between $\Delta_0=\Delta(\nu\rightarrow 0)$, its value at the Fermi level, and its asymptotic value, that equals the absolute value of the bare interaction times the condensate fraction $\langle\psi_{\downarrow}\psi_{\uparrow}\rangle=\int_\mathbf{k}\langle \psi_{-\mathbf{k},\downarrow}\psi_{\mathbf{k},\uparrow}\rangle$. $\Delta_0$ also represents the gap between the upper and lower Bogoliubov band. Magnetic and charge fluctuations above the critical scale significantly renormalize the gap with respect to the Hartree-Fock calculation ($\widetilde{V}=-U$ in Eq.~\eqref{eq_fRG+MF: gap equation fermionic}), that in the present case coincides with Bardeen-Cooper-Schrieffer (BCS) theory. This effect is reminiscent of the Gor'kov-Melik-Barkhudarov correction in weakly coupled superconductors~\cite{Gorkov1961}. The computed frequency dependence of the gap compares qualitatively well with Ref.~\cite{Eberlein2013}, where a more sophisticated truncation of the flow equations has been carried out.
    
    Since $\Delta$ is a spin singlet superfluid gap, and we have chosen $\alpha$ to be real, it obeys 
    \begin{equation}
        \Delta(\nu) = \Delta(-\nu) = \Delta^*(-\nu),
    \end{equation}
    where the first equality comes from the spin singlet nature and the second one from the time reversal symmetry of the effective action. Therefore, the imaginary part of the gap is always zero. By contrast, a magnetic gap would gain, in general, a finite (and antisymmetric in frequency) imaginary part.

    \begin{figure}[t]
        \centering
        \includegraphics[width=0.6\textwidth]{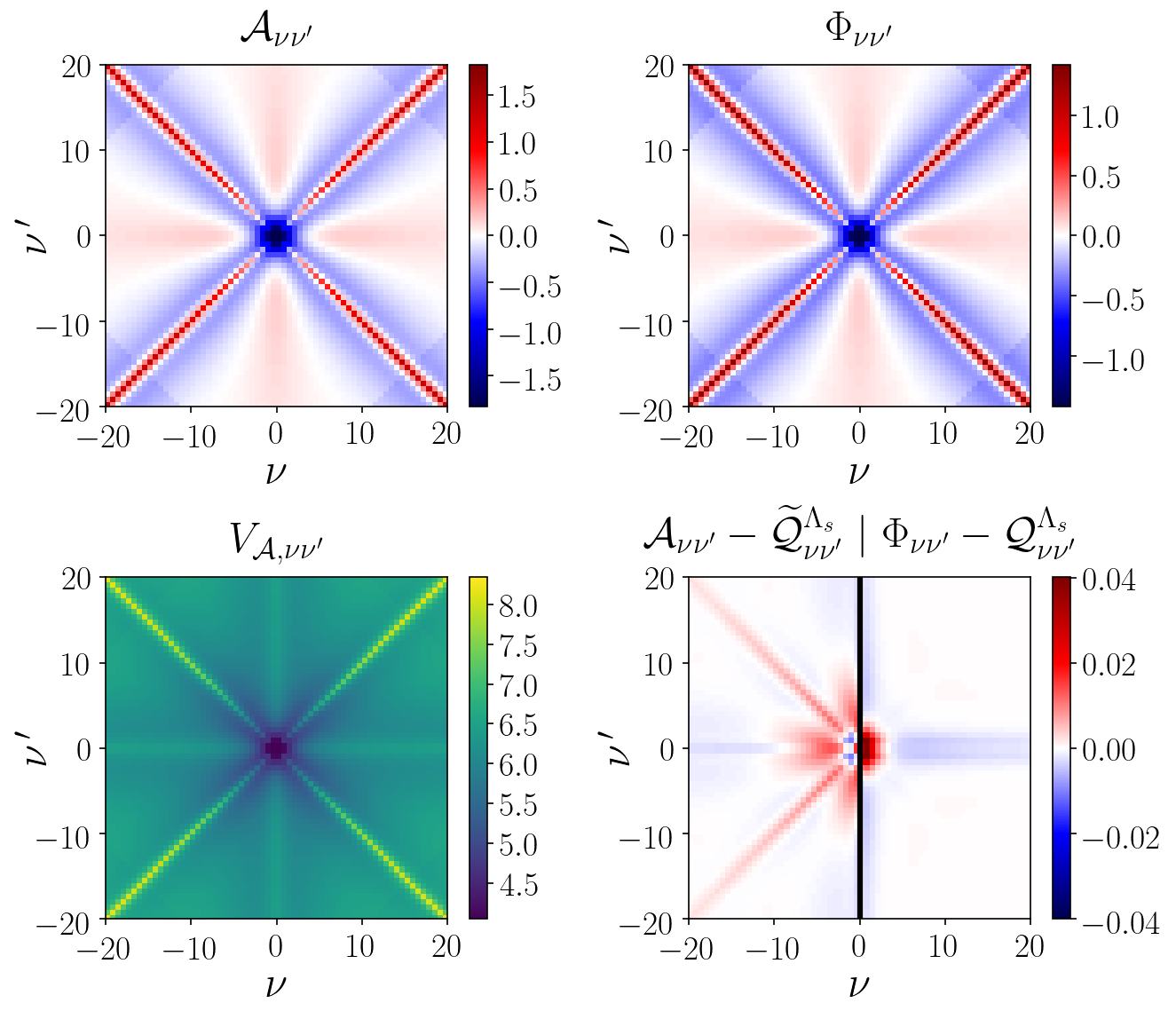}
        \caption{Effective interactions calculated in the SSB phase as functions of Matsubara frequencies.\\
        \textit{Upper left}: longitudinal residual two fermion interaction $\mathcal{A}$.\\
        \textit{Upper right}: transverse residual two fermion interaction $\Phi$.\\
        \textit{Lower left}: longitudinal effective two fermion interaction $V_\mathcal{A}$.\\
        \textit{Lower right}: longitudinal residual two fermion interaction $\mathcal{A}$ with its reduced counterpart $\widetilde{\mathcal{Q}}$ at the critical scale subtracted (left), and transverse longitudinal residual two fermion interaction $\Phi$ minus its equivalent, $\mathcal{Q}$, at $\Lamc$ (right). Both quantities exhibit very small values, showing that $\mathcal{A}$ and $\Phi$ do not deviate significantly from $\widetilde{\mathcal{Q}}$ and $\mathcal{Q}$, respectively.}
        \label{fig_fRG+MF: final vertices}
    \end{figure}
\begin{figure}[t]
        \centering
        \includegraphics[width=0.65\textwidth]{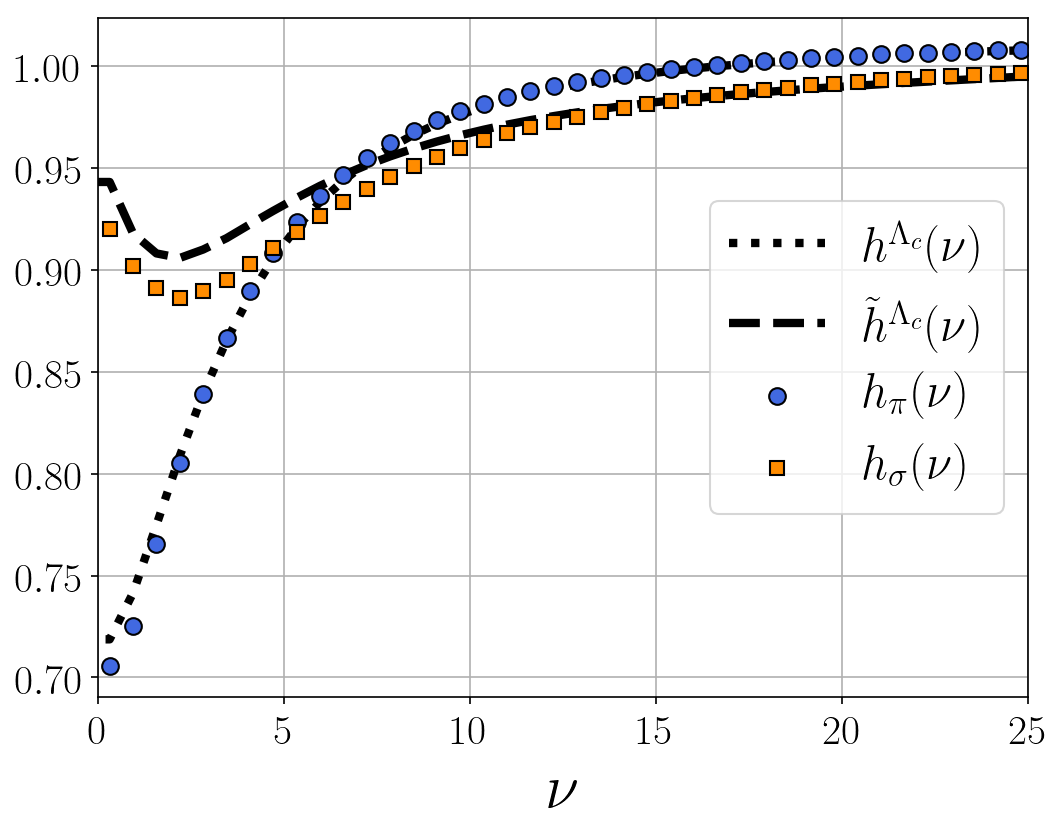}
        \caption{Frequency dependence of Yukawa couplings both at the critical scale $\Lamc$ and in the SSB phase. While $h_\pi$ coincides with $h^{\Lamc}$, the longitudinal coupling $h_\sigma$ does not differ significantly from the reduced one at the critical scale, $\widetilde{h}^{\Lamc}$. The continuous lines for $h^{\Lamc}$ and $\widetilde{h}^{\Lamc}$ are an interpolation through the data calculated on the Matsubara frequencies.}
        \label{fig_fRG+MF: hs}
    \end{figure}

    In Fig.~\ref{fig_fRG+MF: final vertices} we show the results for the residual two fermion interactions in the longitudinal and transverse channels, together with the total effective interaction in the longitudinal channel, defined as 
    \begin{equation}
        V_{\mathcal{A},\nu\nu'}=\frac{h_\sigma(\nu)h_\sigma(\nu')}{m_\sigma}+\mathcal{A}_{\nu\nu'}.
        \label{eq_fRG+MF: VA SSB bosonic}
    \end{equation}
    The analog of Eq.~\eqref{eq_fRG+MF: VA SSB bosonic} for the transverse channel cannot be computed, because the transverse mass $m_\pi$ is zero, in agreement with the Goldstone theorem. The key result is that the residual interactions $\mathcal{A}_{\nu\nu'}$ and $\Phi_{\nu\nu'}$ inherit the frequency structures of $\widetilde{\mathcal{Q}}^{\Lamc}_{\nu\nu'}$ and $\mathcal{Q}^{\Lamc}_{\nu\nu'}$, respectively, and they are also close to them in values (compare with Fig.~\ref{fig_fRG+MF: vertices Lambda s}). The same occurs for the Yukawa couplings, as shown in Fig.~\ref{fig_fRG+MF: hs}. Indeed, the calculated transverse coupling $h_\pi$ does not differ at all from the Yukawa coupling at the critical scale $h^{\Lamc}$. In other words, if instead of solving the self consistent equations, one runs a flow in the SSB phase, the transverse Yukawa coupling will stay the same from $\Lamc$ to $\Lfin$. Furthermore, the longitudinal coupling $h_\sigma$ develops a dependence on the frequency which does not differ significantly from the one of $\widetilde{h}^{\Lamc}$. This feature, at least for our choice of parameters, can lead to some simplifications in the flow equations of Sec.~\ref{sec_fRG+MF: bosonic flow and integration}. Indeed, when running a fRG flow in the SSB phase, one might let flow only the bosonic inverse propagators by keeping the Yukawa couplings and residual interactions fixed at their values, reduced or not, depending on the channel, at the critical scale. This simplifications can be crucial to make computational costs lighter when including bosonic fluctuations of the order parameter, which, similarly, do not significantly renormalize Yukawa couplings in the SSB phase~\cite{Obert2013,Bartosch2009}.
    %

\cleardoublepage
    \rhead[\fancyplain{}{\bfseries Single boson exchange decomposition of the two-particle vertex}]{\fancyplain{}{\bfseries\thepage}}
    \lhead[\fancyplain{}{\bfseries\thepage}]{\fancyplain{}{\bfseries Single boson exchange decomposition}}
    \chapter{Single boson exchange decomposition of the two-particle vertex}
    \label{chap: Bos Fluct Norm}
    In this chapter, we introduce a reformulation of the fRG equations that exploits the \emph{single boson exchange} (SBE) representation of the vertex function, introduced in Ref.~\cite{Krien2019_I}. The latter is based on a diagrammatic decomposition classifying the contributions to the vertex function in terms of their reducibility with respect to removing a bare interaction vertex. This idea is implemented in the fRG by writing each physical channel in terms of a single boson process and a residual two-particle interaction. On the one hand, the present decomposition offers numerical advantages, substantially reducing the computational complexity of the vertex function. On the other hand, it provides a physical insight into the collective fluctuations of the correlated system. We apply the SBE decomposition to the strongly interacting Hubbard model, by combining it with the DMF\textsuperscript{2}RG (see Chap.~\ref{chap: methods}), both at half filling and at finite doping. The results presented in this chapter can be found in Ref.~\cite{Bonetti2021}. 
    \section{Single boson exchange decomposition}
    \begin{figure}[b]
        \centering
        \includegraphics[width= 0.7\textwidth]{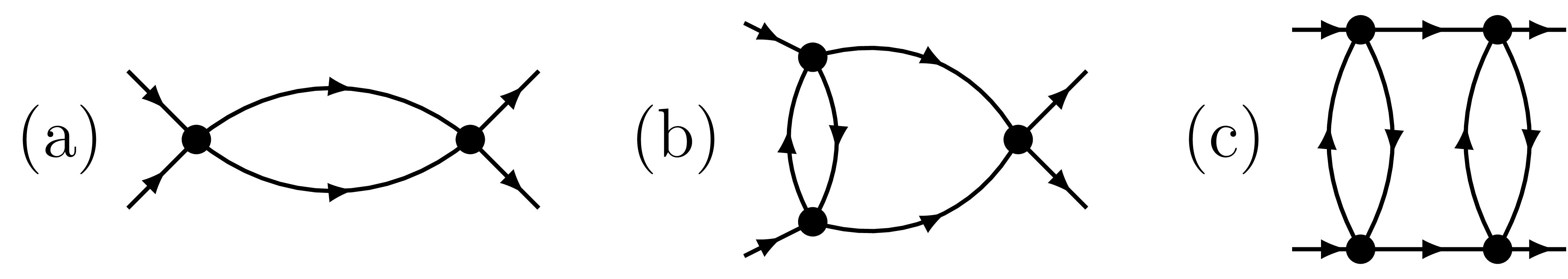}
        \caption{Representative diagrams of the $U$-reducibility. While all three diagrams are two-particle-$pp$ reducible, diagram (a) and (b) are also $U$-$pp$ reducible, while (c) is $U$-irreducible.}
        \label{fig_SBE_fRG: SBE diagrams}
    \end{figure}
    \begin{figure}[t]
        \centering
        \includegraphics[width= 0.5\textwidth]{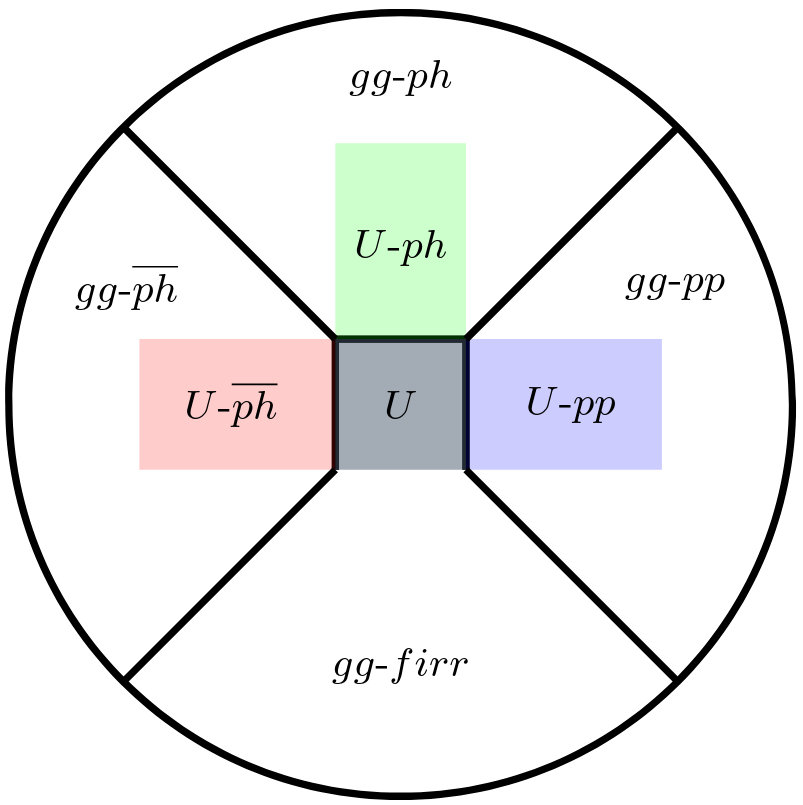}
        \caption{Venn diagram showing the differences between the $U$- and two-particle reducibility (here denoted as $gg$). We notice that in a given channel ($pp$, $ph$ or $\phx$) the $U$-reducible diagrams are a subset of those which are two-particle reducible. The diagram consisting of the bare interaction $U$ is the only diagram that is $U$-reducible but two-particle irreducible. Taken from Ref.~\cite{Krien2019_I}.}
        \label{fig_SBE_fRG: SBE venn diagramm}
    \end{figure}
    In this section, we introduce the SBE decomposition, and we refer to Ref.~\cite{Krien2019_I} for further details. The SBE decomposition relies on the concept of $U$-reducibility~\cite{GiulianiVignale}. The diagrams contributing to the two-particle vertex $V$ can be classified as two-particle reducible or irreducible, depending on whether they can be cut into two disconnected parts by removing a pair of fermionic propagators. The $U$-reducibility sets in as an alternative criterion to classify these diagrams. A diagram is called $U$-reducible (irreducible) if it can (cannot) be cut in two by the removal of a bare interaction vertex. Furthermore, similarly to what happens for the two-particle reducibility, a diagram can be classified as $U$-$pp$ (particle-particle), $U$-$ph$ (particle-hole), or $U$-$\phx$ (particle-hole-crossed) reducible, depending on how the fermionic Green's functions are connected to the removed interaction. Moreover, since the bare vertex has always two pairs of fermionic legs attached, a $U$-reducible diagram is always also two-particle reducible in the same channel, while the opposite is in general not true, as shown by the exemplary diagrams in Fig.~\ref{fig_SBE_fRG: SBE diagrams}. The only exception to this rule is the diagram consisting of a single bare interaction, which as a convention we choose to be $U$-reducible, but it is two-particle irreducible (see Fig.~\ref{fig_SBE_fRG: SBE venn diagramm}).  
    
    Switching from diagrammatic to physical channels, one can re-write the vertex decomposition in Eq.~\eqref{eq_methods: channel decomp physical} as
    \begin{equation}
        \begin{split}
            V(k_1',k_2',k_1) = \,\,&\Lambda_{U\mathrm{irr}}(k_1',k_2',k_1) -2U \\
            &+ \frac{1}{2}\phi^{m,\mathrm{SBE}}_{k_{ph},k_{ph}'}(k_1-k_1') 
            + \frac{1}{2}\phi^{c,\mathrm{SBE}}_{k_{ph},k_{ph}'}(k_1-k_1') \\
            &+ \phi^{m,\mathrm{SBE}}_{k_{\phx},k_{\phx}'}(k_2'-k_1) \\
            &+ \phi^{p,\mathrm{SBE}}_{k_{pp},k_{pp}'}(k_1'+k_2'),
        \end{split}
        \label{eq_SBE_fRG: channel decomp SBE}
    \end{equation}
    with $k_{ph}$, $k'_{ph}$, $k_{\phx}$, $k'_{\phx}$, $k_{pp}$, and $k'_{pp}$ defined as in Eq.~\ref{eq_methods: k k' pp ph phx}. Here, $\Lambda_{U\mathrm{irr}}$ is given by the $U$-irreducible diagrams, and $[\phi^{m,\mathrm{SBE}}+\phi^{c,\mathrm{SBE}}]/2$, $\phi^{m,\mathrm{SBE}}$, and $\phi^{p,\mathrm{SBE}}$ by the $ph$, $\phx$, and $pp$ $U$-reducible diagrams, respectively. Notice that a term $2U$ has been subtracted to avoid double counting of the bare interaction, present in each of the $\phi^{X,\mathrm{SBE}}$. Every $U$-reducible channel can be then further reduced in more fundamental building blocks. Because of the locality of the Hubbard interaction $U$, its dependence on the fermionic arguments $k$ and $k'$ gets completely factorized, and it can be written as 
    \begin{equation}
        \phi^{X,\mathrm{SBE}}_{k,k'}(q) = h^X_k(q)\,D^X(q)\,h^X_{k'}(q),
        \label{eq_SBE_fRG: phi SBE}
    \end{equation}
    where $X=m$, $c$ or $p$, and $h^X$ are referred to as Yukawa (or sometimes Hedin) couplings and $D^X$ as bosonic propagators of screened interactions. The former are related to the three point Green's functions $G^{(3)X}$ via
    \begin{subequations}
        \begin{align}
            &h^m_k(q)=\frac{G^{(3)m}_k(q)}{\chi^{0,ph}_k(q)\left[1+U\chi^m(q)\right]},\\
            &h^c_k(q)=\frac{G^{(3)c}_k(q)+\beta n G(k)\delta_{q,0}}{\chi^{0,ph}_k(q)\left[1-U\chi^c(q)\right]}, \label{eq_SBE_fRG: G3c}\\
            &h^p_k(q)=\frac{G^{(3)p}_k(q)}{\chi^{0,pp}_k(q)\left[1-U\chi^p(q)\right]},
        \end{align}
        \label{eq_SBE_fRG: yukawas from G3}
    \end{subequations}
    where $G(k)$ is the fermionic Green's function, $\chi^X(q)$ the magnetic, charge or pairing susceptibility, $n$ the particle density, and the generalized bare bubbles are defined as
    \begin{subequations}
        \begin{align}
            &\chi^{0,ph}_k(q)=G\left(k+\rnddo{q}\right)G\left(k-\rndup{q}\right), \\
            &\chi^{0,pp}_k(q)=G\left(\rnddo{q}+k\right)G\left(\rndup{q}-k\right).     
        \end{align}
    \end{subequations}
    The three point Green's functions are then related to the four point one $G^{(4)}$ via
    \begin{subequations}
        \begin{align}
            &G^{(3)m}_k(q)=\sum_{\sigma=\up,\down}\int_{k'}\mathrm{sgn}(\sigma)\, G^{(4)}_{\up\sigma\up\sigma}\left(k-\rnddo{q},k'+\rndup{q},k+\rndup{q}\right),\\
            &G^{(3)c}_k(q)=\sum_{\sigma=\up,\down}\int_{k'} G^{(4)}_{\up\sigma\up\sigma}\left(k-\rnddo{q},k'+\rndup{q},k+\rndup{q}\right),\\
            &G^{(3)p}_k(q)=\int_{k'} G^{(4)}_{\up\down\up\down}\left(\rnddo{q}+k,\rndup{q}-k,\rnddo{q}+k'\right),
        \end{align}
    \end{subequations}
    where $\mathrm{sgn}(\up)=+1$, $\mathrm{sgn}(\down)=-1$, and the definition for $G^{(4)}$ is a straightforward lattice generalization of Eq.~\eqref{eq_methods: G4 DMFT}. Notice that in Eq.~\eqref{eq_SBE_fRG: G3c} a disconnected term has been removed from the definition of the charge Yukawa coupling. 
    
    The screened interactions are related to the susceptibilities through
    \begin{subequations}
        \begin{align}
            &D^m(q) = U + U^2\chi^m(q),\\
            &D^c(q) = U - U^2\chi^c(q),\\
            &D^p(q) = U - U^2\chi^p(q).
        \end{align}
        \label{eq_SBE_fRG: D from chi}
    \end{subequations}
    We therefore see that the division by a term $1\pm U\chi^X(q)$ in Eq.~\eqref{eq_SBE_fRG: yukawas from G3} is necessary to avoid double counting of the diagrams in $\phi^{X,\mathrm{SBE}}$. 
    
    It is then interesting to analyze the limits when the frequencies contained in the variables $k$ and $q$ are sent to infinity. All the susceptibilities decay to zero for large frequency, implying
    \begin{equation}
        \lim_{\Omega\to\infty} D^X(\bq,\Omega)=U.
        \label{eq_SBE_fRG: limit D}
    \end{equation}
    Concerning the Yukawa couplings, with some algebra one can express them in the form~\cite{Krien2019_III}
    \begin{subequations}
        \begin{align}
            &h^m_k(q) = 1 + \int_{k'} \varphi^m_{k,k'}(q) \chi^{0,ph}_{k'}(q),\\
            &h^c_k(q) = 1 - \int_{k'} \varphi^c_{k,k'}(q) \chi^{0,ph}_{k'}(q),\\
            &h^p_k(q) = 1 - \int_{k'} \varphi^p_{k,k'}(q) \chi^{0,pp}_{k'}(q),
        \end{align}
    \end{subequations}
    with 
    \begin{equation}
        \varphi^X_{k,k'}(q)=V^X_{k,k'}(q)-h^X_k(q)D^X(q)h^X_{k'}(q),
        \label{eq_SBE_fRG: varphi def}
    \end{equation}
    and
    \begin{subequations}
        \begin{align}
            &V^m_{k,k'}(q) = V\left(k-\rndup{q},k'+\rndup{q},k'-\rndup{q}\right),\\
            &V^c_{k,k'}(q) = 2V\left(k-\rndup{q},k'+\rndup{q},k+\rnddo{q}\right)-V\left(k-\rndup{q},k'+\rndup{q},k'-\rndup{q}\right),\\
            &V^p_{k,k'}(q)=V\left(\rnddo{q}+k,\rndup{q}-k,\rnddo{q}+k'\right),
        \end{align}
        \label{eq_SBE_fRG: Vertices X}
    \end{subequations}
    \hspace{-1.5mm}where $V=V_{\up\down\up\down}$ is the vertex function defined in Sec.~\ref{subs_methods: vertex flow eq}.
    Combining decomposition~\eqref{eq_SBE_fRG: channel decomp SBE}, Eq.~\eqref{eq_SBE_fRG: limit D} and the fact that $\Lambda_{U\mathrm{irr}}$ decays to zero when sending to infinity one of its frequency arguments (this can be proven diagrammatically), one then sees that
    \begin{equation}
        \lim_{\Omega\to\infty}\varphi^X_{k,k'}(\bq,\Omega)=\lim_{\nu\to\infty}\varphi^X_{(\bk,\nu),k'}(q)=0,
        \label{eq_SBE_fRG: varphi limit}
    \end{equation}
    because the frequencies that are sent to infinity enter as arguments in all the screened interactions present in the definition of $\varphi^X$. This lets us conclude that
    \begin{equation}
        \lim_{\Omega\to\infty}h^X_{k}(\bq,\Omega)=\lim_{\nu\to\infty}h^X_{(\bk,\nu)}(q)=1.
        \label{eq_SBE_fRG: h limit}
    \end{equation}
    The limits here derived can be also proven by means of diagrammatic arguments, as shown in Ref.~\cite{Wentzell2016}, where a different notation has been used. 
    
    The SBE decomposition offers several advantages. In first place, it allows for a substantial reduction of the computational complexity. Indeed, the calculation of the SBE terms, accounting for the asymptotic frequency dependencies of the vertex, reduces to two functions, namely $D^X$ and $h^X$, that depend on at least one less collective variable $k$ than the full channel functions $\phi^X$. Furthermore, since the rest functions are fast decaying in all frequency directions, the number of Masubara frequencies required for their calculation can be kept small. Approximations where the $\mathcal{R}^X$ are fully neglected are also possible, based on the choice of selecting only the class of $U$-reducible diagrams. Secondly, the SBE offers a clear physical interpretation of the processes that generate correlations between the electrons, allowing, for example, to diagnose which kind of collective fluctuations give the largest contribution to a given physical observable. Finally, the clear identification of bosonic fluctuations allows for a better treatment of the Goldstone modes when spontaneous symmetry breaking occurs. 
    \section{SBE representation of the fRG}
    In this section, we implement the SBE decomposition in the 1-loop fRG equations. Generalizations to other truncations (Katanin, 2-loop, multiloop~\cite{Gievers2022}) are also possible. To keep the notation light, we omit the $\L$-dependence of the quantities at play.
    
    We start by recasting the channel flow equations, derived in Sec.~\ref{sec_methods: instability analysis}, in the following form
    \begin{equation}
        \deL \phi^X_{k,k'}(q) = \int_{p}V^X_{k,p}(q)\left[\widetilde{\partial}_\L \chi^{0,X}_p(q)\right]V^X_{p,k'}(q),
    \end{equation}
    where, according to the definitions in Sec.~\ref{sec_methods: instability analysis}, we have defined $\phi^m=U+\mathcal{M}$, $\phi^c=U-\mathcal{C}$, and $\phi^p=U-\mathcal{P}$. The bare bubbles are given by
    \begin{subequations}
        \begin{align}
            &\chi^{0,m}_k(q)=-\chi^{0,ph}_k(q),\\
            &\chi^{0,c}_k(q)=\chi^{0,ph}_k(q),\\
            &\chi^{0,p}_k(q)=-\chi^{0,pp}_k(q).
        \end{align}
    \end{subequations}
    In essence, the $\phi^X$ represent the collection of all two-particle reducible diagrams in a given (physical) channel plus the bare interaction. We can express them in the form
    \begin{equation}
        \phi^X_{k,k'}(q) =
        \phi^{X,\mathrm{SBE}}_{k,k'}(q) + \mathcal{R}^X_{k,k'}(q) - U, 
    \end{equation}
    where $\phi^{X,\mathrm{SBE}}$ is $U$-reducible and can be written as in Eq.~\eqref{eq_SBE_fRG: phi SBE}, and $\mathcal{R}^X$ is $U$-\emph{irreducible} but two particle \emph{reducible} in the given channel. The rest function $\mathcal{R}^X$ decays to zero when \emph{any} of the three frequencies on which it depends is sent to infinity~\cite{Wentzell2016}. With the help of Eqs.~\eqref{eq_SBE_fRG: varphi limit} and \eqref{eq_SBE_fRG: h limit}, one can therefore prove that
    \begin{subequations}
        \begin{align}
            &\lim_{\nu'\to\infty}\phi^X_{k,(\bk',\nu')}(q)
            =\lim_{\nu'\to\infty}V^X_{k,(\bk',\nu')}(q) 
            = h^X_k(q)D^X(q),\\
            &\lim_{\substack{\nu\to\infty\\\nu'\to\infty}}\phi^X_{(\bk,\nu),(\bk',\nu')}(q)
            = \lim_{\substack{\nu\to\infty\\\nu'\to\infty}}V^X_{(\bk,\nu),(\bk',\nu')}(q)
            = D^X(q).
        \end{align}
    \end{subequations}
    The flow equations for the screened interactions, Yukawa couplings, and rest functions immediately follow
    \begin{subequations}
        \begin{align}
            &\deL D^X(q) = \left[D^X(q)\right]^2\int_p h^X_p(q)\left[\widetilde{\partial}_\L \chi^{0,X}_p(q)\right]h^X_p(q),\\
            &\deL h^X_k(q) = \int_p \varphi^X_{k,p}(q)\left[\widetilde{\partial}_\L \chi^{0,X}_p(q)\right]h^X_p(q),\\
            &\deL \mathcal{R}^X_{k,k'}(q) = \int_p \varphi^X_{k,p}(q)\left[\widetilde{\partial}_\L \chi^{0,X}_p(q)\right]\varphi^X_{p,k'}(q),
        \end{align}
    \end{subequations}
    where $\varphi^X$ has been defined in Eq.~\eqref{eq_SBE_fRG: varphi def}. In Appendix~\ref{app: symm V} one can find the symmetry properties of the screened interactions, Yukawa couplings, and rest functions. The above flow equations can be alternatively derived by introducing three bosonic fields in the Hubbard action via as many Hubbard-Stratonovich transformations, and running an fRG flow for a mixed boson-fermion system (for more details see Appendix~\ref{app: SBE_fRG_app}).
    \subsection{Plain fRG}
    For the plain fRG, the initial condition $V^\Lini=U$ translates into $\phi^{X}_{k,k'}(q)=U$, which implies
    \begin{subequations}
        \begin{align}
            &D^{X,\Lini}(q)=U,\\
            &h^{X,\Lini}_k(q)=1,\\
            &\mathcal{R}^{X,\Lini}_{k,k'}(q)=0.
        \end{align}
    \end{subequations}
    Furthermore, in the 1-loop, Katanin, 2-loop, and multiloop approximations, the fully $U$-irreducible term $\Lambda_{U\mathrm{irr}}$ is set to the sum of the three rest functions, lacking any fully two-particle irreducible contribution.
    \subsection{\texorpdfstring{DMF\textsuperscript2RG}{DMF2RG}}
    \label{eq_SBE_fRG: DMF2RG initial conditions}
    Within the DMF\textsuperscript{2}RG, one has to apply the parametrization in Eq.~\eqref{eq_SBE_fRG: channel decomp SBE} also to the impurity vertex, that is
    \begin{equation}
        \begin{split}
            V^\mathrm{imp}(\nu_1',\nu_2',\nu_1) = &\Lambda_{U\mathrm{irr}}^\mathrm{imp}(\nu_1',\nu_2',\nu_1) -2U \\
            &+ \frac{1}{2}\phi^{m,\mathrm{SBE},\mathrm{imp}}_{\nu_{ph},\nu_{ph}'}(\nu_1-\nu_1') 
            + \frac{1}{2}\phi^{c,\mathrm{SBE},\mathrm{imp}}_{\nu_{ph},\nu_{ph}'}(\nu_1-\nu_1') \\
            &+ \phi^{m,\mathrm{SBE},\mathrm{imp}}_{\nu_{\phx},\nu_{\phx}'}(\nu_2'-\nu_1) \\
            &+ \phi^{p,\mathrm{SBE},\mathrm{imp}}_{\nu_{pp},\nu_{pp}'}(\nu_1'+\nu_2'),
        \end{split}
    \end{equation}
    where the definitions of the frequencies $\nu_{ph}$, $\nu_{ph}'$, $\nu_{\phx}$, $\nu_{\phx}'$, $\nu_{pp}$, and $\nu_{pp}'$ can be read from the frequency components of Eq.~\eqref{eq_methods: k k' pp ph phx}. The impurity $U$-reducible terms can be written as
    \begin{equation}
        \phi^{X,\mathrm{SBE},\mathrm{imp}}_{\nu\nu'}(\Omega)=h^{X,\mathrm{imp}}_\nu(\Omega) \, D^{X,\mathrm{imp}}(\Omega) \, h^{X,\mathrm{imp}}_{\nu'}(\Omega),
    \end{equation}
    where the impurity Yukawa couplings and screened interactions can be computed from the momentum independent version of Eqs.~\eqref{eq_SBE_fRG: yukawas from G3} and \eqref{eq_SBE_fRG: D from chi}, after the DMFT self-consistent loop has converged. The $U$-irreducible contribution is then obtained by subtracting the $\phi^{X,\mathrm{SBE},\mathrm{imp}}$ from the impurity vertex. In principle, one can invert three Bethe-Salpeter equations to extract the local rest functions from $\Lambda^\mathrm{imp}_{U\mathrm{irr}}$. However, this can be avoided assigning to the flowing $k$-dependent rest functions only those contributions arising on top the local ones. 
    
    The DMF\textsuperscript{2}RG initial conditions thus read as
    \begin{subequations}
        \begin{align}
            &D^{X,\Lini}(q)=D^{X,\mathrm{imp}}(\Omega),\\
            &h^{X,\Lini}_k(q)=h^{X,\mathrm{imp}}_\nu(\Omega),\\
            &\mathcal{R}_{k,k'}^{X,\Lini}(q)=0.
        \end{align}
    \end{subequations}
    Within the 1-loop, Katanin, 2-loop, and multiloop approximations, the DMF\textsuperscript{2}RG $U$-irreducible vertex consists of two terms: a non-flowing one, accounting \emph{also} for the local fully two-particle irreducible contributions, and a flowing one, given by the sum of the three rest functions, consisting of nonlocal two-particle reducible but $U$-irreducible corrections. 
    \subsection{Results at half filling}
    \begin{figure}[t]
        \centering
        \includegraphics[width=0.65\textwidth]{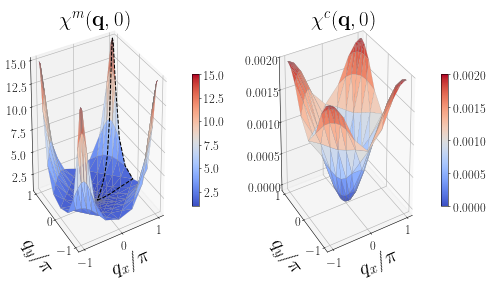}
        \caption{Magnetic (left) and charge (right) susceptibilities at zero frequency as functions of the spatial momentum $\bq$ for $U=16t$ and $T=0.286t$. In the left panel the black dashed line indicates the values taken by $\chi^m(\bq,0)$ along the path in the Brillouin zone considered in Fig.~\ref{fig_SBE_fRG: chim rest no rest}.}
        \label{fig_SBE_fRG: 3D chis HF}
    \end{figure}
    In this section we test the validity of the SBE decomposition on the Hubbard model from moderate to strong coupling by means of the DMF\textsuperscript{2}RG. In order to further simplify the numerics, we project the Yukawa couplings and rest functions dependencies on secondary momenta $\bk$ and $\bk'$ onto $s$-wave form factors, so that
    \begin{subequations}
        \begin{align}
            &h^X_{(\bk,\nu)}(q)\simeq h^X_\nu(q),\\
            &\mathcal{R}^X_{(\bk,\nu),(\bk',\nu')}(q)\simeq \mathcal{R}^X_{\nu\nu'}(q).
        \end{align}
    \end{subequations}
    The flow equations therefore simplify to
    \begin{subequations}
        \begin{align}
            &\deL D^X(q) = \left[D^X(q)\right]^2\,T\sum_\omega h^X_\omega(q)\left[\widetilde{\partial}_\L \chi^{0,X}_\omega(q)\right]h^X_\omega(q),\\
            &\deL h^X_\nu(q) = T\sum_\omega \varphi^X_{\nu\omega}(q)\left[\widetilde{\partial}_\L \chi^{0,X}_\omega(q)\right]h^X_\omega(q),\\
            &\deL \mathcal{R}^X_{\nu\nu'}(q) = T\sum_\omega \varphi^X_{\nu\omega}(q)\left[\widetilde{\partial}_\L \chi^{0,X}_\omega(q)\right]\varphi^X_{\omega\nu'}(q),
        \end{align}
    \end{subequations}
    where we have projected $\varphi^X$ and the bubbles onto $s$-wave form factors, that is
    \begin{subequations}
        \begin{align}
            &\chi_{\nu}^{0,X}(q)=\int_\bk \chi^{0,X}_{(\bk,\nu)}(q),\label{eq_SBE_fRG: chi0 nu}\\
            &\varphi_{\nu\nu'}^X(q)=\int_{\bk,\bk'} \varphi^X_{(\bk,\nu),(\bk',\nu')}(q).
        \end{align}
    \end{subequations}
    We notice that in some parameter ranges the Yukawa couplings and, more importantly, the rest functions may acquire a strong dependence on $\bk$ and $\bk'$. In this case, the $s$-wave approximation is no longer justified. However, in this section we will focus on the half-filled Hubbard model at fairly high temperature, where the dependencies of the vertices on secondary momenta are expected to be weak. In all the rest of the chapter we will neglect the flow of the self-energy, which we keep fixed at the DMFT value. 
    
    As far as the computation of the DMFT initial conditions is concerned, we use ED with 4 bath sites as impurity solver. After the self-consistent loop has converged, we calculate the impurity three- and four-point Green's functions as well as the susceptibilities from their Lehmann representation~\cite{Tagliavini2018}, and extract the respective Yukawa couplings, screened interactions, and the $U$-irreducible DMFT vertex. 
    
    In this section we focus on the half-filled Hubbard model with only nearest neighbor hoppings ($t'=t''=0$) for different couplings and temperatures. For the present choice of parameters particle-hole symmetry is realized. In the results below, the flow of the rest functions has been neglected, when not explicitly stated otherwise. We take the hopping $t$ as energy unit. 
    \subsubsection{Susceptibilities}
    \begin{figure}[t]
        \centering
        \begin{subfigure}[t]{0.495\textwidth}
            \centering
            \includegraphics[width=\textwidth]{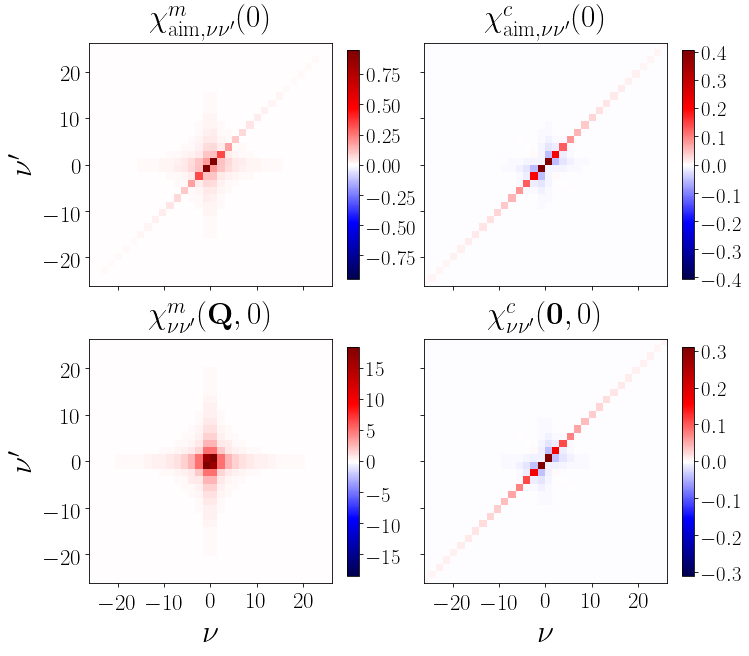}
            \caption{}
            \label{fig_SBE_fRG: gen chis HF U4}
        \end{subfigure}
        \begin{subfigure}[t]{0.495\textwidth}
            \centering
            \includegraphics[width=\textwidth]{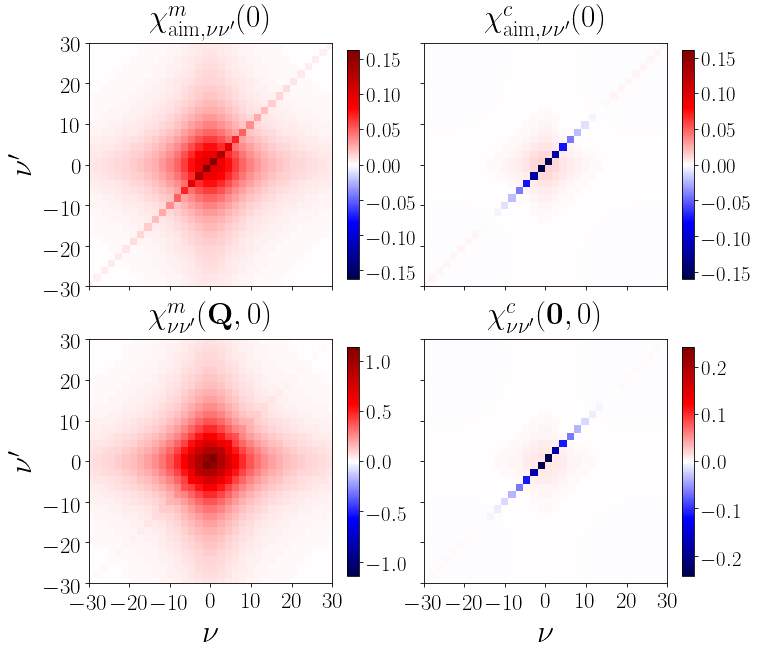}
            \caption{}
            \label{fig_SBE_fRG: gen chis HF U16}
        \end{subfigure}
        \caption{\emph{Panel (a)}: Generalized magnetic (left column) and charge (right) susceptibilities at $U=4t$, and $T=0.250t$, obtained from the impurity model (upper row) and from the DMF\textsuperscript{2}RG (lower row). \emph{Panel (b)}: same as (a), with $U=16t$ and $T=0.286t$.}
        \label{fig_SBE_fRG: gen chis HF}
    \end{figure}
    We start by testing the validity of the SBE decomposition at strong coupling, focusing on the physical response functions.
    
    In Fig.~\ref{fig_SBE_fRG: 3D chis HF}, we show the zero frequency magnetic and charge susceptibilities, extracted from the computed screened interactions $D^m$ and $D^c$, as functions of the lattice momentum for $U=16t$ and $T=0.286t$, that is, slightly above the N\'eel temperature predicted by DMFT at this coupling (see also leftmost panel of Fig.~\ref{fig_SBE_fRG: chim rest no rest}). We notice that particle-hole symmetry implies $D^p(\bq,\Omega)=D^c(\bq+\bQ,\Omega)$, with $\bQ=(\pi,\pi)$. The 1-loop truncation of the DMF\textsuperscript{2}RG does not substantially suppress the N\'eel temperature $T_N$ predicted by the DMFT, resulting in large peaks of $\chi^m(\bq,0)$ at $\bq=\bQ$. It is remarkable, however, that within the DMF\textsuperscript{2}RG $T_N$ is much smaller than the one that plain fRG would give for the present coupling, that is, $T_N\propto U$ for large $U$. The charge susceptibility $\chi^c$ is strongly suppressed at strong coupling, vanishing at $\bq=\bzero$, due to the fully insulating nature of the system at the coupling here considered. Indeed, $U=16t$ lies far above the critical coupling at which the Mott metal to insulator occurs in DMFT ($U_\mathrm{MIT}(T=0)\simeq12t$). 
    
    Following the analysis in Ref.~\cite{Chalupa2021}, it is instructive to analyze the evolution of the generalized susceptibilities, introduced in Sec.~\ref{subs_methods: DMFT susceptibilities}, as the coupling is tuned across the Mott transition. They are in general defined as
    \begin{subequations}
        \begin{align}
            &\chi^m_{k,k'}(q) = -\chi^{0,ph}_k(q)\delta_{k,k'} + \chi^{0,ph}_k(q)V^m_{k,k'}(q)\chi^{0,ph}_{k'}(q),\\
            &\chi^c_{k,k'}(q) = -\chi^{0,ph}_k(q)\delta_{k,k'} - \chi^{0,ph}_k(q)V^c_{k,k'}(q)\chi^{0,ph}_{k'}(q),\\
            &\chi^p_{k,k'}(q) = \chi^{0,pp}_k(q)\delta_{k,k'} + \chi^{0,pp}_k(q)V^c_{k,k'}(q)\chi^{0,pp}_{k'}(q),
        \end{align}
        \label{eq_SBE_fRG: gen chi k}
    \end{subequations}
    \hspace{-1.5mm}where $\delta_{k,k'}=\beta\delta_{\nu\nu'}\delta(\bk-\bk')$, and $V^X$ defined as in Eq.~\eqref{eq_SBE_fRG: Vertices X}. The physical susceptibilities are then obtained from $\chi^X(q)=\int_{k,k'}\chi^X_{k,k'}(q)$. We notice that in a conserving approximation (such as the multiloop fRG) the $\chi^X(q)$ calculated with the above "post-processing" formula coincide with the ones extracted from the screened interactions $D^X(q)$. However, for the 1-loop truncation here employed, the two calculations might yield different results. In the following, we project the $\bk$ and $\bk'$ dependencies of the generalized susceptibility onto $s$-wave form factors, that is, we consider
    \begin{subequations}
        \begin{align}
            &\chi^m_{\nu\nu'}(q) = -\beta\chi^{0,ph}_\nu(q)\delta_{\nu\nu'} + \chi^{0,ph}_\nu(q)V^m_{\nu\nu'}(q)\chi^{0,ph}_{\nu'}(q),\\
            &\chi^c_{\nu\nu'}(q) = -\beta\chi^{0,ph}_\nu(q)\delta_{\nu\nu'} - \chi^{0,ph}_\nu(q)V^c_{\nu\nu'}(q)\chi^{0,ph}_{\nu'}(q),\\
            &\chi^p_{\nu\nu'}(q) = +\beta\chi^{0,pp}_\nu(q)\delta_{\nu\nu'} + \chi^{0,pp}_\nu(q)V^p_{\nu\nu'}(q)\chi^{0,pp}_{\nu'}(q),
        \end{align}
        \label{eq_SBE_fRG: gen chi nu}
    \end{subequations}
    with $\chi^{0,X}_\nu(q)$ as defined in Eq.~\eqref{eq_SBE_fRG: chi0 nu}, and $V^X_{\nu\nu'}(q)=\int_{\bk,\bk'}V_{k,k'}^X(q)$.
    
    In the following, we will focus on the generalized susceptibilities at zero bosonic frequency and for two coupling values, $U=4t$, and $U=16t$, below and above the Mott transition. The corresponding temperatures are chosen to be close to the DMFT N\'eel temperature for the given coupling, that is, $T=0.250t$ and $T=0.286t$. The corresponding results are shown in Fig.~\ref{fig_SBE_fRG: gen chis HF}, where we also plot the corresponding generalized susceptibilities for the self-consistent impurity problem, denoted as $\chi^X_{\mathrm{aim},\nu\nu'}(\Omega)$. At moderate coupling (Fig.~\ref{fig_SBE_fRG: gen chis HF U4}), the leading structure of the charge susceptibility, both for the AIM and DMF\textsuperscript{2}RG results, is given by a \emph{positive} diagonal decaying to zero for large $\nu=\nu'$, arising from the bubble term $-\chi^{0,ph}_\nu(q)$, built upon a metallic Green's function. At the AIM level, the role of vertex corrections appears to be marginal in both channels, with small negative (positive) off-diagonal elements, leading to an overall mild suppression (enhancement) of the physical charge (magnetic) susceptibility.
    While for the charge channel the nonlocal DMF\textsuperscript{2}RG corrections are essentially irrelevant, in the magnetic one, they lead to a strong enhancement of $\chi^m_{\nu\nu'}(\bQ,0)$, signaling strong antiferromagnetic correlations. In the Mott phase, the picture changes drastically, due to large vertex corrections. In the magnetic channel, they strongly enhance the physical susceptibility even at the AIM level overtaking the diagonal term. This is a clear hallmark of the formation of local magnetic moments, resulting in a large magnetic response at zero frequency, following the Curie-Weiss law. Differently, in the charge channel, the vertex strongly suppresses the physical response, flipping the sign of the diagonal entries up to frequencies $|\nu=\nu'|\sim U$. In more detail, these negative values are responsible for the freezing of charge fluctuations in the deep insulating regime~\cite{Gunnarsson2016,Chalupa2021}. This observation can be interpreted as the charge counterpart of the local moment formation in the magnetic sector. The negative diagonal entries are in general related to negative eigenvalues of the generalized susceptibility. Increasing the coupling $U$, when one of the eigenvalues flips its sign, the matrix $\chi^c_{\nu\nu'}(0)$ becomes non-invertible, leading to divergences of the irreducible vertex function~\cite{Schaefer2013,Gunnarsson2016,Chalupa2020,Springer2020}, which are in turn related to the multivaluedness of the Luttinger-Ward functional~\cite{Kozik2014,Vucicevic2018}.

    \subsubsection{Role of the rest functions}
    \begin{figure}[t]
        \centering
        \includegraphics[width=\textwidth]{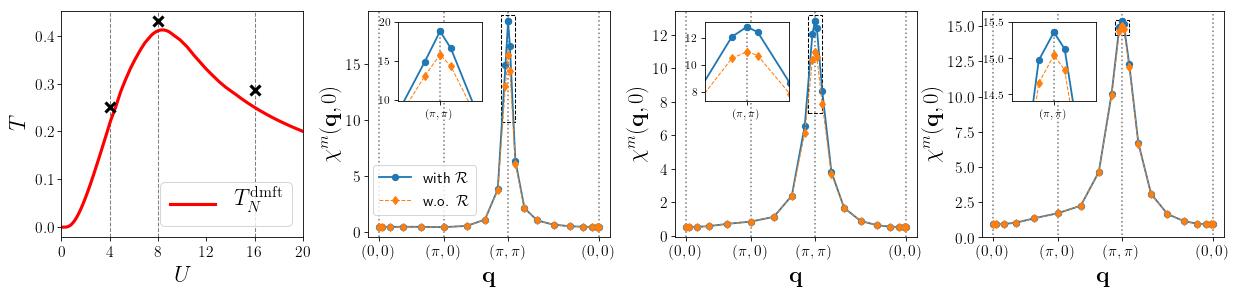}
        \caption{Left panel: DMFT N\'eel temperature and location of the parameters considered in the remaining three panels. Other panels: magnetic susceptibility at zero frequency in a path in the Brillouin zone (see Fig.~\ref{fig_SBE_fRG: 3D chis HF} for its definition) calculated with and without considering the flow of the rest functions. The coupling values considered are, from left to right, $U=4t$, $8t$, and $16t$.}
        \label{fig_SBE_fRG: chim rest no rest}
    \end{figure}
    \begin{figure}[b!]
        \centering
        \includegraphics[width=0.6\textwidth]{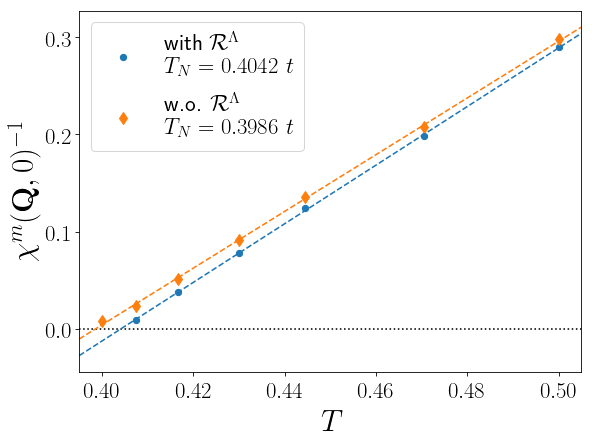}
        \caption{Inverse magnetic susceptibility at zero frequency and $\bq=\bQ$ as a function of temperature for a coupling value of $U=8t$. The dashed lines correspond to linear fits of the data, whose extrapolation yields the N\'eel temperature. }
        \label{fig_SBE_fRG: xi vs T}
    \end{figure}
    All the results presented so far have been obtained neglecting the flow of the rest functions. While this approximation significantly reduces the computational cost, the extent of its validity has to be verified in different coupling regimes. We recall that by considering the flow of $\mathcal{R}^X$ we recover the results obtained by conventional implementations of the fRG (see, for example Refs.~\cite{Vilardi2017,Vilardi2019}). 
    
    In Fig.~\ref{fig_SBE_fRG: chim rest no rest}, we analyze the impact of the inclusion/neglection of the rest functions on the magnetic susceptibilities at coupling values of $U=4t$, $8t$, and $16t$ and temperatures close to the corresponding $T_N$ as obtained in the DMFT, namely $T=0.250t$, $0.444t$, and $0.286t$, respectively (see leftmost panel for the location of these points in the $(U,T)$ phase diagram). The corrections due to the inclusion of the $\mathcal{R}^X$ are rather marginal for all couplings considered, resulting in only a slight enhancement of magnetic correlations. As a consequence of this, the N\'eel temperature, which is finite as the 1-loop truncation here considered violates the Mermin-Wagner theorem, is very mildly affected by the rest functions. This can be observed in Fig.~\ref{fig_SBE_fRG: xi vs T}, where we plot the inverse magnetic susceptibility at $\bq=\bQ$ and zero frequency as a function of the temperature for $U=8t$. We notice that the inclusion of the $\mathcal{R}^X$ yields a N\'eel temperature of $T_N=0.4042t$, and the one obtained without rest function lies very close to it, $T_N=0.3986t$. The effects of the rest functions on the charge and pairing susceptibilities (not shown) are negligible. 
    
\begin{figure}[t]
        \centering
        \includegraphics[width=0.95\textwidth]{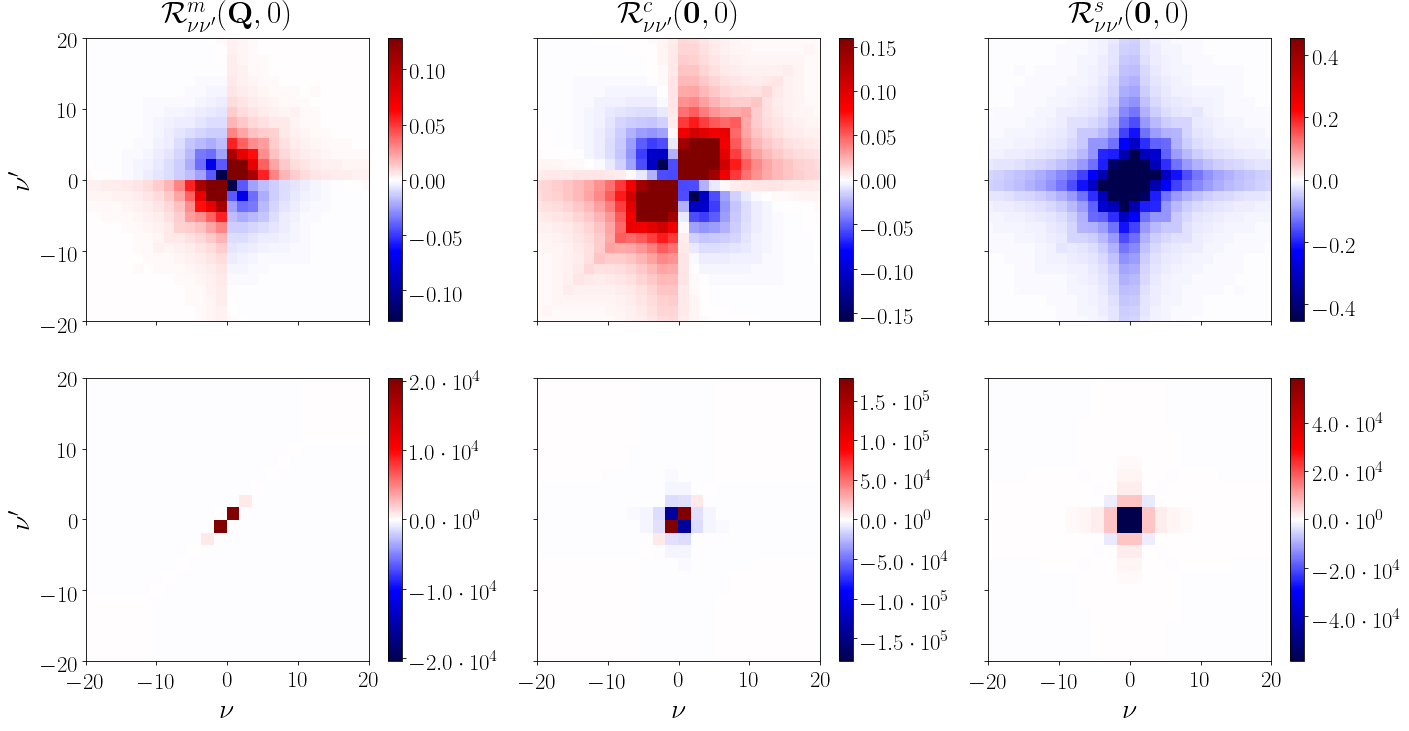}
        \caption{Fermionic frequency dependence of the magnetic (left column), charge (central column), and pairing (right column) rest functions for zero bosonic frequency, and for couplings $U=4t$ (upper row) and $U=16t$ (lower row). The temperatures are the same as in Fig.~\ref{fig_SBE_fRG: chim rest no rest}.}
        \label{fig_SBE_fRG: rests}
    \end{figure}

    In Fig.~\ref{fig_SBE_fRG: rests}, we plot the frequency structure of the rest functions for the three channels at zero bosonic frequency, for $U=4t$ and $U=16t$ and for the same temperatures considered in Fig.~\ref{fig_SBE_fRG: chim rest no rest}. The decay to zero for large frequencies $\nu$, $\nu'$ is clear, particularly at strong coupling (lower row), where the $\mathcal{R}^X$ take extremely large values at the lowest Matsubara frequencies. However, in the insulating regime the Green's function is suppressed at the smallest Matsubara frequencies, strongly reducing the effect of the large values of the rest functions on the physical observables. 
    \subsection{Finite doping: fluctuation diagnostics of \texorpdfstring{$d$-wave}{d-wave} correlations}
    In this section, we show results for the doped Hubbard model at fairly low temperature. The parameter set we consider is a hole doping of $p=1-n=0.18$, $T=0.044t$, a next-to-nearest neighbor hopping of $t'=-0.2t$, and $U=8t$. Since at finite doping and low temperatures the Hubbard model is expected to display sizable $d$-wave correlations, we improve our form factor expansion to include them. Considering the pairing channel, we notice that the $U$-reducible term can have a finite $d$-wave coefficient thanks to the Yukawa coupling
    \begin{equation}
        h^{p}_{(\bk,\nu)}(q)\sim  h^{p,s}_{\nu}(q) + h^{p,d}_{\nu}(q) d_\bk,
    \end{equation}
    with $d_\bk=\cos k_x-\cos k_y$. However, due to the locality of the bare interaction, the function $h_\nu^{p,d}(\bq,\Omega)$ identically vanishes for $\bq=\bzero$, therefore not contributing to an eventual $d$-wave pairing state. For this reason we retain only the $s$-wave contribution to the pairing Yukawa coupling. What would really drive the formation of a $d$-wave superconducting gap is the rest function $\mathcal{R}^p$. We expand the latter as 
    \begin{equation}
        \mathcal{R}^p_{k,k'}(q)\simeq\mathcal{R}^{p}_{\nu\nu'}(q)-\mathcal{D}_{\nu\nu'}(q)d_{\bk} d_{\bk'}, 
        \label{eq_SBE_fRG: expansion of Rp}
    \end{equation}
    where we have neglected possible $s$-$d$-wave mixing terms, and the minus sign has been chosen for convenience. In essence, the function $\mathcal{D}_{\nu\nu'}(q)$, which we refer to as $d$-wave pairing channel, is given by diagrams that are two-particle-$pp$ reducible but $U$-irreducible and, at the same time, exhibit a $d$-wave symmetry in the dependence on $\bk$ and $\bk'$. The flow equation for $\mathcal{D}_{\nu\nu'}(q)$ reads as
    \begin{equation}
        \deL \mathcal{D}_{\nu\nu'}(q)= T\sum_\omega V^{d}_{\nu\omega}(q)\left[\widetilde{\partial}_\L\chi^{0,d}_\omega(q)\right]V^{d}_{\omega\nu'}(q),
        \label{eq_SBE_fRG: flow eq D}
    \end{equation}
    with 
    \begin{equation}
        \chi^{0,d}_\nu(q)=\int_{\bk}d_\bk^2\,\chi^{0,pp}_{(\bk,\nu)}(q),
    \end{equation}
    and 
    \begin{equation}
        V^{d}_{\nu\nu'}(q)=\int_{\bk,\bk'}d_\bk d_{\bk'}\,V^p_{(\bk,\nu),(\bk',\nu')}(q).
    \end{equation}
    The flow equations of the other quantities remain unchanged, except that the contribution in Eq.~\eqref{eq_SBE_fRG: expansion of Rp} has to be considered in the calculation of the functions $V^X_{\nu\nu'}(q)$, with $X=m$, $c$, or $p$. In this section we neglect the flow of all the rest functions but $\mathcal{D}$.

    \begin{figure}[t]
        \centering
        \includegraphics[width= 0.95\textwidth]{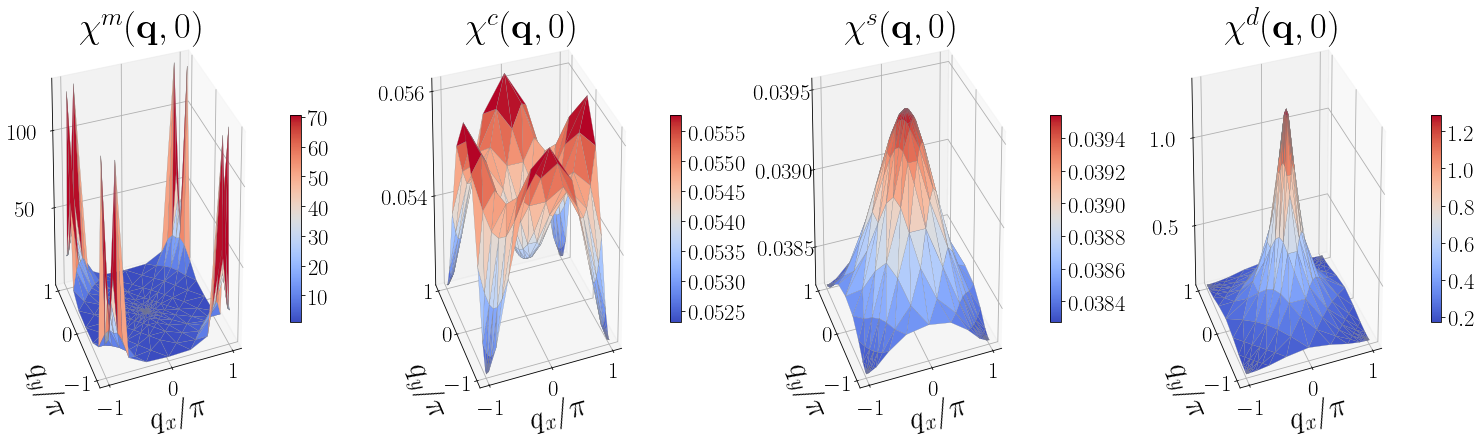}
        \caption{From left to right: magnetic, charge, $s$- and $d$-wave pairing susceptibilities at zero bosonic frequency as functions of the lattice momentum $\bq$, for $p=0.18$, $t'=-0.2t$, $T=0.044t$, and $U=8t$, determined at the stopping scale $\Lamc=0.067t$.}
        \label{fig_SBE_fRG: dwave 3d chis}
    \end{figure}
    In the parameter regime considered, and within the 1-loop truncation employed, the system is unstable under the formation of incommensurate magnetic order. Thus the flow needs to be stopped due to the divergence of $D^m(\bq,0)$ at $\bq=(\pi-2\pi\eta,\pi)$ (and symmetry related). We therefore arbitrarily define the stopping scale $\Lamc$ as the one at which $D^m$ exceeds the value of $8\times 10^3t$, corresponding to a magnetic susceptibility of $\sim 120t^{-1}$. We obtain $\Lamc=0.067t$. While the choice of a parameter regime close to a magnetic instability is crucial to detect \emph{sizable} $d$-wave pairing correlations, we expect that an improved truncation (as, for example, the multiloop extension) would remove the divergence in the magnetic channel, allowing to continue the flow down to $\L=0$, thereby probably enhancing the $d$-wave susceptibility. 
    
    In Fig.~\ref{fig_SBE_fRG: dwave 3d chis}, we show the magnetic, charge, $s$- and $d$-wave pairing susceptibilities, computed at the stopping scale. While the first three have been extracted from the bosonic propagators $D^X$ (see Eq.~\eqref{eq_SBE_fRG: D from chi}), the $d$-wave pairing susceptibility has been calculated with the "post-processing" formula 
    \begin{equation}
        \chi^d(q) = T\sum_\nu\chi^{0,d}_\nu(q) + T^2\sum_{\nu,\nu'}\chi^{0,d}_\nu(q)\,V^d_{\nu\nu'}(q) \,\chi^{0,d}_{\nu'}(q).
    \end{equation}
    The magnetic susceptibility displays very large values in the form of peaks at wave vectors $(\pi-2\pi\eta,\pi)$ and symmetry related ($\eta\simeq0.08$) due to the incommensurate antiferromagnetic instability. Differently, the charge and $s$-wave pairing response functions are rather suppressed, with $\chi^c(q)$ exhibiting peaks at $\bq=(\pi,0)$ (and symmetry related), signaling \emph{very mild} charge stripe correlations. Finally, $\chi^d(q)$, although being not excessively large, presents a well-defined peak at $\bq=\bzero$ and is by far the second largest response function.

    \begin{figure}[t]
        \centering
        \includegraphics[width= 0.65\textwidth]{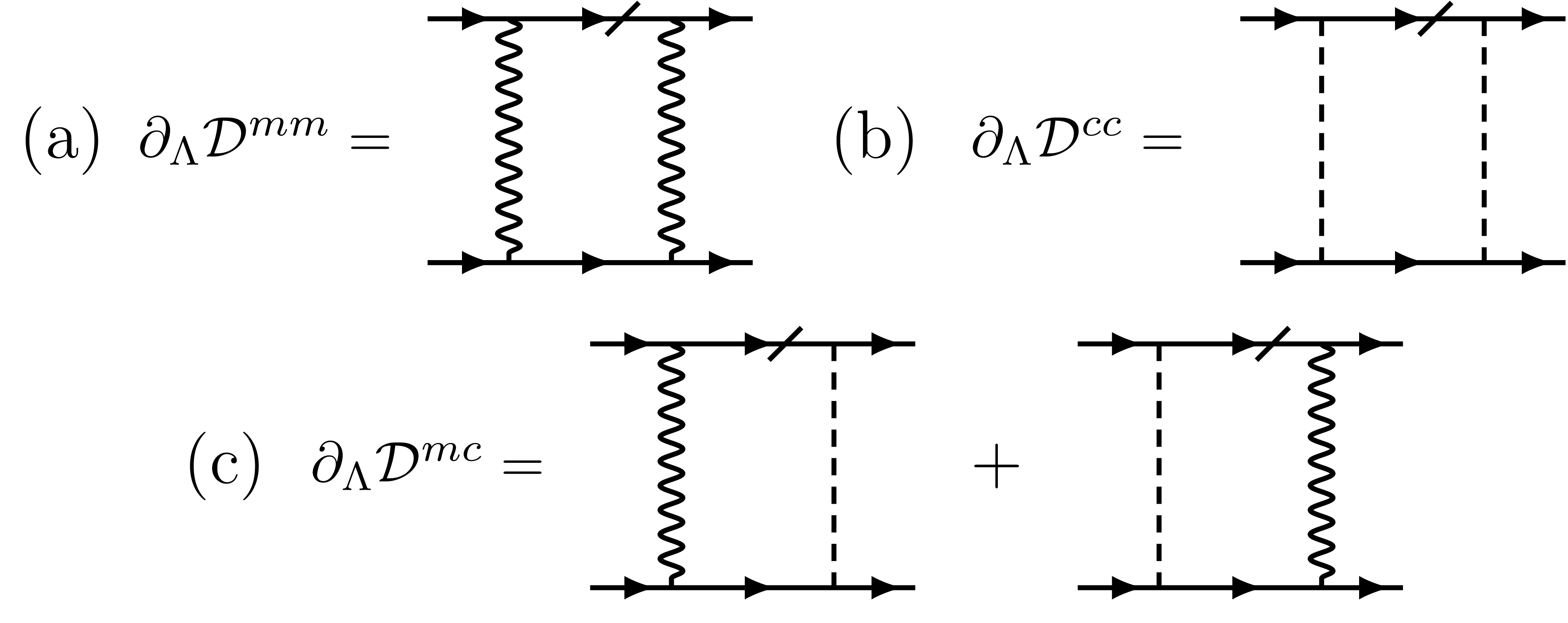}
        \caption{Diagrammatic representation of the two boson contributions to the flow equation of $\mathcal{D}_{\nu\nu'}(q)$. Wavy (dashed) lines represent magnetic (charge) screened interactions, and solid lines fermionic Green's functions. The ticked lines indicate single scale propagators.}
        \label{fig_SBE_fRG: dwave diagrams}
    \end{figure}
\begin{figure}[t]
        \centering
        \includegraphics[width= 0.65\textwidth]{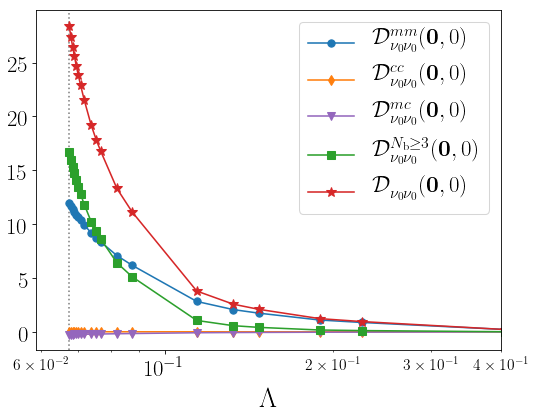}
        \caption{Different terms contributing to the $d$-wave pairing channel $\mathcal{D}_{\nu\nu'}(q)$, as defined in Eq.~\eqref{eq_SBE_fRG: D contributions}, at $q=(\bzero,0)$ and $\nu=\nu'=\nu_0\equiv\pi T$ as functions of the scale $\L$. The total flow of $\mathcal{D}$ is also reported for comparison (red stars).}
        \label{fig_SBE_fRG: dwave diagnostics flow}
    \end{figure}

    Similarly to what has been done in the \emph{fluctuation diagnostics} for the self-energy~\cite{Gunnarsson2015,Krien2020}, it is instructive to analyze the different bosonic fluctuations contributing to the formation of a \emph{sizable} $d$-wave pairing channel $\mathcal{D}$. The function $V^d_{\nu\nu'}(q)$ entering the flow equation of $\mathcal{D}$ can be written as
    \begin{equation}
        \begin{split}
            V^d_{\nu\nu'}(\bq,\Omega)=
            -L^{d,m}_{\nu\nu'}(\Omega)
            -L^{d,c}_{\nu\nu'}(\Omega)
            -\mathcal{D}_{\nu\nu'}(\bq,\omega),
        \end{split}
    \end{equation}
    where we have defined 
    \begin{subequations}
        \begin{align}
            L^{d,m}_{\nu\nu'}(\Omega)=&\phantom{+}\frac{1}{2} \phi^{m,\mathrm{SBE},d}\left(\rnddo{\Omega}+\rndup{\nu+\nu'},\rndup{\Omega}-\rndup{\nu+\nu'};\nu'-\nu\right)\nonumber\\
            &+\phi^{m,\mathrm{SBE},d}\left(\rndup{\nu-\nu'+\Omega},\rndup{\nu'-\nu+\Omega};-\nu-\nu' + \Omega\,\mathrm{m}\,2\right), \\
            L^{d,c}_{\nu\nu'}(\Omega)=&\phantom{+}\frac{1}{2} \phi^{c,\mathrm{SBE},d}\left(\rnddo{\Omega}+\rndup{\nu+\nu'},\rndup{\Omega}-\rndup{\nu+\nu'};\nu'-\nu\right),
        \end{align}
    \end{subequations}
    with
    \begin{equation}
        \phi^{X,\mathrm{SBE},d}(\nu,\nu';\Omega) = -\int_\bq\frac{\cos q_x+\cos q_y}{2}\, \phi^{X,\mathrm{SBE}}_{\nu\nu'}(\bq,\Omega).
    \end{equation}
    We can then split the different terms contributing to~\eqref{eq_SBE_fRG: flow eq D} as
    \begin{subequations}
        \begin{align}
            &\deL\mathcal{D}^{mm}_{\nu\nu'}(q)=T\sum_\omega L^{d,m}_{\nu\omega}(q)\left[\widetilde{\partial}_\L \chi^{0,d}_\omega(q)\right] L^{d,m}_{\omega\nu'}(q),
            \label{eq_SBE_fRG: D mm}\\
            &\deL\mathcal{D}^{cc}_{\nu\nu'}(q)=T\sum_\omega L^{d,c}_{\nu\omega}(q)\left[\widetilde{\partial}_\L \chi^{0,d}_\omega(q)\right] L^{d,c}_{\omega\nu'}(q),
            \label{eq_SBE_fRG: D cc}\\
            &\deL\mathcal{D}^{mc}_{\nu\nu'}(q)=T\sum_\omega L^{d,m}_{\nu\omega}(q)\left[\widetilde{\partial}_\L \chi^{0,d}_\omega(q)\right] L^{d,c}_{\omega\nu'}(q) 
            +T\sum_\omega L^{d,c}_{\nu\omega}(q)\left[\widetilde{\partial}_\L \chi^{0,d}_\omega(q)\right] L^{d,m}_{\omega\nu'}(q),
            \label{eq_SBE_fRG: D mc}\\
            &\deL\mathcal{D}^{N_b\geq3}_{\nu\nu'}(q)=\deL\mathcal{D}_{\nu\nu'}(q)-\deL\mathcal{D}^{mm}_{\nu\nu'}(q)-\deL\mathcal{D}^{cc}_{\nu\nu'}(q)-\deL\mathcal{D}^{mc}_{\nu\nu'}(q)\nonumber\\
            &\phantom{\deL\mathcal{D}^{N_b\geq3}_{\nu\nu'}(q)}=T\sum_\omega \mathcal{D}_{\nu\omega}(q)\left[\widetilde{\partial}_\L \chi^{0,d}_\omega(q)\right] \mathcal{D}_{\omega\nu'}(q)
            +\sum_{X=m,c}
            T\sum_\omega \mathcal{D}_{\nu\omega}(q)\left[\widetilde{\partial}_\L \chi^{0,d}_\omega(q)\right] L^{d,X}_{\omega\nu'}(q)
            \nonumber\\
            &\hskip 2.5cm
            +\sum_{X=m,c}
            T\sum_\omega L^{d,X}_{\nu\omega}(q)\left[\widetilde{\partial}_\L \chi^{0,d}_\omega(q)\right] \mathcal{D}_{\omega\nu'}(q).
            \label{eq_SBE_fRG: D Nb>=3}
        \end{align}
        \label{eq_SBE_fRG: D contributions}
    \end{subequations}
    A diagrammatic representation of the flow equations of the first three terms, $\mathcal{D}^{mm}$, $\mathcal{D}^{cc}$, and $\mathcal{D}^{mc}$ is given in Fig.~\ref{fig_SBE_fRG: dwave diagrams}. They represent two boson processes, also known as Aslamazov-Larkin diagrams. Inspecting Eqs.~\eqref{eq_SBE_fRG: D mm},~\eqref{eq_SBE_fRG: D cc}, and~\eqref{eq_SBE_fRG: D mc}, one can notice that they are not fully reconstructed by the flow, as the functions $L^{d,m}$ and $L^{d,c}$ (and the self-energy) also depend on the fRG scale. This is a feature of the 1-loop truncation and is not present in the framework of the multiloop extension. It is nonetheless reasonable to interpret these contributions as two boson processes, and the remainder $\mathcal{D}^{N_b\geq3}$ as a higher order contribution in the number of exchanged bosons. In Fig.~\ref{fig_SBE_fRG: dwave diagnostics flow}, we plot the different contributions to $\mathcal{D}_{\nu\nu'}$ at $q=0$ and $\nu=\nu'=\nu_0\equiv\pi T$ as functions of the scale $\L$. We notice that in the early stages of the flow the largest contribution to $\mathcal{D}$ comes from magnetic two boson process, confirming that magnetic fluctuations provide the seed for the formation of $d$-wave pairing in the 2D Hubbard model, as found in other fRG studies~\cite{Halboth2000_PRL,Husemann2009,Katanin2009,Vilardi2017,Vilardi2019}.
    Moreover, the multiboson term ($\mathcal{D}^{N_b\geq3}$) develops at smaller scales, compared to the two boson ones. At the same time, it increases considerably when approaching the stopping scale $\Lamc$, overtaking the other contributions. In general, arbitrarily close to a thermodynamic instability towards a $d$-wave pairing state, the $N_b\geq\bar{N}$ boson contribution is always larger than all the $N_b<\bar{N}$ ones, for every finite $\bar{N}$~\cite{Bonetti2021}. In the present parameter region we indeed observe $\mathcal{D}^{N_b\geq3}>\mathcal{D}^{mm},\mathcal{D}^{cc},\mathcal{D}^{mc}$, which means that an important precondition for the onset of a thermodynamic superconducting instability has already been realized.
    %

\cleardoublepage
    \rhead[\fancyplain{}{\bfseries Low-energy physics of metallic spiral magnets}]{\fancyplain{}{\bfseries\thepage}}
    \lhead[\fancyplain{}{\bfseries\thepage}]{\fancyplain{}{\bfseries Low-energy physics of metallic spiral magnets}}
    \chapter{Collective modes of metallic spiral magnets}
    \label{chap: low energy spiral}
    In this chapter, we present a detailed analysis of the low-energy magnons of a spiral magnet. In particular, we show that, differently than a N\'eel antiferromagnet, the SU(2) spin symmetry is broken down to $\mathbb{Z}_2$, giving rise to three  Goldstone modes, corresponding to three gapless magnon branches. We focus on the case of \emph{coplanar} spiral order, that implies that two of the three magnon modes have the same dispersion. In particular, one finds one \emph{in-plane} and two \emph{out-of-plane} modes. We perform a low energy expansion of the magnetic susceptibilities in the spiral magnetic state, and derive general expressions for the spin stiffnesses and spectral weights of the magnon excitations. We also show that they can be alternatively computed from the response to a gauge field. We prove that the equivalence of this approach with a low-energy expansion of the susceptibilities is enforced by some \emph{Ward identities}. Moreover, we analyze the size and the low-momentum and frequency dependence of the Landau damping of the Goldstone modes. 
    The understanding of the low energy physics of a spiral magnet will be of fundamental importance for the next chapter, where a model for the pseudogap phase is presented in terms of short-range spiral order.
    
    This chapter is organized as it follows. In Sec.~\ref{sec_low_spiral: Ward Identities}, we derive the local Ward identities that enforce the equality of the spin stiffnesses and spectral weights computed expanding the susceptibilities near their Goldstone pole and from the response to a gauge field. In this Section, besides the spiral magnet, we also analyze the case of a superconductor and of a N\'eel antiferromagnet. In Secs.~\ref{sec_low_spiral: MF spiral} and \ref{sec_low_spiral: RPA spiral}, we present the mean-field and random phase approximation (RPA) approaches to the spiral magnetic state and its collective excitations. In Sec.~\ref{sec_low_spiral: properties of Goldstones} we expand the RPA magnetic susceptibilities around their Goldstone poles and derive expressions for the spin stiffnesses, spectral weights and Landau dampings of the three magnon modes. In Sec.~\ref{sec_low_spiral: explicit WIs} we compute the spin stiffnesses and spectral weights as response coefficients of the spiral magnet to a SU(2) gauge field and show the equivalence with the formulas of Sec.~\ref{sec_low_spiral: properties of Goldstones}. In Sec.~\ref{sec_low_spiral: Neel limit} we analyze the N\'eel limit, and in Sec.~\ref{sec_low_spiral: numerical results} we show a numerical evaluation of the results of the previous sections. 
    The content of this chapter has been published in Refs.~\cite{Bonetti2022} and \cite{Bonetti2022_II}. 
    \section{Local Ward identities for spontaneously broken symmetries}
    \label{sec_low_spiral: Ward Identities}
    In this section we derive and discuss the Ward identities connected with a specific gauge symmetry which gets globally broken due to the onset of long-range order in the fermionic system. We focus on two specific symmetry groups: the (abelian) U(1) charge symmetry and the (nonabelian) SU(2) spin symmetry. All over the chapter we employ Einstein's notation, that is, a sum over repeated indices is implicit. 
    \subsection{U(1) symmetry}
    We consider the generating functional of susceptibilities of the superconducting order parameter and gauge kernels, defined as:
    \begin{equation}
        \begin{split}
            \mathcal{G}\left[A_\mu,J,J^*\right]=
            -\ln \int \!\mathcal{D}\psi\mathcal{D}\psibar e^{-\mathcal{S}\left[\psi,\psibar,A_\mu\right]+(J^*,\psi_\down\psi_\up)+(J,\psibar_\up\psibar_\down)},
        \end{split}
        \label{eq_low_spiral: G-functional U(1)}
    \end{equation}
    where $\psi=(\psi_\up,\psi_\down)$ ($\psibar=(\psibar_\up,\psibar_\down)$) are Grassmann spinor fields corresponding to the annihilation (creation) of a fermion, $A_\mu$ is the electromagnetic field, $J$ ($J^*$) is a source field that couples to the superconducting order parameter $\psibar_\up\psibar_\down$ ($\psi_\down\psi_\up$), and $\mathcal{S}[\psi,\psibar,A_\mu]$ is the action of the system. The index $\mu=0,1,\dots,d$, with $d$ the system dimensionality, runs over temporal ($\mu=0$) and spatial ($\mu=1,\dots,d$) components. In the above equation and from now on, the expression $(A,B)$ has to be intended as $\int_x A(x)B(x)$, where $x$ is a collective variable consisting of a spatial coordinate $\bs{x}$ (possibly discrete, for lattice systems), and an imaginary time coordinate $\tau$, and $\int_x$ is a shorthand for $\int d^d\bs{x}\int_0^\beta d\tau$, with $\beta$ the inverse temperature. Even in the case of a lattice system, we define the gauge field over a \emph{continuous} space time, so that expressions involving its gradients are well defined. 
    
    We let the global U(1) charge symmetry be broken by an order parameter that, to make the treatment simpler, we assume to be local ($s$-wave)
    \begin{equation}
        \langle \psi_\down(x)\psi_\up(x)\rangle 
        = \langle \psibar_\up(x)\psibar_\down(x)\rangle= \varphi_0,
    \end{equation}
    where the average is computed at zero source and gauge fields, and, without loss of generality, we choose $\varphi_0\in \mathbb{R}$. A generalization to systems with nonlocal order parameters, such as $d$-wave superconductors, is straightforward. 
    
    The functional~\eqref{eq_low_spiral: G-functional U(1)} has been defined such that its second derivative with respect to $J$ and $J^*$ at zero $J$, $J^*$ and $A_\mu$ gives (minus) the susceptibility of the order parameter $\chi(x,x')$, while (minus) the gauge kernel $K_{\mu\nu}(x,x')$ can be extracted differentiating twice with respect to the gauge field. In formulas
    \begin{subequations}
        \begin{align}
            &\chi(x,x')=-\frac{\delta^2\mathcal{G}}{\delta J(x)\delta J^*(x')}\bigg\rvert_{J=J^*=A_\mu=0},\\
            &K_{\mu\nu}(x,x')=-\frac{\delta^2\mathcal{G}}{\delta A_\mu(x)\delta A_\nu(x')}\bigg\rvert_{J=J^*=A_\mu=0}.
        \end{align}
    \end{subequations}
    Let us now consider the constraints that the U(1) gauge invariance imposes on the functional $\mathcal{G}$. Its action on the fermionic fields is
    \begin{subequations}\label{eq_low_spiral: U(1) gauge transf on psis}
        \begin{align}
            &\psi(x) \to e^{i\theta(x)}\psi(x),\\
            &\psibar(x) \to e^{-i\theta(x)}\psibar(x),
        \end{align}
    \end{subequations}
    \noindent with $\theta(x)$ a generic function. Similarly, the external fields transform as
    \begin{subequations}\label{eq_low_spiral: ext fields U(1) transformation}
        \begin{align}
            &J(x) \to J^\prime(x)= e^{2i\theta(x)}J(x),\\
            &J^*(x) \to [J^{\prime}(x)]^*= e^{-2i\theta(x)}J^*(x),\\
            & A_\mu(x) \to A_\mu^\prime(x)=A_\mu(x)-\partial_\mu \theta(x),
        \end{align}
    \end{subequations}
    where $\partial_\mu =(i\partial_\tau,\nabla)$. In Eqs.~\eqref{eq_low_spiral: U(1) gauge transf on psis} and \eqref{eq_low_spiral: ext fields U(1) transformation} the spatial coordinate $\bs{x}$ of the spinors $\psi$ and $\psibar$, as well as the sources $J$ and $J^*$ may be a lattice one, while the gauge field $A_\mu$ and the parameter $\theta$ are always defined over a continuous space. To keep the notation lighter, we always indicate the space-time coordinate as $x$, keeping in mind that its spatial component could have a different meaning depending on the field it refers to. 
    
    For $\mathcal{G}$ to be invariant under a U(1) gauge transformation, it must not to depend on $\theta(x)$:
    \begin{equation}
        \frac{\delta}{\delta \theta(x)}\mathcal{G}[A_\mu',J',(J^{\prime})^*]=0.
        \label{eq_low_spiral: dG/dtheta(x) U(1)}
    \end{equation}
    Considering an infinitesimal transformation, that is $|\theta(x)|\ll 1$, from Eqs.~\eqref{eq_low_spiral: ext fields U(1) transformation} and~\eqref{eq_low_spiral: dG/dtheta(x) U(1)}, we obtain 
    \begin{equation}
        \partial_\mu \left(\frac{\delta\mathcal{G}}{\delta A_\mu(x)}\right)+2i\left[\frac{\delta\mathcal{G}}{J(x)}J(x)-\frac{\delta\mathcal{G}}{J^*(x)}J^*(x)\right]=0.
        \label{eq_low_spiral: Ward identity G}
    \end{equation}
    We now consider the change of variables 
    \begin{subequations}
        \begin{align}
            & J(x)=J_{1}(x)+iJ_{2}(x),\\
            & J^*(x)=J_{1}(x)-iJ_{2}(x),
        \end{align}
    \end{subequations}
    such that $J_{1}(x)$ ($J_{2}(x)$) is a source field coupling to longitudinal (transverse) fluctuations of the order parameter, and the functional $\Gamma$, defined as the Legendre transform of $\mathcal{G}$,
    \begin{equation}
        \begin{split}
            \Gamma[A_\mu,\phi_{1},\phi_{2}]=\sum_{a=1,2}&\int_x\phi_{a}(x) J_{a}(x)
            +\mathcal{G}[A_\mu,J_{1},J_{2}],
        \end{split}
    \end{equation}
    where $\phi_{a}(x)=\frac{\delta\mathcal{G}[A_\mu,J_{1},J_{2}]}{\delta J_{a}(x)}$. The gauge kernel can be computed from $\Gamma$ as well:
    \begin{equation}
        K_{\mu\nu}(x,x')=-\frac{\delta^2 \Gamma}{\delta A_\mu(x) \delta A_\nu(x')}\Big\rvert_{\vec{\phi}=A_\mu=0},
    \end{equation}
    because, thanks to the Legendre transform properties, $\delta\Gamma/\delta A_\mu(x) = \delta\mathcal{G}/\delta A_\mu(x)$. Differently, differentiating $\Gamma$ twice with respect to the fields $\phi_{a}$ returns the inverse correlator
    \begin{equation}
        C^{ab}(x,x')=-\frac{\delta^2\Gamma}{\delta\phi_{a}(x)\delta\phi_{b}(x')}\bigg\rvert_{\vec{\phi}=A_\mu=0},
    \end{equation}
    which obeys a reciprocity relation~\cite{NegeleOrland}
    \begin{equation}
        \int_{x^{\prime\prime}}C^{ac}(x,x^{\prime\prime})\chi^{cb}(x^{\prime\prime},x')=\delta_{ab}\delta(x-x'),
        \label{eq_low_spiral: reciprocity relation}
    \end{equation}
    with the generalized susceptibility $\chi^{ab}(x,x')$, defined as
    \begin{equation}
        \chi^{ab}(x,x')=-\frac{\delta^2\mathcal{G}}{\delta J_{a}(x)\delta J_{b}(x')}\bigg\rvert_{J_{a}=A_\mu=0}.
    \end{equation}
    Eq.~\eqref{eq_low_spiral: Ward identity G} can be expressed in terms of $\Gamma$ as
    \begin{equation}
        \begin{split}
            &\partial_\mu \left(\frac{\delta\Gamma}{\delta A_\mu(x)}\right)
            -2\left[\frac{\delta\Gamma}{\delta\phi_{1}(x)}\phi_{2}(x)
            -\frac{\delta\Gamma}{\delta\phi_{2}(x)}\phi_{1}(x)\right]
            =0.
        \end{split}
        \label{eq_low_spiral: Ward identity Gamma}
    \end{equation}
    Eq.~\eqref{eq_low_spiral: Ward identity Gamma} is an identity for the generating functional $\Gamma$ stemming from U(1) gauge invariance of the theory. Taking derivatives with respect to the fields, one can derive an infinite set of Ward identities. 
    
    We are interested in the relation between the gauge kernel and the transverse inverse susceptibility $C^{22}(x,x')$.
    For this purpose, we differentiate Eq.~\eqref{eq_low_spiral: Ward identity Gamma} once with respect to $\phi_{2}(x')$ and once with respect to $A_\nu(x')$, and then set the fields to zero. We obtain the set of equations
    \begin{subequations}
        \begin{align}
            &-\partial_\mu \mathcal{C}_\mu^2(x,x')=2\varphi_0\, C^{22}(x,x'),\label{eq_low_spiral: WI 1}\\ 
            &-\partial_\mu K_{\mu\nu}(x,x')=2\varphi_0\, \mathcal{C}_\nu^{2}(x,x'),
            \label{eq_low_spiral: WI 2}
        \end{align}
    \end{subequations}
    where $\varphi_0=\langle\phi(x)\rangle=\langle\phi_{1}(x)\rangle=\langle\psi_\down(x)\psi_\up(x)\rangle$, and we have defined the quantity
    \begin{equation}
        \mathcal{C}_\mu^{a}(x,x')=-\frac{\delta^2\Gamma}{\delta A_\mu(x)\delta\phi_{a}(x')}\bigg\rvert_{\vec{\phi}=A_\mu=0}.
    \end{equation}
    Combining \eqref{eq_low_spiral: WI 1} and \eqref{eq_low_spiral: WI 2}, we obtain
    \begin{equation}
        \partial_\mu\partial_\nu K_{\mu\nu}(x,x')=4\varphi_0^2\,C^{22}(x,x').
        \label{eq_low_spiral: WI for SC}
    \end{equation}
    Fourier transforming Eq.~\eqref{eq_low_spiral: WI for SC} and rotating to real frequencies, we have
    \begin{equation}
        -q_\mu q_\nu K_{\mu\nu}(q)=4\varphi_0^2 C^{22}(q),
        \label{eq_low_spiral: WI for SC in q space}
    \end{equation}
    with $q=(\bq,\omega)$ a collective variable combining momentum and real frequency. 
    
    
    We now define the superfluid stiffness $J_{\alpha\beta}$ and the uniform density-density susceptibility $\chi_n$\footnote{\label{note_low_spiral: Note_def_stiffnesses}The spin stiffnesses and dynamical susceptibilities (or density-density uniform susceptibility, for the superconductor) can be equivalently defined as the coefficients of a low-energy expansion of the transverse susceptibilities. Here, we choose to define them from the gauge kernels and show that, within a conserving approximation, the two definitions are equivalent.}
    as
    \begin{subequations}
        \begin{align}
            &J_{\alpha\beta}\equiv-\lim_{\bq\to\bzero}K_{\alpha\beta}(\bq,\omega=0) \label{eq_low_spiral: Jsc definition},\\
            &\chi_n\equiv\lim_{\omega\to 0}K_{00}(\bq=\bzero,\omega),\label{eq_low_spiral: den-den chi SC}
        \end{align}
    \end{subequations}
    where the minus sign in \eqref{eq_low_spiral: Jsc definition} has been introduced so that $J_{\alpha\beta}$ is positive definite. Notice that, even though the limits $\bq\to\bzero$ and $\omega\to 0$ in Eq.~\eqref{eq_low_spiral: den-den chi SC} have been taken in the opposite order compared to what it is conventionally done, they commute in a $s$-wave superconductor because of the absence of gapless fermionic excitations. In the above equation and from now on, we employ the convention that the indices labeled as $\mu$, $\nu$ include temporal and spatial components, whereas $\alpha$ and $\beta$ only the latter. Taking the second derivative with respect to $q$ on both sides of \eqref{eq_low_spiral: WI for SC in q space}, we obtain
    \begin{subequations}\label{eq_low_spiral: J and chi from K22 SC}
        \begin{align}
            &J_{\alpha\beta}=2\varphi_0^2  \partial^2_{q_\alpha q_\beta}C^{22}(\bq,\omega=0)\big\rvert_{\bq\to\bzero},\\
            &\chi_n=-2\varphi_0^2  \partial^2_{\omega}C^{22}(\bq=\bzero,\omega)\big\rvert_{\omega\to0},
        \end{align}
    \end{subequations}
    where $\partial^2_{q_\alpha q_\beta}$ and $\partial^2_\omega$ are shorthands for $\frac{\partial^2}{\partial q_\alpha q_\beta}$ and $\frac{\partial^2}{\partial\omega^2} $, respectively. Moreover, we have made use of the Goldstone theorem, reading $C^{22}(\bzero,0)=0$. To derive Eq.~\eqref{eq_low_spiral: J and chi from K22 SC} from \eqref{eq_low_spiral: WI for SC in q space} we have exploited the finiteness of the gauge kernel $K_{\mu\nu}(q)$ in the $\bq\to\bzero$ and $\omega\to 0$ limits. 
    Eq.~\eqref{eq_low_spiral: J and chi from K22 SC} states that the superfluid stiffness and the uniform density-density correlation function are not only the zero momentum and frequency limit of the gauge kernel, but also the coefficients of the inverse transverse susceptibility when expanded for small $\bq$ and $\omega$, respectively. Inverting Eq.~\eqref{eq_low_spiral: reciprocity relation}, $C^{22}(q)$ can be expressed in terms of $\chi^{ab}(q)$ as 
    \begin{equation}
        C^{22}(q)= \frac{1}{\chi^{22}(q)-\chi^{21}(q)\frac{1}{\chi^{11}(q)}\chi^{12}(q)}.
    \end{equation}
    In the limit $q\to 0=(\bzero,0)$, $\chi^{22}(q)$ diverges for the Goldstone theorem, while the second term in the denominator vanishes like some power of $q$. This implies that, for small $q$,
    \begin{equation}
        C^{22}(q)\simeq \frac{1}{\chi^{22}(q)}. 
    \end{equation}
    From this consideration, together with \eqref{eq_low_spiral: J and chi from K22 SC}, we can deduce that the transverse susceptibility can be written as 
    \begin{equation}
        \chi^{22}(\bq,\omega)\simeq \frac{4\varphi_0^2}{-\chi_n \omega^2+J_{\alpha\beta}q_\alpha q_\beta},
        \label{eq_low_spiral: chi22 SC small q}
    \end{equation}
    for small $\bq$ and $\omega$. 
    
    The above form of the $\chi^{22}(q)$ can be also deduced from a low energy theory for the phase fluctuations of the superconducting order parameter. Setting $J$ and $J^*$ to zero in \eqref{eq_low_spiral: G-functional U(1)}, and integrating out the Grassmann fields, one obtains an effective action for the gauge fields. The quadratic contribution in $A_\mu$ is 
    \begin{equation}
        \mathcal{S}_\mathrm{eff}^{(2)}[A_\mu]=-\frac{1}{2}\int_q K_{\mu\nu}(q) A_\mu(-q) A_\nu(q),
    \end{equation}
    where $\int_q$ is a shorthand for $\int \frac{d\omega}{2\pi}\int \frac{d^d \bq}{(2\pi)^d}$. Since we are focusing only on slow and long-wavelength fluctuations of $A_\mu$, we replace $K_{\mu\nu}(q)$ with $K_{\mu\nu}(0)$. Considering a pure gauge field, $A_\mu(x)=-\partial_\mu\theta(x)$, where $\theta(x)$ is (half) the phase of the superconducting order parameter ($\phi(x)=\varphi_0 e^{-2i\theta(x)}$), we obtain 
    \begin{equation}
        \mathcal{S}_\mathrm{eff}[\theta]=\frac{1}{2}\int_x \left\{-\chi_n\left[\partial_t\theta(x)\right]^2+J_{\alpha\beta}\partial_\alpha\theta(x)\partial_\beta\theta(x)\right\},
        \label{eq_low_spiral: phase field action}
    \end{equation}
    with $\theta(x)\in[0,2\pi]$ a periodic field. The above action is well known to display a Berezinskii-Kosterlitz-Thouless (BKT) transition~\cite{Berezinskii1971,Kosterlitz1973} for $d=1$ (at $T=0$) and $d=2$ (at $T>0)$, while for $d=3$ ($T\geq 0$) or $d=2$ ($T=0$), it describes a gapless phase mode known as Anderson-Bogoliubov phonon~\cite{Anderson1958}.
    
    From~\eqref{eq_low_spiral: phase field action}, we can extract the propagator of the field $\theta(x)$
    \begin{equation}
        \langle \theta(-q)\theta(q)\rangle =  \frac{1}{-\chi_n \omega^2+J_{\alpha\beta}q_\alpha q_\beta},
    \end{equation}
    where we have neglected the fact that $\theta(x)$ is defined modulo $2\pi$. 
    Writing $\phi_{2}(x)=(\phi(x)-\phi^*(x))/(2i)=-\varphi_0 \sin(2\theta(x))\simeq -2\varphi_0 \theta(x)$, $\chi^{22}(q)$ can be expressed as
    \begin{equation}
        \begin{split}
            \chi^{22}(q)=\langle \phi_{2}(-q)\phi_{2}(q)\rangle \simeq 4\varphi_0^2 \langle \theta(-q)\theta(q)\rangle
            = \frac{4\varphi_0^2}{-\chi_n \omega^2+J_{\alpha\beta}q_\alpha q_\beta},
        \end{split}
    \end{equation}
    which is in agreement with Eq.~\eqref{eq_low_spiral: chi22 SC small q}. 
    \subsection{SU(2) symmetry}
    In this Section, we repeat the same procedure we have applied in the previous one to derive the Ward identities connected to a SU(2) gauge invariant system. We consider the functional
    \begin{equation}
        \mathcal{G}[A_\mu,\vec{J}]=-\ln \int \!\mathcal{D}\psi\mathcal{D}\psibar e^{-\mathcal{S}\left[\psi,\psibar,A_\mu\right]+(\vec{J},\frac{1}{2}\psibar\vec{\sigma}\psi)},
    \end{equation}
    where $A_\mu(x)=A_\mu^a(x)\frac{\sigma^a}{2}$ is a SU(2) gauge field, $\vec{\sigma}$ are the Pauli matrices, and $\vec{J}(x)$ is a source field coupled to the fermion spin operator $\frac{1}{2}\psibar(x)\vec{\sigma}\psi(x)$. Similarly to the previous section, derivatives of $\mathcal{G}$ with respect to $A_\mu$ and $\vec{J}$ at zero external fields give minus the gauge kernels and spin susceptibilities, respectively. In formulas,
    \begin{subequations}
        \begin{align}
            &\chi^{ab}(x,x')=-\frac{\delta^2\mathcal{G}}{\delta J_{a}(x)\delta J_{b}(x')}\bigg\rvert_{\vec{J}=A_\mu=0},\\
            &K_{\mu\nu}^{ab}(x,x')=-\frac{\delta^2\mathcal{G}}{\delta A^a_\mu(x)\delta A^b_\nu(x')}\bigg\rvert_{\vec{J}=A_\mu=0}.\label{eq_low_spiral: gauge kernel definition}
        \end{align}
    \end{subequations}
    We let the SU(2) symmetry be broken by a (local) order parameter of the form 
    \begin{equation}\label{eq_low_spiral: magnetic order parameter}
        \left\langle \frac{1}{2}\psibar(x)\vec{\sigma}\psi(x) \right\rangle = m \hat{v}(\bs{x}),
    \end{equation}
    with $\hat{v}(\bs{x})$ a position-dependent unit vector pointing along the local direction of the magnetization. 
    
    A SU(2) gauge transformation on the fermionic fields reads
    \begin{subequations}
        \begin{align}
            &\psi(x) \to R(x)\psi(x),\\
            &\psibar(x) \to \psibar(x)R^\dagger(x),
        \end{align}
        \label{eq_low_spiral: SU(2) gauge symm}
    \end{subequations}
    where $R(x)\in\mathrm{SU(2)}$ is a matrix acting on the spin indices of $\psi$ and $\psibar$. The external fields 
    transform as
    \begin{subequations}
        \begin{align}
            J_{a}(x) \to J^{\prime}_a(x)= &\mathcal{R}^{ab}(x)J_{b}(x),\\
            A_\mu(x) \to A_\mu^\prime(x)=&R^\dagger(x)A_\mu(x)R(x)
            +iR^\dagger(x)\partial_\mu R(x), \label{eq_low_spiral: SU(2) Amu transformation}
        \end{align}
    \end{subequations}
    where $\mathcal{R}(x)$ is the adjoint representation of $R(x)$
    \begin{equation}
        \mathcal{R}^{ab}(x)\sigma^b = R(x)\sigma^a R^\dagger(x).
        \label{eq_low_spiral: mathcalR definition}
    \end{equation}
    The SU(2) gauge invariance of $\mathcal{G}$ can be expressed as
    \begin{equation}
        \frac{\delta}{\delta R(x)}\mathcal{G}[A_\mu',\vec{J}^\prime]=0.
    \end{equation}
    Writing $R(x)=e^{i \theta_a(x)\frac{\sigma^a}{2}}$, $R^\dagger(x)=e^{-i \theta_a(x)\frac{\sigma^a}{2}}$, and considering an infinitesimal transformation $|\theta_a(x)|\ll1$, we obtain the functional identity
    \begin{equation}
        \begin{split}
            \partial_\mu \left(\frac{\delta\Gamma}{\delta A_\mu^a(x)}\right)
            -\varepsilon^{a\ell m}\bigg[
            \frac{\delta\Gamma}{\delta\phi^\ell(x)}\phi^m(x)
            -\frac{\delta\Gamma}{\delta A_\mu^\ell(x)}A_\mu^m(x)
            \bigg]
            =0,
        \end{split}
        \label{eq_low_spiral: Ward identity SU(2)}
    \end{equation}
    where $\varepsilon^{abc}$ is the Levi-Civita tensor. $\Gamma[A_\mu,\vec{\phi}]$ is the Legendre transform of $\mathcal{G}$, defined as
    \begin{equation}
        \Gamma[A_\mu,\vec{\phi}]=\int_x\vec{\phi}(x)\cdot\vec{J}(x) + \mathcal{G}[A_\mu,\vec{J}],
    \end{equation}
    with $\phi_{a}(x)=\frac{\delta\mathcal{G}[A_\mu,\vec{J}]}{\delta J_{a}(x)}$. The inverse susceptibilities $C^{ab}(x,x')$, defined as,
    \begin{equation}
        C^{ab}(x,x')=-\frac{\delta^2\Gamma}{\delta\phi_{a}(x)\delta\phi_{b}(x')}\bigg\rvert_{\vec{\phi}=A_\mu=0},
    \end{equation}
    obey a reciprocity relation with the spin susceptibilities $\chi^{ab}(x,x')$ similar to \eqref{eq_low_spiral: reciprocity relation}.
    
    Defining the quantities
    \begin{subequations}
        \begin{align}
            &\mathcal{C}_\mu^{ab}(x,x')=-\frac{\delta^2\Gamma}{\delta A^a_\mu(x)\delta\phi_{b}(x')}\bigg\rvert_{\vec{\phi}=A_\mu=0},\\
            &\mathcal{B}_{\mu}^{a}(x)=-\frac{\delta\Gamma}{\delta A^a_\mu(x)}\bigg\rvert_{\vec{\phi}=A_\mu=0},
        \end{align}    
    \end{subequations}
    we obtain from~\eqref{eq_low_spiral: Ward identity SU(2)} the set of equations
    \begin{subequations}
        \begin{align}
            -\partial_\mu  \mathcal{C}_\mu^{ab}(x,x')&=m \varepsilon^{a\ell m}C^{\ell b}(x,x')v_m(\bs{x}),\label{eq_low_spiral: WI SU(2) I}\\
            -\partial_\mu  K_{\mu\nu}^{ab}(x,x')&=m \varepsilon^{a\ell m}\mathcal{C}^{b\ell}_\nu(x,x')v_m(\bs{x})
            -\varepsilon^{a\ell b}\mathcal{B}_\nu^\ell(x)\delta(x-x')\label{eq_low_spiral: WI SU(2) II},\\
            \partial_\mu  \mathcal{B}_{\mu}^{a}(x)&=0\label{eq_low_spiral: WI SU(2) III},
        \end{align}
    \end{subequations}
    where \eqref{eq_low_spiral: WI SU(2) I}, \eqref{eq_low_spiral: WI SU(2) II}, have been obtained differentiating \eqref{eq_low_spiral: Ward identity SU(2)} with respect to $\phi_b(x')$ and $A_\nu(x')$, respectively, and setting the fields to zero. Eq.~\eqref{eq_low_spiral: WI SU(2) III} simply comes from \eqref{eq_low_spiral: Ward identity SU(2)} computed at zero $A_\mu$, $\phi_{a}$. According to Eq.~\eqref{eq_low_spiral: magnetic order parameter}, the expectation value of $\vec{\phi}(x)$ takes the form $\langle \vec{\phi}(x)\rangle=m \hat{v}(\bs{x})$. Combining \eqref{eq_low_spiral: WI SU(2) I}, \eqref{eq_low_spiral: WI SU(2) II}, and \eqref{eq_low_spiral: WI SU(2) III}, we obtain the Ward identity
    \begin{equation}
        \partial_\mu\partial_\nu K_{\mu\nu}^{ab}(x,x')=m^2 \varepsilon^{a\ell m}\varepsilon^{bnp} v^\ell(
        \bs{x}) v^n(\bs{x}') C^{mp}(x,x'),
        \label{eq_low_spiral: WI gauge Kab SU(2)}
    \end{equation}
    which connects the gauge kernels with the inverse susceptibilities. 
    
    In the following, we analyze two concrete examples where the above identity applies, namely the N\'eel antiferromagnet and the spiral magnet. We do not consider ferromagnets or, in general, systems with a net average magnetization, as in this case the divergence of the transverse components of the kernel $K_{00}^{ab}(q)$ for $q\to 0$ leads to changes in the form of the Ward identities. In this case, one can talk of type-II Goldstone bosons~\cite{Wilson2020}, characterized by a non-linear dispersion. 
    \subsubsection{N\'eel order}
    We now consider the particular case of antiferromagnetic (or N\'eel) ordering for a system on a $d$-dimensional bipartite lattice. In this case $\hat{v}(\bs{x})$ takes the form $(-1)^\bs{x}\hat{v}$, with $(-1)^\bs{x}$ being 1 ($-1$) on the sites of sublattice A (B), and $\hat{v}$ a constant unit vector. In the following, without loss of generality, we consider $\hat{v}=(1,0,0)$. Considering only the diagonal ($a=b$) components of \eqref{eq_low_spiral: WI gauge Kab SU(2)}, we have
    \begin{subequations}
        \begin{align}
            \partial_\mu\partial_\nu  K_{\mu\nu}^{11}(x,x')&=0,\label{eq_low_spiral: WI K11}\\
            \partial_\mu\partial_\nu  K_{\mu\nu}^{22}(x,x')&=m^2 (-1)^{\bs{x}-\bs{x}'} C^{33}(x,x')\label{eq_low_spiral: WI K22},\\
            \partial_\mu\partial_\nu  K_{\mu\nu}^{33}(x,x')&=m^2 (-1)^{\bs{x}-\bs{x}'} C^{22}(x,x')\label{eq_low_spiral: WI K33}.
        \end{align}
    \end{subequations}
    Despite N\'eel antiferromagnetism breaking the lattice translational symmetry, the components of the gauge Kernel considered above depend only on the difference of their arguments $x-x'$, and thus have a well-defined Fourier transform.
    Eq.~\eqref{eq_low_spiral: WI K11} implies $q_\mu q_\nu K_{11}^{\mu\nu}(\bq,\omega)=0$, as expected due to the residual U(1) gauge invariance in the N\'eel state. In particular, one obtains $\lim_{\bq\to\bzero}K^{11}_{\alpha\beta}(\bq,0)=0$, and $\lim_{\omega\to 0}K^{11}_{00}(\bzero,\omega)=0$. Eqs.~\eqref{eq_low_spiral: WI K22} and \eqref{eq_low_spiral: WI K33} are the same equation as we have $K_{22}(x,x')=K_{33}(x,x')$, again because of the residual symmetry. If we rotate them onto the real time axis and perform the Fourier transform, we get
    \begin{subequations}\label{eq_low_spiral: chi and J Neel}
        \begin{align}
            &J_{\alpha\beta}\equiv-\lim_{\bq\to\bzero}K^{22}_{\alpha\beta}(\bq,0)=\frac{1}{2}m^2 \partial^2_{q_\alpha q_\beta} C^{33}(\bq,0)\Big\rvert_{\bq\to\bQ},\label{eq_low_spiral: spin stiffness Neel}\\
            &\chi_\mathrm{dyn}^\perp\equiv\lim_{\omega\to 0}K^{22}_{00}(\bzero,\omega)=-\frac{1}{2}m^2 \partial^2_{\omega} C^{33}(\bQ,\omega)\Big\rvert_{\omega\to 0},\label{eq_low_spiral: chi perp Neel}
        \end{align}
    \end{subequations}
    where $J_{\alpha\beta}$ is the spin stiffness, $\bQ=(\pi/a_0,\dots,\pi/a_0)$, with $a_0$ the lattice spacing, and we name $\chi_\mathrm{dyn}^\perp$ as transverse dynamical susceptibility\footnote{See footnote\ref{note_low_spiral: Note_def_stiffnesses}}. In the above equations we have made use of the Goldstone theorem, which in the present case reads
    \begin{equation}\label{eq_low_spiral: Goldstone Neel}
        C^{22}(\bQ,0)=C^{33}(\bQ,0)=0.
    \end{equation}
    %
    Furthermore, to derive Eq.~\eqref{eq_low_spiral: chi and J Neel} from \eqref{eq_low_spiral: WI K22}, we have used the finiteness of the $\bq\to\bzero$ and $\omega\to 0$ limits of the gauge kernels. 
    Following the argument given in the previous section, for $q=(\bq,\omega)$ close to $Q=(\bQ,0)$, we can replace $C^{33}(q)$ by $1/\chi^{33}(q)$ in \eqref{eq_low_spiral: chi and J Neel}, implying
    \begin{equation}
        \begin{split}
            \chi^{22}(q\simeq Q)=\chi^{33}(q\simeq Q)
            \simeq\frac{m^2}{-\chi_\mathrm{dyn}^\perp \omega^2 + J_{\alpha\beta}(q-Q)_\alpha(q-Q)_\beta}.
        \end{split}
        \label{eq_low_spiral: chi22 and chi33 Neel small q}
    \end{equation}
    Notice that in Eq.~\eqref{eq_low_spiral: chi22 and chi33 Neel small q} we have neglected the imaginary parts of the susceptibilities, that, for doped antiferromagnets, can lead to \emph{Landau damping} of the Goldstone modes~\cite{Bonetti2022}.
    
    Also for N\'eel ordering, form \eqref{eq_low_spiral: chi22 and chi33 Neel small q} of the transverse susceptibilities can be deduced from a low energy theory for the gauge field $A_\mu(x)$, that is,
    \begin{equation}
        \begin{split}
            \mathcal{S}_\mathrm{eff}[A_\mu]=-\frac{1}{2}\int_q 
            \Big[
            K_{00}^{ab}(\bzero,\omega\to0) A^a_0(-q)
            A^b_0(q)
            +K_{\alpha\beta}^{ab}(\bq\to\bzero,0) A^a_\alpha(-q)
            A^b_\beta(q)
            \Big].
        \end{split}
    \end{equation}
    Considering a pure gauge field
    \begin{equation}
        A_\mu(x)=iR^\dagger(x)\partial_\mu R(x),
        \label{eq_low_spiral: pure gauge Amu SU(2)}
    \end{equation}
    with $R(x)$ a SU(2) matrix, we obtain the action
    \begin{equation}
        \mathcal{S}_\mathrm{eff}[\hat{n}]=\frac{1}{2}\int_x \left\{-\chi_\mathrm{dyn}^\perp \left|\partial_t\hat{n}(x)\right|^2+J_{\alpha\beta} \partial_\alpha\hat{n}(x)\cdot\partial_\beta\hat{n}(x)\right\},
        \label{eq_low_spiral: SU(2)/U(1) NLsM}
    \end{equation}
    where $\hat{n}(x)=(-1)^\bs{x}\mathcal{R}(x)\hat{v}(\bs{x})$, with $\mathcal{R}(x)$ defined as in Eq.~\eqref{eq_low_spiral: mathcalR definition}, and $|\hat{n}(x)|^2=1$. Eq.~\eqref{eq_low_spiral: SU(2)/U(1) NLsM} is the well-known $\mathrm{O(3)/O(2)}$ non-linear sigma model (NL$\sigma$M) action, describing low-energy properties of quantum antiferromagnets~\cite{Haldane1983,AuerbachBook1994}.
    
    Writing $R(x)=e^{i\theta_a(x)\frac{\sigma^a}{2}}$, and expanding to first order in $\theta_a(x)$, $\hat{n}(x)$ becomes $\hat{n}(x)\simeq(1,\theta_2(x),-\theta_3(x))$.
    Considering the expression $\vec{\phi}(x)=(-1)^\bs{x} m \hat{n}(x)$ for the order parameter field, we see that small fluctuations in $\hat{n}(x)$ only affect the 2- and 3-components of $\vec{\phi}(x)$. The transverse susceptibilities can be therefore written as
    \begin{equation}
        \begin{split}
            \chi^{22}(q)&=\chi^{33}(q)=\langle \phi_{2}(q)\phi_{2}(-q)\rangle
             \simeq 
            m^2 \langle n_2(q+Q) n_2(-q-Q)\rangle \\
            &= \frac{m^2}{-\chi_\mathrm{dyn}^\perp  \omega^2+J_{\alpha\beta}(q-Q)_\alpha(q-Q)_\beta}, 
        \end{split}
        \label{eq_low_spiral: goldstone chis neel}
    \end{equation}
    which is the result of Eq.~\eqref{eq_low_spiral: chi22 and chi33 Neel small q}. In Eq.~\eqref{eq_low_spiral: goldstone chis neel} we have made use of the propagator of the $\hat{n}$-field dictated by the action of Eq.~\eqref{eq_low_spiral: SU(2)/U(1) NLsM}, that is, 
    \begin{equation}
        \langle n_a(q) n_a(-q)\rangle = \frac{1}{-\chi_\mathrm{dyn}^\perp \omega^2+J_{\alpha\beta}q_\alpha q_\beta}.
    \end{equation}
    Eq.~\eqref{eq_low_spiral: goldstone chis neel} predicts two degenerate magnon branches with linear dispersion for small $\bq-\bQ$. In the case of an isotropic antiferromagnet ($J_{\alpha\beta}=J\delta_{\alpha\beta}$), we have $\omega_\bq = c_s |\bq|$, with the spin wave velocity given by $c_s=\sqrt{J/\chi_\mathrm{dyn}^\perp}$.
    \subsubsection{Spiral magnetic order}
    \label{subsec_low_spiral: WI spiral}
    We now turn our attention to the case of spin spiral ordering, described by the magnetization direction 
    \begin{equation}
        \hat{v}(\bs{x})=\cos(\bQ\cdot\bs{x})\hat{v}_1+\sin(\bQ\cdot\bs{x})\hat{v}_2,
        \label{eq_low_spiral: spiral magnetization}
    \end{equation}
    where $\hat{v}_1$ and $\hat{v}_2$ are two generic constant unit vectors satisfying $\hat{v}_1\cdot\hat{v}_2=0$, and at least one component of $\bQ$ is neither 0 nor $\pi/a_0$. Without loss of generality, we choose $\hat{v}_1=\hat{e}_1=(1,0,0)$ and $\hat{v}_2=\hat{e}_2=(0,1,0)$. It is convenient to rotate the field $\vec{\phi}(x)$ to a basis in which $\hat{v}(\bs{x})$ is uniform. This is achieved by the transformation~\cite{Kampf1996}
    \begin{equation}
        \vec{\phi}^\prime(x) = \mathcal{M}(x) \vec{\phi}(x),
        \label{eq_low_spiral: spiral rotation field}
    \end{equation}
    with
    \begin{equation}
        \mathcal{M}(x)=\left(
        \begin{array}{ccc}
            \cos(\bQ\cdot\bs{x}) & \sin(\bQ\cdot\bs{x}) & 0 \\
            -\sin(\bQ\cdot\bs{x}) & \cos(\bQ\cdot\bs{x}) & 0 \\
            0 & 0 & 1
        \end{array}
        \right). 
        \label{eq_low_spiral: spiral rotation matrix}
    \end{equation}
    In this way, the inverse susceptibilities are transformed into
    \begin{equation}
        \begin{split}
            \widetilde{C}^{ab}(x,x')=&- \frac{\delta^2\Gamma}{\delta \phi^{\prime}_a(x)\delta \phi^{\prime}_b(x')}\bigg\rvert_{\vec{\phi}'=A_\mu=0}
            =[\mathcal{M}^{-1}(x)]^{ac}[\mathcal{M}^{-1}(x')]^{bd} C^{cd}(x,x').
        \end{split}
        \label{eq_low_spiral: rotated Ks spiral}
    \end{equation}
    If we now apply the Ward identity \eqref{eq_low_spiral: WI gauge Kab SU(2)}, we obtain
    \begin{subequations}\label{eq_low_spiral: WI real space spiral}
        \begin{align}
            \partial_\mu\partial_\nu  K_{\mu\nu}^{11}(x,x')&=m^2 \sin(\bQ\cdot\bs{x})\sin(\bQ\cdot\bs{x}') \widetilde{C}^{33}(x,x'),\\
            \partial_\mu\partial_\nu  K_{\mu\nu}^{22}(x,x')&=m^2 \cos(\bQ\cdot\bs{x})\cos(\bQ\cdot\bs{x}') \widetilde{C}^{33}(x,x'),\\
            \partial_\mu\partial_\nu  K_{\mu\nu}^{33}(x,x')&=m^2\widetilde{C}^{22}(x,x'),
        \end{align}
    \end{subequations}
    with 
    \begin{subequations}
        \begin{align}
            \widetilde{C}^{33}(x,x')&=C^{33}(x,x'),\\
            \widetilde{C}^{22}(x,x')&=\sin(\bQ\cdot\bs{x})\sin(\bQ\cdot\bs{x}') C^{11}(x,x')
            +\cos(\bQ\cdot\bs{x})\cos(\bQ\cdot\bs{x}') C^{22}(x,x')\nonumber\\
            &\phantom{=}-\sin(\bQ\cdot\bs{x})\cos(\bQ\cdot\bs{x}')C^{12}(x,x')
            -\cos(\bQ\cdot\bs{x})\sin(\bQ\cdot\bs{x}')C^{21}(x,x').
        \end{align}
    \end{subequations}
    We remark that an order parameter of the type~\eqref{eq_low_spiral: spiral magnetization} completely breaks the SU(2) spin symmetry, which is why none of the right hand sides of the equations above vanishes. We have considered a \emph{coplanar} spiral magnetic order, that is, we have assumed all the spins to lie in the same plane, so that out of the three Goldstone modes, two are degenerate and correspond to out-of-plane fluctuations, and one to in-plane fluctuations of the spins.
    Furthermore, translational invariance is broken, so the Fourier transforms of the gauge kernels $K_{\mu\nu}^{ab}(\bq,\bq',\omega)$ and inverse susceptibilities $C^{ab}(\bq,\bq',\omega)$ are nonzero not only for $\bq-\bq'=\bzero$ but also for $\bq-\bq'=\pm\bQ$ or $\pm2\bQ$. Time translation invariance is preserved, and the gauge kernels and the inverse susceptibilities depend on one single frequency. However, in the basis obtained with transformation~\eqref{eq_low_spiral: spiral rotation matrix}, translational invariance is restored, so that the Fourier transform of $\widetilde{C}^{ab}(x,x')$ only depends on one spatial momentum. With this in mind, we can extract expressions for the spin stiffnesses and dynamical susceptibilities from~\eqref{eq_low_spiral: WI real space spiral}. After rotating to real frequencies, and using the property that for a spiral magnet the gauge kernels are finite in the limits $\bq=\bq'\to\bzero$ and $\omega\to 0$, we obtain\footnote{See footnote\ref{note_low_spiral: Note_def_stiffnesses}}
    \begin{subequations}\label{eq_low_spiral: spin stiff spiral def}
        \begin{align}
            J^{\perp,1}_{\alpha\beta}\equiv&-\lim_{\bq\to\bzero}K^{11}_{\alpha\beta}(\bq,0)
            =\frac{1}{8}m^2\partial^2_{q_\alpha q_\beta}\sum_{\eta=\pm}\widetilde{C}^{33}(\bq+\eta\bQ,0)\bigg\rvert_{\bq\to\bzero},\label{eq_low_spiral: expression J1 spiral}\\
            J^{\perp,2}_{\alpha\beta}\equiv&-\lim_{\bq\to\bzero}K^{22}_{\alpha\beta}(\bq,0)
            =\frac{1}{8}m^2\partial^2_{q_\alpha q_\beta}\sum_{\eta=\pm}\widetilde{C}^{33}(\bq+\eta\bQ,0)\bigg\rvert_{\bq\to\bzero},\label{eq_low_spiral: expression J2 spiral}\\
            J^\smsqr_{\alpha\beta}\equiv&-\lim_{\bq\to\bzero}K^{33}_{\alpha\beta}(\bq,0)
            =\frac{1}{2}m^2\partial^2_{q_\alpha q_\beta}\widetilde{C}^{22}(\bq,0)\bigg\rvert_{\bq\to\bzero},\label{eq_low_spiral: expression J3 spiral}
        \end{align}
    \end{subequations}
    and
    \begin{subequations}\label{eq_low_spiral: Z factors spiral def}
        \begin{align}
            \chi_\mathrm{dyn}^{\perp,1}\equiv&\lim_{\omega\to 0}K^{11}_{00}(\bzero,\omega)
            =-\frac{1}{8}m^2\partial^2_\omega \sum_{\eta=\pm}\widetilde{C}^{33}(\eta\bQ,\omega)\bigg\rvert_{\omega\to0},\label{eq_low_spiral: expression chi1 spiral}\\
            \chi_\mathrm{dyn}^{\perp,2}\equiv&\lim_{\omega\to, 0}K^{22}_{00}(\bzero,\omega)
            =-\frac{1}{8}m^2\partial^2_\omega \sum_{\eta=\pm}\widetilde{C}^{33}(\eta\bQ,\omega)\bigg\rvert_{\omega\to0},\label{eq_low_spiral: expression chi2 spiral}\\
            \chi_\mathrm{dyn}^\smsqr\equiv&\lim_{\omega\to 0}K^{33}_{00}(\bzero,\omega)
            =-\frac{1}{2}m^2\partial^2_\omega \widetilde{C}^{22}(\bzero,\omega)\bigg\rvert_{\omega\to0}\label{eq_low_spiral: expression chi3 spiral},
        \end{align}
    \end{subequations}
    where the labels $\perp$ and $\smsqr$ denote out-of-plane and in-plane quantities, respectively. In the equations above, we have defined $K^{ab}_{\mu\nu}(\bq,\omega)$ as the prefactors of the components of the gauge kernels $K^{ab}_{\mu\nu}(\bq,\bq',\omega)$ which are proportional to $(2\pi)^d\delta^d(\bq-\bq')$.
    From Eqs.~\eqref{eq_low_spiral: spin stiff spiral def} and \eqref{eq_low_spiral: Z factors spiral def} it immediately follows that $J^{\perp,1}_{\alpha\beta}=J^{\perp,2}_{\alpha\beta}\equiv J^{\perp}_{\alpha\beta}$, and $\chi_\mathrm{dyn}^{\perp,1}=\chi_\mathrm{dyn}^{\perp,2}\equiv \chi_\mathrm{dyn}^{\perp}$, as expected in the case of coplanar order~\cite{Azaria1990}. To derive the equations above, we have made use of the Goldstone theorem, which for spiral ordering reads (see for example Refs.~\cite{Chubukov1995,Kampf1996})
    \begin{subequations}
        \begin{align}
            &\widetilde{C}^{33}(\pm\bQ,0)=0,\\
            &\widetilde{C}^{22}(\bzero,0)=0.
        \end{align}
    \end{subequations}
    Notice that the above relations can be also derived from a functional identity similar to~\eqref{eq_low_spiral: Ward identity SU(2)} but descending from the \emph{global} SU(2) symmetry.
    Moreover, close to their respective Goldstone points ($(\bzero,0)$ for $\widetilde{C}^{22}$, and $(\pm\bQ,0)$ for $\widetilde{C}^{33}$), $\widetilde{C}^{22}(q)$ can be replaced with $1/\chit^{22}(q)$, and $\widetilde{C}^{33}(q)$ with $1/\chit^{33}(q)$, with the rotated susceptibilities defined analogously to~\eqref{eq_low_spiral: rotated Ks spiral}.   
    If the spin spiral state occurs on a lattice that preserves parity, we have $\widetilde{C}^{aa}(\bq,\omega)=\widetilde{C}^{aa}(-\bq,\omega)$, from which we obtain
    \begin{subequations}\label{eq_low_spiral: stiffenss and chi spiral final}
        \begin{align}
            &J_{\alpha\beta}^\perp=\frac{1}{4}m^2 \partial^2_{q_\alpha q_\beta}\left( \frac{1}{\chit^{33}(\bq,0)}\right)\bigg\rvert_{\bq\to \pm \bQ},
            \label{eq_low_spiral: stiffenss and chi spiral final J2}\\
            &J^\smsqr_{\alpha\beta}=\frac{1}{2}m^2 \partial^2_{q_\alpha q_\beta}\left( \frac{1}{\chit^{22}(\bq,0)}\right)\bigg\rvert_{\bq\to \bzero},
            \label{eq_low_spiral: stiffenss and chi spiral final J3}\\
            &\chi_\mathrm{dyn}^\perp=-\frac{1}{4}m^2 \partial^2_{\omega}\left( \frac{1}{\chit^{33}(\pm\bQ,\omega)}\right)\bigg\rvert_{\omega\to 0},
            \label{eq_low_spiral: stiffenss and chi spiral final chi2}\\
            &\chi_\mathrm{dyn}^\smsqr=-\frac{1}{2}m^2 \partial^2_{\omega}\left( \frac{1}{\chit^{22}(\bzero,\omega)}\right)\bigg\rvert_{\omega\to 0}.
            \label{eq_low_spiral: stiffenss and chi spiral final chi3}
        \end{align}
    \end{subequations}
    Neglecting the imaginary parts of the susceptibilities, giving rise to dampings of the Goldstone modes~\cite{Bonetti2022}, from Eq.~\eqref{eq_low_spiral: stiffenss and chi spiral final} we can obtain expressions for the susceptibilities near their Goldstone points
    \begin{subequations}\label{eq_low_spiral: low energy chis spiral}
        \begin{align}
            &\chit^{22}(q\simeq(\bzero,0))\simeq \frac{m^2}{-\chi^\smsqr_\mathrm{dyn}\omega^2+J^\smsqr_{\alpha\beta}q_\alpha q_\beta},\\
            &\chit^{33}(q\simeq(\pm\bQ,0))\simeq \frac{m^2/2}{-\chi^\perp_\mathrm{dyn}\omega^2+J_{\alpha\beta}^\perp(q\mp Q)_\alpha (q\mp Q)_\beta}.
        \end{align}
    \end{subequations}

    Expressions \eqref{eq_low_spiral: low energy chis spiral} can be deduced from a low energy model also in the case of spin spiral ordering. Similarly to what we have done for the N\'eel case, we consider a pure gauge field, giving the non-linear sigma model action
    \begin{equation}
        \mathcal{S}_\mathrm{eff}[\mathcal{R}]=\frac{1}{2}\int_x \tr\left[\mathcal{P}_{\mu\nu}\partial_\mu \mathcal{R}(x)\partial_\nu \mathcal{R}^T(x)\right],
        \label{eq_low_spiral: O(3)xO(2)/O(2) NLsM}
    \end{equation}
    where $\mathcal{R}(x)\in\mathrm{SO(3)}$ is defined as in Eq.~\eqref{eq_low_spiral: mathcalR definition}, and now $\partial_\mu$ denotes $(-\partial_t,\vec{\nabla})$. The matrix $\mathcal{P}_{\mu\nu}$ is given by
    \begin{equation}
        \mathcal{P}_{\mu\nu}=
        \left(
        \begin{array}{ccc}
            \frac{1}{2}J^\smsqr_{\mu\nu} & 0 & 0 \\
            0 & \frac{1}{2}J^\smsqr_{\mu\nu} & 0 \\
            0 & 0 & J^\perp_{\mu\nu}-\frac{1}{2}J^\smsqr_{\mu\nu}
        \end{array}
        \right),
    \end{equation}
    with 
    \begin{equation}
        J^a_{\mu\nu}=
        \left(
        \begin{array}{c|c}
            -\chi_\mathrm{dyn}^a & 0 \\ \hline
            0 & J^a_{\alpha\beta}
        \end{array}
        \right),
    \end{equation}
    for $a\in\{\smsqr,\perp\}$.
    Action~\eqref{eq_low_spiral: O(3)xO(2)/O(2) NLsM} is a NL$\sigma$M describing low energy fluctuations around a spiral magnetic ordered state. It has been introduced and studied in the context of frustrated antiferromagnets~\cite{Azaria1990,Azaria1992,Azaria1993,Azaria1993_PRL}.
    
    We now write the field $\vec{\phi}^\prime(x)$ as $\vec{\phi}^\prime(x)=m \mathcal{M}(x)\mathcal{R}(x)\hat{v}(\bs{x})$, and consider an  $\mathcal{R}(x)$ stemming from a SU(2) matrix $R(x)=e^{i\theta_a(x)\frac{\sigma^a}{2}}$ with $\theta_a(x)$ infinitesimal, that is, 
    \begin{equation}
        \mathcal{R}_{ab}(x)\simeq\delta_{ab}-\varepsilon^{abc}\theta_c(x),
        \label{eq_low_spiral: mathcal R small theta}
    \end{equation}
    we get
    \begin{equation}
        \begin{split}
            \vec{\phi}^\prime(x)&\simeq m \mathcal{M}(x)[\hat{v}(\bs{x})-\hat{v}(\bs{x})\times\vec{\theta}(x)]\\
            &=m[\hat{e}_1-\hat{e}_1\times\vec{\theta}^\prime(x)],
        \end{split}
    \end{equation}
    with $\hat{e}_1=(1,0,0)$, and $\vec{\theta}^\prime(x)=\mathcal{M}(x)\vec{\theta}(x)$. Inserting \eqref{eq_low_spiral: mathcal R small theta} into \eqref{eq_low_spiral: O(3)xO(2)/O(2) NLsM}, we obtain 
    \begin{equation}
        \begin{split}
            \mathcal{S}_\mathrm{eff}[\vec{\theta}]=\frac{1}{2}\int_x \bigg\{J^\perp_{\mu\nu}\sum_{a=1,2}\left[\partial_\mu \theta_a(x)\partial_\nu \theta_a(x)\right]
            +J^\smsqr_{\mu\nu}\partial_\mu \theta_3(x)\partial_\nu \theta_3(x)\bigg\}.
        \end{split}
        \label{eq_low_spiral: NLsM spiral linearized}
    \end{equation}
    We are finally in the position to extract the form of the susceptibilities for small fluctuations
    \begin{subequations}\label{eq_low_spiral: low energy susceptibilities}
        \begin{align}
            \chit^{22}(q)=&\langle\phi^{\prime}_2(q)\phi^{\prime}_2(-q)\rangle\simeq m^2\langle\theta^{\prime}_3(q)\theta^{\prime}_3(-q)\rangle
            =\frac{m^2}{-\chi_\mathrm{dyn}^\smsqr\omega^2+J^\smsqr_{\alpha\beta}q_\alpha q_\beta},\\
            \chit^{33}(q)=&\langle\phi^{\prime}_3(q)\phi^{\prime}_3(-q)\rangle\simeq m^2\langle\theta^{\prime}_2(q)\theta^{\prime}_2(-q)\rangle
            =\sum_{\eta=\pm}\frac{m^2/2}{-\chi_\mathrm{dyn}^\perp\omega^2+J_{\alpha\beta}^\perp(q-\eta Q)_\alpha (q-\eta Q)_\beta},
        \end{align}
    \end{subequations}
    which is the result of Eq.~\eqref{eq_low_spiral: low energy chis spiral}. 
    In the above equations we have used the correlators of the $\theta$ field descending from action \eqref{eq_low_spiral: NLsM spiral linearized}. Form~\eqref{eq_low_spiral: low energy chis spiral} of the susceptibilities predicts three linearly dispersing Goldstone modes, two of which (the out-of-plane ones) are degenerate and propagate with velocities $c_\perp^{(n)} = \sqrt{\lambda_\perp^{(n)}/\chi^\perp_\mathrm{dyn}}$, where $\lambda^{(n)}_\perp$ are the eigenvalues of $J_{\alpha\beta}^\perp$ and $n=1,\dots,d$. Similarly, the in-plane mode velocity is given by $c_\smsqr^{(n)}=\sqrt{\lambda_\smsqr^{(n)}/\chi_\smsqr^\mathrm{dyn}}$, with $\lambda_\smsqr^{(n)}$ the eigenvalues of $J_{\alpha\beta}^\smsqr$.
    \section{Mean-field treatment of the spiral magnetic state}
    \label{sec_low_spiral: MF spiral}
    For this chapter to be self-contained, we repeat here the some of the basic concepts of spiral magnetism already presented in Chapter~\ref{chap: spiral DMFT}.
    
    A spiral magnetic state is characterized by an average magnetization lying in a plane, which, by rotational invariance, can have \emph{any} orientation. Without loss of generality, we choose it to be the $xy$-plane. We therefore express the expectation value of the spin operator as
    \begin{equation}
        \langle \vec{S}_j \rangle = m \left[\cos(\bQ\cdot\bs{R}_j)\hat{e}_1+\sin(\bQ\cdot\bs{R}_j)\hat{e}_2\right],
        \label{eq_low_spiral: spiral magn}
    \end{equation}
    where $m$ is the magnetization amplitude, $\bs{R}_j$ is the spatial coordinate of lattice site $j$, $\hat{e}_a$ is a unit vector pointing along the $a$-direction, and $\bQ$ is a fixed wave vector. In an itinerant electron system, the spin operator is given by
    \begin{equation}
        S^a_j = \frac{1}{2}\sum_{s,s'=\up,\down}c^\dagger_{j,s}\PauliMat^a_{ss'}c_{j,s'},
        \label{eq_low_spiral: spin operator}
    \end{equation}
    where $\sigma^a$ ($a=1,2,3$) are the Pauli matrices, and $c^\dagger_{j,s}$ ($c_{j,s}$) creates (annihilates) an electron at site $j$ with spin projection $s$. Fourier transforming Eq.~\eqref{eq_low_spiral: spiral magn}, we find that the magnetization amplitude is given by the momentum integral
    \begin{equation}
        \int_\bk \langle c^\dagger_{\bk,\up} c_{\bk+\bQ,\down}\rangle,
        \label{eq_low_spiral: spiral magn k-space}
    \end{equation}
    where $c^\dagger_{\bk,\sigma}$ ($c_{\bk,\sigma}$) is the Fourier transform of $c^\dagger_{j,s}$ ($c_{j,s}$), $\int_\bk=\int\frac{d^d\bk}{(2\pi)^d}$ denotes a $d$-dimensional momentum integral, with $d$ the system dimensionality. From Eq.~\eqref{eq_low_spiral: spiral magn k-space}, we deduce that spiral magnetism only couples the electron states $(\bk,\up)$ and $(\bk+\bQ,\down)$. It is therefore convenient to use a rotated spin reference frame~\cite{Kampf1996}, corresponding to transformation
    \begin{equation}
         \tilde c_j =
         e^{-\frac{i}{2}\bQ\cdot\bs{R}_j} e^{\frac{i}{2}\bQ\cdot\bs{R}_j \sigma^3} c_j \, , \quad
         \tilde c_j^\dagger =
         c_j^\dagger \, e^{-\frac{i}{2}\bQ\cdot\bs{R}_j \sigma^3} e^{\frac{i}{2}\bQ\cdot\bs{R}_j}.
         \label{eq_low_spiral: ctildes}
    \end{equation}
    In this basis, the Fourier transform of the spinor $\tilde{c}_j$ is given by $\tilde{c}_\bk=(c_{\bk,\up},c_{\bk+\bQ,\down})$, and the magnetization~\eqref{eq_low_spiral: spiral magn} points along the $\hat{e}_1$ axis:
    \begin{equation}
        \langle\widetilde{S}^\alpha_j\rangle =
        \frac{1}{2} \big\langle \tilde c^\dagger_j \sigma^a \tilde c_j \big\rangle =
        m \delta_{a,1}.
    \end{equation}
    With the help of transformation~\eqref{eq_low_spiral: ctildes}, we can express the mean-field Green's function in Matsubara frequencies as
    \begin{equation} 
        \widetilde{\mathbf{G}}_\bk(i\nu_n) = 
        \left( \begin{array}{cc}
        i\nu_n-\xi_{\bk} & - \Delta \\ - \Delta & i\nu_n-\xi_{\bk+\bQ}
        \end{array} \right)^{-1} ,
        \label{eq_low_spiral: matrix Gtilde}
    \end{equation}
    where $\nu_n=(2n+1)\pi T$, $\xi_\bk = \eps_\bk - \mu$, with the single-particle dispersion $\eps_\bk$ and the chemical potential $\mu$, while $\Delta$ is the magnetic gap associated with the spiral order. Diagonalizing~\eqref{eq_low_spiral: matrix Gtilde}, one obtains the quasiparticle bands
    \begin{equation}
         E_{\bk}^\pm = g_{\bk} \pm \sqrt{h_{\bk}^2 + \Delta^2},
         \label{eq_low_spiral: QP dispersions}
    \end{equation}
    where $g_\bk = \frac{1}{2}(\xi_\bk + \xi_{\bk+\bQ})$ and
    $h_\bk = \frac{1}{2}(\xi_\bk - \xi_{\bk+\bQ})$. It is convenient to express the Green's function~\eqref{eq_low_spiral: matrix Gtilde} as 
    \begin{equation}
        \widetilde G_\bk(i\nu_n) = \frac{1}{2} \sum_{\ell=\pm} 
        \frac{u^\ell_{\bk}}{i\nu_n-E^\ell_{\bk}},
        \label{eq_low_spiral: spiral Gf comf}
    \end{equation}
    with the coefficients
    \begin{equation} 
         u_\bk^\ell = \sigma^0 + \ell \, \frac{h_\bk}{e_\bk} \sigma^3 +
         \ell \, \frac{\Delta}{e_\bk} \sigma^1, 
         \label{eq_low_spiral: ukl def}
    \end{equation}
    where $\sigma^0$ is the $2\times2$ unit matrix and $e_\bk = \sqrt{h_\bk^2 + \Delta^2}$.
    
    We assume the spiral states to emerge from a lattice model with onsite repulsive interactions (Hubbard model), with imaginary time action
    \begin{equation}\label{eq_low_spiral: Hubbard action}
        \begin{split}
            \mathcal{S}[\psi,\psibar]=\int_0^\beta\!d\tau\left\{\sum_{j,j',\sigma}\psibar_{j,\sigma}\left[(\partial_\tau - \mu)\delta_{jj'} + t_{jj'}\right]\psi_{j',\sigma}
            + U\sum_{j}\psibar_{j,\up}\psibar_{j,\down}\psi_{j,\down}\psi_{j,\up}\right\},
        \end{split}
    \end{equation}
    where $t_{jj'}$ describes the hopping amplitude between the lattice sites labeled by $j$ and $j'$ and $U$ is the Hubbard interaction. 
    The Hartree-Fock or mean-field (MF) gap equation at temperature $T$ reads
    \begin{equation}
         \Delta = - U \int_\bk T\sum_\nu \widetilde G^{\up\down}_\bk(\nu) =
          U \int_\bk \frac{\Delta}{2e_\bk} \left[f(E^-_\bk)-f(E^+_\bk)\right] \, ,
          \label{eq_low_spiral: gap equation}
    \end{equation}
    where $f(x)=(e^{x/T}+1)^{-1}$ is the Fermi function, the magnetization amplitude is related to $\Delta$ via $\Delta = Um$.
    
    Finally, the optimal $\bQ$-vector is obtained minimizing the mean-field free energy
    \begin{equation}\label{eq_low_spiral: MF free energy}
        \begin{split}
            \mathcal{F}_\mathrm{MF}(\bQ)=&-T\sum_{\nu_n}\int_\bk \Tr\ln \widetilde{\mathbf{G}}_\bk(i\nu_n)+\frac{\Delta^2}{U}+\mu n \\
            =&-T\int_\bk \sum_{\ell=\pm}\ln\left(1+e^{-E^\ell_\bk/T}\right)+\frac{\Delta^2}{U}+\mu n, 
        \end{split} 
    \end{equation}
    where $n$ is the fermion density.
    \section{Susceptibilities and Goldstone modes}
    \label{sec_low_spiral: RPA spiral}
    In spin spiral state, the spin and charge susceptibilities are coupled. It is therefore convenient to treat them on equal footing by extending the definition of the spin operator in Eq.~\eqref{eq_low_spiral: spin operator} to
    \begin{equation}
        S^a_j = \frac{1}{2}\sum_{s,s'=\up,\down}c^\dagger_{j,s}\PauliMat^a_{ss'}c_{j,s'},
    \end{equation}
    where now $a$ runs from 0 to 3, with $\sigma^0$ the unit 2$\times$2 matrix. It is evident that for $a=1,2,3$ we recover the usual spin operator, while $a=0$ gives half the density. 
    We then consider the imaginary-time susceptibility 
    \begin{equation}
        \chi^{ab}_{jj'}(\tau) = \langle {\cal T} S_j^a(\tau) S_{j'}^b(0) \rangle,
    \end{equation}
    where $\mathcal{T}$ denotes time ordering. Fourier transforming to Matsubara frequency representation and analytically continuing to the real frequency axis, we obtain the retarded susceptibility $\chi^{ab}_{jj'}(\omega)$. As previously mentioned, $\chi^{ab}_{jj'}(\omega)$ is not invariant under spatial translations. It is therefore convenient to compute the susceptibilities in the rotated reference frame~\cite{Kampf1996} of Eq.~\eqref{eq_low_spiral: spiral rotation matrix}
    \begin{equation}
        \chit_{jj'}(\omega)=\mathcal{M}_j \chi_{jj'}(\omega) \mathcal{M}^T_{j'},
        \label{eq_low_spiral: rotated chis spiral RPA}
    \end{equation}
    where in this case we have
    \begin{equation}
        \mathcal{M}_j=\left(
        \begin{array}{cccc}
            1 & 0 & 0 & 0\\
            0 & \cos(\bQ\cdot\bs{R}_j) & \sin(\bQ\cdot\bs{R}_j) & 0 \\
            0 & -\sin(\bQ\cdot\bs{R}_j) & \cos(\bQ\cdot\bs{R}_j) & 0 \\
            0 & 0 & 0 & 1
        \end{array}
        \right).
        \label{eq_low_spiral: spiral rotation matrix with charge}
    \end{equation}
    The physical susceptibilities $\chi_{jj'}(\omega)$ can be obtained inverting Eq.~\eqref{eq_low_spiral: rotated chis spiral RPA}. Their momentum representation typically involves two distinct spatial momenta $\bq$ and $\bq'$, where $\bq'$ can equal $\bq$, $\bq\pm\bQ$ (only for $a\neq b$), or $\bq\pm 2\bQ$. Inverting Eq.~\eqref{eq_low_spiral: rotated chis spiral RPA} and Fourier transforming, we obtain the following relations between the momentum and spin diagonal components of the physical susceptibilities and those within the rotated reference frame
    \begin{subequations}
        \begin{align}
            \chi^{00}(\bq,\bq,\omega) &= \chit^{00}(\bq,\omega) \\
            \chi^{11}(\bq,\bq,\omega) &= \chi^{22}(\bq,\bq,\omega) \nonumber \\
            &= \frac{1}{4} \big[
            \chit^{11}(\bq+\bQ,\omega) + \chit^{11}(\bq-\bQ,\omega) +
            \chit^{22}(\bq+\bQ,\omega) + \chit^{22}(\bq-\bQ,\omega) \nonumber \\
            & \quad + 2i \, \chit^{12}(\bq+\bQ,\omega) +
                2i \, \chit^{21}(\bq-\bQ,\omega) \big] \nonumber \\
            &= \chit^{-+}(\bq+\bQ,\omega) + \chit^{+-}(\bq-\bQ,\omega) \, ,
            \label{eq_low_spiral: chi11phys} \\
             \chi^{33}(\bq,\bq,\omega) &= \chit^{33}(\bq,\omega) \, ,
        \end{align}
        \label{eq_low_spiral: chi physical mom diagonal}
    \end{subequations}
    where we have used $\chit^{21}(q)=-\chit^{12}(q)$ (see Table \ref{tab_low_spiral: symmetries}), and we have defined 
    \begin{equation}
        \chit^{+-}(\bq,\omega)=\langle \widetilde{S}^+_{-\bq,-\omega}\widetilde{S}^-_{\bq,\omega}\rangle,
    \end{equation}
    with $\widetilde{S}^\pm=(\widetilde{S}^1\pm i\widetilde{S}^2)/2$. For $a=b$ the only momentum off-diagonal components are given by
    \begin{subequations}
        \begin{align}
            \chi^{11}(\bq,\bq \pm 2\bQ,\omega) &=\frac{1}{4} \left[
            \chit^{11}(\bq \mp \bQ,\omega) - \chit^{22}(\bq \mp \bQ,\omega) \right] , \\
            \chi^{22}(\bq,\bq \pm 2\bQ,\omega) &=\frac{1}{4} \left[
            \chit^{22}(\bq \mp \bQ,\omega) - \chit^{11}(\bq \mp \bQ,\omega) \right] .
        \end{align}
        \label{eq_low_spiral: chi physical mom off-diagonal}
    \end{subequations}
    Here, with a slight abuse of notation, we denoted as $\chi^{aa}(\bq,\bq,\omega)$ and $\chi^{aa}(\bq,\bq\pm2\bQ,\omega)$ the prefactors of $(2\pi)^d\delta^d(\bq-\bq')$ and $(2\pi)^d\delta^d(\bq-\bq'\mp2\bQ)$, respectively, in the contributions to the full susceptibilities $\chi^{aa}(\bq,\bq',\omega)$. 
    In the spacial case of N\'eel ordering, the observation $2\bQ\simeq \bzero$ combined with Eqs.~\eqref{eq_low_spiral: chi physical mom diagonal} and \eqref{eq_low_spiral: chi physical mom off-diagonal}, implies $\chi^{11}(\bq,\bq,\omega)\neq \chi^{22}(\bq,\bq,\omega)$, differently than for the spiral state, where $2\bQ\neq\bzero$. 
    
    Within the random phase approximation, the susceptibilities of the Hubbard model are given by
    \begin{equation} \label{eq_low_spiral: chi RPA}
        \chit(q)=\chit_0(q)[\mathbb{1}_4-\Gamma_0\chit_0(q)]^{-1},
    \end{equation}
    where $\mathbb{1}$ is the $4\times 4$ unit matrix, and $\Gamma_0=2U\mathrm{diag}(-1,1,1,1)$ is the bare interaction. The bare bubbles on the real frequency axis are given by 
    \begin{equation}
         \begin{split}
             &\chit^{ab}_0(\bq,\omega) =
             - \frac{1}{4} \int_\bk T \sum_{\nu_n} \tr \big[ 
            \sigma^a\,\widetilde{\mathbf{G}}_\bk(i\nu_n) \,
            \sigma^b\,\widetilde{\mathbf{G}}_{\bk+\bq}(i\nu_n+i\Omega_m) \big] \Big\rvert_{i\Omega_m\to\omega+i0^+},
         \end{split}
         \label{eq_low_spiral: chi0 expression}
    \end{equation}
    where $\Omega_m=2m\pi T$ denotes a bosonic Matsubara frequency. Using \eqref{eq_low_spiral: spiral Gf comf}, one can perform the frequency sum, obtaining
    \begin{equation}
         \chit^{ab}_0(\bq,\omega) = - \frac{1}{8}\sum_{\ell,\ell'=\pm}\int_\bk \mathcal{A}^{ab}_{\ell\ell'}(\bk,\bq)F_{\ell\ell'}(\bk,\bq,\omega),
         \label{eq_low_spiral: chi0 def}
    \end{equation}
    where we have defined 
    \begin{equation}
        F_{\ell\ell'}(\bk,\bq,\omega)=\frac{f(E^\ell_\bk)-f(E^{\ell'}_{\bk+\bq})}{\omega+i0^++E^\ell_\bk-E^{\ell'}_{\bk+\bq}},
        \label{eq_low_spiral: Fll def}
    \end{equation}
    and the coherence factors
    \begin{equation}
        \mathcal{A}^{ab}_{\ell\ell'}(\bk,\bq)=\frac{1}{2}\Tr\left[\sigma^a u^\ell_\bk\sigma^b u^{\ell'}_{\bk+\bq}\right].
        \label{eq_low_spiral: coh fact def}
    \end{equation}
    The coherence factors are either purely real or purely imaginary, depending on $a$ and $b$. The functions $F_{\ell\ell'}(\bk,\bq,\omega)$ have a real part and an imaginary part proportional to a $\delta$-function. To distinguish the corresponding contributions to $\chit^{ab}_0(\bq,\omega)$, we refer to the contribution coming from the real part of $F_{\ell\ell'}(\bk,\bq,\omega)$ as $\chit^{ab}_{0r}(\bq,\omega)$, and to the contribution from the imaginary part of $F_{\ell\ell'}(\bk,\bq,\omega)$ as $\chit^{ab}_{0i}(\bq,\omega)$. Note that $\chit^{ab}_{0r}(\bq,\omega)$ is imaginary and $\chit^{ab}_{0i}(\bq,\omega)$ is real if the corresponding coherence factor is imaginary. 
    \subsection{Symmetries of the bare susceptibilities}
    Both contributions $\chit_{0r}^{ab}$ and $\chit_{0i}^{ab}$ to $\chit_0^{ab}$ have a well defined parity under $\bq \to -\bq$. In Appendix~\ref{app: low en spiral} we show that the diagonal components of $\chit_{0r}^{ab}$ and the off-diagonal ones which do not involve either the 2- or the 3-component of the spin are symmetric, while the other off-diagonal elements are antisymmetric.
    The sign change of $\chit_{0i}^{ab}(q)$ under $\bq \to -\bq$ is the opposite, that is, $\chit_{0i}^{ab}(q)$ is antisymmetric if $\chit_{0r}^{ab}(q)$ is symmetric and vice versa.
    In two spatial dimensions, for a spiral wave vector $\bQ$ of the form $(\pi-2\pi\eta,\pi)$ all the susceptibilities are symmetric under $q_y \to -q_y$. This implies that those susceptibilities which are antisymmetric for $\bq \to -\bq$ are identically zero for $q_y=0$, and vanish in the limit of N\'eel order ($\eta \to 0$). Similarly, for a diagonal spiral $\bQ = (\pi-2\pi\eta,\pi-2\pi\eta)$ all the susceptibilities are symmetric for $q_x \leftrightarrow q_y$ and those which are antisymmetric in $\bq$ vanish for $q_x = q_y$.
    
    The contributions $\chit_{0r}^{ab}$ and $\chit_{0i}^{ab}$ to $\chit_0^{ab}$ are also either symmetric or antisymmetric under the transformation $\omega \to -\omega$.
    In Appendix~\ref{app: low en spiral} we show that among the functions $\chit_{0r}^{ab}$ all the diagonal parts and the off-diagonal ones which do not involve the 3-component of the spin are symmetric in $\omega$. The off-diagonal terms involving the 3-component of the spin are antisymmetric. $\chit_{0i}^{ab}(q)$ is antisymmetric under $\omega \to -\omega$ if $\chit_{0r}^{ab}(q)$ is symmetric and vice versa.
    
    In Table \ref{tab_low_spiral: symmetries} we show a summary of the generic (for arbitrary $\bQ$) symmetries of the bare susceptibilities. Susceptibilities with real (imaginary) coherence factors are symmetric (antisymmetric) under the exchange $a \leftrightarrow b$.
    \begin{table}[ht!]
    \centering
    \begin{tabular}{|c|c|c|c|c|}
            \hline
            $a,b$ & 0 & 1 & 2 & 3  \\
            \hline
            0 & $+,+,+$ & $+,+,+$ & $-,+,-$ & $-,-,+$ \\
            \hline
            1 & $+,+,+$ & $+,+,+$ & $-,+,-$ & $-,-,+$ \\
            \hline
            2 & $-,+,-$ & $-,+,-$ & $+,+,+$ & $+,-,-$ \\
            \hline
            3 & $-,-,+$ & $-,-,+$ & $+,-,-$ & $+,+,+$ \\
            \hline
    \end{tabular}
    \caption{Symmetries of the bare susceptibilities. The first sign in each field represents the sign change of $\chit_{0r}^{ab}(q)$ under $\bq \to -\bq$. The second one represents the sign change of ${\chit_{0r}^{ab}}(q)$ under $\omega \to -\omega$. The sign changes of ${\chit_{0i}^{ab}}(q)$ under $\bq \to -\bq$ or $\omega \to -\omega$ are just the opposite. The third sign in each field is the sign change of $\chit_0^{ab}(q)$ under the exchange $a \leftrightarrow b$.}
    \label{tab_low_spiral: symmetries}
    \end{table}
    
    \subsection{Location of Goldstone modes}
    We now identify the location of Goldstone modes in the spiral magnet by analyzing divergences of the rotated susceptibilities $\chit(q)$.
    \subsubsection{In-plane mode}
    For $q=0=(\bzero,0)$, all the off-diagonal components of the bare bubbles $\chit_0(q)$ involving the 2-component of the spin vanish: $\chit_0^{20}(q)$ and $\chit_0^{21}(q)$ because they are odd in momentum, and $\chit_0^{23}(q)$ because it is antisymmetric for $\omega\to-\omega$. We also remark that all the $\chit_{0i}^{ab}$ vanish at zero frequency. The RPA expression for the 22-component of the rotated susceptibility therefore takes the simple form
    \begin{equation}
         \chit^{22}(0) =
         \frac{\chit^{22}_0(0)}{1 - 2 U\chit^{22}_0(0)}.
         \label{eq_low_spiral: chi22(0) RPA}
    \end{equation}
    Notice that the limits $\bq\to\bzero$ and $\omega\to0$ commute for $\chit_0^{22}$ as the intraband coherence factor $A^{22}_{\ell\ell}(\bk,\bq)$ vanishes for $\bq=\bzero$ (see Appendix~\ref{app: low en spiral}). Eq.~\eqref{eq_low_spiral: chi0 def} yields
    \begin{equation}
        \chit_0^{22}(0)=\int_\bk \frac{f(E^-_\bk)-f(E^+_\bk)}{4e_\bk}.
    \end{equation}
    The denominator of Eq.~\eqref{eq_low_spiral: chi22(0) RPA} vanishes if the gap equation~\eqref{eq_low_spiral: gap equation} is fulfilled. Thus, $\chit^{22}(0)$ is divergent. From Eq.~\eqref{eq_low_spiral: chi11phys}, we see that this makes the momentum diagonal part of the physical susceptibilities $\chi^{11}(\bq,\bq,0)$ and $\chi^{22}(\bq,\bq,0)$ divergent at $\bq=\pm \bQ$. These divergences are associated with a massless Goldstone mode corresponding to fluctuations of the spins within the $xy$ plane~\cite{Chubukov1995}, in which the magnetization is aligned. By contrast, $\chit^{11}(\bq,0)$ is always finite and corresponds to a massive amplitude mode. 
    \subsubsection{Out-of-plane modes}
    By letting $\omega\to 0$, all the off-diagonal components of the bare susceptibilities involving the 3-component of the spin vanish as they are odd in $\omega$. Hence, we can express $\chit^{33}(\bq,0)$ as
    \begin{equation}
         \chit^{33}(\bq,0) =
         \frac{\chit^{33}_0(\bq,0)}{1 - 2 U \chit^{33}_0(\bq,0)} .
          \label{eq_low_spiral: chi33(q,0)}
    \end{equation}
    In Appendix~\ref{app: low en spiral}, we show that
    \begin{equation}
         \chit_0^{33}(\pm\bQ,0) = \int_\bk \frac{f(E^-_\bk)-f(E^+_\bk)}{4e_\bk} =
         \chit_0^{22}(0) ,
          \label{eq_low_spiral: chi33_0(q,0)}
    \end{equation}
    so that the denominator of \eqref{eq_low_spiral: chi33(q,0)} vanishes if the gap equation is fulfilled. Therefore, the 33-component of the susceptibility is divergent at $q=(\pm \bQ,0)$ due to two degenerate Goldstone modes corresponding to fluctuations of the spins out of the $xy$ plane~\cite{Chubukov1995}.
    \section{Properties of the Goldstone modes}
    \label{sec_low_spiral: properties of Goldstones}
    As we have already anticipated in Sec.~\ref{subsec_low_spiral: WI spiral}, the susceptibilities containing a Goldstone mode can be expanded around their zero-energy pole as (cf.~Eq.~\eqref{eq_low_spiral: low energy susceptibilities})
    \begin{subequations}
        \begin{align}
            &\chit^{22}(\bq,\omega) \sim \frac{m^2}
            {J^{\smsqr}_{\alpha\beta} \, q_\alpha q_\beta
            - \chi_\mathrm{dyn}^{\smsqr} \,\omega^2 + iD^\smsqr(\bq,\omega)} \, ,\\
            &\chit^{33}(\bq,\omega) \sim \frac{m^2/2}
            {J^{\perp}_{\alpha\beta} \, \big( q_\alpha \mp Q_{\alpha} \big)
            \big( q_\beta\mp Q_{\beta} \big)
            - \chi_\mathrm{dyn}^{\perp} \,\omega^2 + iD^\perp(\bq,\omega)} \, ,
        \end{align}
    \end{subequations}
    where $m$ is the magnetization amplitude as defined before, $J^{a}_{\alpha\beta}$ ($a\in\{\smsqr,\perp\}$) are the spin stiffnesses, and $\chi_\mathrm{dyn}^{a}$ the dynamical susceptibilities. The ratios $m^2/\chi_\mathrm{dyn}^{a}$ define the spectral weights of the Goldstone modes and $[J^{a}_{\alpha\beta}/\chi_\mathrm{dyn}^{a}]^{1/2}$ their velocity tensors. Compared with Eq.~\eqref{eq_low_spiral: low energy susceptibilities}, we have also considered an imaginary part $iD^{a}(\bq,\omega)$ in the denominator of the susceptibilities, due to \emph{Landau damping} of the collective excitations due to their decay into particle-hole pairs, which has been instead neglected in Sec.~\ref{subsec_low_spiral: WI spiral}. The structure of this term will be discussed below. 
    
    In the following we will discuss how to extract the spin stiffnesses, dynamical susceptibilities and Landau dampings from the RPA expressions for $\chit^{22}(\bq,\omega)$ and $\chit^{33}(\bq,\omega)$.
    \subsection{In-plane mode}
    Using \eqref{eq_low_spiral: chi RPA}, the in-plane susceptibility can be conveniently written as 
    \begin{equation}
        \chit^{22}(q)=\frac{\overline{\chi}_0^{22}(q)}{1-2U\overline{\chi}_0^{22}(q)}, 
        \label{eq_low_spiral: chi22 RPA}
    \end{equation}
    with
    \begin{equation}
        \overline{\chi}_0^{22}(q)=\chit_0^{22}(q) + \sum_{a,b\in\{0,1,3\}}\chit_0^{2a}(q)\Gammat_{2}^{ab}(q)\chit_0^{b2}(q).
    \end{equation}
    $\Gammat_{2}(q)$ is given by
    \begin{equation}
        \Gammat_{2}(q)=\left[\mathbb{1}_3-\Gamma_{0,2}\chit_{0,2}(q)\right]^{-1}\Gamma_{0,2},
    \end{equation}
    where $\Gamma_{0,2}^{ab}$ and $\chit_{0,2}^{ab}(q)$ are matrices obtained from $\Gamma_0^{ab}$ and $\chit_0^{ab}(q)$ removing the components where $a=2$ and/or $b=2$, and $\mathbb{1}_3$ denotes the $3\times3$ identity matrix.
    For later convenience, we notice that for $q=0$, all the off-diagonal elements $\chit_0^{2a}(q)$ and $\chit_0^{a2}(q)$ vanish, so that $\Gammat_{2}(0)$ can be obtained from the full expression
    \begin{equation}
        \Gammat(q)=\left[\mathbb{1}-\Gamma_{0}\chit_{0}(q)\right]^{-1}\Gamma_{0},
        \label{eq_low_spiral: Gamma(q) definition}
    \end{equation}
    selecting only the components in which the indices take the values 0,1, or 3. 
    
    \subsubsection{Spin stiffness}
    Setting $\omega=0$, the bare susceptibilities $\chit_0^{23}(q)$ and $\chit_0^{32}(q)$ vanish as they are odd in $\omega$. Moreover, in the limit $\bq\to\bzero$, $\chit_0^{2a}(\bq,0)$ and $\chit_0^{a2}(\bq,0)$, with $a=0,1$, are linear in $\bq$ as they are odd under $\bq\to-\bq$. The in-plane spin stiffness can be therefore written as (cf.~Eq.~\eqref{eq_low_spiral: stiffenss and chi spiral final J3})
    \begin{equation}
        \begin{split}
            J^\smsqr_{\alpha\beta}=&
            -2\Delta^2\partial^2_{q_\alpha q_\beta}\overline{\chi}_0^{22}(0)\\
            =&-2\Delta^2\bigg[
            \partial^2_{q_\alpha q_\beta}\chit_0^{22}(0)+
            2\sum_{a,b\in\{0,1\}}\partial_{q_\alpha}\chit_0^{2a}(0)\,\Gammat^{ab}(\bq\to\bzero,0)\,\partial_{q_\alpha}\chit_0^{b2}(0)
            \bigg],
        \end{split}
        \label{eq_low_spiral: J3 RPA}
    \end{equation}
    where we have used $\overline{\chi}_0^{22}(0)=\chit_0^{22}(0)=1/(2U)$, descending from the gap equation, and $\partial_{q_\alpha}f(0)$ is a shorthand for $\partial f(\bq,0)/\partial q_\alpha |_{\bq\to\bzero}$, and similarly for $\partial^2_{q_\alpha q_\beta}f(0)$.
    
    \subsubsection{Dynamical susceptibility}
    In a similar way, if we set $\bq$ to $\bzero$ and consider the limit of small $\omega$, the terms where $a$ and/or $b$ are 0 or 1 vanish as $\chit_0^{2a}(q)$ and $\chit_0^{a2}(q)$ for a=0,1 are odd in $\bq$. On the other hand, $\chit_0^{23}(q)$ and $\chit_0^{32}(q)$ are linear in $\omega$ for small $\omega$. With these considerations, the in-plane dynamical susceptibility is given by (see Eq.~\eqref{eq_low_spiral: stiffenss and chi spiral final chi3})
    \begin{equation}
        \begin{split}
            \chi_\mathrm{dyn}^\smsqr=
            &2\Delta^2\partial^2_{\omega}\overline{\chi}_0^{22}(0)\\
            =&2\Delta^2\Big[
            \partial^2_{\omega}\chit_0^{22}(0)
            +2\partial_{\omega}\chit_0^{23}(0)\,\Gammat^{33}(\bzero,\omega\to 0)\,\partial_{\omega}\chit_0^{32}(0)
            \Big],
        \end{split}
        \label{eq_low_spiral: chi3 RPA}
    \end{equation}
    where $\partial^n_{\omega}f(0)$ is a shorthand for $\partial^n f(\bzero,\omega)/\partial \omega^n |_{\omega\to 0}$, and $\Gammat^{33}(\bzero,\omega\to 0)$ can be cast in the simple form
    \begin{equation}
        \Gammat^{33}(\bzero,\omega\to 0)=\frac{2U}{1-2U \chit_0^{33}(\bzero,\omega\to 0)}.
    \end{equation}
    \subsubsection{Landau damping}
    We now analyze the leading imaginary term, describing the damping of the in-plane Goldstone mode for small $\bq$ and $\omega$. Imaginary contributions to the bare susceptibilities arise from the $\delta$-function term in $\chit_{0i}^{ab}$. For small $\bq$ and $\omega$ only intraband ($\ell=\ell'$) terms contribute since $E^+_\bk-E^-_\bk > 2\Delta$. We expand the imaginary part of $1/\chit^{22}(\bq,\omega)$ for small $\bq$ and $\omega$ by keeping the ratio $\hat{\omega}=\omega/|\bq|$ fixed. The coupling to the 3-component can be neglected, since the intraband coherence factor $A^{23}(\bk-\bq/2,\bq)$ is already of order $|\bq|^2$ for small $\bq$. Hence at order $|\bq|^2$, we obtain
    \begin{equation} \label{eq_low_spiral: chit22ominv}
        \begin{split}
             {\rm Im} \, \frac{1}{\chit^{22}(\bq,\hat{\omega}|\bq|)} = - 4U^2 \bigg[
             \chit_{0i}^{22}(\bq,\hat{\omega}|\bq|) + {\rm Im} \! \sum_{a,b=0,1}
             \chit_0^{2a}(\bq,\hat{\omega}|\bq|) \Gammat^{ab}(\bzero,0) \, \chit_0^{b2}(\bq,\hat{\omega}|\bq|)
             \bigg].
        \end{split}
    \end{equation}
    Note that $\Gammat^{ab}(\bzero,0)=\lim_{|\bq|\to0}\Gammat^{ab}(\bq,\hat{\omega}|\bq|)$ depends on $\hat{\omega}$ and the direction of $\hat{q}=\bq/|\bq|$. We now show that both terms in Eq.~\eqref{eq_low_spiral: chit22ominv} are of order $|\bq|^2$ at fixed $\hat{\omega}$. 
    
    Shifting the integration variable $\bk$ in Eq.~\eqref{eq_low_spiral: chi0 def} by $-\bq/2$, $\chit_{0i}^{22}$ becomes
    \begin{equation}
        \chit_{0i}^{22}(\bq,\omega)=\frac{i\pi}{8}\int_\bk \sum_{\ell,\ell'}A^{22}_{\ell\ell'}(\bk-\bq/2,\bq) \big[ f(E_{\bk-\bq/2}^\ell) - f(E_{\bk+\bq/2}^{\ell'}) \big]
        \delta(\omega + E_{\bk-\bq/2}^\ell - E_{\bk+\bq/2}^{\ell'}).
    \end{equation}
    For small $\omega$, only the intraband terms contribute. The intraband coherence factor 
    \begin{equation}
        A^{22}_{\ell\ell}(\bk-\bq/2,\bq)=1-\frac{h_{\bk-\bq/2}h_{\bk+\bq/2}+\Delta^2}{e_{\bk-\bq/2} e_{\bk+\bq/2}},
    \end{equation}
    is of order $|\bq|^2$ for small $\bq$. Expanding $E_{\bk+\bq/2}^\ell - E_{\bk-\bq/2}^{\ell}=\bq\cdot\nabla_\bk E^\ell_\bk+O(|\bq|^3)$ and $f(E_{\bk-\bq/2}^\ell) - f(E_{\bk+\bq/2}^{\ell})=-f^\prime(E^\ell_\bk) (\bq\cdot\nabla_\bk E^\ell_\bk)+O(|\bq|^3)$, with $f^\prime(x)=df(x)/dx$, and using $\delta(|\bq|x)=|\bq|^{-1}\delta(x)$, we obtain
    \begin{equation}\label{eq_low_spiral: chi0i22}
        \chit_{0i}^{22}(\bq,\omega)=-\frac{i\pi}{16}\hat{\omega}q_\alpha q_\beta \int_\bk\sum_{\ell} \left[\partial^2_{q_\alpha q_\beta}A^{22}_{\ell\ell}(\bk-\bq/2,\bq)\big\rvert_{\bq=\bzero}\right]f^\prime(E_\bk^\ell)\, \delta(\hat{\omega}-\hat{\bq}\cdot\nabla_\bk E^\ell_\bk)+O(|\bq|^3).
    \end{equation}
    We thus conclude that $\chit_{0i}^{22}(\bq,\omega)$ is of order $\hat{\omega}|\bq|^2$ for small $\bq$ and $\omega=\hat{\omega} |\bq|$. Since at low temperatures $T\ll\Delta$ the term $f^\prime(E^\ell_\bk)$ in Eq.~\eqref{eq_low_spiral: chi0i22} behaves as $-\delta(E^\ell_\bk)$, we deduce that the only presence of Fermi surfaces (that is, $\bk$-points where $E^\ell_\bk=0$) is sufficient to induce a finite $\chit_{0i}^{22}(\bq,\omega)$ at small $\bq$.
    
    Since the effective interaction $\Gammat^{ab}(\bzero,0)$ is real, the second term in Eq.~\eqref{eq_low_spiral: chit22ominv} receives contribution only from the cross terms $\chit_{0r}^{2a}(\bq,\omega) \Gammat^{ab}(\bzero,0) \, \chit_{0i}^{b2}(\bq,\omega)$ and
    $\chit_{0i}^{2a}(\bq,\omega) \Gammat^{ab}(\bzero,0)
    \cdot$ $\chit_{0r}^{b2}(\bq,\omega)$. For small $\omega$ only intraband terms contribute to $\chit_{0i}^{2a}(\bq,\omega)$ and $\chit_{0i}^{b2}(\bq,\omega)$. Both are of order $\hat{\omega}\bq$ for small $\bq$ at fixed $\hat{\omega}$ because the intraband coherence factors $A_{\ell\ell}^{02}(\bk,\bq) = - A_{\ell\ell}^{20}(\bk,\bq)$ and $A_{\ell\ell}^{12}(\bk,\bq) = - A_{\ell\ell}^{21}(\bk,\bq)$ are of order $\bq$. Moreover, $\chit_{0r}^{2a}(\bq,\omega)$ and $\chit_{0r}^{b2}(\bq,\omega)$ are antisymmetric in $\bq$ and thus of order $\bq$, too. Hence, the second term in Eq.~\eqref{eq_low_spiral: chit22ominv} is of order $\hat{\omega}|\bq|^2$.
    
    We thus have shown that the damping term of the in-plane mode has the form
    \begin{equation} \label{eq_low_spiral: damping2}
        {\rm Im} \, \frac{m^2}{\chit^{22}(\bq,\omega)} =
        - \hat{\omega}|\bq|^2 \gamma(\hat\bq,\hat\omega) + O(|\bq|^3) \,,
    \end{equation}
    where the scaling function is symmetric in $\hat{\omega}$ and finite for $\hat{\omega}=0$. The Landau damping of the in-plane mode has the same form of the two Goldstone modes in a N\'eel antiferromagnet~\cite{Sachdev1995}. It is of the same order of the leading real terms near the pole, implying that the damping of the Goldstone mode is of the same order as its excitation energy, that is, $|\bq|$. This implies that the in-plane mode of a spin spiral state and the Goldstone modes of a N\'eel antiferromagnet are not \emph{asymptotically stable} quasiparticles, as this would require a damping rate vanishing faster than the excitation energy when approaching the pole of the susceptibility. 
    %
    \subsection{Out-of-plane modes}
    Similarly to the in-plane mode, one can write the out-of-plane susceptibility in the form
    \begin{equation}
        \chit^{33}(q)=\frac{\overline{\chi}_0^{33}(q)}{1-2U\overline{\chi}_0^{33}(q)}, 
        \label{eq_low_spiral: chi33 RPA}
    \end{equation}
    with
    \begin{equation}{\label{eq_low_spiral: chit33(q)}}
        \overline{\chi}_0^{33}(q)=\chit_0^{33}(q) + \sum_{a,b\in\{0,1,2\}}\chit_0^{3a}(q)\Gammat_{3}^{ab}(q)\chit_0^{b3}(q),
    \end{equation}
    where $\Gammat_{3}(q)$ is defined similarly to $\Gammat_{2}(q)$, that is, removing the components that involve the index 3 instead of the index 2. We also notice that 
    \begin{equation}
        \Gammat_{3}^{ab}(\bq,0)=\Gammat^{ab}(\bq,0),
    \end{equation}
    for $a$, $b=0,1,2$, because all the off-diagonal components $\chit_0^{3a}(q)$ and $\chit_0^{a3}(q)$ vanish for zero frequency. 
    \subsubsection{Spin stiffness}
    Using Eq.~\eqref{eq_low_spiral: stiffenss and chi spiral final J2} and $\overline{\chi}_0^{33}(\pm Q)=\chit_0^{33}(\pm Q)=1/(2U)$, we obtain the out-of-plane spin stiffness
    \begin{equation}
        J_{\alpha\beta}^\perp=
            -\Delta^2\partial^2_{q_\alpha q_\beta}\overline{\chi}_0^{33}(\pm Q)=
            -\Delta^2\partial^2_{q_\alpha q_\beta}\chit_0^{33}(\pm Q),
    \end{equation}
    where $\partial^2_{q_\alpha q_\beta} f(\pm Q)$ stands for $\partial^2f(\bq,0)/\partial q_\alpha\partial q_\beta |_{\bq\to\pm\bQ}$. 
    \subsubsection{Dynamical susceptibility}
    In the limit $\omega\to 0$, all the $\chit_0^{3a}(q)$ and $\chit_0^{a3}(q)$, with $a=0,1,2$, are linear in $\omega$, and the dynamical susceptibility is given by (see Eq.~\eqref{eq_low_spiral: stiffenss and chi spiral final chi2})
    \begin{equation}
        \begin{split}
            \chi_\mathrm{dyn}^\perp&=\Delta^2 \partial^2_\omega \overline{\chi}_0^{33}(\pm Q)\\ 
            &=\Delta^2 \bigg[
            \partial^2_\omega \chit^{33}_0(\pm Q)
            + 2\sum_{a,b\in\{0,1,2\}}\partial_\omega \chit_0^{3a}(\pm Q) \Gammat^{ab}(\pm Q) \partial_\omega \chit_0^{b3}(\pm Q)
            \bigg],
        \end{split}
    \end{equation}
    with $\partial^n_{\omega} f(\pm Q)$ a shorthand for $\partial^n f(\pm\bQ,\omega)/\partial \omega^n |_{\omega\to 0}$. We remark that for $\Gammat^{ab}(q)$ the limits $\bq\to\bQ$ and $\omega\to 0$ commute if $\bQ$ is not a high-symmetry wave vector, that is, if $E^\ell_{\bk+\bQ}\neq E^\ell_\bk$. 
    \subsubsection{Landau damping}
    We now analyze the $\bq$- and $\omega$-dependence of the imaginary part of $1/\chit^{33}$ for small $\omega$ and for $\bq$ near $\pm\bQ$. We discuss the case $\bq\sim\bQ$. The behavior for $\bq\sim-\bQ$ is equivalent. 
    
    We first fix $\bq=\bQ$ and study the $\omega$-dependence of the damping term. Since all the off-diagonal bare susceptibilities $\chit_0^{3a}(q)$ and $\chit_0^{b3}(q)$ in Eq.~\eqref{eq_low_spiral: chit33(q)} vanish for $\omega=0$, we obtain the following expansion of $1/\chit^{33}(\bQ,\omega)$ for small $\omega$
    \begin{equation} \label{eq_low_spiral: chit33inv}
         \frac{1}{\chit^{33}(\bQ,\omega)} = 2U \bigg[
         1 - 2U \chit_0^{33}(\bQ,\omega) - 2U \!\!\! \sum_{a,b\in\{0,1,2\}}
         \chit_0^{3a}(\bQ,\omega) \Gammat^{ab}(\bQ,0) \, 
         \chit_0^{b3}(\bQ,\omega) \bigg]
         + O(\omega^3) \, .
    \end{equation}
    The first contribution to the imaginary part of $1/\chit^{33}$ comes from the imaginary part of the bare susceptibility $\chit^{33}_{0i}(\bQ,\omega)$:
    \begin{equation} \label{eq_low_spiral: chit0i33}
         \chit_{0i}^{33}(\bQ,\omega) = \frac{i\pi}{8} \int_\bk \sum_{\ell,\ell'}
         A_{\ell\ell'}^{33}(\bk,\bQ)
         \big[ f(E_\bk^\ell) - f(E_{\bk+\bQ}^{\ell'}) \big]
         \delta(\omega + E_\bk^\ell - E_{\bk+\bQ}^{\ell'}) \, .
    \end{equation}
    For small $\omega$, only momenta for which \emph{both} $E^\ell_\bk$ and $E^{\ell'}_{\bk+\bQ}$ are $O(\omega)$ contribute to the integral. These momenta are restricted to a small neighborhood of $\emph{hot spots}$ $\bk_H$, defined by the relations
    \begin{equation}\label{eq_low_spiral: hot spots}
        E^\ell_{\bk_H}=E^{\ell'}_{\bk_H+\bQ}=0. 
    \end{equation}
    In most cases, only intraband ($\ell=\ell'$) hot spots appear. While the existence of interband hot spots cannot be excluded in general, we restrict our analysis to intraband contributions. 
    
    For $\ell=\ell'$, Eq.~\eqref{eq_low_spiral: hot spots} is equivalent to
    \begin{equation}
         E_{\bk_H}^\ell = 0 \quad \mbox{and} \quad \xi_{\bk_H} = \xi_{\bk_H+2\bQ} \, .
    \end{equation}
    In the N\'eel state $2\bQ$ is a reciprocal lattice vector and the second equation is always satisfied, so that all momenta on the Fermi surfaces are hot spots. The condition $\xi_{\bk_H} = \xi_{\bk_H+2\bQ}$ implies $h_{\bk_H+\bQ}=-h_{\bk_H}$, which in turn translates into $A^{33}_{\ell\ell}(\bk_H,\bQ)=0$ and also $\nabla_\bk A^{33}_{\ell\ell}(\bk,\bQ)|_{\bk\to\bk_H}=0$. For small frequencies and temperatures, the momenta contributing to the integral~\eqref{eq_low_spiral: chit0i33} are situated at a distance of order $\omega$ from the hot spots. For these momenta, the coherence factor is of order $\omega^2$ as both $A^{33}_{\ell\ell}(\bk,\bQ)$ and its gradient vanish at the hot spots. Multiplying this result with the usual factor $\propto\omega$ coming from the difference of the Fermi functions, we obtain
    \begin{equation}
        \chit^{33}_{0i}(\bQ,\omega) \propto \omega^3,
    \end{equation}
    for small $\omega$.
    
    We now consider the contribution to the imaginary part of $1/\chit^{33}$ coming from the second term in Eq.~\eqref{eq_low_spiral: chit33inv}. Since $\Gammat^{ab}(\bQ,0)$ is real, only the cross terms $\sum_{a,b}\chit_{0i}^{3a}(\bQ,\omega)$ $\Gammat^{ab}(\bQ,0)$ $\chit_{0r}^{b3}(\bQ,\omega)$ and $\sum_{a,b}\chit_{0r}^{3a}(\bQ,\omega)$ $\Gammat^{ab}(\bQ,0)$ $\chit_{0i}^{b3}(\bQ,\omega)$ contribute to the damping of the out-of-plane modes. The real parts of the bare susceptibilities $\chit^{3a}_{0r}(\bQ,\omega)$ and $\chit^{b3}_{0r}(\bQ,\omega)$ are antisymmetric in $\omega$ and of order $\omega$ for small frequencies. Their coherence factor vanish at $\bk=\bk_H$ but they have a finite gradient there. Hence, following the arguments given before for $\chit^{33}_{0i}(\bQ,\omega)$, we deduce
    \begin{equation}
        \chit^{3a}_{0i}(\bQ,\omega)\propto\omega^2
    \end{equation}
    for $a\in\{0,1,2\}$ and small $\omega$. The imaginary part of the second term in Eq.~\eqref{eq_low_spiral: chit33inv} is therefore of order $\omega^3$. We thus have shown that the Landau damping of the out-of-plane modes at $\bq=\bQ$ obeys
    \begin{equation}\label{eq_low_spiral: damping33 q=Q}
        {\rm Im}\frac{m^2}{\chit^{33}(\bQ,\omega)}\propto\omega^3.
    \end{equation}

    For $\bq\neq\bQ$, the hot spots are determined by $E^\ell_{\bk_H}=E^{\ell'}_{\bk_H+\bq}$ and the coherence factors remain finite there, so that
    \begin{equation}
        \chit^{3a}_{0i}(\bq,\omega)\simeq -p^{3a}(\bq)\omega
    \end{equation}
    for $a=0,1,2,3$ and small $\omega$. For $a=3$ both the coherence factor $A_{\ell\ell}^{33}(\bk,\bq)$ and its gradient vanish at the hot spots, implying $p^{33}(\bq)\propto(\bq-\bQ)^2$ for $\bq\sim\bQ$. Differently, for $a\neq3$ the gradient of the coherence factor remains finite, so that $p^{3a}(\bq)\propto|\bq-\bQ|$ for $\bq\sim\bQ$ (and $a\neq3$). For $\omega\to 0$ the contribution coming fror $\chit^{33}_{0i}$ is leading and we can generalize Eq.~\eqref{eq_low_spiral: damping33 q=Q} as 
    \begin{equation}\label{eq_low_spiral: damping outofplane2}
        {\rm Im}\frac{m^2}{\chit^{33}(\bq,\omega)}=-\gamma(\bq)\omega+O(\omega^2),
    \end{equation}
    with $\gamma(\bq)\propto(\bq-\bQ)^2$ for $\bq\to\bQ$. The contributions to the damping coming from the off-diagonal susceptibilities are of order $\omega^2$ with a prefactor linear in $|\bq-\bQ|$. Considering the limit $\omega\to 0$, $\bq\to\bQ$ at fixed $\hat{\omega}=\omega/|\bq-\bQ|$, both diagonal and off-diagonal terms are of order $|\bq-\bQ|^3$. 
    
    The above results are strongly dependent on the existence of hot spots. If Eq.~\eqref{eq_low_spiral: hot spots} has no solutions, the out-of-plane modes are not damped at all, at least within the RPA. 
    \section{Explicit evaluation of the Ward identities}
    \label{sec_low_spiral: explicit WIs}
    In this section, we explicitly evaluate the Ward identities derived in Sec.~\ref{sec_low_spiral: Ward Identities} for a spiral magnet and explicitly show that the expressions for the spin stiffnesses and dynamical susceptibilities obtained from the response to a SU(2) gauge field coincide (within the RPA, which is a conserving approximation in the sense of Baym and Kadanoff~\cite{Baym1961,KadanoffBaym}) with those derived within the low-energy expansion of the susceptibilities, carried out in Sec.~\ref{sec_low_spiral: properties of Goldstones}. 
    \subsection{Gauge kernels}
    We begin by setting up the formalism to compute the response to a SU(2) gauge field $A_\mu$ within the Hubbard model. We couple our system to $A_\mu$ via a Peierls substitution in the quadratic part of the Hubbard action~\eqref{eq_low_spiral: Hubbard action}:
\begin{equation}\label{eq_low_spiral: S0 coupled to SU(2) gauge field}
    \begin{split}
        \mathcal{S}_0[\psi,\psibar,A_\mu]=\int_0^\beta \!d\tau \sum_{jj'}&\psibar_j \Big[
        (\partial_\tau - A_{0,j} + \mu)\delta_{jj'}
        +t_{jj'}e^{-\bs{r}_{jj'}\cdot(\nabla-i \bs{A}_j)}
    \Big]\psi_j,
    \end{split}
\end{equation}
where $e^{-\bs{r}_{jj'}\cdot\nabla}$ is the translation operator from site $j$ to site $j'$, with $\bs{r}_{jj'}=\bs{r}_j-\bs{r}_{j'}$. Notice that under the transformation $\psi_j\to R_j\psi_j$, with $R_j\in\mathrm{SU(2)}$, the interacting part of the action~\eqref{eq_low_spiral: Hubbard action} is left unchanged, while the gauge field transforms according to \eqref{eq_low_spiral: SU(2) Amu transformation}. Since the gauge kernels correspond to correlators of two gauge fields, we expand \eqref{eq_low_spiral: S0 coupled to SU(2) gauge field} to second order in $A_\mu$. After a Fourier transformation one obtains
\begin{equation}
    \begin{split}
        \mathcal{S}_0[\psi,\psibar,A_\mu]=
        &-\int_k \psibar_k \left[i\nu_n+\mu-\eps_\bk\right]\psi_k\\
        &+\frac{1}{2}\int_{k,q}A_\mu^a(q) \gamma^\mu_{\bk}\, \psibar_{k+q}\sigma^a\psi_k
        -\frac{1}{8}\int_{k,q,q'}A^a_\alpha(q-q')A^a_\beta(q') \gamma^{\alpha\beta}_\bk \psibar_{k+q}\psi_k,
    \end{split}
    \label{eq_low_spiral: S0 coupled with Amu SU(2) momentum space}
\end{equation}
where the first order coupling is given by $\gamma^\mu_\bk=(1,\nabla_\bk\eps_\bk)$, and the second order one is $\gamma^{\alpha\beta}_\bk=\partial^2_{k_\alpha k_\beta}\eps_{\bk}$. In the equation above,the symbol $\int_k=\int_\bk T\sum_{\nu_n}$ ($\int_q=\int_\bq T\sum_{\Omega_m}$) denotes an integral over spatial momenta and a sum over fermionic (bosonic) Matsubara frequencies. Analyzing the coupling of the temporal component of the gauge field to the fermions in \eqref{eq_low_spiral: S0 coupled to SU(2) gauge field} and \eqref{eq_low_spiral: S0 coupled with Amu SU(2) momentum space}, we notice that the temporal components of the gauge kernel (see definition~\eqref{eq_low_spiral: gauge kernel definition}) are nothing but the susceptibilities in the original (unrotated) spin basis
\begin{equation}
    K_{00}^{ab}(\bq,\bq',\omega)=\chi^{ab}(\bq,\bq',\omega),
\end{equation}
where $\omega$ is a real frequency. 

\begin{figure*}
    \centering
    \includegraphics[width=1.\textwidth]{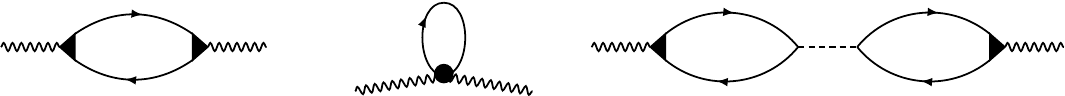}
    \caption{Diagrams contributing to the spin stiffnesses. The wavy line represents the external SU(2) gauge field, the solid lines the electronic Green's functions, the black triangles the paramagnetic vertex $\gamma^\mu_\bk \sigma^a$, the black circle the diamagnetic one $\gamma^{\alpha\beta}_\bk \sigma^0$, and the dashed line the effective interaction $\Gamma(\bq,\bq',\omega)$.}
    \label{fig: fig1}
\end{figure*}
The spatial components of the gauge kernel can be expressed in the general form (see Fig.~\ref{fig: fig1})
\begin{equation}
    \begin{split}
        K_{\alpha\beta}^{ab}(\bq,\bq',\omega)=
        &K_{\mathrm{para},\alpha\beta}^{ab}(\bq,\bq',\omega)
        +\delta_{ab}\,K_{\alpha\beta}^{\mathrm{dia}}\\
        &+\int_{\bq^{\prime\prime},\bq^{\prime\prime\prime}}\sum_{c,d}K_{\mathrm{para},\alpha 0}^{ac}(\bq,\bq^{\prime\prime},\omega)
        \Gamma^{cd}(\bq^{\prime\prime},\bq^{\prime\prime\prime},\omega)K_{\mathrm{para},0\beta}^{db}(\bq^{\prime\prime\prime},\bq^{\prime},\omega),
    \end{split}
    \label{eq_low_spiral: gauge kernel formula}
\end{equation}
where $\Gamma(\bq',\bq^{\prime\prime},\omega)$ is the effective interaction~\eqref{eq_low_spiral: Gamma(q) definition} expressed in the unrotated basis. Within the RPA, the paramagnetic terms are given by
\begin{equation}
    \begin{split}
        &K_{\mathrm{para},\mu\nu}^{ab}(\bq,\bq',\omega)=
        -\frac{1}{4}\int_{\bk,\bk'}T\sum_{\nu_n}
    \gamma^\mu_\bk\gamma^\nu_{\bk'+\bq'}\tr\Big[\sigma^a\bs{G}_{\bk,\bk'}(i\nu_n)\sigma^b
    \bs{G}_{\bk'+\bq',\bk+\bq}(i\nu_n+i\Omega_m)\Big],
    \end{split}
    \label{eq_low_spiral: paramagnetic contr Kernel}
\end{equation}
with the replacement $i\Omega_m\to\omega+i0^+$.
The Green's function in the unrotated basis takes the form
\begin{equation}\label{eq_low_spiral: G unrotated}
    \bs{G}_{\bk,\bk'}(i\nu_n)=\left(
    \begin{array}{cc}
        G_\bk(i\nu_n)\delta_{\bk,\bk'} & F_{\bk}(i\nu_n)\delta_{\bk,\bk'-\bQ} \\
        F_{\bk-\bQ}(i\nu_n)\delta_{\bk,\bk'+\bQ} & \overline{G}_{\bk-\bQ}(i\nu_n)\delta_{\bk,\bk'}
    \end{array}
    \right),
\end{equation}
where $\delta_{\bk,\bk'}$ is a shorthand for $(2\pi)^d\delta^d(\bk-\bk')$, and
\begin{subequations}
    \begin{align}
        &G_\bk(i\nu_n)=\frac{i\nu_n-\xi_{\bk+\bQ}}{(i\nu_n-\xi_{\bk})(i\nu_n-\xi_{\bk+\bQ})-\Delta^2},\label{eq_low_spiral: Gk up unrotated}\\
        &\overline{G}_\bk(i\nu_n)=\frac{i\nu_n-\xi_{\bk}}{(i\nu_n-\xi_{\bk})(i\nu_n-\xi_{\bk+\bQ})-\Delta^2},\label{eq_low_spiral: Gk down unrotated}\\
        &F_\bk(i\nu_n)=\frac{\Delta}{(i\nu_n-\xi_{\bk})(i\nu_n-\xi_{\bk+\bQ})-\Delta^2}.
    \end{align}
\end{subequations}
The diamagnetic term does not depend on $\bq$, $\bq'$ and $\omega$, and is proportional to the unit matrix in the gauge indices. It evaluates to 
\begin{equation}
    K_{\alpha\beta}^{\mathrm{dia}}=-\frac{1}{4}\int_{\bk,\bk'} T\sum_{\nu_n} (\partial^2_{k_\alpha k_\beta}\eps_\bk)\tr\left[\bs{G}_{\bk,\bk'}(i\nu_n)\right].
\end{equation}
We can now compute the spin stiffnesses and dynamical susceptibilities from the gauge kernels. 
\subsubsection{In-plane mode}
The in-plane spin stiffness is defined as
\begin{equation}
    J^\smsqr_{\alpha\beta}=-\lim_{\bq\to\bzero}K_{\alpha\beta}^{33}(\bq,0),
\end{equation}
where we have defined as $K_{\mu\nu}(\bq,\omega)$ the prefactors of those components of the gauge kernels $K_{\mu\nu}(\bq,\bq',\omega)$ which are proportional to $\delta_{\bq,\bq'}$.
In addition to the bare term 
\begin{equation}
    J_{\alpha\beta}^{0,\smsqr}=-\lim_{\bq\to\bzero}K_{\alpha\beta}^{\mathrm{para},33}(\bq,0)-K_{\alpha\beta}^{\mathrm{dia}},
    \label{eq_low_spiral: J inplane bare}
\end{equation}
we find nonvanishing paramagnetic contributions that mix spatial and temporal components. They involve
\begin{subequations}\label{eq_low_spiral: off_diagonal kalphat_3a}
    \begin{align}
        &\lim_{\bq\to\bzero}K_{0\alpha}^{30}(\bq,\bq',0)=\kappa_\alpha^{30}(\bzero)\delta_{\bq',\bzero},\\
        &\lim_{\bq\to\bzero}K_{0\alpha}^{31}(\bq,\bq',0)=\kappa_\alpha^{31}(\bzero)\frac{\delta_{\bq',\bQ}+\delta_{\bq',-\bQ}}{2}, \label{eq_low_spiral: k31 def}\\
        &\lim_{\bq\to\bzero}K_{0\alpha}^{32}(\bq,\bq',0)=\kappa_\alpha^{32}(\bzero)\frac{\delta_{\bq',\bQ}-\delta_{\bq',-\bQ}}{2i}\label{eq_low_spiral: k32 def},
    \end{align}
\end{subequations}
    where $\kappa_\alpha^{32}(\bzero)=\kappa_\alpha^{31}(\bzero)$. Noticing that for $a=0,1,2$, we have $\lim_{\bq\to\bzero}K^{3a}_{\alpha 0}(\bq,0)=\lim_{\bq\to\bzero}K^{a3}_{0\alpha}(\bq,0)$, and inserting this result into \eqref{eq_low_spiral: gauge kernel formula}, we obtain
    \begin{equation}
        J_{\alpha\beta}^{\smsqr}=J_{\alpha\beta}^{0,\smsqr}-\sum_{a,b\in\{0,1\}}\kappa_\alpha^{3a}(\bzero)\Gammat^{ab}(\bq\to\bzero,0)\kappa_\beta^{3b}(\bzero),\label{eq_low_spiral: J3 gauge}
    \end{equation}
    where $\Gammat(\bq\to\bzero,0)$ is the effective interaction in the rotated spin basis, defined in Eq.~\eqref{eq_low_spiral: Gamma(q) definition}. Notice that the delta functions in Eq.~\eqref{eq_low_spiral: off_diagonal kalphat_3a} convert the unrotated $\Gamma$ to $\Gammat$ and, together with the equality $\kappa_\alpha^{32}(\bzero)=\kappa_\alpha^{31}(\bzero)$, they remove the terms where $a$ or $b$ equal 2 in the sum. 
    The dynamical susceptibility is defined as
    \begin{equation}
        \chi_\mathrm{dyn}^\smsqr=\lim_{\omega\to0}K_{00}^{33}(\bzero,\omega)=\lim_{\omega\to0}\chi^{33}(\bzero,\omega).
    \end{equation}
    From Eq.~\eqref{eq_low_spiral: spiral rotation matrix with charge} we deduce that
    \begin{equation}
        \chi^{33}(\bq,\omega)=\chit^{33}(\bq,\omega).
    \end{equation}
    Remarking that for $\omega=0$ all the off-diagonal elements of the bare susceptibilities with one (and only one) of the two indices equal to 3 vanish, we obtain the RPA expression for $\chi_\mathrm{dyn}^\smsqr$
    \begin{equation}
        \begin{split}
            \chi_\mathrm{dyn}^\smsqr = \lim_{\omega\to 0}
            \frac{\chit_0^{33}(\bzero,\omega)}{1-2U\chit_0^{33}(\bzero,\omega)}.
        \end{split}
        \label{eq_low_spiral: chi3 gauge}
    \end{equation}
    \subsubsection{Out-of-plane modes}
    To compute the the out-of-plane stiffness, that is, 
    \begin{equation}
        J_{\alpha\beta}^\perp=-\lim_{\bq\to\bzero}K_{\alpha\beta}^{22}(\bq,0),
    \end{equation}
    we find that all the paramagnetic contributions to the gauge kernel that mix temporal and spatial components vanish in the $\omega\to0$ and $\bq=\bq'\to\bzero$\footnote{The terms $K^{\alpha 0}_{13}(\mathbf{0},\pm\mathbf{Q},0)$ and $K^{\alpha 0}_{23}(\mathbf{0},\pm\mathbf{Q},0)$ are zero only if $\mathbf{Q}$ is chosen such that it minimizes the free energy (see Eq.~\eqref{eq_low_spiral: MF free energy}). In fact, if this is not the case, they can be shown to be proportional to $\partial_{q_\alpha}\chit^0_{33}(\pm Q)$, which is finite for a generic $\mathbf{Q}$. } limits. Moreover, the $\bq\to\bzero$ limit of the momentum diagonal paramagnetic contribution can be written as
    \begin{equation}
        \begin{split}
            \lim_{\bq\to\bzero}K_{\mathrm{para},\alpha\beta}^{22}(\bq,0)=
            &-\frac{1}{4}\int_{\bk,\bk'} T\sum_{\substack{\nu_n\\\zeta=\pm}}
            \gamma^\alpha_\bk\gamma^\beta_{\bk'}\tr\left[\sigma^\zeta\bs{G}_{\bk,\bk'}(i\nu_n)\sigma^{-\zeta}\bs{G}_{\bk',\bk}(i\nu_n)\right]\\
            =&-\frac{1}{2}\int_{\bk}T\sum_{\nu_n}\gamma^\alpha_\bk\gamma^\beta_{\bk}\,G_\bk(i\nu_n)\overline{G}_{\bk-\bQ}(i\nu_n),
        \end{split}
    \end{equation}
    where we have defined $\sigma^\pm=(\sigma^1\pm i\sigma^2)/2$. The out-of-plane spin stiffness is thus given by
    \begin{equation}
        \begin{split}
            J^\perp_{\alpha\beta} = &-\frac{1}{2}\int_{\bk}T\sum_{\nu_n}\gamma^\alpha_\bk\gamma^\beta_{\bk}\,G_\bk(i\nu_n)\overline{G}_{\bk-\bQ}(i\nu_n)
            -\frac{1}{4}\int_{\bk,\bk'} T\sum_{\nu_n} (\partial^2_{k_\alpha k_\beta}\eps_\bk)\tr\left[\bs{G}_{\bk,\bk'}(i\nu_n)\right]
        \end{split}
    \end{equation}

    Finally, we evaluate the dynamical susceptibility of the out-of-plane modes. This is defined as
    \begin{equation}
        \chi_\mathrm{dyn}^\perp=\lim_{\omega\to 0}K_{00}^{22}(\bzero,\omega)=\lim_{\omega\to 0}\chi^{22}(\bzero,\omega).
    \end{equation}
    Applying transformation~\eqref{eq_low_spiral: rotated chis spiral RPA}, we can express the momentum-diagonal component of $\chi^{22}(\bq,\bq',\omega)$ in terms of the susceptibilities in the rotated basis as
    \begin{equation}
        \begin{split}
            \chi^{22}(\bq,\omega)=\frac{1}{4}\sum_{\zeta=\pm}\left[
            \chit^{11}(\bq+\zeta\bQ,\omega)
            +\chit^{22}(\bq+\zeta\bQ,\omega)
            +2i\zeta\chit^{12}(\bq+\zeta\bQ,\omega)
            \right],
            \label{eq_low_spiral: uniform chi22 spiral}
        \end{split}
    \end{equation}
    where we have used (see Table~\ref{tab_low_spiral: symmetries}) $\chit^{12}(q)=-\chit^{21}(q)$. Sending $\bq$ to $\bzero$ in \eqref{eq_low_spiral: uniform chi22 spiral}, and using the symmetry properties of the susceptibilities for $\bq\to-\bq$ (see Table~\ref{tab_low_spiral: symmetries}), we obtain
    \begin{equation}
        \begin{split}
            \chi^{22}(\bzero,\omega)&=\frac{1}{2}\left[\chit^{11}(\bQ,\omega)+\chit^{22}(\bQ,\omega)+2i\chit^{12}(\bQ,\omega)\right]
            =2\chit^{-+}(\bQ,\omega),
        \end{split}
    \end{equation}
    with $\chit^{-+}(q)=\langle S^-(-q)S^+(q)\rangle$, and $S^\pm(q)=(S^1(q)\pm iS^2(q))/2$.
    It is convenient to express $\chit^{-+}(\bQ,\omega)$ as 
    \begin{equation}
        \begin{split}
            \chit^{-+}(\bQ,\omega)&=\chit^{-+}_0(\bQ,\omega)
            +\sum_{a,b\in\{0,1,2,3\}}\chit^{-a}_0(\bQ,\omega)\Gammat^{ab}(\bQ,\omega)\chit^{b+}_0(\bQ,\omega),
        \end{split}
        \label{eq_low_spiral: chit pm RPA}
    \end{equation}
    where we have defined
    \begin{subequations}
        \begin{align}
            &\chit_0^{-a}(q)=\frac{1}{2}\left[\chit_0^{1a}(q)-i\chit_0^{2a}(q)\right],\\
            &\chit_0^{a+}(q)=\frac{1}{2}\left[\chit_0^{a1}(q)+i\chit_0^{a2}(q)\right].
        \end{align}
    \end{subequations}
    In the limit $\omega\to 0$, $\chit_0^{-3}(\bQ,\omega)$ and $\chit_0^{3+}(\bQ,\omega)$ vanish as they are odd in frequency (see Table~\ref{tab_low_spiral: symmetries}). We can now cast the dynamical susceptibility in the form
    \begin{equation}
        \begin{split}
            \chi_\mathrm{dyn}^\perp&=2\chit^{-+}_0(Q)
            +2\sum_{a,b\in\{0,1,2\}}\chit^{-a}_0(Q)\Gammat^{ab}(Q)\chit^{b+}_0(Q),
        \end{split}
        \label{eq_low_spiral: chi22 gauge}
    \end{equation}
    or, equivalently,
    \begin{equation}
        \begin{split}
            \chi_\mathrm{dyn}^\perp&=2\chit^{+-}_0(-Q)
            +2\sum_{a,b\in\{0,1,2\}}\chit^{+a}_0(-Q)\Gammat^{ab}(-Q)\chit^{b-}_0(-Q).
        \end{split}
    \end{equation}
    We remark that in the formulas above we have not specified in which order the limits $\bq\to\pm\bQ$ and $\omega\to0$ have to be taken as they commute. 
    \subsection{Equivalence of RPA and gauge theory approaches}
    In this Section, we finally prove that the expressions for the spin stiffnesses and dynamical susceptibilities obtained from a low energy expansion of the susceptibilities (Sec.~\ref{sec_low_spiral: properties of Goldstones}) coincide with those computed via the SU(2) gauge response of the previous section via a direct evaluation.
    \subsubsection{In-plane mode}
    We start by computing the first term in Eq.~\eqref{eq_low_spiral: J3 RPA}. The second derivative of the 22-component of the bare susceptibility can be expressed as
    \begin{equation}
        \begin{split}
            -2\Delta^2\partial^2_{q_\alpha q_\beta}\chit_0^{22}(0)=
            -\Delta^2\int_\bk \gamma_{\bk}^\alpha\gamma_{\bk+\bQ}^\beta\left[\frac{f(E^-_\bk)-f(E^+_\bk)}{4e_\bk^3}
            +\frac{f'(E^+_\bk)+f'(E^-_\bk)}{4e_\bk^2}\right],
        \end{split}
        \label{eq_low_spiral: J03 RPA}
    \end{equation}
    where $f'(x)=df/dx$ is the derivative of the Fermi function. On the other hand, the bare contribution to $J^\smsqr_{\alpha\beta}$ (Eq.~\eqref{eq_low_spiral: J inplane bare}) reads
    \begin{equation}
        \begin{split}
            J^{0,\smsqr}_{\alpha\beta}=&\frac{1}{4}\int_\bk T\sum_{\nu_n}\left[
            G_\bk(i\nu_n)^2\gamma_{\bk}^\alpha\gamma_{\bk}^\beta+\overline{G}_\bk(i\nu_n)^2\gamma_{\bk+\bQ}^\alpha\gamma_{\bk+\bQ}^\beta
            -2F_\bk(i\nu_n)^2\gamma_{\bk}^\alpha\gamma_{\bk+\bQ}^\beta
            \right]
            \\
            &+\frac{1}{4}\int_\bk T\sum_{\nu_n} \Big[
            G_\bk(i\nu_n)\gamma^{\alpha\beta}_\bk
            +\overline{G}_\bk(i\nu_n)\gamma^{\alpha\beta}_{\bk+\bQ}
            \Big].
        \end{split}
    \end{equation}
    %
    The second (diamagnetic) term can be integrated by parts, giving
    \begin{equation}
        \begin{split}
            -\frac{1}{4}\int_\bk T\sum_{\nu_n} \left[
            G^2_\bk(i\nu_n)\gamma_{\bk}^\alpha\gamma_{\bk}^\beta+\overline{G}^2_\bk(i\nu_n)\gamma_{\bk+\bQ}^\alpha\gamma_{\bk+\bQ}^\beta
            +2F^2_\bk(i\nu_n)\gamma_{\bk}^\alpha\gamma_{\bk+\bQ}^\beta
            \right],
        \end{split}
    \end{equation}
    where we have used the properties
    \begin{subequations}
        \begin{align}
            &\partial_{k_\alpha}G_\bk(i\nu_n)=G^2_\bk(i\nu_n)\gamma^\alpha_\bk+F^2_\bk(i\nu_n)\gamma^\alpha_{\bk+\bQ},\\      
            &\partial_{k_\alpha}\overline{G}_\bk(i\nu_n)=\overline{G}^2_\bk(i\nu_n)\gamma^\alpha_{\bk+\bQ}+F^2_\bk(i\nu_n)\gamma^\alpha_{\bk}.
        \end{align}
        \label{eq_low_spiral: derivatives of G}
    \end{subequations}
    Summing up both terms, we obtain
    \begin{equation}
        \begin{split}
            J_{0,\smsqr}^{\alpha\beta}=
        -\int_\bk T\sum_{\nu_n}\gamma_{\bk}^\alpha\gamma_{\bk+\bQ}^\beta &F^2_\bk(i\nu_n).
        \end{split}
        \label{eq_low_spiral: J03 explicit}
    \end{equation}
    Performing the Matsubara sum, we arrive at
    \begin{equation}\label{eq_low_spiral: J0inplane}
        \begin{split}
            J^{0,\smsqr}_{\alpha\beta}=
            -\Delta^2\int_\bk \gamma_{\bk}^\alpha\gamma_{\bk+\bQ}^\beta\left[\frac{f(E^-_\bk)-f(E^+_\bk)}{4e_\bk^3}
            +\frac{f'(E^+_\bk)+f'(E^-_\bk)}{4e_\bk^2}\right],
        \end{split}
    \end{equation}
    which is the same result as in \eqref{eq_low_spiral: J03 RPA}. Furthermore, one can show that
    \begin{subequations}
        \begin{align}
            &2i\Delta\partial_{q_\alpha}\chit_0^{20}(0)=-2i\Delta\partial_{q_\alpha}\chit_0^{02}(0)=\kappa^{30}_{\alpha}(\bzero),\\
            &2i\Delta\partial_{q_\alpha}\chit_0^{21}(0)=-2i\Delta\partial_{q_\alpha}\chit_0^{12}(0)=\kappa^{31}_{\alpha}(\bzero).
        \end{align}
        \label{eq_low_spiral: RPA vs gauge inplane J, offdiagonal}
    \end{subequations}
    Inserting results~\eqref{eq_low_spiral: J03 RPA}, \eqref{eq_low_spiral: J03 explicit}, and \eqref{eq_low_spiral: RPA vs gauge inplane J, offdiagonal} into \eqref{eq_low_spiral: J3 RPA} and \eqref{eq_low_spiral: J3 gauge}, we prove that these two expressions give the same result for the in-plane stiffness. Explicit expressions for $\kappa^{30}_{\alpha}(\bzero)$ and $\kappa^{31}_{\alpha}(\bzero)$ are given in Appendix~\ref{app: low en spiral}.
    
    If we now consider the dynamical susceptibility, it is straightforward to see that
    \begin{equation}
        \begin{split}
            2\Delta^2\partial^2_\omega\chit_0^{22}(0)=&2i\Delta\partial_\omega\chit^{23}_0(0)=\lim_{\omega\to 0}\chit_0^{33}(\bzero,\omega)
            =\Delta^2\int_\bk \frac{f(E^-_\bk)-f(E^+_\bk)}{4e^3_\bk},
        \end{split}
        \label{eq_low_spiral: RPA=gauge for chi3}
    \end{equation}
    which, if inserted into Eqs.~\eqref{eq_low_spiral: chi3 RPA} and \eqref{eq_low_spiral: chi3 gauge}, proves that the calculations of $\chi_\mathrm{dyn}^\smsqr$ via gauge kernels and via the low-energy expansion of the susceptibilities provide the same result.
    \subsubsection{Out-of-plane modes}
    With the help of some lengthy algebra, one can compute the second momentum derivative of the bare susceptibility $\chit_0^{33}(q)$, obtaining
    \begin{equation}
        \begin{split}
            -\Delta^2\partial^2_{q_\alpha q_\beta}\chit_0^{33}(Q)
            =&\frac{1}{8}\int_\bk\sum_{\ell,\ell'=\pm}\left(1-\ell\frac{h_\bk}{e_\bk}\right)\left(1+\ell\frac{h_{\bk+\bQ}}{e_{\bk+\bQ}}\right)\gamma^\alpha_{\bk+\bQ}\gamma^\beta_{\bk+\bQ}
            \frac{f(E^\ell_\bk)-f(E^{\ell'}_{\bk+\bQ})}{E^\ell_\bk-E^{\ell'}_{\bk+\bQ}}\\
            &-\frac{1}{8}\int_\bk \sum_{\ell=\pm}\left[\left(1-\ell\frac{h_\bk}{e_\bk}\right)^2\gamma^{\alpha}_{\bk+\bQ}+\frac{\Delta^2}{e_\bk^2}\gamma^{\alpha}_{\bk}\right]\gamma^\beta_{\bk+\bQ} f'(E^\ell_\bk)\\
            &-\frac{1}{8}\int_\bk \sum_{\ell=\pm}\left[\frac{\Delta^2}{e_\bk^2}(\gamma^{\alpha}_{\bk+\bQ}-\gamma^{\alpha}_{\bk})\right]\gamma^\beta_{\bk+\bQ} \frac{f(E^\ell_\bk)-f(E^{-\ell}_{\bk})}{E^\ell_\bk-E^{-\ell}_{\bk}}.
        \end{split}
    \end{equation}
    Similarly to what we have done for the in-plane mode, we integrate by parts the diamagnetic contribution to the gauge kernel. Its sum with the paramagnetic one gives
    \begin{equation}
        \begin{split}
            -\lim_{\bq\to\bzero}K^{22}_{\alpha\beta}&(\bq,\bq,0)=\\
            =&\frac{1}{2}\int_\bk T\sum_{\nu_n}\Big\{
            \gamma^\alpha_{\bk+\bQ}\gamma^\beta_{\bk+\bQ}\overline{G}_\bk(i\nu_n)\left[G_{\bk+\bQ}(i\nu_n)-\overline{G}_\bk(i\nu_n)\right]
            -\gamma^\alpha_{\bk}\gamma^\beta_{\bk+\bQ}F^2_\bk(i\nu_n)
            \Big\}.
        \end{split}
    \end{equation}
    Performing the Matsubara sums, one can prove the equivalence of the RPA and gauge theory approach for the calculation of $J_{\alpha\beta}^{\perp}$. 
    
    Similarly, we obtain for the second frequency derivative of the bubble $\chit_0^{33}(q)$
    \begin{equation}
        \begin{split}
            \Delta^2\partial^2_\omega\chit_0^{33}(Q)=-\frac{1}{8}\int_\bk\sum_{\ell,\ell'=\pm}
            \left(1-\ell\frac{h_\bk}{e_\bk}\right)\left(1+\ell\frac{h_{\bk+\bQ}}{e_{\bk+\bQ}}\right)
            \frac{f(E^\ell_\bk)-f(E^{\ell'}_{\bk+\bQ})}{E^\ell_\bk-E^{\ell'}_{\bk+\bQ}}=2\chit^{-+}_0(Q).
        \end{split}
        \label{eq_low_spiral: proof Z2, diagonal component}
    \end{equation}
    Furthermore, one can prove that
    \begin{equation}
        \begin{split}
            \Delta\partial_\omega \chit_0^{3a}(Q)=\Delta[\partial_\omega \chit_0^{a3}(Q)]^*
            =\chit_0^{-a}(Q)=[\chit_0^{a+}(Q)]^*,
        \end{split}
        \label{eq_low_spiral: proof Z2, off diagonal components}
    \end{equation}
    for $a=0,1,2$. Inserting results \eqref{eq_low_spiral: proof Z2, diagonal component} and \eqref{eq_low_spiral: proof Z2, off diagonal components} into Eqs.~\eqref{eq_low_spiral: chi22 RPA} and \eqref{eq_low_spiral: chi22 gauge}, one sees that the RPA and gauge theory approaches are equivalent for the calculation of $\chi_\mathrm{dyn}^\perp$. In Appendix~\ref{app: low en spiral} we provide explicit expressions for the off diagonal bare susceptibilities $\chit^{-a}_0(Q)$.
    \subsubsection{Remarks on more general models}
    We remark that in the more general case of an interaction of the type
    \begin{equation}
        \mathcal{S}_\mathrm{int}=\int_{k,k',q}\!U_{k,k'}(q)[\psibar_{k+q}\vec{\sigma}\psi_k]\cdot[\psibar_{k'-q}\vec{\sigma}\psi_{k'}],
    \end{equation}
    producing, in general, a $k$-dependent gap, the identities we have proven above do not hold anymore within the RPA, as additional terms in the derivative of the inverse susceptibilities emerge, containing expressions involving first and second derivatives of the gap with respect to the spatial momentum and/or frequency. In fact, in the case of nonlocal interactions, gauge invariance requires additional couplings to the gauge field in $\mathcal{S}_\mathrm{int}$, complicating our expressions for the gauge kernels. Similarly, even for action~\eqref{eq_low_spiral: Hubbard action}, approximations beyond the RPA produce in general a $k$-dependent $\Delta$, and vertex corrections in the kernels are required to obtain the same result as the one obtained expanding the susceptibilities. 
    \section{N\'eel limit}
    \label{sec_low_spiral: Neel limit}
    In this Section, we analyze the N\'eel limit, that is, $\bQ=(\pi/a_0,\dots,\pi/a_0)$. In this case, it is easy to see that, within the RPA, the bare susceptibilities in the rotated basis obey the identities
    \begin{subequations}
        \begin{align}
            &\chit_0^{22}(\bq,\omega)=\chit_0^{33}(\bq+\bQ,\omega),\\
            &\chit_0^{20}(\bq,\omega)=\chit_0^{21}(\bq,\omega)=0,\\
            &\chit_0^{30}(\bq,\omega)=\chit_0^{31}(\bq,\omega)=0.
        \end{align}
    \end{subequations}
    Furthermore, we obtain for the mixed gauge kernels (see Appendix~\ref{app: low en spiral})
    \begin{equation}
        K^{ab}_{\mathrm{para},\alpha 0}(\bq,\bq',\omega)=K^{ab}_{\mathrm{para},0\alpha}(\bq,\bq',\omega)=0.
    \end{equation}
    We also notice that $K^{11}_{\alpha\beta}(\bq,\bq',0)$ and $K^{22}_{\alpha\beta}(\bq,\bq',0)$ have (different) momentum off-diagonal contributions for which $\bq'=\bq\pm 2\bQ$. If $\bQ=(\pi/a_0,\dots,\pi/a_0)$, these terms become diagonal in momentum, as $2\bQ\sim\bzero$, such that
    \begin{subequations}
        \begin{align}
            &\lim_{\bq\to\bzero}K^{11}_{\alpha\beta}(\bq,0)=0,\\
            &K^{22}_{\alpha\beta}(\bq,0)=K^{33}_{\alpha\beta}(\bq,0).
        \end{align}
    \end{subequations}

    From the above relations, we can see that $J_{\alpha\beta}^\perp=J^\smsqr_{\alpha\beta}\equiv J_{\alpha\beta}$, and $\chi_\mathrm{dyn}^\perp=\chi_\mathrm{dyn}^\smsqr\equiv \chi_\mathrm{dyn}^\perp$, as expected for the N\'eel state.
    
    From these considerations, we obtain for the spin stiffness
    \begin{equation}
        \begin{split}
            J_{\alpha\beta}=&-\lim_{\bq\to\bzero}K^{22}_{\alpha\beta}(\bq,0)=-\lim_{\bq\to\bzero}K^{33}_{\alpha\beta}(\bq,0)
            =-2\Delta^2\partial^2_{q_\alpha q_\beta}\chit_0^{22}(0)
            =-2\Delta^2\partial^2_{q_\alpha q_\beta}\chit_0^{33}(Q),
        \end{split}
    \end{equation}
    which implies that $J_{\alpha\beta}$ is given by Eq.~\eqref{eq_low_spiral: J0inplane}. If the underlying lattice is $C_4$-symmetric, the spin stiffness is isotropic in the N\'eel state, that is, $J_{\alpha\beta}=J\delta_{\alpha\beta}$. 
    Similarly, for the dynamical susceptibility, we have
    \begin{equation}
        \begin{split}
            \chi_\mathrm{dyn}^\perp=&\lim_{\omega\to 0}\chi^{22}(\bzero,\omega)=\lim_{\omega\to 0}\chi^{33}(\bzero,\omega)
            =2\Delta^2\partial^2_\omega \chit_0^{22}(0)=2\Delta^2\partial^2_\omega \chit_0^{33}(Q),
        \end{split}
    \end{equation}
    which, combined with \eqref{eq_low_spiral: RPA=gauge for chi3}, implies
    \begin{equation}
        \chi_\mathrm{dyn}^\perp=\lim_{\omega\to 0}\frac{\chit_0^{33}(\bzero,\omega)}{1-2U\chit_0^{33}(\bzero,\omega)},
    \end{equation}
    with $\chit_0^{33}(\bzero,\omega\to 0)$ given by Eq.~\eqref{eq_low_spiral: RPA=gauge for chi3}.
    
    We notice that the dynamical susceptibility is obtained from the susceptibility by letting $\bq\to\bzero$ \emph{before} $\omega\to 0$. This order of the limits removes the intraband terms (that is, the $\ell=\ell'$ terms in Eq.~\eqref{eq_low_spiral: chi0 def}), which instead would yield a finite contribution to the \emph{uniform transverse susceptibility} $\chi^\perp\equiv\lim_{\bq\to\bzero}\chi^{22}(\bq,0)$. In the special case of an insulator at low temperature $T\ll\Delta$, the intraband contributions vanish and one has the identity $\chi_\mathrm{dyn}^\perp=\chi^\perp$, leading to the hydrodynamic relation for the spin wave velocity~\cite{Halperin1969} $c_s=\sqrt{J/\chi^\perp}$ (in an isotropic antiferromagnet). As noticed in Ref.~\cite{Sachdev1995}, in a doped antiferromagnet this hydrodynamic expression does not hold anymore, and one has to replace the uniform transverse susceptibility with the dynamical susceptibility. Since $J=0$ and $\chi_\mathrm{dyn}^\perp=0$ in the symmetric phase due to SU(2) gauge invariance, the expression $c_s=\sqrt{J/\chi_\mathrm{dyn}^\perp}$ yields a finite value $c_s$ at the critical point $\Delta\to 0$, provided that $J$ and $\chi_\mathrm{dyn}^\perp$ scale to zero with the same power of $\Delta$, as it happens within mean-field theory. Note that in the symmetric phase SU(2) gauge invariance does not pose any constraint on $\chi^\perp$, which is generally finite. 
    
    In the simpler case of perfect nesting, that is, when $\xi_\bk=-\xi_{\bk+\bQ}$, corresponding to the half-filled particle-hole symmetric Hubbard model, and at zero temperature, expressions for $J$ and $\chi^\perp$ have been derived in Refs.~\cite{Schulz1995,Borejsza2004} for two spatial dimensions, and it is straightforward to check that our results reduce to these in this limit. Moreover,  Eqs.~31-34 in Ref.~\cite{Schulz1995} are similar to our Ward identities but no derivation is provided. 
    
    We finally analyze the Landau damping of the Goldstone modes for a N\'eel antiferromagnet. Using the decoupling of the sector 0 and 1 from sectors 2 and 3, one obtains
    \begin{equation}
        {\rm Im}\frac{1}{\chit^{22}(\bq,\omega)}=-4U^2\chit^{22}_{0i}(\bq,\omega)+O(|\bq|^3)
    \end{equation}
    for small $\bq$, and 
    \begin{equation}
        {\rm Im}\frac{1}{\chit^{33}(\bq,\omega)}=-4U^2\chit^{22}_{0i}(\bq,\omega)+O(|\bq-\bQ|^3)
    \end{equation}
    for $\bq\sim\bQ$. Because of $\chit^{22}(\bq,\omega)=\chit^{33}(\bq+\bQ,\omega)$, the damping of the two Goldstone modes is identical. Returning to the susceptibilities in the unrotated basis, where $\chi^{22}(\bq,\omega)=\chit^{22}(\bq+\bQ,\omega)$ and $\chi^{33}(\bq,\omega)=\chit^{33}(\bq,\omega)$, one has
    \begin{equation}
        {\rm Im}\frac{1}{\chi^{22}(\bq',\omega)}={\rm Im}\frac{1}{\chi^{33}(\bq',\omega)}=-|\bq'|^2\hat{\omega}\gamma(\hat{\bq}',\hat{\omega})+O(|\bq|^3),
    \end{equation}
    for small $\bq'=\bq-\bQ$ and at fixed $\hat{\omega}=\omega/|\bq'|$. This form of the Landau damping in the N\'eel state has already been derived by Sachdev \emph{et al.} in Ref.~\cite{Sachdev1995}.
    \section{Numerical results in two dimensions}
    \label{sec_low_spiral: numerical results}
    \begin{figure}[h!]
        \centering
        \includegraphics[width=0.6\textwidth]{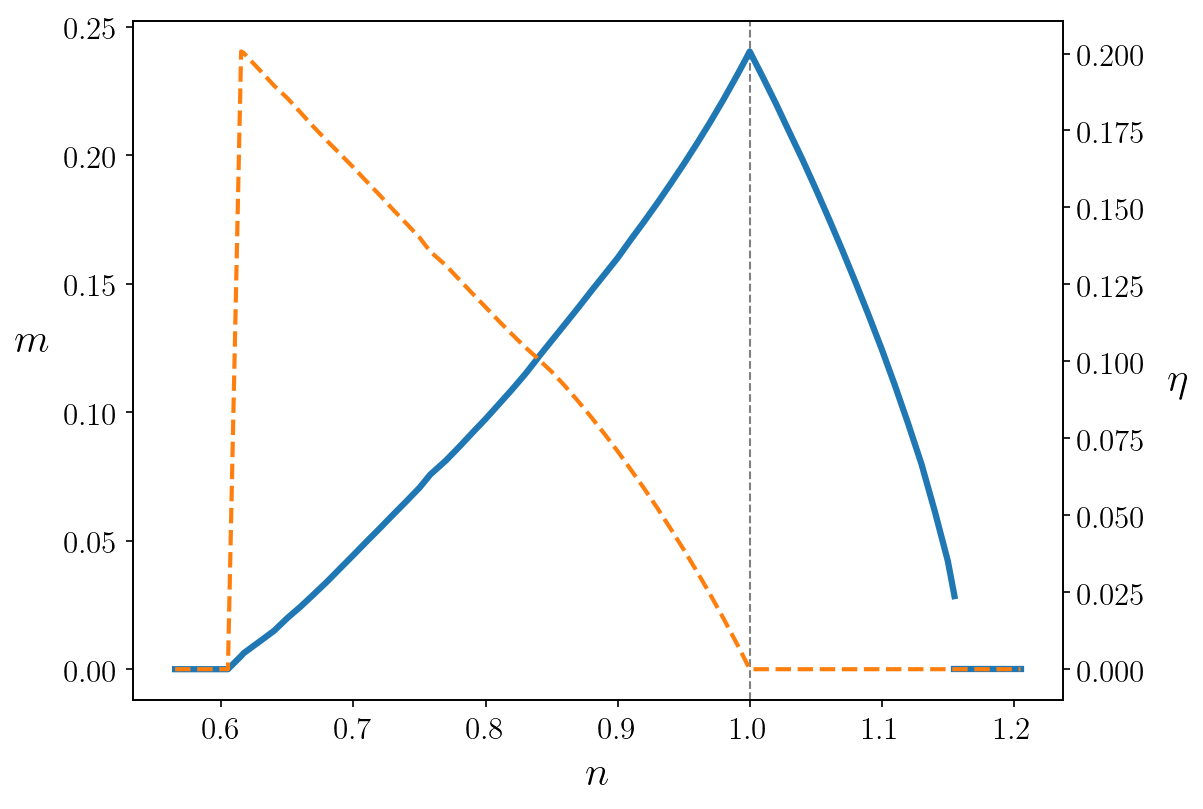}
        \caption{Magnetization $m$ (left axis, solid line) and incommensurability $\eta$ (right axis, dashed line) as functions of the electron density in the mean-field ground state of the two-dimensional Hubbard model for $t'=-0.16t$, $U=2.5t$.}
        \label{fig_low_en_sp: fig1}
    \end{figure}
    \begin{figure}[h!]
        \centering
        \includegraphics[width=1.\textwidth]{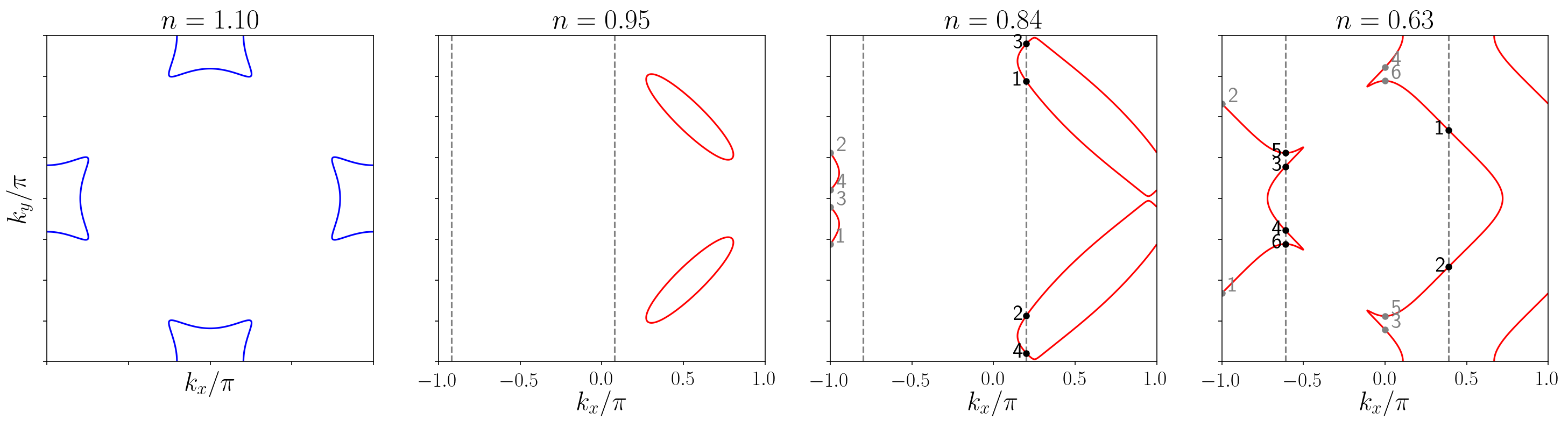}
        \caption{Quasiparticle Fermi surfaces in the magnetic ground state at various densities. The blue (red) lines correspond to solutions of the equation $E^+_\bk=0$ ($E^-_\bk=0$). The gray dashed lines are solutions of $E^\ell_\bk=E^\ell_{\bk+\bQ}$ (or, equivalently, $\xi_\bk=\xi_{\bk+2\bQ}$) for $\bQ\neq(\pi,\pi)$. For the densities $n=0.84$ and $n=0.63$ these lines intersect the Fermi surfaces at the hotspots (black dots), that is the points that are connected to other Fermi surface points (gray dots) by a momentum shift $\bQ$. The numbers indicate a pairwise connection. In the N\'eel state obtained for $n\geq 1$ all $\bk$ points satisfy $E^\ell_\bk=E^\ell_{\bk+\bQ}$. }
        \label{fig_low_en_sp: fig2}
    \end{figure}
    \begin{figure}[h!]
        \centering
        \includegraphics[width=0.5\textwidth]{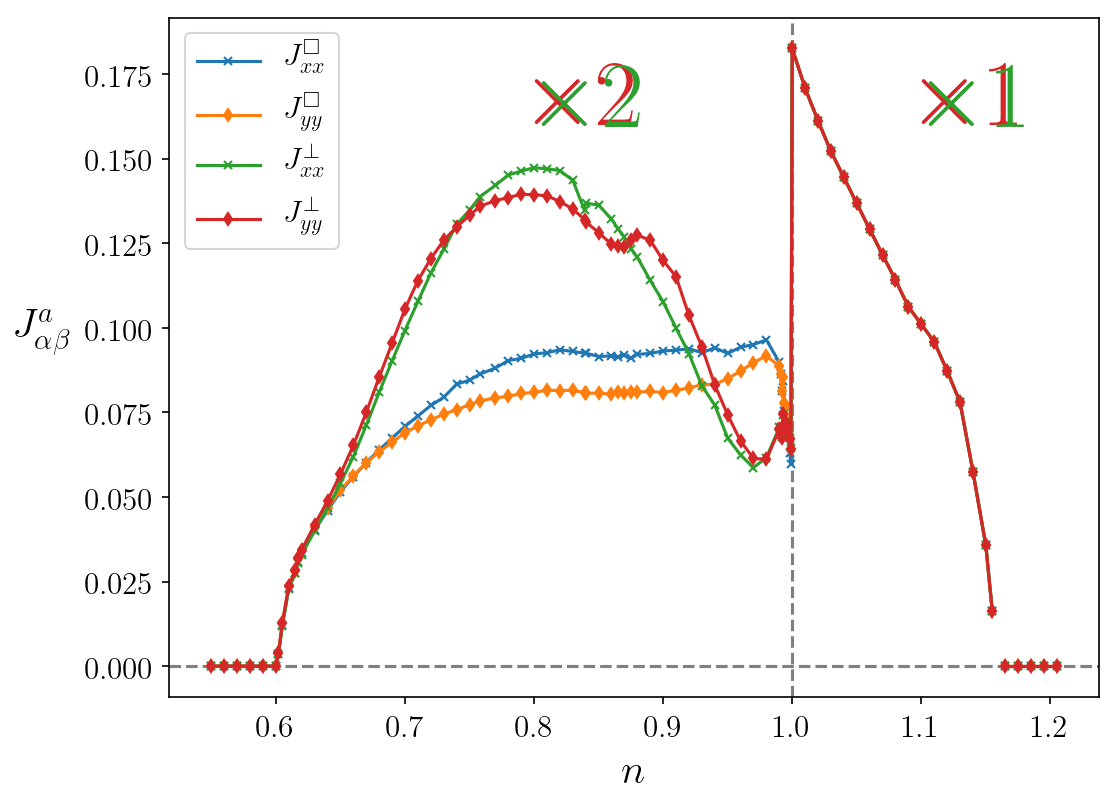}
        \caption{In-plane and out-of-plane spin stiffnesses as functions of the electron density. In the N\'eel state for $n\geq1$ all the stiffnesses take the same value. Notice that for the spiral magnetic state for $n<1$ we have multiplied the out-of-plane spin stiffnesses $J^\perp_{xx}$ and $J^\perp_{yy}$ by a factor of 2.}
        \label{fig_low_en_sp: fig3}
    \end{figure}
    \begin{figure}[h!]
        \centering
        \includegraphics[width=0.5\textwidth]{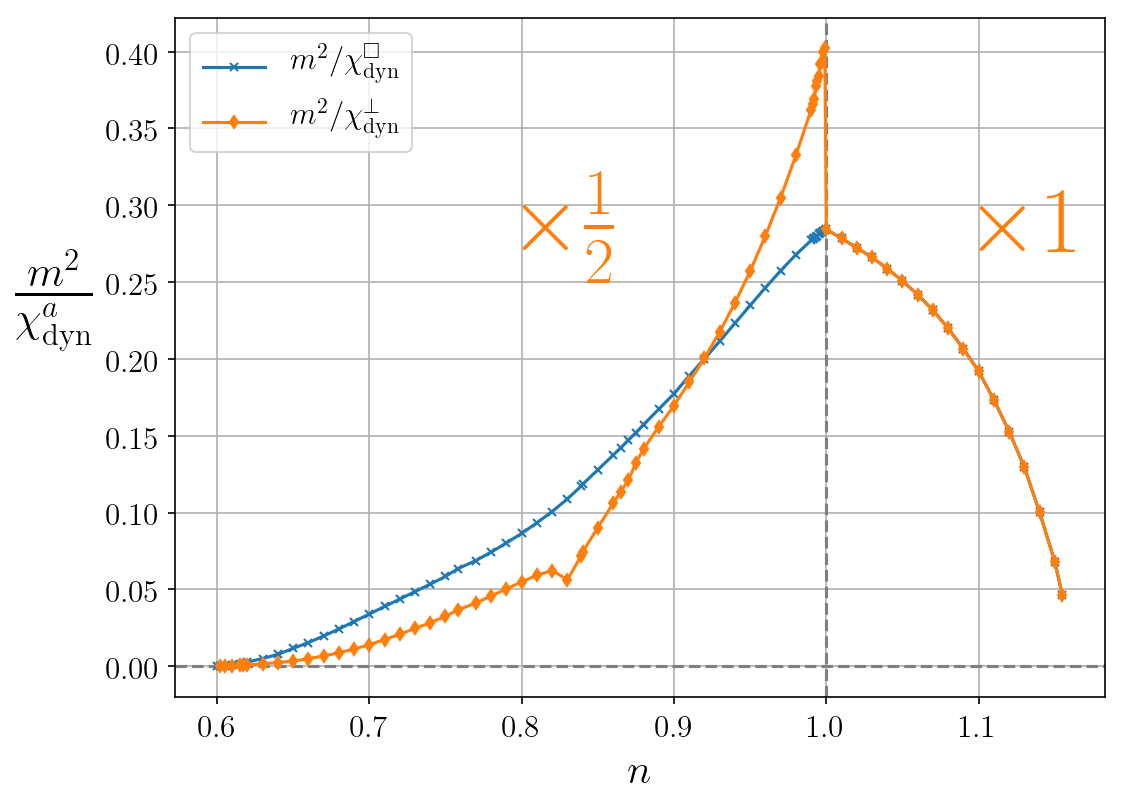}
        \caption{Spectral weights of the in-plane and out-of-plane Goldstone modes as functions of the electron density $n$. In the N\'eel state for $n\geq1$ both weights take the same value. Notice that in the spiral magnetic state for $n<1$ we have multiplied the out-of-plane spectral weight by a factor $1/2$.}
        \label{fig_low_en_sp: fig4}
    \end{figure}
    \begin{figure}[h!]
        \centering
        \includegraphics[width=0.5\textwidth]{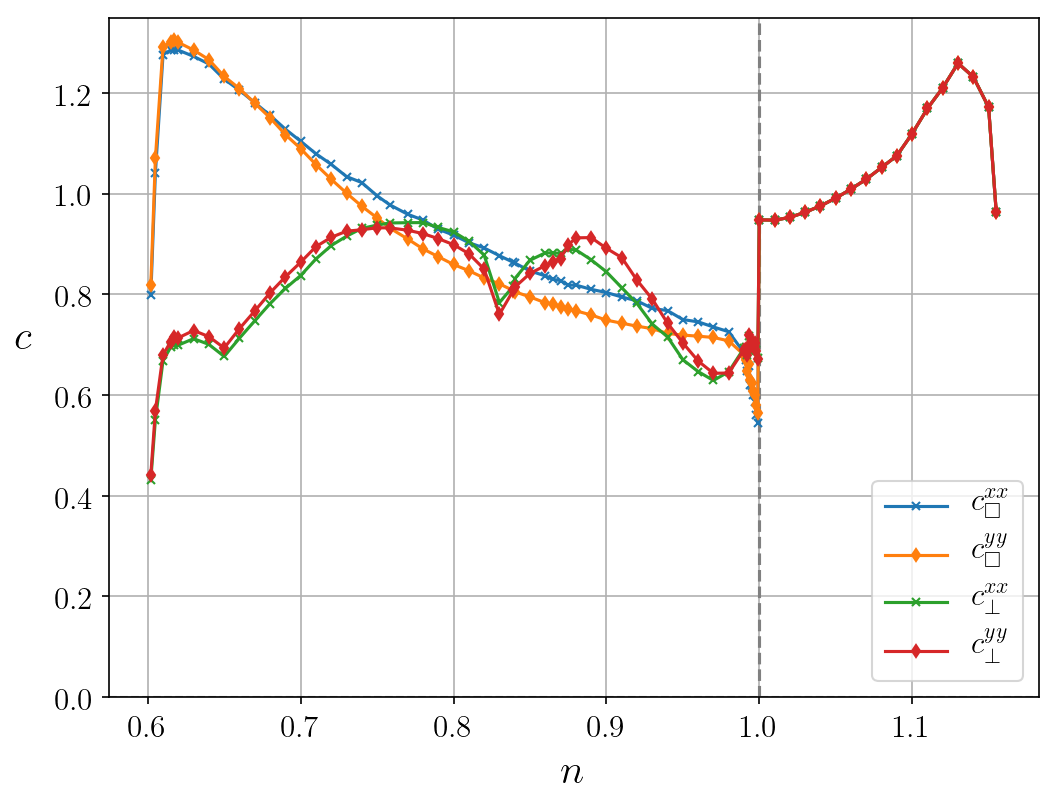}
        \caption{In-plane and out-of-plane magnon velocities $c^a_{\alpha\alpha}=\sqrt{J^a_{\alpha\alpha}/\chi_\mathrm{dyn}^a}$ as functions of the electron density.}
        \label{fig_low_en_sp: fig5}
    \end{figure}
    \begin{figure}[h!]
        \centering
        \includegraphics[width=0.5\textwidth]{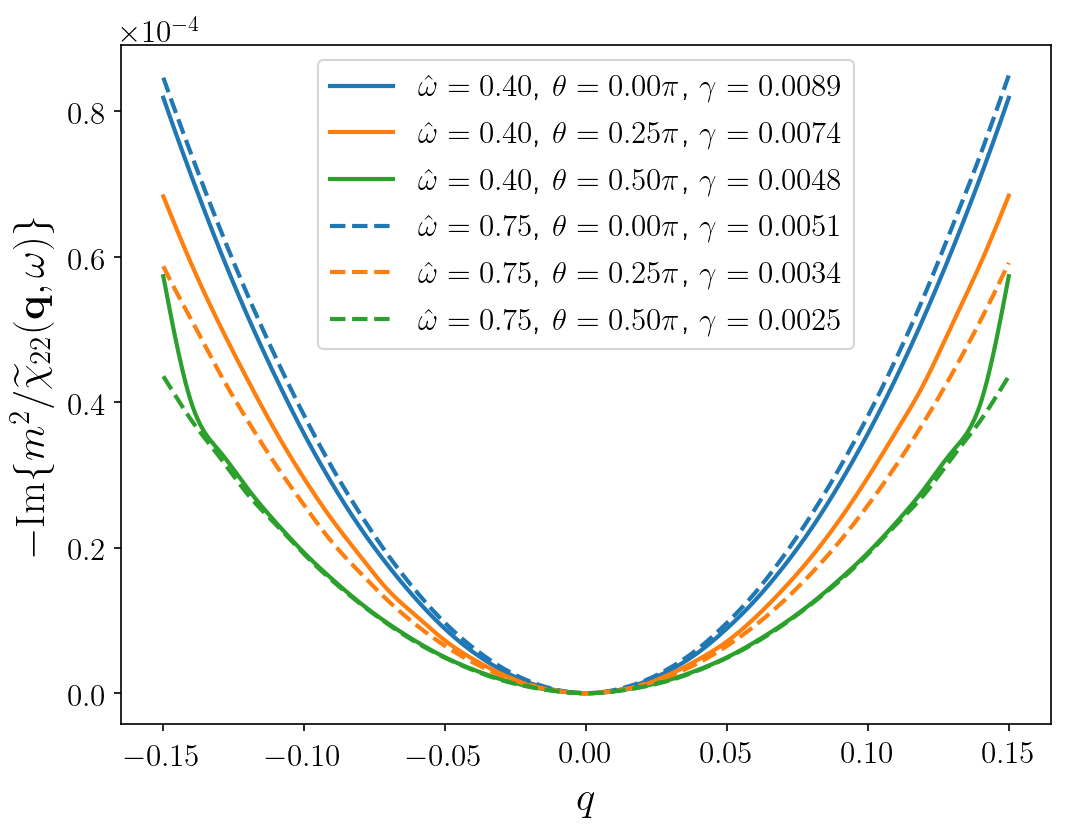}
        \caption{Damping term of the in-plane Goldstone mode as a function of $|\bq|$ at $n=0.84$ for two fixed values of $\hat{\omega}=\omega/|\bq|$ and three fixed directions. The parameter $\theta$ parameterizes the angle between $\bq$ and the $q_x$ axis. The values of the prefactor $\gamma$ of the leading dependence on $\hat{\omega}|\bq|^2$ are shown in the inset.}
        \label{fig_low_en_sp: fig6}
    \end{figure}
    \begin{figure}[h!]
        \centering
        \includegraphics[width=0.5\textwidth]{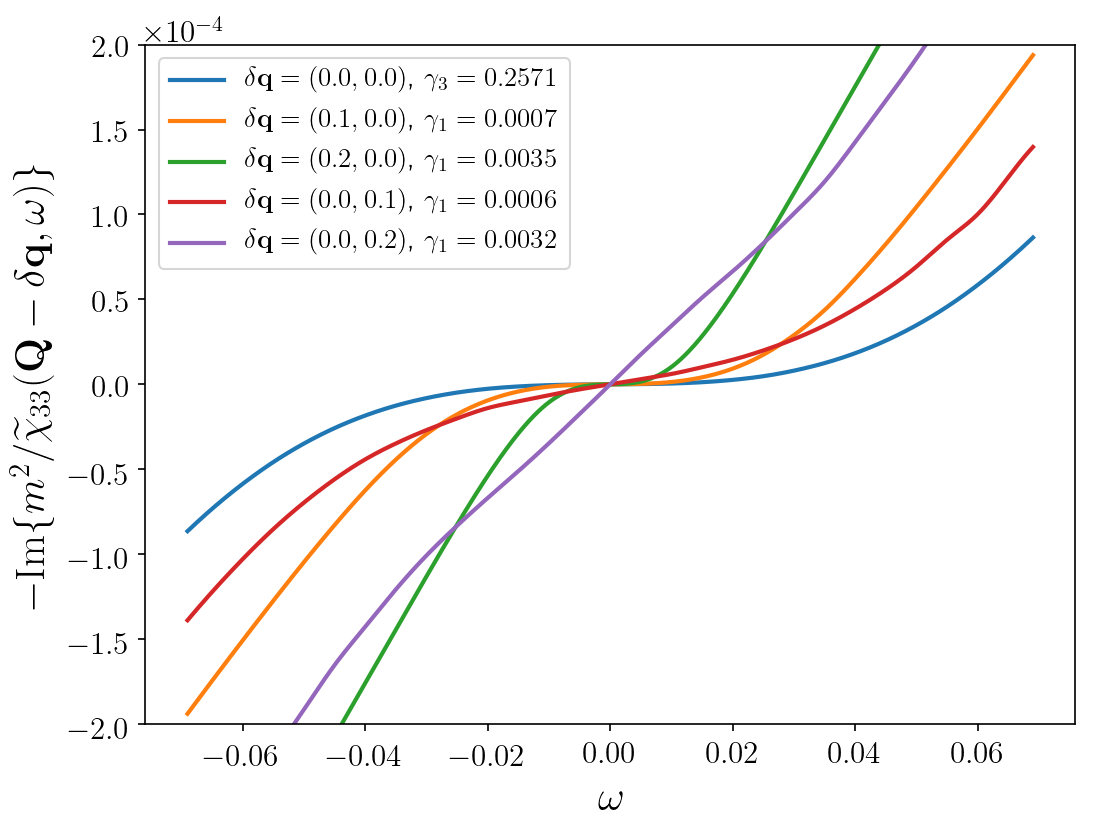}
        \caption{Damping term of the out-of-plane Goldstone mode as a
                 function of $\omega$ for various fixed wave vectors $\bq$ near $\bQ = (0.82\pi, \pi )$ and fixed density $n$ = 0.84. The prefactors $\gamma_1$ of the linear frequency dependence for $\bq\neq\bQ$ and the prefactor $\bQ$ of the cubic frequency dependence for $\bq=\bQ$ are shown in the inset.}
        \label{fig_low_en_sp: fig7}
    \end{figure}
    In this section, we present numerical results for the spin stiffnesses and Landau dampings of the Goldstone modes in a spiral magnetic state obtained from the two-dimensional Hubbard model on a square lattice with neatest and next-t-nearest neighbor hopping amplitudes $t$ and $t'$. All over this section we employ $t$ as energy unit, that is, we set $t=1$. 
    
    All calculations have been performed in the ground state, that is, at $T=0$, and with fixed values of the Hubbard interaction $U=2.5t$ and nearest neighbor hopping $t'=-0.16t$. For this choice of parameters mean-field theory yields a spiral magnetic state for densities ranging from $n\simeq 0.61$ to half filling ($n=1$), with ordering wave vector of the form $\bQ=(\pi-2\pi\eta,\pi)$ and symmetry related. The incommensurability $\eta$ changes increases monotonically upon reducing the density, and vanishes continuously for $n\to1$. For filling factors between $n=1$ and $n\simeq 1.15$, we obtain a N\'eel state. The transition between the paramagnetic and the spiral state at $n\simeq0.61$ is continuous, while the one occurring at $n\simeq1.15$ is of first order with a relatively small jump of the order parameter. The magnetization and incommensurability curves as functions of the electron density are shown in Fig.~\ref{fig_low_en_sp: fig1}.
    
    In Fig.~\ref{fig_low_en_sp: fig2}, we show the quasiparticle Fermi surfaces in the ground state for different densities. In the electron-doped region ($n>1$), these are given by solutions of $E^+_\bk=0$ (cf.~Eq.~\eqref{eq_low_spiral: QP dispersions}), while in the hole-doped regime ($n<1$) by momenta satisfying $E^-_\bk=0$. In principle, for sufficiently small gaps $\Delta$, both equations can have solutions, but we do not find such a case within our choice of parameters. In the ground state, the lines defined by $E^+_\bk=0$ ($E^-_\bk=0$) enclose doubly occupied (empty) states, and we therefore refer to them as electron (hole) pockets. In Fig.~\ref{fig_low_en_sp: fig2}, we also shown the lines along which the equality $E^\ell_\bk=E^\ell_{\bk+\bQ}$ (with $\bQ\neq(\pi,\pi)$) is satisfied. We notice that for small doping these lines never intersect the Fermi surfaces, implying that the out-of-plane modes are not Landau damped at all in this parameter region. 
    
    In Fig.~\ref{fig_low_en_sp: fig3}, we show the in-plane and out-of-plane spin stiffnesses $J^a_{\alpha\beta}$ as functions of the electron density. Both in the spiral state for $n<1$ and in the N\'eel state for $n\geq1$ only the diagonal components $J^a_{xx}$ and $J^a_{yy}$ are nonzero. In the N\'eel state, the stiffnesses are isotropic ($J^a_{xx}=J^a_{yy}$) and degenerate, as dictated by symmetry. In the spiral state the in-plane and the out-of-plane spin stiffnesses differ significantly from each other. Both exhibit a slight nematicity ($J_{xx}^a\neq J^a_{yy}$), coming from the difference between $Q_x$ and $Q_y$\footnote{Note that for a spiral state with $\bQ=(\pi-2\pi\eta,\pi-2\pi\eta)$, or symmetry related, one would have $J^a_{xx}=J^a_{yy}$ but $J^a_{xy}=J^a_{yx}\neq0$.}. Both in-plane and out-of-plane stiffnesses exhibits a sudden and sharp jump upon approaching half filling from the hole-doped side. This discontinuity is due to the sudden appearance of hole pockets, which allow for intraband excitation processes with small energies. Conversely, on the electron-doped side no discontinuity is found as the contributions from the electron pockets are suppressed by vanishing prefactors (see Eq.~\eqref{eq_low_spiral: J0inplane}) at the momenta $(\pi,0)$ and $(0,\pi)$, where they pop up. We remark that we find positive stiffnesses all over the density range in which a magnetic state appears. This proves the stability of the spiral magnetic state over smooth and small deformations of the order parameter, including variations of the wave vector $\bQ$. 
    
    In Fig.~\ref{fig_low_en_sp: fig4}, we plot the spectral weights of the magnon modes as functions of the electron density $n$. The discontinuity in $m^2/\chi_\mathrm{dyn}^\perp$ is due to the intraband terms coming from the emergence of the hole pockets. By contrast, $m^2/\chi_\mathrm{dyn}^\smsqr$ is finite as it only gets contributions from interband processes. The spectral weights vanish near the critical fillings beyond which the magnetic state disappears, indicating that the dynamical susceptibilities vanish slower than $m^2$. The dip in $m^2/\chi^\perp_\mathrm{dyn}$ at $n\approx0.84$ is due to the merging of two electron pockets.
    
    In Fig.~\ref{fig_low_en_sp: fig5}, we plot the magnon velocities $c^a_{\alpha\alpha}=\sqrt{J^a_{\alpha\alpha}/\chi_\mathrm{dyn}^a}$. They only exhibit a mild density dependence and they are always of order $t$ in the entire magnetized regime. It is worthwhile to remark that they remain finite at the critical fillings beyond which the paramagnetic state appears. 
    
    The in-plane damping term ${\rm Im}[m^2/\chit^{22}(\bq,\omega)]$ is plotted in Fig.~\ref{fig_low_en_sp: fig6} as a function of $|\bq|$ for two fixed values of $\hat{\omega}=\omega/|\bq|$ and three fixed directions $\hat{\bq}=\bq/|\bq|$. The density is set to $n=0.84$. The characteristic quadratic behavior of Eq.~\eqref{eq_low_spiral: damping2} is clearly visible, with the prefactors $\gamma(\hat{\bq},\hat{\omega})$ shown in the inset. 
    
    In Fig.~\ref{fig_low_en_sp: fig7}, we plot the frequency dependence of the damping term of the out-of-plane mode ${\rm Im}[m^2/\chit^{33}(\bq,\omega)]$ for various fixed momenta $\bq$ at and near $\bQ$. For $\bq=\bQ$ the damping is proportional to $\omega^3$ in agreement with Eq.~\eqref{eq_low_spiral: damping33 q=Q}. For $\bq\neq\bQ$, one can see the linear frequency dependence predicted by Eq.~\eqref{eq_low_spiral: damping outofplane2}. The prefactors of the leading cubic and linear terms are listed in the inset. 
     

\cleardoublepage
    \rhead[\fancyplain{}{\bfseries SU(2) gauge theory of the pseudogap phase}]{\fancyplain{}{\bfseries\thepage}}
    \lhead[\fancyplain{}{\bfseries\thepage}]{\fancyplain{}{\bfseries SU(2) gauge theory of the pseudogap phase}}
    \chapter{SU(2) gauge theory of the pseudogap phase}
    \label{chap: pseudogap}
    In this Chapter, we derive an effective theory for the pseudogap phase by fractionalizing the electron field into a fermionic chargon, carrying the original electron charge, and a charge neutral spinon. The latter is a SU(2) matrix describing position- and time-dependent (bosonic) fluctuations of the local spin orientation. The fractionalization brings in a SU(2) gauge redundancy, which is why we dub this theory as SU(2) gauge theory. We then consider a magnetically ordered state for the chargons, which leads to a reconstruction of the Fermi surface. We remark that symmetry breaking is nonetheless prevented at finite temperature by the spinon fluctuations, in agreement with Mermin-Wagner theorem. We compute the magnetic state properties of the chargons, starting from the 2D Hubbard model, employing the fRG+MF method described in Chapter~\ref{chap: fRG+MF}. We subsequently integrate out the fermionic degrees of freedom, obtaining an effective non-linear sigma model (NL$\sigma$M) for the spinons. The NL$\sigma$M is described by few parameters, namely spin stiffnesses $J$ and dynamical susceptibilities $\chi_\mathrm{dyn}$, which we compute following the formalism derived in Chapter~\ref{chap: low energy spiral}. A large-$N$ expansion returns a finite temperature pseudogap regime in the hole-doped and electron-doped regions of the phase diagram. On the hole-doped side, we also find a nematic phase at low temperatures, in agreement with the experimentally observed nematicity in cuprate materials~\cite{Ando2002,Cyr-Choiniere2015}. Within our moderate coupling calculation, the spinon fluctuations are found not to be sufficiently strong to destroy long range order in the ground state. The spectral function in the hole doped pseudogap regime has the form of hole pockets with suppressed weight on their backsides, leading to Fermi arcs. The content of this chapter appears in Ref.~\cite{Bonetti2022_III}. 
    \section{SU(2) gauge theory}
    \label{sec_pg: SU(2) gauge theory}
    \subsection{Fractionalizing the electron field}
    We consider the Hubbard model on a square lattice with lattice spacing $a=1$. The action in imaginary time reads
    \begin{eqnarray}
     \mathcal{S}[c,c^*] &=& 
     \int_0^\beta\!d\tau \bigg\{ \sum_{j,j',\sigma} c^*_{j,\sigma}
     \left[ \left( \partial_\tau - \mu\right)\delta_{jj'} + t_{jj'} \right] c_{j',\sigma}
     + \; U \sum_j n_{j,\up}n_{j,\down} \bigg\} , 
     \label{eq_pg: Hubbard action}
    \end{eqnarray}
    where $c_{j,\sigma} = c_{j,\sigma}(\tau)$ and $c^*_{j,\sigma} = c^*_{j,\sigma}(\tau)$ are Grassmann fields corresponding to the annihilation and creation, respectively, of an electron with spin orientation $\sigma$ at site $j$, and $n_{j,\sigma} = c^*_{j,\sigma}c_{j,\sigma}$. The chemical potential is denoted by $\mu$, and $U > 0$ is the strength of the (repulsive) Hubbard interaction. To simplify the notation, we write the dependence of the fields on the imaginary time $\tau$ only if needed for clarity.
    
    The action in \eqref{eq_pg: Hubbard action} is invariant under \emph{global} SU(2) rotations acting on the Grassmann fields as 
    \begin{equation}
     c_j \to \mathcal{U} c_j, \quad\quad
     c^*_j \to c^*_j \, \mathcal{U}^\dagger,
    \label{eq_pg: SU(2) transf. electrons}
    \end{equation}
    where $c_j$ and $c^*_j$ are two-component spinors composed from $c_{j,\sigma}$ and $c^*_{j,\sigma}$, respectively, while $\mathcal{U}$ is a SU(2) matrix acting in spin space.
    
    To separate collective spin fluctuations from the charge degrees of freedom, we fractionalize the electronic fields as~\cite{Schulz1995, Borejsza2004, Scheurer2018, Wu2018}
    \begin{equation}
     c_j = R_j \, \psi_j , \quad\quad
     c^*_j = \psi^*_j \, R^\dagger_j ,
    \label{eq_pg: electron fractionaliz.}
    \end{equation}
    where $R_j \in \mbox{SU(2)}$, to which we refer as ``spinon'', is composed of bosonic fields, and the components of the ``chargon'' $\psi_j$ are fermionic. According to \eqref{eq_pg: SU(2) transf. electrons} and \eqref{eq_pg: electron fractionaliz.} the spinons transform under the global SU(2) spin rotation by a \emph{left} matrix multiplication, while the chargons are left invariant. Conversely, a U(1) charge transformation acts only on $\psi_j$, leaving $R_j$ unaffected.
    The transformation in Eq.~\eqref{eq_pg: electron fractionaliz.} introduces a redundant SU(2) gauge symmetry, acting as 
    \begin{subequations}
    \begin{align}
     & \psi_j \to \mathcal{V}_j\,\psi_j , \qquad\quad
       \psi^*_j \to \psi^*_j \, \mathcal{V}^\dagger_j , \\
     & R_j \to R_j \, \mathcal{V}_j^\dagger, \qquad\quad
       R^\dagger_j \to \mathcal{V}_j \, R^\dagger_{j} ,
    \end{align}
    \end{subequations}
    with $\mathcal{V}_j\in \mbox{SU(2)}$.
    Hence, the components $\psi_{j,s}$ of $\psi_j$ carry an SU(2) gauge index $s$, while the components $R_{j,\sigma s}$ of $R_j$ have two indices, the first one ($\sigma$) corresponding to the global SU(2) symmetry, and the second one ($s$) to SU(2) gauge transformations. 
    
    We now rewrite the Hubbard action in terms of the spinon and chargon fields. The quadratic part of \eqref{eq_pg: Hubbard action} can be expressed as \cite{Borejsza2004}
    \begin{eqnarray}
     \mathcal{S}_0[\psi,\psi^*,R] &=& \int_0^\beta\!d\tau
     \bigg\{ \sum_j \psi^*_j \left[ \partial_\tau - \mu - A_{0,j} \right]
     \psi_{j}
     + \, \sum_{j,j'}t_{jj'}\,\psi^*_{j}\, e^{-\mathbf{r}_{jj'} \cdot \left(\boldsymbol{\nabla} - i\mathbf{A}_j \right)} \, \psi_j \bigg\},
    \label{eq_pg: S0 chargons spinons}
    \end{eqnarray}
    where we have introduced a SU(2) gauge field, defined as 
    \begin{equation}
        A_{\mu,j} = (A_{0,j},\mathbf{A}_j) = i R^\dagger_j \dmu R_j,
        \label{eq_pg: gauge field definition}
    \end{equation}
    with $\dmu = (i\partial_\tau,\boldsymbol{\nabla})$. Here, the nabla operator $\boldsymbol{\nabla}$ is defined as generator of translations on the lattice, that is,
    $e^{-\br_{jj'}\cdot\boldsymbol{\nabla}}$ with $\br_{jj'} = \br_j - \br_{j'}$
    is the translation operator from site $j$ to site $j'$.
    
    To rewrite the interacting part in \eqref{eq_pg: Hubbard action}, we use the decomposition~\cite{Weng1991,Schulz1995,Borejsza2004}
    \begin{equation}
     n_{j,\up}n_{j,\down} = \frac{1}{4}(n_j)^2 - \frac{1}{4}(c^*_j \,
     \vec{\sigma}\cdot\hat{\Omega}_j \, c_j)^2,
    \label{eq_pg: interaction decomposition}
    \end{equation}
    where $n_j = n_{j,\up} + n_{j,\down}$ is the charge density operator,
    $\vec{\sigma} = (\sigma^1,\sigma^2,\sigma^3)$ are the Pauli matrices, and $\hat{\Omega}_j$ is an arbitrary time- and site-dependent unit vector. The interaction term of the Hubbard action can therefore be written in terms of spinon and chargon fields as
    \begin{equation}
     \mathcal{S}_\mathrm{int}[\psi,\psi^*,R] =
     \int_0^\beta\!d\tau \, U \sum_j \left[\frac{1}{4} (n_j^\psi)^2 -
     \frac{1}{4}(\vec{S}^\psi_j\cdot\Hat{\Omega}^R_j)^2 \right] ,
    \end{equation}
    where $n^\psi_j = \psi^*_j\psi_j$ is the chargon density operator, $\vec{S}^\psi_j = \frac{1}{2} \psi^*_j\vec{\sigma}\psi_j$ is the chargon spin operator, and 
    \begin{equation}
     \vec{\sigma}\cdot\hat{\Omega}^R_j =
     R^\dagger_j \, \vec{\sigma} \cdot \hat{\Omega}_j \, R_j .
    \label{eq_pg: Omega and Omega^R}
    \end{equation}
    Using \eqref{eq_pg: interaction decomposition} again, we obtain 
    \begin{equation}
     \mathcal{S}_\mathrm{int}[\psi,\psi^*,R] =
     \int_0^\beta\!d\tau \, U \sum_j n^\psi_{j,\up}n^\psi_{j,\down},
    \end{equation}
    with $n^\psi_{j,s} = \psi^*_{j,s} \psi_{j,s}$.
    Therefore, the final form of the action $\mathcal{S} = \mathcal{S}_0 + \mathcal{S}_\mathrm{int}$ is nothing but the Hubbard model action where the physical electrons have been replaced by chargons coupled to a SU(2) gauge field.
    
    Since the chargons do not carry any spin degree of freedom, a \emph{global} breaking of their SU(2) gauge symmetry ($\langle \vec{S}^\psi_j \rangle \neq 0$) does not necessarily imply long range order for the physical electrons.
    The matrices $R_j$ describe directional fluctuations of the order parameter $\langle \vec{S}_j \rangle$, where the most important ones vary slowly in time and space. 
    \subsection{Non-linear sigma model}
    \label{sec_pg: NLsM}
    We now derive a low energy effective action for the spinon fields $R_j$ by integrating out the chargons,
    \begin{equation}
     e^{-\mathcal{S}_\mathrm{eff}[R]} =
     \int\! \mathcal{D} \psi \mathcal{D} \psi^* \,
      e^{-\mathcal{S}[\psi,\psi^*,R]} . 
    \label{eq_pg: integral over psi}
    \end{equation}
    Since the action $\mathcal{S}$ is quartic in the fermionic fields, the functional integral must be carried out by means of an approximate method. In previous works~\cite{Schulz1995,Dupuis2002,Borejsza2004} a Hubbard-Stratonovich transformation has been applied to decouple the chargon interaction, together with a saddle point approximation on the auxiliary bosonic (Higgs) field. We will employ an improved approximation, which we describe in Sec.~\ref{sec_pg: fRG+MF}. 
    
    The effective action for the spinons can be obtained by computing the response functions of the chargons to a fictitious SU(2) gauge field. Since we assign only low energy long wave length fluctuations to the spinons in the decomposition \eqref{eq_pg: electron fractionaliz.}, the spinon field $R_j$ is slowly varying in space and time. Hence, we can perform a gradient expansion. To second order in the gradient $\dmu R_j$, the effective action $\mathcal{S}_\mathrm{eff}[R]$ has the general form
    \begin{equation}
     \mathcal{S}_\mathrm{eff}[R] = \int_\mathcal {T} \! dx \Big[
     \cB^a_\mu A_{\mu}^a(x) +
     \textstyle{\frac{1}{2}} \cJ^{ab}_\munu  
     A_{\mu}^a(x) A_{\nu}^b(x) \Big] ,
    \label{eq_pg: effective action Amu}
    \end{equation}
    where $\mathcal{T} = [0,\beta] \times \mathbb{R}^2$, repeated indices are summed, and we have expanded the gauge field $A_\mu$ in terms of the SU(2) generators,
    \begin{equation}
     A_\mu(x) = A_\mu^a(x) \, \frac{\sigma^a}{2} ,
    \label{eq_pg: Amu SU(2) generators}
    \end{equation}
    with $a$ running from 1 to 3. 
    In line with the gradient expansion, the gauge field is now defined over a \emph{continuous} space-time. The coefficients in~\eqref{eq_pg: effective action Amu} do not depend on the spatio-temporal coordinates $x = (\tau,\mathbf{r})$ and are given by
    \begin{eqnarray} \label{eq_pg: def Ba}
     \cB^a_\mu &=& \frac{1}{2} \sum_{j,j'} \gamma^{(1)}_{\mu}(j,j')
     \langle \psi^*_j(0) \sigma^a \psi_{j'}(0) \rangle , \\
     \label{eq_pg: def Jab}
     \cJ_\munu^{ab} &=& \frac{1}{4} \sum_{j,j'} \sum_{l,l'}
     \gamma^{(1)}_{\mu}(j,j') \gamma^{(1)}_{\nu}(l,l')
     \nonumber \int_0^\beta d\tau \,
     \big\langle \left( \psi^*_j(\tau) \sigma^a \psi_{j'}(\tau) \right)
     \left( \psi^*_l(0) \sigma^b \psi_{l'}(0) \right) \big\rangle_c
     \nonumber \\
     &&- \frac{1}{4} \sum_{j,j'} \gamma^{(2)}_{\mu\nu}(j,j') 
     \langle \psi^*_j(0) \psi_{j'}(0) \rangle \, \delta_{ab} ,
    \label{eq_pg: spin stiff definitions}
    \end{eqnarray}
    where $\langle\bullet\rangle$ ($\langle\bullet\rangle_c$) denotes the (connected) average with respect to the chargon Hubbard action. The first and second order current vertices have been defined as
    \begin{subequations}
    \begin{align}
    \label{eq_pg: gamma1}
     \gamma^{(1)}(j,j') =& \phantom{-} \left(
     \delta_{jj'}, i\,x_{jj'} \, t_{jj'}, i\,y_{jj'} \, t_{jj'} \right) , \hskip 1cm \\
    \label{eq_pg: gamma2}
     \gamma^{(2)}(j,j') =& - \left( \begin{array}{ccc}
     0 & 0 & 0 \\
     0 & x_{jj'} x_{jj'} \, t_{jj'} & x_{jj'} y_{jj'} \, t_{jj'}\\
     0 & y_{jj'} x_{jj'} \, t_{jj'} & y_{jj'} y_{jj'} \, t_{jj'}\\
     \end{array} \right) , \hskip -5mm
    \end{align}
    \end{subequations}
    where $x_{jj'}$ and $y_{jj'}$ are the $x$ and $y$ components, respectively of $\br_{jj'} = \br_j - \br_{j'}$.
    
    In Sec.~\ref{sec_pg: linear term} we will see that the linear term in~\eqref{eq_pg: effective action Amu} vanishes. We therefore consider only the quadratic contribution to the effective action. 
    
    We now derive an effective theory for the spinon fluctuations, which can be more convenitently expressed in terms of their adjoint representation
    \begin{equation}
    R^\dagger \, \sigma^\a R = \cR^{ab} \sigma^\b .
    \label{eq_pg: R to mathcal R}
    \end{equation}
    We start by proving the identity
    \begin{equation} \label{eq_pg: dmu R identity}
     \dmu\cR = -i \cR\, \Sigma^a A_\mu^a,
    \end{equation}
    $\Sigma^a$ are the generators of the SU(2) in the adjoint representation,
    \begin{equation}
     \Sigma^a_{bc} = -i \varepsilon^{abc},
    \end{equation}
    with $\varepsilon^{abc}$ the Levi-Civita tensor. 
    Rewriting Eq.~\eqref{eq_pg: R to mathcal R} as 
    \begin{equation}
     \cR^{ab} = \frac{1}{2} 
     \Tr\left[ R^\dagger \sigma^a R^{\phantom{\dagger}} \sigma^b \right] \, ,
    \end{equation}
    we obtain the derivative of $\cR$ in the form,
    \begin{equation}
     \dmu\cR^{ab} = \Tr\left[ R^\dagger \sigma^a \, (\dmu R) \sigma^b \right] =\Tr\left[R^\dagger \sigma^a R R^\dagger(\dmu R) \sigma^b \right] 
     = -i \cR^{ac} \Sigma^d_{cb} A_\mu^d,
    \end{equation}
    which is the identity in \eqref{eq_pg: dmu R identity}.
    
    We now aim to express the object $\frac{1}{2} \cJ^{ab}_\munu A_\mu^\a A^b_\nu$ in terms of the matrix field $\cR$. We write the stiffness matrix in terms of a new matrix $\cP_\munu$ via
    \begin{equation} \label{eq_pg: J to P}
     \cJ^{ab}_\munu =\Tr[ \cP_\munu ]\delta_{ab} - \cP^{ab}_\munu  = 
     \Tr \left[ \cP_\munu \Sigma^a \Sigma^b \right] .
    \end{equation}
    Using $\cR^T \cR = \mathbb{1}$, we obtain
    \begin{equation}
        \frac{1}{2} \cJ^{ab}_\munu A_\mu^a A^b_\nu = 
        \frac{1}{2} \Tr \left[ \cP_\munu \,
        \Sigma^a \, \cR^T \cR \, \Sigma^b \right] A_\mu^a A^b_\nu 
        = \frac{1}{2} \Tr \left[ \cP_\munu (\dmu\cR^T)(\dnu\cR) \right] ,
    \end{equation}
    where we have used Eq.~\eqref{eq_pg: dmu R identity} in the last line. The above equation yields
    Eq.~\eqref{eq_pg: general NLsM}. Relation~\eqref{eq_pg: J to P} can be easily inverted using $\Tr[\cJ_\munu ] = 2\Tr[\cP_\munu]$.
    
    We have therefore obtained the non-linear sigma model (NL$\sigma$M) action for the directional fluctuations 
    \begin{equation}
     \mathcal{S}_\mathrm{NL\sigma M} = \int_\mathcal{T}\!dx \,
     \frac{1}{2}\Tr\left[\cP_\munu (\dmu\cR^T)(\dnu\cR)\right],
    \label{eq_pg: general NLsM}
    \end{equation}
    where $\cP_\munu = \frac{1}{2} \Tr[\cJ_\munu] \mathbb{1} - \cJ_\munu $.
    
    The structure of the matrices $\cJ_\munu $ and $\cP_\munu $ depends on the magnetically ordered chargon state. In the trivial case $\langle \vec{S}^\psi_j \rangle = 0$ all the stiffnesses vanish and no meaningful low energy theory for $R$ can be derived. A well-defined low-energy theory emerges, for example, when N\'eel antiferromagnetic order is realized in the chargon sector, that is,
    \begin{equation}
     \langle \vec{S}^\psi_j \rangle \propto (-1)^{\boldsymbol{r}_j} \hat{u},
    \end{equation}
    where $\hat{u}$ is an arbitrary fixed unit vector. Choosing $\hat{u} = \hat{e}_1 = (1,0,0)$, the spin stiffness matrix in the N\'eel state has the form
    \begin{equation}
     \cJ_\munu = \left( \begin{array}{ccc}
     0  & 0 & 0 \\ 0 & J_\munu  & 0 \\ 0 & 0 & J_\munu \end{array} \right) ,
    \end{equation}
    with $(J_{\mu\nu}) = {\rm diag}(-Z,J,J)$.
    In this case the effective theory reduces to the well-known ${\rm O(3)/O(2)} \simeq S_2$ non-linear sigma model \cite{Haldane1983_I,Haldane1983_II}
    \begin{equation}
     \mathcal{S}_\mathrm{NL\sigma M} =
     \frac{1}{2} \int_\mathcal {T} dx \, \left(
     Z |\partial_\tau\hat{\Omega}|^2 + J |\vec{\nabla}\hat{\Omega}|^2
     \right) ,
    \end{equation}
    where $\hat{\Omega}^\a=\cR^{\a1}$, and $|\hat{\Omega}|^2=1$.
    
    Another possibility is spiral magnetic ordering of the chargons,
    \begin{equation}
     \langle\vec{S}^\psi_j\rangle \propto
     \cos(\bQ \cdot \br_j)\hat{u}_1 +
     \sin(\bQ \cdot \br_j)\hat{u}_2,
    \end{equation}
    where $\bQ$ is a fixed wave vector as obtained by minimizing the chargon free energy, while $\hat{u}_1$ and $\hat{u}_2$ are two arbitrary mutually orthogonal unit vectors. The special case $\bQ = (\pi,\pi)$ corresponds to the N\'eel state. Fixing $\hat{u}_1$ to $\hat{e}_1$ and $\hat{u}_2$ to $\hat{e}_2\equiv(0,1,0)$, the spin stiffness matrix takes the form
    \begin{equation}
      \cJ_\munu = \left( \begin{array}{ccc}
      J_\munu^\perp & 0 & 0 \\
      0 & J_\munu^\perp & 0 \\
      0 & 0 & J_\munu^\Box
      \end{array} \right),
    \label{eq_pg:spiral stiffness matrix}
    \end{equation}
    where
    \begin{equation}
     (J_{\mu\nu}^a) =
     \left( \begin{array}{ccc}
     -Z^a & 0 & 0 \\  0 & J_{xx}^a & J_{xy}^a \\ 0 & J_{yx}^a & J_{yy}^a
     \end{array} \right) .
    \end{equation}
    for $a \in \{ \perp,\Box \}$.
    In this case, the effective action maintains its general form~\eqref{eq_pg: general NLsM} and it describes the O(3)$\times$O(2)/O(2) symmetric NL$\sigma$M, which has been previously studied in the context of geometrically frustrated antiferromagnets~\cite{Azaria1990,Azaria1992,Azaria1993_PRL,Azaria1993,Klee1996}. This theory has three independent degrees of freedom, corresponding to one \emph{in-plane} and two \emph{out-of-plane} Goldstone modes. 
    
    Antiferromagnetic N\'eel or spiral orders have been found in the two-dimensional Hubbard model over broad regions of the parameter space by several approximate methods, such as Hartree-Fock~\cite{Igoshev2010}, slave boson mean-field theory~\cite{Fresard1991}, expansion in the hole density~\cite{Chubukov1995}, moderate coupling fRG~\cite{Yamase2016}, and dynamical mean-field theory~\cite{Vilardi2018,Bonetti2020_I}. In our theory the mean-field order applies only to the chargons, while the physical electrons are subject to order parameter fluctuations.
    \section{Computation of parameters}
    \label{sec_pg: fRG+MF}
    In this section, we describe how we evaluate the chargon integral in Eq.~\eqref{eq_pg: integral over psi} to compute the magnetic order parameter and the stiffness matrix $\cJ_{\mu\nu}$. The advantage of the way we formulated our theory in Sec.~\ref{sec_pg: SU(2) gauge theory} is that it allows arbitrary approximations on the chargon action. One can employ various techniques to obtain the order parameter and the spin stiffnesses in the magnetically ordered phase. We use a renormalized mean-field (MF) approach with effective interactions obtained from a functional renormalization group (fRG) flow. In the following we briefly describe our approximation of the (exact) fRG flow, and we refer to Refs.~\cite{Berges2002, Metzner2012, Dupuis2021} and to Chapter~\ref{chap: methods} for the fRG, and to Refs.~\cite{Wang2014, Yamase2016, Bonetti2020_II, Vilardi2020} and to Chapter~\ref{chap: fRG+MF} for the fRG+MF method. 
    
    
    \subsection{Symmetric regime}
    We evaluate the chargon functional integral by using an fRG flow equation \cite{Berges2002, Metzner2012, Dupuis2021}, choosing the temperature $T$ as flow parameter \cite{Honerkamp2001}. Temperature can be used as a flow parameter after rescaling the chargon fields as $\psi_j \to T^\frac{3}{4}\psi_j$, and defining a rescaled bare Green's function,
    \begin{equation}
     G_0^T(\bk,i\nu_n) = \frac{T^{\frac{1}{2}}}{i\nu_n - \epsilon_\bk + \mu} ,
    \end{equation}
    where $\nu_n = (2n+1)\pi T$ the fermionic Matsubara frequency, and $\epsilon_\bk$ is the Fourier transform of the hopping matrix in~\eqref{eq_pg: Hubbard action}.
    
    We approximate the exact fRG flow by a second order (one-loop) flow of the two-particle vertex $V^T$, discarding self-energy feedback and contributions from the three-particle vertex \cite{Metzner2012}. In an SU(2) invariant system the two-particle vertex has the spin structure
    \begin{equation*}
    \begin{split}
     V^T_{\sigma_1\sigma_2\sigma_3\sigma_4}(k_1,k_2,k_3,k_4) &=
     V^T(k_1,k_2,k_3,k_4) \, \delta_{\sigma_1\sigma_3}\,\delta_{\sigma_2\sigma_4} \\
     & - V^T(k_2,k_1,k_3,k_4) \, \delta_{\sigma_1\sigma_4}\,\delta_{\sigma_2\sigma_3} , 
    \end{split}
    \end{equation*}
    where $k_\alpha = (\bk_\alpha,i\nu_{\alpha n})$ are combined momentum and frequency variables. Translation invariance imposes momentum conservation so that $k_1 + k_2 = k_3 + k_4$.
    We perform a static approximation, that is, we neglect the frequency dependency of the vertex. To parametrize the momentum dependence, we use the channel decomposition~\cite{Husemann2009, Husemann2012, Vilardi2017, Vilardi2019}
    \begin{eqnarray}
      V^T(\bk_1,\bk_2,\bk_3,\bk_4) &=& U
     - \phi^{p,T}_{\frac{\bk_1-\bk_2}{2},\frac{\bk_3-\bk_4}{2}}(\bk_1+\bk_2) \nonumber \\
     && + \, \phi^{m,T}_{\frac{\bk_1+\bk_4}{2},\frac{\bk_2+\bk_3}{2}}(\bk_2-\bk_3)
       + \frac{1}{2}\phi^{m,T}_{\frac{\bk_1+\bk_3}{2},\frac{\bk_2+\bk_4}{2}}(\bk_3-\bk_1) 
       \nonumber \\
     && - \frac{1}{2}\phi^{c,T}_{\frac{\bk_1+\bk_3}{2},\frac{\bk_2+\bk_4}{2}}(\bk_3-\bk_1) ,
    \label{eq_pg: vertex parametrization}
    \end{eqnarray}
    where the functions $\phi^{p,T}$, $\phi^{m,T}$, and $\phi^{c,T}$ capture fluctuations in the pairing, magnetic, and charge channel, respectively.
    The dependences of these functions on the linear combination of momenta in the brackets are typically much stronger than those in the subscripts. Hence, we expand the latter dependencies in form factors \cite{Husemann2009,Lichtenstein2017}, keeping only the lowest order s-wave, extended s-wave, p-wave and d-wave contributions.
    
    We run the fRG flow from the initial temperature $T_\mathrm{ini} = \infty$, at which $V^{T_\mathrm{ini}} = U$, down to a critical temperature $T^*$ at which $V^T$ diverges, signaling the onset of spontaneous symmetry breaking (SSB). If the divergence of the vertex is due to $\phi^{m,T}$, the chargons develop some kind of magnetic order. 
    \subsection{Order parameter}
    \label{sec_pg: order parameter and Q}
    In the magnetic phase, that is, for $T < T^*$, we assume an order parameter of the form
    $\langle \psi^*_{\bk,\up} \psi_{\bk+\bQ,\down} \rangle$,  which corresponds to N\'eel antiferromagnetism if $\bQ = (\pi,\pi)$, and to spiral order otherwise. 
    
    For $T < T^*$ we simplify the flow equations by decoupling the three channels $\phi^{P,T}$, $\phi^{M,T}$, and $\phi^{C,T}$. The flow equations can then be formally integrated, and the formation of an order parameter can be easily taken into account \cite{Wang2014}. In the magnetic channel one thus obtains the magnetic gap equation~\cite{Yamase2016}
    \begin{equation}\label{eq_pg: gap equation fRG+MF}
     \Delta_{\bk} = \int_{\bk'} \overline{V}^m_{\bk,\bk'}(\bQ)\,
     \frac{f(E^-_{\bk'}) - f(E^+_{\bk'})}{E^+_{\bk'} - E^-_{\bk'}} \, \Delta_{\bk'} ,
    \end{equation}
    where $f(x)=(e^{x/T}+1)^{-1}$ is the Fermi function, $\int_\bk$ is a shorthand notation for $\int\!\frac{d^2\bk}{(2\pi)^2}$, and $E^\pm_\bk$ are the quasiparticle dispersions
    \begin{equation}
     E^\pm_\bk = \frac{\epsilon_\bk+\epsilon_{\bk+\bQ}}{2}
     \pm\sqrt{\frac{1}{4} \left( \epsilon_\bk-\epsilon_{\bk+\bQ} \right)^2
     + \Delta_\bk^2} \, -\mu .
    \end{equation}
    The effective coupling $\overline{V}^m_{\bk,\bk'}(\bQ)$ is the particle-hole irreducible part of $V^{T^*}$ in the magnetic channel, which can be obtained by inverting a Bethe-Salpeter equation at the critical scale, 
    \begin{equation}
    \begin{split}
     V^{m,T^*}_{\bk,\bk'}(\bq) &= \overline{V}^m_{\bk,\bk'}(\bq)
      - \int_{\bk''} \overline{V}^m_{\bk,\bk''}(\bq) \, \Pi^{T^*}_{\bk''}(\bq) \,
     V^{m,T^*}_{\bk'',\bk'}(\bq) ,
    \end{split}
    \label{eq_pg: V phx Bethe-Salpeter}
    \end{equation}
    where $V^{m,T}_{\bk,\bk'}(\bq) = V^{T}(\bk-\bq/2,\bk'+\bq/2,\bk'-\bq/2,\bk+\bq/2)$, and the particle-hole bubble is given by
    \begin{equation}
     \Pi^{T}_{\bk}(\bq) = \sum_{\nu_n} G_0^{T}\left(\bk-\bq/2,i\nu_n\right)
     G_0^{T}\left(\bk+\bq/2,i\nu_n\right) . 
    \end{equation}
    Although $V^{m,T^*}_{\bk,\bk'}(\bq)$ diverges at certain wave vectors $\bq = \bQ_c$, the irreducible coupling $\overline{V}^m_{\bk,\bk'}(\bq)$ is finite for all $\bq$.
    
    The dependence of $\overline{V}^m_{\bk,\bk'}(\bq)$ on $\bk$ and $\bk'$ is rather weak and of no qualitative importance. Hence, to simplify the calculations, we discard the $\bk$ and $\bk'$ dependencies of the effective coupling by taking the momentum average
    $\overline{V}^m(\bq) = \int_{\bk,\bk'} \overline{V}^m_{\bk,\bk'}(\bq)$.
    The magnetic gap then becomes momentum independent, that is, $\Delta_\bk = \Delta$.
    While the full vertex $V^{m,T}_{\bk,\bk'}(\bq)$ depends very strongly on $\bq$, the dependence of its irreducible part $\overline{V}_{\bk,\bk'}(\bq)$ on $\bq$ is rather weak. The calculation of the stiffnesses in the subsequent section is considerably simplified approximating $\overline{V}^m(\bq)$ by a momentum independent effective interaction $U_{\rm eff}^m = \overline{V}^m(\bQ_c)$.  The gap equation~\eqref{eq_pg: gap equation fRG+MF} therefore simplifies to
    \begin{equation}\label{eq_pg: simplified gap equation}
        1 = U_{\rm eff}^m \int_\bk \frac{f(E^-_{\bk}) - f(E^+_{\bk})}{E^+_{\bk} - E^-_{\bk}}.
    \end{equation}
    The optimal ordering wave vector $\bQ$ is found by minimizing the mean-field free energy of the system
    \begin{equation} \label{eq_pg: MF theromdynamic potential}
     F(\bQ) = - T \int_\bk\sum_{\ell=\pm} \ln\left(1+e^{-E^\ell_\bk(\bQ)/T}\right)
     + \frac{\Delta^2}{2U_{\rm eff}^m} + \mu n ,
    \end{equation}
    where the chemical potential $\mu$ is determined by keeping the density
    $n = \int_\bk \sum_{\ell=\pm} f(E^\ell_\bk)$ fixed and the gap equation~\eqref{eq_pg: gap equation fRG+MF} fulfilled for each value of $\bQ$. The optimal wave vectors $\bQ$ at temperatures $T < T^*$ generally differ from the wave vectors $\bQ_c$ at which $V^{T^*}_{\bk,\bk'}(\bq)$ diverges.
    
    Eq.~\eqref{eq_pg: gap equation fRG+MF} has the form of a mean-field gap equation with a renormalized interaction that is reduced compared to the bare Hubbard interaction $U$ by fluctuations in the pairing and charge channels. This reduces the critical doping beyond which magnetic order disappears, compared to the unrealistically large values obtained already for weak bare interactions in pure Hartree-Fock theory (see e.g.\ Ref.~\cite{Igoshev2010}). 
    
    \subsection{Spin stiffnesses}
    \label{sec_pg: spin stiff formalism}
    
    The NL$\sigma$M parameters, that is, the spin stiffnesses $\cJ_{\mu\nu}^{ab}$, are obtained by evaluating Eq.~\eqref{eq_pg: spin stiff definitions}. These expressions can be viewed as the response of the chargon system to an external SU(2) gauge field in the low energy and long wavelength limit, and they are equivalent to the stiffnesses defined by an expansion of the inverse susceptibilities to quadratic order in momentum and frequency around the Goldstone poles (see Chapter~\ref{chap: low energy spiral}).
    The following evaluation is obtained as a simple generalization of the RPA formula derived in Chapter~\ref{chap: low energy spiral} to a renormalized RPA with effective interaction
    \begin{equation}
     \widetilde{\Gamma}^{ab}_{0}(\bq) = \Gamma^{ab}_{0}(\bq) =
     2 \, {\rm diag} \left[
     -U_{\rm eff}^c(\bq),U_{\rm eff}^m,U_{\rm eff}^m,U_{\rm eff}^m \right] ,
    \end{equation}
    where $U_{\rm eff}^m$ has been defined before, and the effective charge interaction is given by
    $U_{\rm eff}^c(\bq) = \int_{\bk,\bk'} \overline V^c_{\bk,\bk'}(\bq)$,
    where the irreducible coupling $\overline V^c_{\bk,\bk'}(\bq)$ is obtained by inverting a Bethe-Salpeter equation similar to Eq.~\eqref{eq_pg: V phx Bethe-Salpeter},
    \begin{equation}\label{eq_pg: V ph Bethe-Salpeter}
    \begin{split}
     V^{c,T^*}_{\bk,\bk'}(\bq) &= \overline{V}^c_{\bk,\bk'}(\bq) 
     + \int_{\bk''} \overline{V}^c_{\bk,\bk''}(\bq) \, \Pi^{T^*}_{\bk''}(\bq) \,
     V^{c,T^*}_{\bk'',\bk'}(\bq) ,
    \end{split}
    \end{equation}
    with
    \begin{eqnarray}
     V^{c,T}_{\bk,\bk'}(\bq) &=& 2V^{T}(\bk\!-\!\bq/2,\bk'\!+\!\bq/2,\bk\!+\!\bq/2,\bk'\!-\!\bq/2)
     \nonumber \\
     &-& V^{T}(\bk\!-\!\bq/2,\bk'\!+\!\bq/2,\bk'\!-\!\bq/2,\bk\!+\!\bq/2) . \nonumber
    \end{eqnarray}
    Here we keep the dependence on $\bq$ since it does not complicate the calculations. We remark that the temporal stiffnesses $Z^a$ of this chapter conincide with the dynamic susceptibilities $\chi_\mathrm{dyn}^a$ of chapter~\ref{chap: low energy spiral}. 
    \subsection{Linear term in the gauge field}
    \label{sec_pg: linear term}
    We now show that the linear term in Eq.~\eqref{eq_pg: effective action Amu} vanishes. Fourier transforming the vertex and the expectation value, the coefficient $\mathcal{B}_\mu^a$ can be written as
    \begin{equation}
     \mathcal{B}_\mu^a = \frac{1}{2} \int_\bk T \sum_{\nu_n} \gamma_\mu^{(1)}(\bk)
     \Tr \left[ \sigma^a \mathbf{G}_{\bk,\bk}(i\nu_n) \right] \, .
    \end{equation}
    Inserting $\mathbf{G}_{\bk,\bk}(\nu_n) $ from Eq.~\eqref{eq_low_spiral: G unrotated} (see Chapter~\ref{chap: low energy spiral}) one immediately sees that $\mathcal{B}_\mu^1 = \mathcal{B}_\mu^2 = 0$ for $\mu = 0,1,2$, and $\mathcal{B}_0^3 = 0$, too. 
    Performing the Matsubara sum for $\mathcal{B}^3_\alpha$ with $\alpha = 1,2$, we obtain
    \begin{equation}
     \mathcal{B}^3_\alpha = \frac{1}{2} \int_\bk
     \sum_{\ell=\pm} \left[
     (\partial_{k_\alpha}\epsilon_\bk) u^\ell_\bk f(E^\ell_\bk) + 
     (\partial_{k_\alpha}\epsilon_{\bk+\bQ}) u^{-\ell}_\bk f(E^\ell_\bk) \right] \, . 
        \label{eq_pg: linear term in Amu expression}
    \end{equation}
    One can see by direct calculation that this term vanishes if $\partial F(\bQ)/\partial\bQ$ with $F(\bQ)$ given by Eq.~\eqref{eq_pg: MF theromdynamic potential} vanishes. Hence, $\mathcal{B}^3_\alpha$ vanishes if $\bQ$ minimizes the free energy.
    A similar result has been obtained in Ref.~\cite{Klee1996}. 
    \section{Evaluation of sigma model}

    To solve the NL$\sigma$M~\eqref{eq_pg: general NLsM}, we resort to a saddle point approximation in the $\text{CP}^{N-1}$ representation, which becomes exact in the large $N$ limit \cite{AuerbachBook1994,Chubukov1994}.
    
    
    \subsection{\texorpdfstring{CP$^{\bf 1}$}{CP} representation}
    
    The matrix $\cR$ can be expressed as a triad of orthonormal unit vectors:
    \begin{equation} \label{eq_pg:mathcal R to Omegas}
     \cR=\big( \hat\Omega_1,\hat\Omega_2,\hat\Omega_3 \big),
    \end{equation}
    where $\hat\Omega_i\cdot\hat\Omega_j = \delta_{ij}$. We represent these vectors in terms of two complex Schwinger bosons $z_\up$ and $z_\down$ \cite{Sachdev1995}
    \begin{subequations} \label{eq_pg:mathcal R to z}
    \begin{align}
     & \hat\Omega_- = z(i\sigma^2\vec{\sigma})z, \\
     & \hat\Omega_+ = z^*(i\sigma^2\vec{\sigma})^\dagger z^*, \\
     & \hat\Omega_3 = z^*\vec{\sigma}z,
    \end{align}
    \end{subequations}
    with $z = (z_\up,z_\down)$ and $\hat{\Omega}_\pm=\hat{\Omega}_1\mp i \hat{\Omega}_2$.
    The Schwinger bosons obey the non-linear constraint
    \begin{equation} \label{eq_pg:z boson constraint}
     z^*_\up z_\up + z^*_\down z_\down = 1 \, .
    \end{equation}
    The parametrization~\eqref{eq_pg:mathcal R to z} is equivalent to setting
    \begin{equation} \label{eq_pg: R to z}
     R = \left( \begin{array}{cc}
     z_\up &  -z_\down^* \\ z_\down & \phantom{-} z_\up^* 
     \end{array} \right),
    \end{equation}
    in Eq.~\eqref{eq_pg: R to mathcal R}. Inserting the expressions \eqref{eq_pg:mathcal R to Omegas} and \eqref{eq_pg:mathcal R to z} into Eq.~\eqref{eq_pg: general NLsM} and assuming a stiffness matrix $\cJ_\munu $ of the form~\eqref{eq_pg:spiral stiffness matrix}, we obtain the $\rm CP^1$ action
    \begin{equation} \label{eq_pg:CP1 action}
     \mathcal{S}_{\text{CP}^1}[z,z^*] = \int_\mathcal {T} dx \,
     \Big[ 2J^\perp_\munu (\dmu z^*)(\dnu z) 
     - \, 2(J^\perp_\munu - J^\Box_\munu) j_\mu j_\nu \Big] \, ,
    \end{equation}
    with sum convention for the spin indices of $z$ and $z^*$ and the current operator
    \begin{equation}
     j_\mu = \frac{i}{2}\left[z^*(\dmu z)-(\dmu z^*)z\right] \, .
    \end{equation}
    We recall that $x = (\tau,\br)$ comprises the imaginary time and space variables, and
    $\mathcal{T} = [0,\beta] \times \mathbb{R}^2$. 
    
    
    \subsection{Large {\em N} expansion}
    
    The current-current interaction in Eq.~\eqref{eq_pg:CP1 action} can be decoupled by a Hubbard-Stratonovich transformation, introducing a U(1) gauge field $\mathcal{A}_\mu$, and implementing the constraint~\eqref{eq_pg:z boson constraint} by means of a Lagrange multiplier $\lambda$. The resulting form of the action describes the so-called massive $\rm CP^1$ model~\cite{Azaria1995}
    \begin{equation} \label{eq_pg:massive CP1 model}
     \cS_{\text{CP}^1}[z,z^*,\mathcal{A}_\mu,\lambda] = \int_\mathcal {T} dx
     \Big[ 2J^\perp_\munu (D_\mu z)^* (D_\nu z) 
     + \frac{1}{2} M_{\mu\nu} \mathcal{A}_\mu \mathcal{A}_\nu + i\lambda(z^*z-1)
     \Big] \, ,
    \end{equation}
    where $D_\mu = \dmu - i\mathcal{A}_\mu$ is the covariant derivative. The numbers $M_{\mu\nu}$ are the matrix elements of the mass tensor of the U(1) gauge field,
    \begin{equation}
     {\rm M} = 4 \big[ 1 - {\rm J}^\Box ({\rm J}^\perp)^{-1} \big]^{-1} {\rm J}^\Box \, ,
    \end{equation}
    where ${\rm J}^\Box$ and ${\rm J}^\perp$ are the stiffness tensors built from the matrix elements
    $J_{\mu\nu}^\Box$ and $J_{\mu\nu}^\perp$, respectively.
    
    To perform a large $N$ expansion, we extend the two-component field $z = (z_\up,z_\down)$ to an $N$-component field $z = (z_1,\dots,z_N)$, and rescale it by a factor $\sqrt{N/2}$ so that it now satisfies the constraint
    \begin{equation}
     z^*z = \sum_{\alpha=1}^N z^*_\alpha z_\alpha = \frac{N}{2} \, .
    \end{equation}
    To obtain a nontrivial limit $N \to \infty$, we rescale the stiffnesses $J^\perp_{\mu\nu}$ and $J^\Box_{\mu\nu}$ by a factor $2/N$, yielding the action
    \begin{equation} \label{eq_pg:massive CPN1 model}
     \cS_{\text{CP}}^{N-1}[z,z^*,\mathcal{A}_\mu,\lambda] = \int_\mathcal {T} dx
     \Big[ 2J^\perp_\munu (D_\mu z)^* (D_\nu z) 
     + \! \frac{N}{4} M_{\mu\nu} \mathcal{A}_\mu \mathcal{A}_\nu +
     i\lambda \Big( z^*z - \frac{N}{2} \Big)
     \Big] . 
    \end{equation}
    This action describes the massive ${\rm CP}^{N-1}$ model~\cite{Campostrini1993}, which in $d>2$ dimensions displays two distinct critical points~\cite{Azaria1995,Chubukov1994,Chubukov1994_II}. The first one belongs to the pure ${\rm CP}^{N-1}$ class, where $M_{\mu\nu} \to 0$ ($J^\Box_{\mu\nu} = 0$), which applies, for example, in the case of N\'eel ordering of the chargons, and the U(1) gauge invariance is preserved. The second is in the O(2N) class, where $M_{\mu\nu} \to \infty$ ($J^\perp_{\mu\nu} = J^\Box_{\mu\nu}$) and the gauge field does not propagate. At the leading order in $N^{-1}$, the saddle point equations are the same for both fixed points, so that we can ignore this distinction in the following. 
    
    At finite temperatures $T > 0$ the non-linear sigma model does not allow for any long-range magnetic order, in agreement with the Mermin-Wagner theorem. The spin correlations decay exponentially and the spin excitations are bounded from below by a spin gap
    $m_s = \sqrt{i\langle\lambda\rangle/Z^\perp}$.
    
    Integrating out the $z$-bosons from Eq.~\eqref{eq_pg:massive CPN1 model}, we obtain the effective action \cite{AuerbachBook1994}
    \begin{equation}
     \cS[\cA_\mu,\lambda] = N \int_\mathcal{T} dx \Big[
     \ln \left( -2J^\perp_{\mu\nu} D_\mu D_\nu + i \lambda \right)
     - \frac{i}{2} \lambda
      + \,\frac{1}{4} M_{\mu\nu} \cA_\mu \cA_\nu \Big] \, . 
    \end{equation}
    In the large $N$ limit the functional integral for its partition function
    is dominated by its saddle point, which is determined by the stationarity equations
    \begin{equation}
     \frac{\delta\mathcal{S}}{\delta\mathcal{A}_\mu}= 
     \frac{\delta\mathcal{S}}{\delta\lambda} = 0 \, .
    \end{equation}

    The first condition implies $\mathcal{A}_\mu=0$, that is, in the large $N$ limit the U(1) gauge field fluctuations are totally suppressed. The variation with respect to $\lambda$ gives, assuming a spatially uniform average value for $\lambda$, 
    \begin{equation}
     T \sum_{\omega_n} \int_\bq
     \frac{1}{Z^\perp \omega_n^2 + J^\perp_{\alpha\beta} q_\alpha q_\beta +
     i\langle\lambda\rangle} = 1 \, . 
    \end{equation}
    Performing the sum over the bosonic Matsubara frequencies $\omega_n = 2n\pi T$, inserting the identity 
    \begin{equation}
     1 = \int_0^\infty \! d\epsilon \,
     \delta\Big( \epsilon - \sqrt{ J^\perp_{\alpha\beta} q_\alpha q_\beta/Z^\perp } \, \Big) ,
    \end{equation}
    and performing the $\bq$-integral, we obtain a self-consistent equation for the spin gap
    \begin{equation} \label{eq_pg:large N equation}
     \frac{1}{4\pi J}
     \int_0^{c_s\Luv} \!\frac{\epsilon\,d\epsilon}{\sqrt{\epsilon^2+m_s^2}} \,\mathrm{coth}\left(\frac{\sqrt{\epsilon^2+m_s^2}}{2T}\right) = 1 \, ,
    \end{equation}
    where $\Luv$ is an ultraviolet momentum cutoff. The constant $J$ is an ``average'' spin stiffness given by
    \begin{equation}
     J = \sqrt{ \mathrm{det} \left( \begin{array}{cc}
     J^\perp_{xx} & J^\perp_{xy} \\ J^\perp_{yx} & J^\perp_{yy}
     \end{array} \right) } \, ,
    \end{equation}
    and $c_s = \sqrt{J/Z^\perp}$ is the corresponding average spin wave velocity.
    In Sec.~\ref{sec_pg: cutoff}, we shall discuss how to choose the value of $\Luv$. 
    For $m_s \ll c_s\Luv$, and $T \ll c_s\Luv$, the magnetic correlation length $\xi_s = \frac{1}{2} c_s/m_s$, behaves as
    \begin{equation}
     \xi_s = 
     \frac{c_s}{4T \, \sinh^{-1} \! \left[
     \frac{1}{2} e^{-\frac{2\pi}{T}(J - J_c)} \right] } \, ,
    \end{equation}
    with the critical stiffness
    \begin{equation} \label{eq_pg:Jc}
     J_c = \frac{c_s\Luv}{4\pi} \, .
    \end{equation}
    The correlation length is finite at each $T > 0$. For $J > J_c$, $\xi_s$ diverges exponentially for $T \to 0$, while for $J < J_c$ it remains finite in the zero temperature limit.
    
    At $T=0$, Eq.~\eqref{eq_pg:large N equation} may not have a solution for any value of $m_s$. This is due to the Bose-Einstein condensation of the Schwinger bosons $z$. One therefore has to account for this effect by adding a \emph{condensate fraction $n_0$} to the left hand side of Eq.~\eqref{eq_pg:large N equation}. For later convenience, we assume that only $z$ bosons with spin index $\up$ condense. We obtain
    \begin{equation} \label{eq_pg:large N eq at T=0}
     n_0 + \frac{1}{4\pi J}
     \int_0^{c_s\Luv} \! \frac{\epsilon \, d\epsilon}{\sqrt{\epsilon^2 + m_s^2}} = 1 \, ,
    \end{equation}
    where $n_0 = |\langle z_\up \rangle|^2$. Eq.~\eqref{eq_pg:large N eq at T=0} can be easily solved, yielding (if $m_s \ll \Luv$)
    \begin{subequations}
    \begin{align}
     &\begin{cases}
     & m_s=0 \\ & n_0 = 1 - \frac{J_c}{J}
     \end{cases}
     \hskip 5mm \text{for } J>J_c \, , \\ 
     &\begin{cases}
     & n_0 = 0 \\ & m_s = 2\pi J\left[\left( J_c/J \right)^2 - 1 \right]
     \end{cases}
     \hskip 5mm \text{for } J<J_c.
    \end{align}
    \end{subequations}

    The Mermin-Wagner theorem is thus respected already in the saddle-point approximation to the ${\rm CP}^{N-1}$ representation of the non-linear sigma model, that is, there is no long-range order at $T > 0$. In the ground state, long-range order (corresponding to a $z$ boson condensation) is obtained for a sufficiently large spin stiffness, while for $J < J_c$ magnetic order is destroyed by quantum fluctuations even at $T = 0$, giving rise to a paramagnetic state with a spin gap.
    
    \subsection{Choice of ultraviolet cutoff}
    \label{sec_pg: cutoff}
    The impact of spin fluctuations described by the non-linear sigma model depends strongly on the ultraviolet cutoff $\Luv$. In particular, the critical stiffness $J_c$ separating a ground state with magnetic long-range order from a disordered ground state is directly proportional to $\Luv$. The need for a regularization of the theory by an ultraviolet cutoff is a consequence of the gradient expansion. While the expansion coefficients (the stiffnesses) are determined by the microscopic model, there is no systematic way of computing $\Luv$.
    
    A pragmatic choice for the cutoff is given by the ansatz
    \begin{equation} \label{eq_pg: Luv}
     \Luv = C/\xi_A \, ,
    \end{equation}
    where $C$ is a dimensionless number, and $\xi_A$ is the magnetic coherence length, which is the characteristic length scale of spin amplitude correlations. This choice may be motivated by the observation that local moments with a well defined spin amplitude are not defined at length scales below $\xi_A$ \cite{Borejsza2004}.
    The constant $C$ can be fixed by matching results from the non-linear sigma model to results from a microscopic calculation in a suitable special case (see below).
    
    The coherence length $\xi_A$ can be obtained from the connected spin amplitude correlation function $\chi_A(\mathbf{r}_{j},\mathbf{r}_{j'}) = \big\langle (\hat{n}_j \cdot \vec{S}^\psi_j)(\hat{n}_{j'} \cdot \vec{S}^\psi_{j'}) \big\rangle_c$, where
    $\hat{n}_j = \langle \vec{S}^\psi_j \rangle / |\langle \vec{S}^\psi_j \rangle|$.
    At long distances between $\br_j$ and $\br_{j'}$ this function decays exponentially with an exponential dependence $e^{-r/\xi_A}$ of the distance $r$.
    Fourier transforming and using the rotated spin frame introduced in Chapter~\ref{chap: low energy spiral}, the long distance behavior of $\chi_A(\mathbf{r}_{j},\mathbf{r}_{j'})$ can be related to the momentum dependence of the static correlation function $\chit^{ab}(\bq,0)$ in the amplitude channel $a=b=1$ for small $\bq$, which has the general form
    \begin{equation}
     \chit^{11}(\bq,0) \propto \frac{1}{J^A_{\alpha\beta} q_\alpha q_\beta + m_A^2} \, .
    \end{equation}
    The magnetic coherence length is then given by
    \begin{equation}
     \xi_A = \sqrt{J_A}/(2 m_A) \, ,
    \end{equation}
    where $J_A = \left( J_{xx}^A J_{yy}^A - J_{xy}^A J_{yx}^A \right)^\frac{1}{2}$.
    
    The constant $C$ in Eq.~\eqref{eq_pg: Luv} can be estimated by considering the Hubbard model with pure nearest neighbor hopping at half filling. At strong coupling (large $U$) the spin degrees of freedom are then described by the antiferromagnetic Heisenberg model, which exhibits a N\'eel ordered ground state with a magnetization reduced by a factor $n_0 \approx 0.6$ compared to the mean-field value \cite{Manousakis1991}. On the other hand, evaluating the RPA expressions for the Hubbard model in the strong coupling limit, one recovers the mean-field results for the spin stiffness and spin wave velocity of the Heisenberg model with an exchange coupling $J_H = 4t^2/U$, namely $J = J_H/4$ and $c_s = \sqrt{2} J_H$. Evaluating the RPA spin amplitude correlation function yields $\xi_A = 1/\sqrt{8}$ in this limit. With the ansatz \eqref{eq_pg: Luv}, one then obtains $n_0 = 1 - 4C/\pi$. Matching this with the numerical result $n_0 \approx 0.6$ yields $C \approx 0.3$ and $\Luv \approx 0.9$.
    \section{Results}
    In this section we present and discuss results obtained from our theory for the two-dimensional Hubbard model, both in the hole- ($n<1$) and electron-doped ($n>1$) regime. We fix the ratio of hopping amplitudes as $t'/t = -0.2$, and we choose a moderate interaction strength $U=4t$. The energy unit is $t$ in all plots. 
    \subsection{Chargon mean-field phase diagram}
    \begin{figure}[h!]
        \centering
        \includegraphics[width=0.6\textwidth]{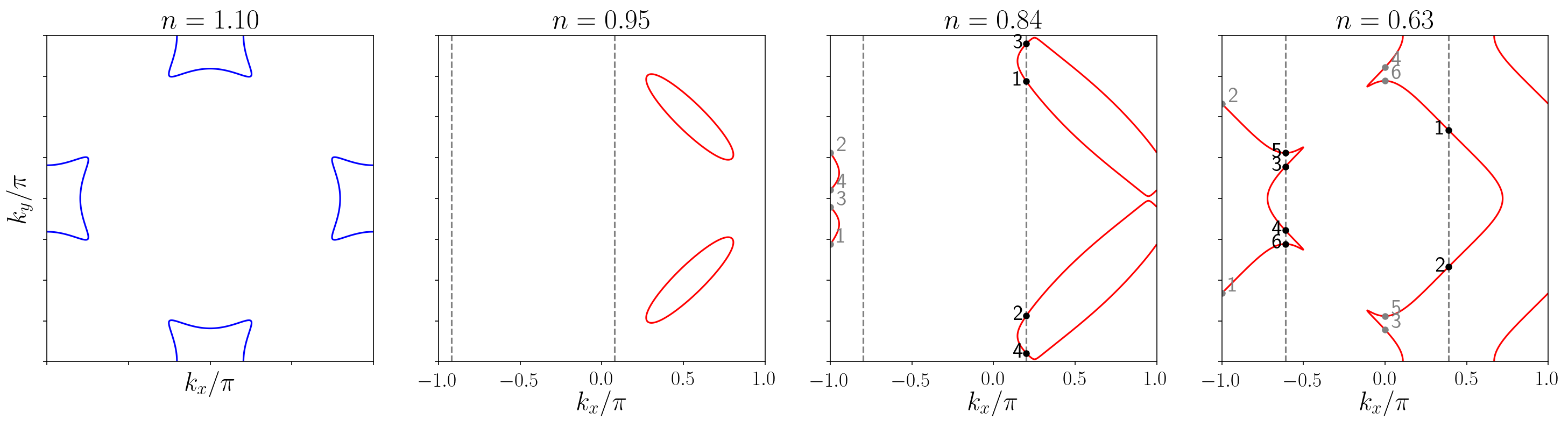}
        \caption{Pseudocritical temperature $T^*$ and nematic temperature $T_\mathrm{nem}$ functions of the density $n$. The labels $T^*_m$ and $T^*_p$ indicate whether the effective interaction diverges in the magnetic or in the pairing channel, respectively. The black solid line indicates the magnetic transition temperature if the pairing instability is ignored (see main text). The labels "N\'eel" and "Spiral" refer to the type of chargon order. The dashed black line refers indicates a topological transition of the quasiparticle Fermi surface within the spiral regime.
        Inset: irreducible magnetic effective interaction $U_{\rm eff}^m$ as a function of density $n$.}
        \label{fig_pg: fig1}
    \end{figure}
    \begin{figure}[h!]
        \centering
        \includegraphics[width=0.6\textwidth]{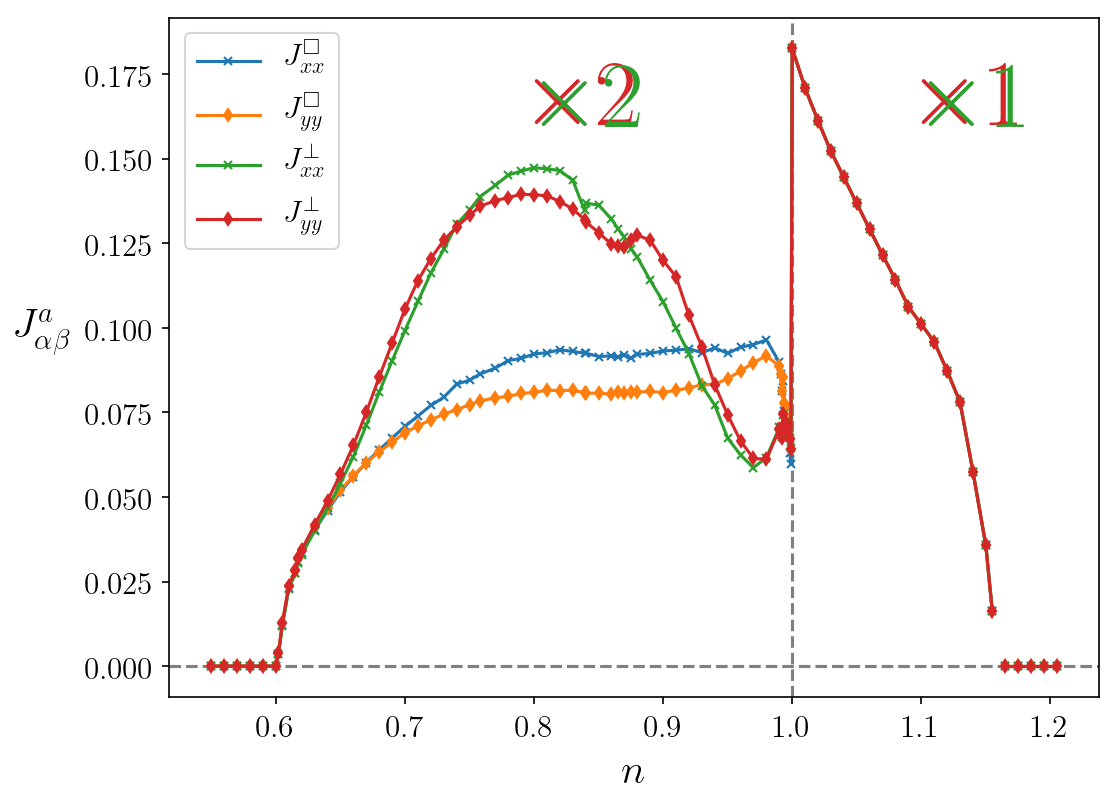}
        \caption{Magnetic gap $\Delta$ (left axis) and incommensurability $\eta$ (right axis) at $T=0$ as functions of the density.}
        \label{fig_pg: fig2}
    \end{figure}
    In Fig.~\ref{fig_pg: fig1}, we plot the critical temperature $T^*$ at which the vertex $V^T(\bk_1,\bk_2,\bk_3,\bk_4)$ diverges. In a wide filling window, from $n=0.84$ to $n=1.08$, the divergence of the vertex is due to a magnetic instability. At the edges of the magnetic dome in the $n$-$T$ phase diagram, the leading instability occurs in the $d$-wave pairing channel. Pairing is expected to extend into the magnetic regime as a secondary instability \cite{Wang2014,Yamase2016}. Vice versa, magnetic order is possible in the regime where pairing fluctuations dominate. In Fig.~\ref{fig_pg: fig1}, we also show the magnetic pseudocritical temperature ($T^*$) obtained by neglecting the onset of pairing (black solid line). This can be determined by setting the magnetic order parameter $\Delta$ to zero in the gap equation~\eqref{eq_pg: gap equation fRG+MF} and solving for the temperature. In the hole-doped part of the $n$-$T$ phase diagram where the $d$-wave superconducting instability is the dominating one, the magnetic critical temperature is only slightly smaller than $T_p^*$. Conversely, in the same region on the electron doped side it vanishes. In Fig.~\ref{fig_pg: fig1}, we also plot the \emph{nematic} temperature $T_\mathrm{nem}$, below which the magnetic chargon state transitions from the N\'eel one to the spiral one, breaking the $C_4$ lattice rotational symmetry. At small hole dopings we first find a N\'eel antiferromagnetic phase at higher temperatures and the system undergoes a second transition to the spiral phase at lower $T$. Conversely, for fillings smaller then $n=0.88$, even right below $T_m^*$ we find a nematic state. Within the spiral regime there is a topological transition of the quasiparticle Fermi surface (indicated by a black dashed line in Fig.~\ref{fig_pg: fig1}), where hole pockets merge. The single-particle spectral function develops Fermi arcs on the right hand side of this transition, while it resembles the large bare Fermi surface on the left (see Sec.~\ref{sec_pg: spectral function}).
    
    In the inset in Fig.~\ref{fig_pg: fig1}, we also show the irreducible effective magnetic interaction $U_\mathrm{eff}^m$ defined in Sec.~\ref{sec_pg: order parameter and Q}. The effective interaction $U_\mathrm{eff}^m$ is strongly reduced from its bare value ($U=4t$) by the non-magnetic channels in the fRG flow.
    
    From now on we ignore the pairing instability and focus on magnetic properties. We compute the magnetic order parameter $\Delta$ together with the optimal wave vector $\bQ$ in the ground state (at $T=0$) as described in Sec.~\ref{sec_pg: order parameter and Q}. In Fig.~\ref{fig_pg: fig2}, we show results for $\Delta$ as a function of the filling. We find a stable magnetic solution extending deep into the hole doped regime down to $n \approx 0.73$. On the electron doped side magnetic order terminates abruptly already at $n \approx 1.08$. This pronounced electron-hole asymmetry and the discontinuous transition on the electron doped side has already been observed in previous fRG+MF calculations for a slightly weaker interaction $U = 3t$ \cite{Yamase2016}.
    The magnetic gap reaches its peak at $n=1$, as expected, although the pseudocritical temperature $T^*$ and the irreducible effective interaction $U_{\rm eff}^m$ exhibit their maximum in the hole doped regime slightly away from half-filling.
    
    The magnetic states are either N\'eel type or spiral with a wave vector of the form $\bQ = (\pi-2\pi\eta,\pi)$, or symmetry related, with an ``incommensurability'' $\eta > 0$. 
    In Fig.~\ref{fig_pg: fig2} results for $\eta$ are shown as a function of the density. At half-filling and in the electron doped region only N\'eel order is found, as expected and in agreement with previous fRG+MF studies \cite{Yamase2016}. Hole doping instead immediately leads to a spiral phase with $\eta > 0$. Whether the N\'eel state persists at small hole doping depends on the hopping parameters and the interaction strength. Its instability toward a spiral state is favored by a larger interaction strength \cite{Chubukov1995}. Indeed, in a previous fRG+MF calculation at weaker coupling the N\'eel state was found to survive up to about 10 percent hole doping \cite{Yamase2016}.
    
    \subsection{Spinon fluctuations}
    \begin{figure}[h!]
        \centering
        \includegraphics[width=1.\textwidth]{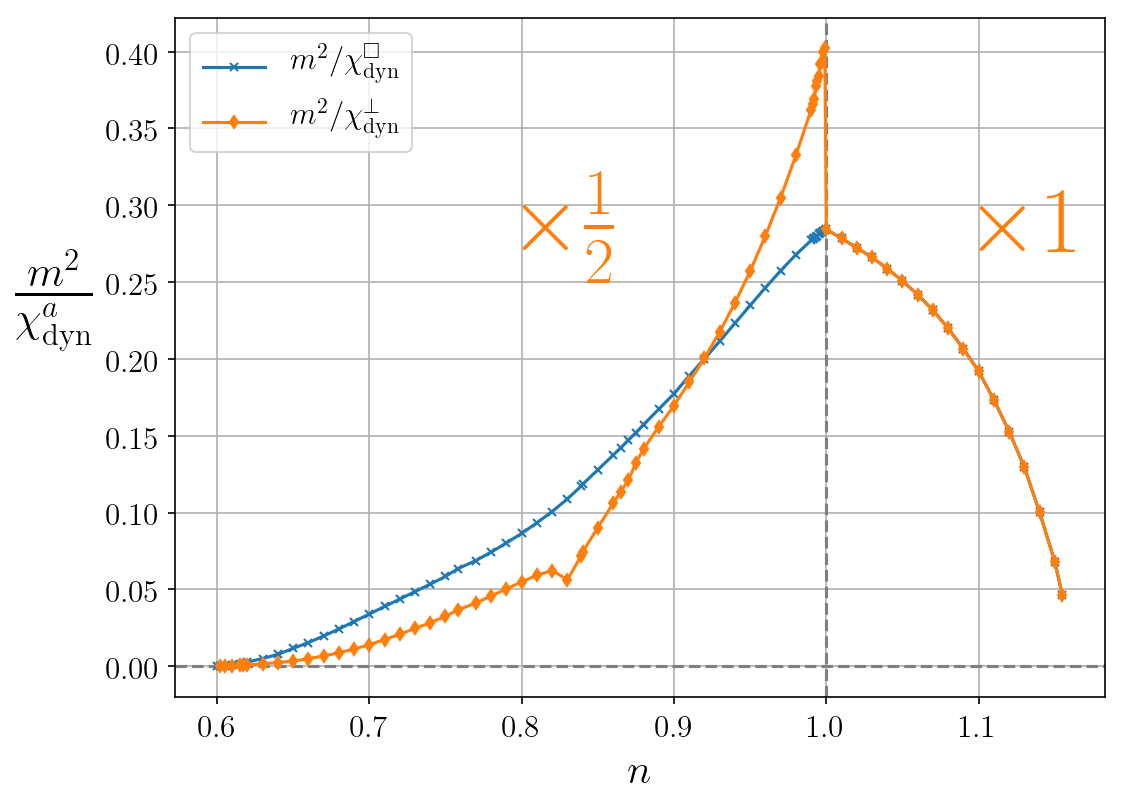}
        \caption{Out-of-plane (left panel) and in-plane (right panel) spatial ($J$) and temporal ($Z$) spin stiffnesses in the ground state ($T=0$) as functions of the filling $n$. In the N\'eel state (for $n \geq 1$) out-of-plane and in-plane stiffnesses coincide.}
        \label{fig_pg: fig3}
    \end{figure}
    \begin{figure}[h!]
        \centering
        \includegraphics[width=0.6\textwidth]{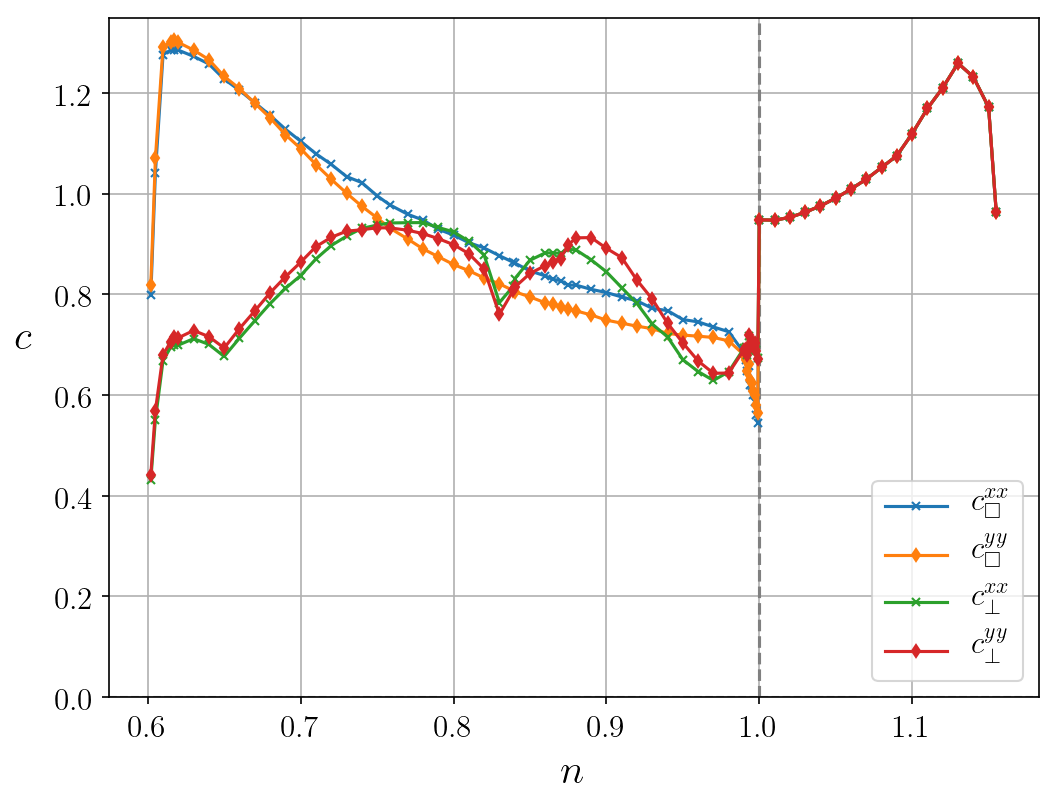}
        \caption{Magnetic coherence length $\xi_A$ (left axis) and average spin wave velocity $c_s$ in the ground state as functions of the filling $n$.}
        \label{fig_pg: fig4}
    \end{figure}
    \begin{figure}[h!]
        \centering
        \includegraphics[width=0.6\textwidth]{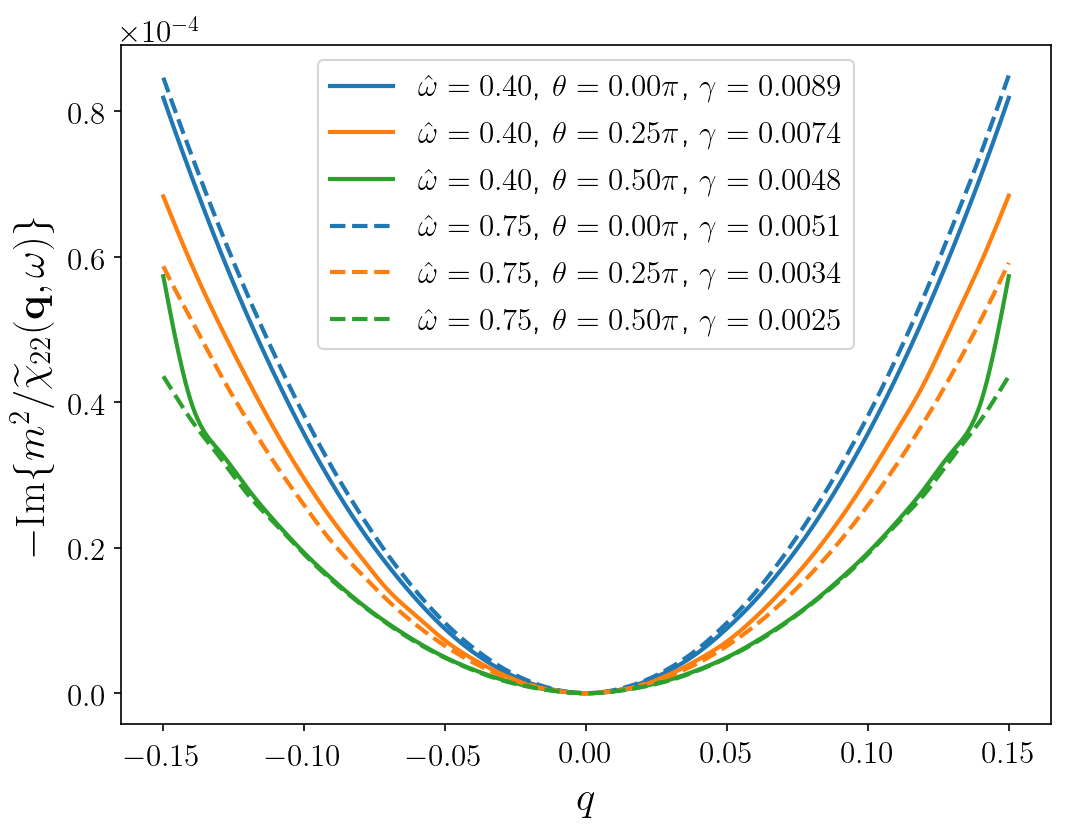}
        \caption{Fraction of condensed $z$-bosons $n_0$ at $T$ = 0 for two distinct choices of the ultraviolet cutoff $\Luv$ as a function of the filling.}
        \label{fig_pg: fig5}
    \end{figure}
    Once the magnetic order parameter $\Delta$ of the chargons and the wave vector $\bQ$ have been computed, we are in the position to calculate the NL$\sigma$M parameters from the expressions presented in Sec.~\ref{sec_pg: spin stiff formalism}. 

    In Fig.~\ref{fig_pg: fig3}, we plot results for the spatial and temporal spin stiffnesses $J^a_{\alpha\alpha}$ and $Z^a$ in the ground state. In the spiral state (for $n < 1$) out-of-plane and in-plane stiffnesses are distinct, while in the N\'eel state (for $n \geq 1$) they coincide. Actually, the order parameter defines an axis, not a plane, in the latter case.
    All the quantities except $Z^\Box$ exhibit pronounced jumps between half-filling and infinitesimal hole-doping. These discontinuities are due to the sudden appearance of hole pockets around the points $(\frac{\pi}{2},\frac{\pi}{2})$ in the Brillouin zone \cite{Bonetti2022}. The spatial stiffnesses are almost constant over a broad range of hole-doping, with a small spatial anisotropy $J^a_{xx} \neq J^a_{yy}$. The temporal stiffnesses $Z^a$ exhibit a stronger doping dependence. The peak of $Z^\perp$ at $n \approx 0.79$ is associated with a van Hove singularity of the quasiparticle dispersion \cite{Bonetti2022}. On the electron doped side all stiffnesses decrease almost linearly with the electron filling. The off-diagonal spin stiffnesses $J^a_{xy}$ and $J^a_{yx}$ vanish both in the N\'eel state and in the spiral state with $\bQ = (\pi-2\pi\eta,\pi)$ and symmetry related.
    
    In Fig.~\ref{fig_pg: fig4}, we show the magnetic coherence length $\xi_A$ and the average spin wave velocity $c_s$ in the ground state. The coherence length is rather short and only weakly doping dependent from half-filling up to 15 percent hole-doping, while it increases strongly toward the spiral-to-paramagnet transition on the hole-doped side. On the electron-doped side it almost doubles from half-filling to infinitesimal electron doping. This jump is due to the sudden appearance of electron pockets upon electron doping. Note that $\xi_A$ does not diverge at the transition to the paramagnetic state on the electron doped side, as this transition is first order.
    The average spin wave velocity exhibits a pronounced jump at half-filling, which is inherited from the jumps of $J_{\alpha\alpha}^\perp$ and $Z^\perp$. Besides this discontinuity it does not vary much as a function of density, remaining finite at the transition points to a paramagnetic state, both on the hole- and electron-doped sides.

    We now investigate whether the magnetic order in the ground state is destroyed by quantum fluctuations or not. To this end we compute the boson condensation fraction $n_0$ as obtained from the large-$N$ expansion of the NL$\sigma$M. This quantity depends on the ultraviolet cutoff $\Luv$. As a reference point, we may use the half-filled Hubbard model at strong coupling, as discussed in Sec.~\ref{sec_pg: cutoff}, which yields $\Luv \approx 0.9$, and the constant in the ansatz Eq.~\eqref{eq_pg: Luv} is thereby fixed to $C \approx 0.3$.
    
    In Fig.~\ref{fig_pg: fig5} we show the condensate fraction $n_0$ computed with two distinct choices of the ultraviolet cutoff: $\Luv = \Luv(n) = C/\xi_A(n)$ and $\Luv = C/\xi_A(n=1)$. For the former choice the cutoff vanishes at the edge of the magnetic region on the hole-doped side, where $\xi_A$ diverges. One can see that $n_0$ remains finite for both choices of the cutoff in nearly the entire density range where the chargons order. Only near the hole-doped edge of the magnetic regime, $n_0$ vanishes slightly above the mean-field transition point, if the ultraviolet cutoff is chosen as density independent. The discontinuous drop of $n_0$ upon infinitesimal hole doping is due to the corresponding drop of the out-of-plane stiffness. In the weakly hole-doped region there is a substantial reduction of $n_0$ below one, for both choices of the cutoff. Except for the edge of the magnetic region on the hole-doped side, the choice of the cutoff has only a mild influence on the results, and the condensate fraction remains well above zero. Hence, we can conclude that the ground state of the Hubbard model with a moderate coupling $U = 4t$ is magnetically ordered over wide density range. The spin stiffness is sufficiently large to protect the magnetic order against quantum fluctuations of the order parameter. 
    
    \subsection{Electron spectral function}
    \label{sec_pg: spectral function}
    \begin{figure}[h!]
        \centering
        \includegraphics[width=0.6\textwidth]{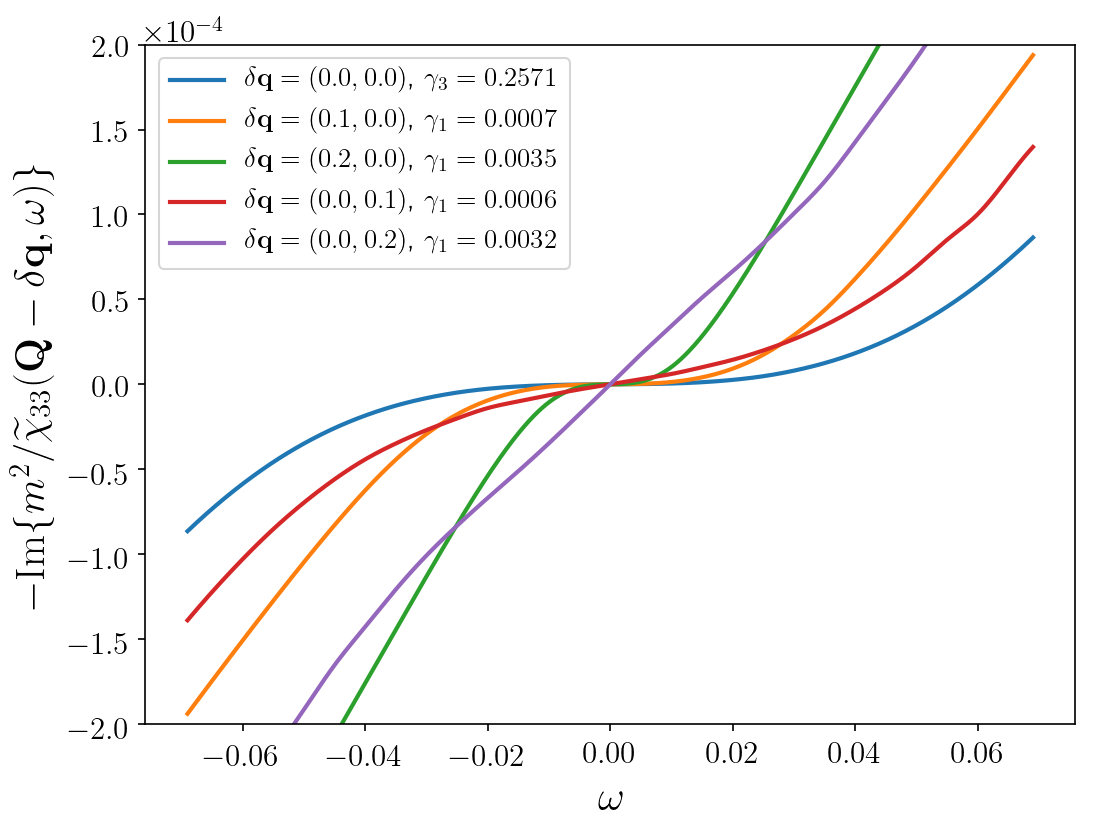}
        \caption{ Quasiparticle Fermi surfaces defined as zeros of
        the chargon quasiparticle energies $E_\bk^\pm$ (left column) and momentum dependence of electron spectral function at zero frequency (right column) for various electron densities. The temperature is $T=0.05t$.
        }
        \label{fig_pg: fig6}
    \end{figure}
    Fractionalizing the electron operators as in Eq.~\eqref{eq_pg: electron fractionaliz.}, the electron Green's function assumes the form
    \begin{eqnarray}
     [\mathcal{G}^e_{jj'}(\tau)]_{\sigma\sigma'} &=&
     - \langle c_{j' \sigma'}(\tau) c^*_{j \sigma}(0) \rangle \nonumber \\
     &=& - \langle [R_{j'}(\tau)]_{\sigma' s'}[R^*_j(0)]_{\sigma s} \,
     \psi_{j' s'}(\tau) \psi^*_{j s}(0) \rangle \, . \nonumber \\
    \end{eqnarray}
    To simplify this expression, one can decouple the average $\langle R R^* \psi\psi^*\rangle$ as $\langle RR^*\rangle\langle\psi\psi^*\rangle$, yielding~\cite{Borejsza2004,Scheurer2018,Wu2018}
    \begin{equation}\label{eq_pg: Gelectron real space}
        [\mathcal{G}^e_{jj'}(\tau)]_{\sigma\sigma'}
        \simeq -\langle[R_{j'}(\tau)]_{\sigma's'}[R_j^*(0)]_{\sigma s}\rangle\,
        \langle \psi_{j's'}(\tau) 
        \psi^*_{js}(0)\rangle.
    \end{equation}
    The spinon Green's function can be computed from the NL$\sigma$M in the continuum limit. Using the Schwinger boson parametrization~\eqref{eq_pg: R to z}, we obtain in the large-$N$ limit
    \begin{equation}
        \langle [R_{j'}(\tau)]_{\sigma'\alpha'}[R^*_j(0)]_{\sigma s}\rangle 
        = -D(\mathbf{r}_j-\mathbf{r}_{j'},\tau)\delta_{\sigma\sigma'}\delta_{ss'}+n_0 \delta_{\sigma s}\delta_{\sigma's'}.
    \end{equation}
    The boson propagator $D(\mathbf{r},\tau)$ is the Fourier transform of
    \begin{equation}
        D(\bq,i\omega_m)=\frac{1/Z_\perp }{\omega_m^2+(J_\perp^{\alpha\beta}q_\alpha q_\beta)/Z_\perp +m_s^2},
    \end{equation}
    with $\omega_m=2\pi m T$ a bosonic Matsubara frequency. Fourier transforming Eq.~\eqref{eq_pg: Gelectron real space}, we obtain the electron Green's function in momentum representation
    \begin{equation}\label{eq_pg: Ge before freq sum}
        \mathcal{G}_{\sigma\sigma'}^e(\bk,\bk',i\nu_n)= -T\sum_{\omega_m}\int_\bq \tr\left[\mathbf{G}_{\bk-\bq,\bk'-\bq}(i\nu_n-i\omega_m)\right] D(\bq,\omega_m)\delta_{\sigma\sigma'}
        +n_0 \,[\mathbf{G}_{\bk,\bk'}(i\nu_n)]_{\sigma\sigma'} ,
    \end{equation}
    where $\mathbf{G}_{\bk,\bk'}(i\nu_n)$ is the mean-field chargon Green's function, given by Eq.~\eqref{eq_low_spiral: G unrotated}. 
    We see that when $n_0=0$, the electron Green's function is diagonal in momentum, that is, it is translational invariant, as the diagonal components of the chargon one entering the trace are nonzero only for $\bk=\bk'$. Furthermore, in this case there is no spontaneous symmetry breaking, because $\mathbf{G}^e$ is proportional to the identity matrix in spin space.
    
    The first term in Eq.~\eqref{eq_pg: Ge before freq sum} describes incoherent excitations and it is the only contribution to the electron Green's function at finite temperature and in the quantum disordered regime, where $n_0=0$. Performing the bosonic Matsubara sum and analytically continuing to real frequencies ($i\nu_n\to\omega+i0^+$), the first term in~\eqref{eq_pg: Ge before freq sum} becomes
    \begin{equation}
        \begin{split}
            G^e(\bk,\omega) = \sum_{\ell,p=\pm}\int_{|\bq|\leq\Luv} \frac{1}{4Z^\perp \omega^{\mathrm{sp}}_\bq}\left(1+\ell\frac{h_{\bk-\bq}}{e_{\bk-\bq}}\right)\frac{f(pE^\ell_{\bk-\bq})+n_B(\omega^{\mathrm{sp}}_\bq)}{\omega+i0^+-E^\ell_{\bk-\bq}+p\,\omega^{\mathrm{sp}}_\bq} + \{\bk\to-\bk\},
        \end{split}
    \end{equation}
    where we have defined the spinon dispersion $\omega^\mathrm{sp}_\bq=\sqrt{(J^\perp_{\alpha\beta} q_\alpha q_\beta)/Z^\perp + m_s^2}$, and $n_B(x)=(e^{x/T}-1)^{-1}$ is the Bose distribution function. The electron spectral function is computed from
    \begin{equation}\label{eq_pg: sp function electron}
        A^e(\bk,\omega)=-\frac{1}{\pi} \mathrm{Im}\left\{G^e(\bk,\omega)+n_0 \left[G_\bk(\omega)+\overline{G}_{\bk-\bQ}(\omega)\right]\right\},
    \end{equation}
    where $G_\bk(\omega)$ and $\overline{G}_{\bk-\bQ}(\omega)$ are obtained by analytically continuing (that is, by replacing $i\nu_n\to\omega+i0^+$) Eqs.~\eqref{eq_low_spiral: Gk up unrotated} and~\eqref{eq_low_spiral: Gk down unrotated}, respectively.
    
    In the right column of Fig.~\ref{fig_pg: fig6}, we show the spectral function $A^e(\bk,\omega)$ at zero frequency as a function of momentum for various electron densities in the hole-doped regime. The temperature $T=0.05t$ is below the chargon ordering temperature $T^*$ in all cases. Because of the finite temperature, only the first term in Eq.~\eqref{eq_pg: sp function electron} contributes to the spectral function. The Fermi surface topology is the same as the one obtained from a mean-field approximation of spiral magnetic order~\cite{Eberlein2016}. At low hole doping, it originates from a superposition of hole pockets (see left column of Fig.~\ref{fig_pg: fig6}), where the spectral weight on the back sides is drastically suppressed by coherence factors, so that only the front sides are visible. The spinon fluctuations lead to a broadening of the spectral function, smearing out the Fermi surface. Since the spinon propagator does not depend on the fermionic momentum, the broadening occurs uniformly in the entire Brillouin zone. Hence, the backbending at the edges of the "arcs" obtained in our theory for $n=0.9$ is more pronounced than experimentally observed in cuprates. This backbending can be further suppressed by including a momentum dependent self-energy or scattering rate with a larger imaginary part in the antinodal region~\cite{Mitscherling2021}. For fillings smaller than $n\approx 0.82$ the chargon hole pockets merge and the spectral function resembles the large bare Fermi surface (see last row of Fig.~\ref{fig_pg: fig6}).
    %

\cleardoublepage
%
\phantomsection
\chapter*{Conclusion}
\label{chap: conclusion}
\rhead[\fancyplain{}{\bfseries Conclusion}]{\fancyplain{}{\bfseries\thepage}}
\lhead[\fancyplain{}{\bfseries\thepage}]{\fancyplain{}{\bfseries Conclusion}}
\addcontentsline{toc}{chapter}{Conclusion}

    %
    In this Thesis, we have dealt with two main problems. The first one was the identification of collective bosonic fluctuations in interacting systems, independent of the coupling strength, where the vertex function may exhibit an intricate dependence on momenta and frequencies. For the symmetric phase, we have combined the single-boson exchange (SBE) parametrization of the vertex function~\cite{Krien2019_I} with the functional renormalization group (fRG) and its fusion with dynamical mean-field theory (DMF\textsuperscript{2}RG). This allows not only for a clear and physically intuitive identification of the bosonic modes at play in the many-particle system, but also for a \emph{substantial} simplification of the complexity of the vertex function. In the symmetry-broken phases, this identification permits the explicit introduction of a bosonic field, describing order parameter fluctuations, and it therefore facilitates the study of fluctuation effects on top of mean-field solutions.
    
    The second problem we dealt with was the development of a theory for the pseudogap phase able to reconcile features typical of a magnetically ordered state, such as Fermi arcs in the spectral function~\cite{Eberlein2016} and charge carrier drop~\cite{Mitscherling2018,Bonetti2020_I}, with the experimentally observed absence of long-range order. This is realized by fractionalizing the electron into a fermionic chargon, carrying the original electron charge, and a bosonic spinon, carrying the electron spin~\cite{Schulz1995}. The resulting theory acquires a SU(2) gauge redundancy~\cite{Scheurer2018}. While the chargon degrees of freedom can be treated within a mean-field-like (MF) approximation, giving some kind of magnetic order (often N\'eel or spiral antiferromagnetism), the computation of the spinon dynamics requires to study the fluctuations on top of the MF. We have therefore analyzed the long-wavelength and low-frequency properties of the directional fluctuations (Goldstone modes) of the spins in an itinerant spiral magnet and studied their damping rates due to their decay into particle-hole pairs. We have also proven that a computation of the low-energy coefficients of the propagators of the Goldstone modes performed by expanding their relative susceptibilities is equivalent to computing the system's response to a fictitious SU(2) gauge field. Finally, we have applied the SU(2) gauge theory to the two-dimensional Hubbard model at moderate coupling and derived an effective non-linear sigma model (NL$\sigma$M) describing the slow and long wavelength dynamics of the spinons, which enabled us to study the pseudogap regime.  
    
    In the following, we summarize the key results of each chapter. 
    \subsection*{Charge carrier drop driven by spiral antiferromagnetism}
    In this chapter, we have performed a dynamical mean-field theory (DMFT) calculation in the magnetically ordered phase of the two-dimensional Hubbard model on a square lattice at strong coupling and finite temperature. We have found that over a broad doping regime spiral magnetic states have a lower energy than the N\'eel solution, and have a wave vector of the form $\bQ=(\pi-2\pi\eta,\pi)$ (or symmetry related) with the incommensurability $\eta$ increasing monotonically as the hole doping $p$ is increased. The magnetic order parameter $\Delta$ decreases with $p$ and vanishes at a critical doping $p^*$. A zero temperature extrapolation gives an approximate linear dependence $\Delta(p)\propto p^*-p$ in a broad doping region below $p^*$. Spiral magnetic ordering leads to a Fermi surface reconstruction for $p<p^*$ that is responsible for the abrupt change in the charge carriers. 
    
    We have computed the longitudinal and Hall conductivities by inserting the magnetic gap $\Delta$, the wave vector $\bQ$, and the quasiparticle renormalization $Z$ (extracted from the diagonal component of the DMFT self-energy) into transport equations for mean-field spin-density wave states with a phenomenological scattering rate~\cite{Mitscherling2018}. Calculations have been performed with band parameters mimicking the real compounds YBa\textsubscript 2Cu\textsubscript 3O\textsubscript y (YBCO), and La\textsubscript{2-x}Sr\textsubscript xCuO\textsubscript4 (LSCO). We found a pronounced drop in both the longitudinal conductivity and the Hall number in a narrow doping range below $p^*$, in agreement with experiments performed at high magnetic fields~\cite{Badoux2016}. For $p>p^*$ the calculated Hall number $n_H(p)$ is close to the na\"ively expected value $1+p$ for YBCO, while for LSCO parameters it deviates significantly. This is due to the fact that in this regime the band structure in the vicinity of the Fermi surface cannot be approximated by a simple parabolic dispersion (in which the $1+p$ behavior has been derived). For $p<p^*$ and sufficiently far away from $p^*$, we find that $n_H(p)\sim p$, in agreement with the fact that the density of charge carriers is given be the volume of the Fermi pockets. The zero temperature extrapolation of our results as functions of the doping yields $p^*=0.21$ for LSCO parameters and $p^*=0.15$ for YBCO. Both values are in the correct range. To better reproduce the experimentally observed critical dopings one would probably need a modeling that goes beyond the single-band Hubbard model. 
    \subsection*{(Bosonized) fRG+MF approach to symmetry broken states}
    In this chapter, we have performed a dynamical fRG analysis of magnetic and superconducting ordering tendencies in the 2D Hubbard model at moderate coupling $U=3t$. We have combined a one-loop flow with coupled charge, magnetic and pairing channels in the symmetric phase above the critical fRG scale $\Lambda_c$ with a mean-field approximation with decoupled channels in the symmetry-broken regime below $\Lambda_c$. All along the calculation, the full frequency dependence of the two-particle vertex has been retained, therefore methodologically improving the results of Ref.~\cite{Yamase2016}. For the parameters chosen, magnetism is the leading instability at $\Lambda_c$ in the hole doping range from half filling to about 20\%. Between 10\% and 20\% hole doping, also a robust $d$-wave pairing gap has been found, allowing for a computation of the superfluid phase stiffness and the Berezinskii-Kosterlitz-Thouless transition temperature $T_\mathrm{KT}$. 
    
    In order to go beyond the mean-field approximation, one needs to account for order parameter fluctuations. This can be conveniently achieved by introducing a bosonic field by means of a Hubbard-Stratonovich transformation. However, this task may become difficult when the two-particle vertex at the critical scale exhibits an intricate dependence on momenta and frequencies. We have therefore devised a technique to factorize the singular part of the vertex at $\Lambda_c$ to introduce a bosonic field. We have subsequently reformulated the fRG+MF equations for a mixed boson-fermion system and proven that they reproduce the results of the "fermionic" framework and they fulfill fundamental constraints such as the Goldstone theorem and the Ward identities associated with global symmetries. As a practical example of the feasibility of the method, we have studied the attractive 2D Hubbard model at half filling, and computed the superconducting order parameter. We have then computed and analyzed frequency dependencies of the longitudinal and transverse Yukawa couplings, describing the interaction between the electron and collective amplitude and phase fluctuations of the order parameter, respectively, as well as those of the so-called residual two-fermion interactions, representing all the non factorizable (bot not singular) contributions to the two-particle vertex.
    \subsection*{SBE decomposition of the fRG}
    In this chapter, we have applied the single-boson exchange (SBE) representation of the vertex function~\cite{Krien2019_I} to the fRG and DMF\textsuperscript{2}RG. This representation relies on a diagrammatic decomposition in contributions mediated by the exchange of a single boson in the different channels. We have recast the fRG flow equations for the two-particle vertex into SBE contributions and a residual four-point vertices, which we label as rest functions. 
    
    This formulation leads to a substantial reduction of the numerical effort required to compute the vertex function. In fact, the SBE contributions consist of one screened interaction, representing the propagator of an effective boson, and two Yukawa couplings, describing the interaction between the electrons and the boson. If on the one hand the vertex function is a challenging object to compute, as it depends on three variables $k_1$, $k_2$ and $k_3$, each of them combining momentum and frequency, on the other hand the Yukawa coupling and the screened interaction require a smaller memory cost, as they depend on two and one variable, respectively. Furthermore, we have shown that the rest functions are localized objects in frequency space, particularly at strong coupling, and one can therefore significantly restrict the total number of frequencies taken into account or even neglect all the non-SBE terms. The reduced numerical effort facilitates the applicability of the fRG and DMF\textsuperscript{2}RG to the physically most interesting regime of low temperatures. 
    
    We have demonstrated the advantage of the implementation of the SBE decomposition by means of DMF\textsuperscript{2}RG calculations for the 2D Hubbard model performed up to very large interactions $U=16t$ at and away from half filling. We have specifically analyzed the impact of neglecting the rest function and observed a marginal effect from weak to strong coupling. Moreover, the SBE decomposition allows for a physical identification of the collective modes at play in the system, and we have therefore employed it to diagnose the mechanism for $d$-wave pairing formation in the doped regime in terms of processes involving the exchange of magnetic and charge bosons. 
    \subsection*{Collective modes of metallic spiral magnets}
    In this chapter, we have derived Ward identities for fermionic systems in which a gauge symmetry is globally broken. In particular, we have shown that the zero-energy and long-wavelength components of the gauge kernels are connected to the transverse susceptibilities of the order parameter by exact relations. We have analyzed several examples, namely a superconductor, a N\'eel antiferromagnet, and a spiral magnet. 
    
    In the latter case, we have performed a random phase approximation (RPA) analysis and identified three Goldstone poles in the susceptibilities, one associated with in-plane, and two associated with out-of-plane fluctuations of the order parameter. Expanding the susceptibilities near their poles, we have derived expressions for the spin stiffness and spectral weights of the magnons (corresponding to the Goldstone modes) and checked that they coincide with those derived by computing the response of the system to a fictitious SU(2) gauge field, as predicted by the Ward identities. Moreover, we have determined the form and the size of the decay rates of the magnons due to Landau damping. The Landau damping of the in-plane mode has the same form as that of a N\'eel antiferromagnet~\cite{Sachdev1995} and is of the same order as the energy $\omega$ of the mode. By contrast, the out-of-plane modes possess a parametrically smaller Landau damping, of the order $\omega^{3/2}$, implying that they are asymptotically stable excitations in the low-energy limit.
    
    In the N\'eel antiferromagnet, we have also shown that the hydrodynamic relation for the magnon velocities $c_s=\sqrt{J/\chi^\perp}$, with $J$ the spin stiffness and $\chi^\perp$ the static transverse susceptibilities, does not hold in presence of gapless fermionic excitations. In fact, it must be replaced by $c_s=\sqrt{J/\chi^\perp_\mathrm{dyn}}$, where $\chi^\perp_\mathrm{dyn}$ is obtained from the transverse susceptibility $\chi^\perp(\bq,\omega)$ by taking the $\omega\to 0$ limit \emph{after} letting $\bq\to\bzero$, that is, the limits are taken in the reverse order compared to $\chi^\perp$. The equality $\chi^\perp=\chi^\perp_\mathrm{dyn}$ only holds for insulating magnets at low temperatures. Similar relations hold for a spiral magnet, too.
    
    We have complemented our analysis with a numerical evaluation of the spin stiffnesses, spectral weights, and decay rates for a specific two-dimensional model system. Some of the quantities exhibit peaks and discontinuities as a function of the electron density which are related to changes of the Fermi surface topology and special contributions in the N\'eel state.
    \subsection*{SU(2) gauge theory of the pseudogap phase}
    In this chapter, we have presented a SU(2) gauge theory of fluctuating magnetic order in the two-dimensional Hubbard model. The theory is based on a fractionalization of the electron field in fermionic chargons and bosonic spinons~\cite{Schulz1995,Borejsza2004,Scheurer2018}. We have treated the chargons within a renormalized mean-field theory with effective interactions obtained by a functional renormalization group flow, as described in Chapter~\ref{chap: fRG+MF}. We have found a broad density range in which they undergo N\'eel or spiral magnetic order below a density-dependent temperature $T^*$. We have treated the spinons, describing fluctuations of the spin orientation, within a gradient expansion, and found that their dynamics is governed by a non-linear sigma model (NL$\sigma$M). The parameters of the NL$\sigma$M, namely the spin stiffnesses, have been computed on top of the magnetically ordered chargon state using a renormalized RPA, closely following the formulas of Chapter~\ref{chap: low energy spiral}. At any finite temperature the spinon fluctuations prevent long-range order, in agreement with the Mermin-Wagner theorem, while at zero temperature they are not strong enough to destroy the magnetic order. Our approximations are valid for a weak or moderate Hubbard interaction $U$. It is possible that at strong coupling spinon fluctuations get enhanced, thus destroying long-range order \emph{even} in the ground state. 
    
    Despite the moderate interaction strength chosen in our calculations, the phase below $T^*$, where the chargon magnetically order, displays all important feature typical of the pseudogap regime in high-$T_c$ cuprates. Even though spinon fluctuations destroy long-range order at any finite $T$, they do not strongly affect the electron spectral function, which remains similar to that of a magnetically ordered state, thus displaying Fermi arcs. They also do not affect charge transport significantly, that is, quantities like longitudinal or Hall conductivities can be computed within the ordered chargon subsystem, yielding~\cite{Eberlein2016,Mitscherling2018,Bonetti2020_I,Storey2016,Storey2017,Chatterjee2017,Verret2017} the drastic charge carrier drop observed at the onset of the psedogap regime in hole-doped cuprates~\cite{Badoux2016,Collignon2017,Proust2019}. Spiral order of the chargons entails nematic order of the electrons. At low hole doping, the chargons form a N\'eel state at $T^*$, and a spiral state below $T_\mathrm{nem}$. Thus, the electrons undergo a nematic phase transition at a critical temperature \emph{below} the pseudogap temperature $T^*$. Evidence for a nematic transition at a temperature $T_\mathrm{nem}<T^*$ has been found recently in slightly underdoped YBCO~\cite{Grissonnanche2022}. For large hole doping, instead, the nematic transition occurs exactly at $T^*$. For electron doping nematic order is completely absent. 
    \section*{Outlook}
    The results presented in this thesis showed methodological advances and raised further questions beyond the scope of this thesis. In the following, we shortly present several paths for extensions. 
    
    First of all, the parametrization of the vertex function in terms of single boson exchange processes and rest functions allows for a substantial reduction of the computational cost. In fact, the Yukawa coupling and the bosonic propagator depend on less arguments than the full two-particle vertex, while the rest function is shown to display a fast decay with respect to all its three frequency variables, especially in the strong coupling regime. The reduced numerical effort facilitates the applicability of the fRG and DMF\textsuperscript{2}RG to the most interesting regime of strong correlations and low temperatures. The SBE decomposition also offers the possibility to \emph{explicitly} introduce bosonic fields and therefore study the flow of mixed boson-fermion systems. This extension is particularly interesting to analyze the impact of bosonic fluctuations on top of mean-field solutions below the (pseudo-) critical scale, where symmetry breaking occurs. The reformulation of the fRG+MF approach with the explicit introduction of a bosonic field offers, in this respect, a convenient starting point. The generalization of the SBE decomposition to other models with different lattices or non-local interactions, where the higher degree of frustration reduces the pseudo-critical temperatures, is also an interesting extension. 
    
    A second path for extensions is given by refinements of the SU(2) gauge theory for the pseudogap regime. In this thesis, we have considered only N\'eel or spiral ordering of the chargons. In the ground state of the two-dimensional Hubbard model, however, there is a whole zoo of possible magnetic ordering patterns, and away from half filling N\'eel or spiral order do not always minimize the energy. One possible competitor is stripe order, where the spins are antiferromagnetically ordered with a unidirectional periodic modulation of the amplitude of the order parameter and of the electron density. If we treat the chargon stripe phase with the same formalism developed in this thesis, magnetic long-range order would be destroyed by directional fluctuations of the spins, while the charge density wave (CDW) may survive. In the, actually quite general, case of \emph{incommensurate} stripe order wave vector, also charge order can become fluctuating due to a soft \emph{sliding} mode that acts as a Goldstone mode and destroys the CDW at finite temperatures, thus explaining the experimental observation of fluctuating charge order within the pseudogap phase~\cite{Frano2020}. Another refinement of our SU(2) gauge theory to make it more quantitative is to circumvent the need of a \emph{ultraviolet cutoff} by formulating it on the lattice, that is, by avoiding the long-wavelength expansion. The weak coupling calculation presented in this thesis has revealed that quantum fluctuations of the spinons are not strong enough to destroy long-range order in the ground state, giving rise to exponentially small spin gaps at low temperatures. This is due to the large value of the magnetic coherence length $\xi_A$ that makes the ultraviolet cutoff $\Luv$ small, thereby weakening quantum fluctuations. At \emph{strong coupling}, the situation might change, as $\xi_A$ gets drastically reduced and $\Luv$ enhanced, thus possibly disordering the ground state. Furthermore, our theory does not take into account topological defects in the spin pattern, as we expect them to be suppressed at low temperatures and deep in the pseudogap phase. However, at higher values of $T$ or near the critical doping at which the chargon order parameter vanishes, they may proliferate, potentially making the sharp metal-to-pseudogap-metal transition more similar to a crossover, similarly to what is observed in experiments. 
    %
    %

\cleardoublepage
%
\addtocontents{toc}{\setcounter{tocdepth}{0}}
\begin{appendices}
\rhead[\fancyplain{}{\bfseries Appendix}]{\fancyplain{}{\bfseries\thepage}}        \lhead[\fancyplain{}{\bfseries\thepage}]{\fancyplain{}{\bfseries Appendix}} 
\addcontentsline{toc}{part}{Appendix}
\part*{Appendix}

    \chapter{Symmetries and flow equation of the vertex function}
    \label{app: symm V}
    In this Appendix, we present the symmetries and the explicit flow equation of the vertex function $V$. 
    \section{Symmetries of \texorpdfstring{$V$}{V}}
    we start by considering the effect of the following symmetries on $V$: SU(2)-spin, lattice-, and time reversal (TRS) symmetries, translational invariance, remnants of anti-symmetry (RAS), and complex conjugation (CC). More detailed discussions can be found in~\cite{Husemann2009,Rohringer2012,Wentzell2016}.
    \subsection{Antisymmetry properties}{}
    The two-particle vertex enters the effective action as 
    \begin{equation}
        \frac{1}{(2!)^2}\sum_{\substack{x_1',x_2',\\x_1,x_2}}V(x_1',x_2',x_1,x_2)\overline{\psi}(x_1')\overline{\psi}(x_2')\psi(x_2)\psi(x_1),
        \label{eq_app_V: gamma4}
    \end{equation}
    where $x=(\bk,\nu,\sigma)$ is a collective variable enclosing the lattice momentum $\bk$, a fermionic Matsubara frequency $\nu$ and the spin quantum number $\sigma$. From now on we label as $k$ the pair $(\bk,\nu)$. From Eq.~\eqref{eq_app_V: gamma4}, we immediately see that exchanging the dummy variables $x_1'$ and $x_2'$ or $x_1$ and $x_2$ the effective action gets a minus sign because of the Grassmann algebra of the fields $\psi$ and $\overline{\psi}$. To keep the effective action invariant, the vertex must therefore obey
    \begin{equation}
        V(x_1',x_2',x_1,x_2) = 
        -V(x_2',x_1',x_1,x_2) =
        -V(x_1',x_2',x_2,x_1).
        \label{eq_app_V: Antisymmetry}
    \end{equation}
    \subsection{SU(2)-spin symmetry}{}
    The SU(2)-spin symmetry acts on the fermionic fields as 
    \begin{subequations}
        \begin{align}
            &\psi_{k,\sigma}\to \sum_{\sigma'}U_{\sigma\sigma'}\,\psi_{k,\sigma}, \\
            &\overline{\psi}_{k,\sigma}\to \sum_{\sigma'}U^\dagger_{\sigma\sigma'}\,\overline{\psi}_{k,\sigma},
            \label{eq_app_V: SU(2) psi}        
        \end{align}
    \end{subequations}
    with $U\in\mathrm{SU(2)}$. A vertex that is invariant under~\eqref{eq_app_V: SU(2) psi} can be expressed as (see also Eq.~\eqref{eq_methods: V SU(2) inv})
    \begin{equation}
        \begin{split}
            V_{\sigma_1'\sigma_2'\sigma_1\sigma_2}(k_1',k_2',k_1,k_2) = 
            V(k_1',k_2',k_1,k_2)\delta_{\sigma_1'\sigma_1}\delta_{\sigma_2'\sigma_2}
            +\overline{V}(k_1',k_2',k_1,k_2)\delta_{\sigma_1'\sigma_2}\delta_{\sigma_2'\sigma_1},
        \end{split}
    \end{equation}
    where Eq.~\eqref{eq_app_V: Antisymmetry} forces the identity
    \begin{equation}
        \overline{V}(k_1',k_2',k_1,k_2)=-V(k_2',k_1',k_1,k_2)=-V(k_1',k_2',k_2,k_1).
    \end{equation}
    From now on we only consider symmetry properties of the vertex function $V=V_{\up\down\up\down}$. 
    \subsection{Time and space translational invariance}{}
    The invariance of the system under time and space translations implies energy and momentum conservation, respectively. If these symmetries are fulfilled, then the vertex function can be written as
    \begin{equation}
        V(k_1',k_2',k_1,k_2)=V(k_1',k_2',k_1)\, \delta\left(k_1'+k_2'-k_1-k_2\right).
    \end{equation}
    \subsection{Remnants of antisymmetry}{}
    The vertex function $V$ is not antisymmetric under the exchange of the pair $(k_1',k_2')$ or $(k_1,k_2)$. It is however invariant under a \emph{simultaneous} exchange of them, that is,
    \begin{equation}
        V(k_1',k_2',k_1,k_2)=V(k_2',k_1',k_2,k_1).
    \end{equation}
    We call this symmetry remnants of antisymmetry (RAS). 
    \subsection{Time reversal symmetry}{}
    A time reversal transformation exchanges the fermionic creation and annihilation operators. It acts on the Grassmann variables as
    \begin{subequations}
        \begin{align}
            &\psi_{k,\sigma}\to i\overline{\psi}_{k,\sigma}, \\
            &\overline{\psi}_{k,\sigma}\to i\psi_{k,\sigma}.       
        \end{align}
    \end{subequations}
    Since this is a symmetry of the bare action, the vertex function must obey
    \begin{equation}
        V(k_1',k_2',k_1,k_2)=V(k_1,k_2,k_1',k_2').
    \end{equation}
    \subsection{Lattice symmetries}{}
    The square lattice considered in this Thesis is invariant under transformations belonging to the discrete group $C_4$. The latter are implemented on the fermionic fields as
    \begin{subequations}
        \begin{align}
            &\psi_{(\bk,\nu),\sigma}\to \psi_{(R\bk,\nu),\sigma}, \\
            &\overline{\psi}_{(\bk,\nu),\sigma}\to \overline{\psi}_{(R\bk,\nu),\sigma}, 
        \end{align}
    \end{subequations}
    with $R\in C_4$. If the lattice symmetries are not spontaneously broken, the vertex function obeys
    \begin{equation}
        V(k_1',k_2',k_1,k_2)=V(R\,k_1',R\,k_2',R\,k_1,R\,k_2),
    \end{equation}
    with $R\,k=(R\bk,\nu)$. For a more detailed discussion see Ref.~\cite{Platt2013}.
    \subsection{Complex conjugation}{}
    The complex conjugation (CC) transformation acts as
    \begin{subequations}
        \begin{align}
            &\psi_{k,\sigma}\to i\mathcal{K}\overline{\psi}_{\widetilde{k},\sigma},\\
            &\overline{\psi}_{k,\sigma}\to i\mathcal{K}\psi_{\widetilde{k},\sigma},
        \end{align}
    \end{subequations}
    where $\widetilde{k}=(\bk,-\nu)$, and the operator $\mathcal{K}$ transforms scalars into their complex conjugate. Since CC is a symmetry of the Hubbard action, the vertex function fulfills
    \begin{equation}
         V(k_1',k_2',k_1,k_2)= \left[V(\widetilde{k}_1',\widetilde{k}_2',\widetilde{k}_1,\widetilde{k}_2)\right]^*.
    \end{equation}
    \subsection{Channel decomposition}{}
    Let us now analyze how the above described symmetries act on the physical channels in which the vertex function can be decomposed (see also Eq.~\eqref{eq_methods: channel decomp physical})
    \begin{equation}
        \begin{split}
            V^\L(k_1',k_2',k_1) = &\lambda(k_1',k_2',k_1) \\
            &+ \frac{1}{2}\mathcal{M}^{\L}_{k_{ph},k_{ph}'}(k_1-k_1') 
            - \frac{1}{2}\mathcal{C}^{\L}_{k_{ph},k_{ph}'}(k_1-k_1') \\
            &+ \mathcal{M}^{\L}_{k_{\phx},k_{\phx}'}(k_2'-k_1) \\
            &- \mathcal{P}^{\L}_{k_{pp},k_{pp}'}(k_1'+k_2'),
        \end{split}
    \end{equation}
    with $k_{ph}$, $k'_{ph}$, $k_{\phx}$, $k'_{\phx}$, $k_{pp}$, and $k'_{pp}$ defined as in Eq.~\eqref{eq_methods: k k' pp ph phx}. 
    Combining RAS, TRS, CC, and, among the lattice symmetries, only the spatial inversion, we can prove that
    \begin{subequations}
        \begin{align}
            &\mathcal{M}_{k,k'}(q)=\mathcal{M}_{k',k}(q),\\
            &\mathcal{M}_{k,k'}(q)=\mathcal{M}_{k,k'}(-q),\\
            &\mathcal{M}_{k,k'}(q) = \left[\mathcal{M}_{-k+q\mathrm{m}2,-k'+q\mathrm{m}2}(q)\right]^*,
        \end{align}
    \end{subequations}
    with $q\mathrm{m}2=(\bzero,2 (j\,\mathrm{mod}\,2)\pi T)$, $j\in\mathbb{Z}$. The same relations can be obtained for the charge channel
    \begin{subequations}
        \begin{align}
            &\mathcal{C}_{k,k'}(q)=\mathcal{C}_{k',k}(q),\\
            &\mathcal{C}_{k,k'}(q)=\mathcal{C}_{k,k'}(-q),\\
            &\mathcal{C}_{k,k'}(q) = \left[\mathcal{C}_{-k+q\mathrm{m}2,-k'+q\mathrm{m}2}(q)\right]^*,
        \end{align}
    \end{subequations}
    Differently, for the pairing channel, we have
    \begin{subequations}
        \begin{align}
            &\mathcal{P}_{k,k'}(q)=\mathcal{P}_{-k+q\mathrm{m}2,-k'+q\mathrm{m}2}(q),\\
            &\mathcal{P}_{k,k'}(q)=\mathcal{P}_{k',k}(q),\\
            &\mathcal{P}_{k,k'}(q) = \left[\mathcal{P}_{k,k'}(-q)\right]^*.
        \end{align}
    \end{subequations}
    \subsection{SBE decomposition}
    It is also useful to apply the symmetries described above to the SBE decomposition of the vertex function, introduced in Chap.~\ref{chap: Bos Fluct Norm}. In more detail, we study the symmetry properties of screened interactions and Yukawa couplings, as the rest functions obey the same relations as their relative channel, as described above.
    \subsubsection{Magnetic channel}
    The symmetries in the magnetic channel read as 
    \begin{subequations}
        \begin{align}
            &h^m_k(q) = h^m_k(-q),\\
            &h^m_k(q) = \left[h^m_{-k+q\mathrm{m}2}(q)\right]^*,\\
            &D^m(q)=D^m(-q),\\
            &D^m(q)=\left[D^m(q)\right]^*.
        \end{align}
    \end{subequations}
    \subsubsection{Charge channel}
    Similarly, in the charge channel we have
    \begin{subequations}
        \begin{align}
            &h^c_k(q) = h^c_k(-q),\\
            &h^c_k(q) = \left[h^c_{-k+q\mathrm{m}2}(q)\right]^*,\\
            &D^c(q)=D^c(-q),\\
            &D^c(q)=\left[D^c(q)\right]^*.
        \end{align}
    \end{subequations}
    \subsubsection{Pairing channel}
    In the pairing channel we obtain
    \begin{subequations}
        \begin{align}
            &h^p_k(q) = h^p_{-k+q\mathrm{m}2}(q),\\
            &h^p_k(q) = \left[h^p_{k}(-q)\right]^*,\\
            &D^p(q)=\left[D^p(-q)\right]^*.
        \end{align}
    \end{subequations}
    \section{Explicit flow equations for physical channels}
    In this section, we explicitly express the flow equations for the physical channels within the form factor expansion introduced in Sec.~\ref{sec_fRG_MF: symmetric regime}.
    The flow equation for the $s$-wave projected magnetic channel reads as 
    \begin{equation}
        \deL \mathcal{M}_{\nu,\nu'}(q)=-T\sum_\omega V^m_{\nu\omega}(q)\left[\widetilde{\partial}_\L \chi^{0,ph}_\omega(q)\right] V^m_{\omega\nu}(q),
    \end{equation}
    with the bubble reading as 
    \begin{equation}
        \chi^{0,ph}_\nu(q)=\int_\bk G\left((\bk,\nu)+\rnddo{q}\right)G\left((\bk,\nu)-\rndup{q}\right),
    \end{equation}
    and the vertex $V^m$ as
    \begin{equation}
        \begin{split}
            V^m_{\nu\nu'}(\bq,\Omega)= 
            &U + \mathcal{M}_{\nu\nu'}(\bq,\Omega)\\
            +&\int_\bk \bigg\{
                -\frac{1}{2}\mathcal{C}_{\rndup{\nu+\nu'}-\rndup{\Omega},\rndup{\nu+\nu'}+\rnddo{\Omega}}(\bk,\nu'-\nu)\\
                &\phantom{\int_\bk \Big\{}
                +\frac{1}{2}\mathcal{M}_{\rndup{\nu+\nu'}-\rndup{\Omega},\rndup{\nu+\nu'}+\rnddo{\Omega}}(\bk,\nu'-\nu)\\
                &\phantom{\int_\bk \Big\{}
                -\mathcal{S}_{\rndup{\nu-\nu'-\Omega},\rndup{\nu'-\nu-\Omega}}(\bk,\nu+\nu'-\Omega\mathrm{m}2)\\
                &\phantom{\int_\bk \Big\{}
                -\mathcal{D}_{\rndup{\nu-\nu'-\Omega},\rndup{\nu'-\nu-\Omega}}(\bk,\nu+\nu'-\Omega\mathrm{m}2) \,\frac{\cos k_x+\cos k_y}{2}
            \bigg\}.
        \end{split}
    \end{equation}
    Similarly, in the charge channel, we have
    \begin{equation}
        \deL \mathcal{C}_{\nu,\nu'}(q)=-T\sum_\omega V^c_{\nu\omega}(q)\left[\widetilde{\partial}_\L \chi^{0,ph}_\omega(q)\right] V^c_{\omega\nu}(q),
    \end{equation}
    with
    \begin{equation}
        \begin{split}
            V^c_{\nu\nu'}(\bq,\Omega)= 
            &U - \mathcal{C}_{\nu\nu'}(\bq,\Omega)\\
            +&\int_\bk \bigg\{
                \frac{1}{2}\mathcal{C}_{\rndup{\nu+\nu'}-\rndup{\Omega},\rndup{\nu+\nu'}+\rnddo{\Omega}}(\bk,\nu'-\nu)\\
                &\phantom{\int_\bk \Big\{}
                +\frac{3}{2}\mathcal{M}_{\rndup{\nu+\nu'}-\rndup{\Omega},\rndup{\nu+\nu'}+\rnddo{\Omega}}(\bk,\nu'-\nu)\\
                &\phantom{\int_\bk \Big\{}
                -2\mathcal{S}_{\rndup{\nu-\nu'-\Omega},\rndup{\nu-\nu'+\Omega}}(\bk,\nu+\nu'-\Omega\mathrm{m}2)\\
                &\phantom{\int_\bk \Big\{}
                +\mathcal{S}_{\rndup{\nu-\nu'-\Omega},\rndup{\nu'-\nu-\Omega}}(\bk,\nu+\nu'-\Omega\mathrm{m}2)\\
                &\phantom{\int_\bk \Big\{}
                -2\mathcal{D}_{\rndup{\nu-\nu'-\Omega},\rndup{\nu-\nu'+\Omega}}(\bk,\nu+\nu'-\Omega\mathrm{m}2) \,\frac{\cos k_x+\cos k_y}{2}\\
                &\phantom{\int_\bk \Big\{}
                +\mathcal{D}_{\rndup{\nu-\nu'-\Omega},\rndup{\nu'-\nu-\Omega}}(\bk,\nu+\nu'-\Omega\mathrm{m}2) \,\frac{\cos k_x+\cos k_y}{2}
            \bigg\}.
        \end{split}
    \end{equation}
    The flow equation for the $s$-wave pairing channel reads as
    \begin{equation}
        \deL \mathcal{S}_{\nu,\nu'}(q)=T\sum_\omega V^p_{\nu\omega}(q)\left[\widetilde{\partial}_\L \chi^{0,pp}_\omega(q)\right] V^p_{\omega\nu}(q),
    \end{equation}
    with the particle-particle bubble given by
    \begin{equation}
        \chi^{0,pp}_\nu(q)=\int_\bk G\left(\rnddo{q}+(\bk,\nu)\right)G\left(\rndup{q}-(\bk,\nu)\right),
    \end{equation}
    and the vertex $V^p$ as
    \begin{equation}
        \begin{split}
            V^p_{\nu\nu'}(\bq,\Omega)= 
            &U - \mathcal{S}_{\nu\nu'}(\bq,\Omega)\\
            +&\int_\bk \bigg\{
                -\frac{1}{2}\mathcal{C}_{\rnddo{\Omega}-\rndup{\nu+\nu'},\rndup{\Omega}-\rndup{\nu+\nu'}}(\bk,\nu'-\nu)\\
                &\phantom{\int_\bk \Big\{}
                +\frac{1}{2}\mathcal{M}_{\rnddo{\Omega}-\rndup{\nu+\nu'},\rndup{\Omega}-\rndup{\nu+\nu'}}(\bk,\nu'-\nu)\\
                &\phantom{\int_\bk \Big\{}
                +\mathcal{M}_{\rndup{\nu-\nu'+\Omega},\rndup{\nu'-\nu+\Omega}}(\bk,-\nu-\nu'+\Omega\mathrm{m}2)
            \bigg\}.
        \end{split}
    \end{equation}
    Finally, in the $d$-wave pairing channel, we have
    \begin{equation}
        \deL \mathcal{D}_{\nu,\nu'}(q)=T\sum_\omega V^d_{\nu\omega}(q)\left[\widetilde{\partial}_\L \chi^{0,pp,d}_\omega(q)\right] V^d_{\omega\nu}(q),
    \end{equation}
    where the $d$-wave pairing bubble is
    \begin{equation}
        \chi^{0,pp,d}_\nu(q)=\int_\bk\,d_\bk^2\, G\left(\rnddo{q}+(\bk,\nu)\right)G\left(\rndup{q}-(\bk,\nu)\right),
    \end{equation}
    with $d_\bk=\cos k_x-\cos k_y$, and the vertex $V^d$ is given by
    \begin{equation}
        \begin{split}
            V^d_{\nu\nu'}(\bq,\Omega)= 
            &-\mathcal{D}_{\nu\nu'}(\bq,\Omega)\\
            +&\int_\bk \frac{\cos k_x+\cos k_y}{2}\bigg\{
                +\frac{1}{2}\mathcal{C}_{\rnddo{\Omega}-\rndup{\nu+\nu'},\rndup{\Omega}-\rndup{\nu+\nu'}}(\bk,\nu'-\nu)\\
                &\phantom{\int_\bk \frac{\cos k_x+\cos k_y}{2}\Big\{}
                -\frac{1}{2}\mathcal{M}_{\rnddo{\Omega}-\rndup{\nu+\nu'},\rndup{\Omega}-\rndup{\nu+\nu'}}(\bk,\nu'-\nu)\\
                &\phantom{\int_\bk \frac{\cos k_x+\cos k_y}{2}\Big\{}
                -\mathcal{M}_{\rndup{\nu-\nu'+\Omega},\rndup{\nu'-\nu+\Omega}}(\bk,-\nu-\nu'+\Omega\mathrm{m}2)
            \bigg\}.
        \end{split}
    \end{equation}
    %
    %

    %
    \chapter{Bosonized flow equations in the SSB phase}
    \label{app: fRG+MF app}
    \section{Derivation of flow equations in the bosonic formalism}
    In this section we will derive the flow equations used in Chap.~\ref{chap: fRG+MF}.
    
    We consider only those terms in which the dependence on the center of mass momentum $q$ is fixed to zero by the topology of the relative diagram or that depend only parametrically on it. These diagrams are the only ones necessary to reproduce the MF approximation.
    
\begin{figure}[t]
        \centering
        \includegraphics[width=0.75\textwidth]{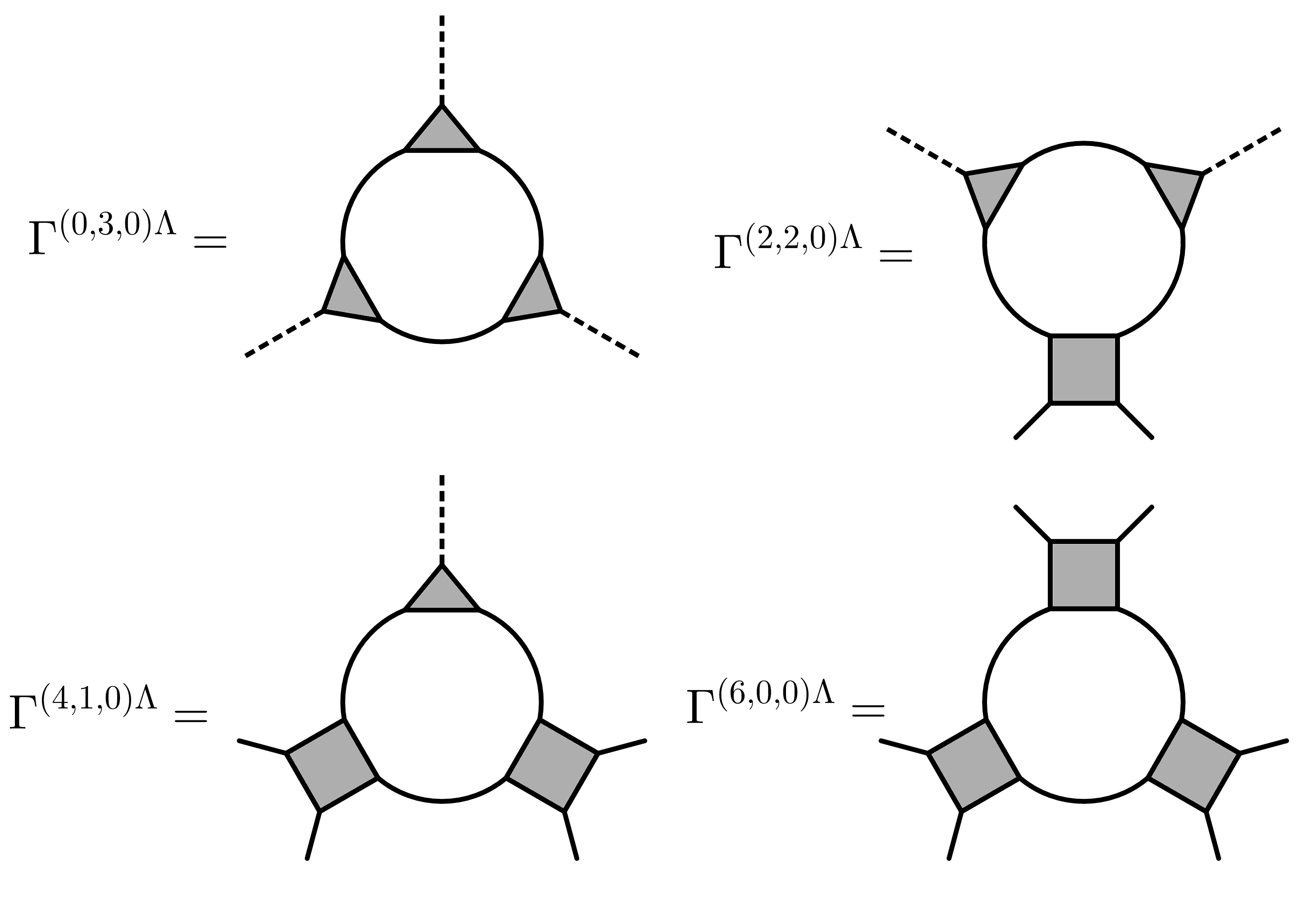}
        \caption{Feynman diagrams describing the Katanin-like approximation higher order correlation functions. The conventions are the same as in Figs.~\ref{fig_fRG+MF: flow eqs} and~\ref{fig_fRG+MF: flow eqs gaps}.}
        \label{fig_app_fRG+MF: katanin}
    \end{figure}

    The flow equations will be derived directly from the Wetterich equation~\eqref{eq_methods: Wetterich eq RL}, with a slight modification, since we have to keep in mind that the bosonic field $\phi$ acquires a scale dependence due to the scale dependence of its expectation value. The flow equation reads (for real $\alpha^\Lambda$):
    \begin{equation}
        \partial_\Lambda\Gamma^\Lambda=\frac{1}{2}\widetilde{\partial}_\Lambda\text{Str}\ln\left[\mathbf{\Gamma}^{(2)\Lambda}+R^\Lambda\right]+\frac{\delta\Gamma^\Lambda}{\delta\sigma_{q=0}} \, \partial_\Lambda \alpha^\Lambda,
        \label{eq_app_fRG+MF: Wetterich eq.}
    \end{equation}
    where $\mathbf{\Gamma}^{(2)\Lambda}$ is the matrix of the second derivatives of the action with respect to the fields, the supertrace Str includes a minus sign when tracing over fermionic variables. The first equation we derive is the one for the flowing expectation value $\alpha^\Lambda$. This is obtained by requiring that the one-point function for $\sigma_q$ vanishes. Taking the $\sigma_q$ derivative in Eq.~\eqref{eq_app_fRG+MF: Wetterich eq.} and setting the fields to zero, we have
    \begin{equation}
        \begin{split}
            \partial_\Lambda\Gamma^{(0,1,0)\Lambda}(q=0)\equiv\partial_\Lambda\frac{\delta\Gamma^\Lambda}{\delta\sigma_{q=0}}\bigg\lvert_{\Psi,\overline{\Psi},\sigma,\pi=0}
            =-\int_k h^\Lambda_\sigma(k;0)\, \widetilde{\partial}_\Lambda F^\Lambda(k)+m^\Lambda_\sigma(0)\,\partial_\Lambda\alpha^\Lambda=0,
        \end{split}
        \label{eq_app_fRG+MF: alpha eq app}
    \end{equation}
    where we have defined
    \begin{equation}
        \Gamma^{(2n_1,n_2,n_3)\Lambda}=\frac{\delta^{(2n_1+n_2+n_3)}\Gamma^\Lambda}{\left(\delta\overline{\Psi}\right)^{n_1}\left(\delta\Psi\right)^{n_1}\left(\delta\sigma\right)^{n_2}\left(\delta\pi\right)^{n_3}}.    
    \end{equation}
    From Eq.~\eqref{eq_app_fRG+MF: alpha eq app} we get the flow equation for $\alpha^\Lambda$.
    \begin{equation}
        \partial_\Lambda\alpha^\Lambda=\frac{1}{m^\Lambda_\sigma(0)}\int_k h^\Lambda_\sigma(k;0)\, \widetilde{\partial}_\Lambda F^\Lambda(k).
        \label{eq_app_fRG+MF: alpha flow}
    \end{equation}
    The MF flow equation for the fermionic gap reads
    \begin{equation}
        \begin{split}
            \partial_\Lambda \Delta^\Lambda(k)= 
            \int_{k'}\mathcal{A}^\Lambda(k,k';0)\,\widetilde{\partial}_\Lambda F^\Lambda(k')
            +\partial_\Lambda\alpha^\Lambda\,h_\sigma^\Lambda(k;0),    
        \end{split}
        \label{eq_app_fRG+MF: gap equation bosonic}
    \end{equation}
    with $\mathcal{A}^\Lambda$ being the residual two fermion interaction in the longitudinal channel. 
    The equation for the inverse propagator of the $\sigma_q$ boson is
    \begin{equation}
        \begin{split}
            \partial_\Lambda m_\sigma^\Lambda(q)=\int_p h_\sigma^\Lambda(p;q)\left[\widetilde{\partial}_\Lambda\Pi^\Lambda_{11}(p;q)\right] h_\sigma^\Lambda(p;q)
            +&\int_p \Gamma^{(2,2,0)\Lambda}(p,0,q)\,\widetilde{\partial}_\Lambda F^\Lambda(p)\\
            +&\partial_\Lambda\alpha^\Lambda\,\Gamma^{(0,3,0)\Lambda}(q,0),
            \label{eq_app_fRG+MF: full P sigma flow}
        \end{split}
    \end{equation}
    where we have defined the bubble at finite momentum $q$ as
    \begin{equation}
        \Pi^\Lambda_{\alpha\beta}(k;q)=-\frac{1}{2}\Tr\left[\tau^\alpha\mathbf{G}^\Lambda(k)\tau^\beta\mathbf{G}^\Lambda(k-q)\right], 
    \end{equation}
    $\Gamma^{(0,3,0)\Lambda}$ is an interaction among three $\sigma$ bosons and $\Gamma^{(2,2,0)\Lambda}$ couples one fermion and 2 longitudinal bosonic fluctuations. 
    The equation for the longitudinal Yukawa coupling is 
    \begin{equation}
        \begin{split}
            \partial_\Lambda h^\Lambda_\sigma(k;q)=\int_p\mathcal{A}^\Lambda(k,p;q)\left[\widetilde{\partial}_\Lambda\Pi^\Lambda_{11}(p;q)\right]h^\Lambda_\sigma(p,q)
            +&\int_{k'}\Gamma^{(4,1,0)\Lambda}(k,p,q,0)\,\widetilde{\partial}_\Lambda F^\Lambda(p)\\
            +&\partial_\Lambda\alpha^\Lambda\,\Gamma^{(2,2,0)\Lambda}(k,q,0),
            \label{eq_app_fRG+MF: full h_sigma flow}
        \end{split}
    \end{equation}
    where $\Gamma^{(4,1,0)\Lambda}$ is a coupling among 2 fermions and one $\sigma$ boson.
    The flow equation for the coupling $\mathcal{A}^\Lambda$ reads instead 
    \begin{equation}
        \begin{split}
            \partial_\Lambda\mathcal{A}^\Lambda(k,k';q)=&\int_p\mathcal{A}^\Lambda(k,p;q)\left[\widetilde{\partial}_\Lambda\Pi^\Lambda_{11}(p;q)\right]\mathcal{A}^\Lambda(p,k';q)\\
            +&\int_{p}\Gamma^{(6,0,0)\Lambda}(k,k',q,p,0)\,\widetilde{\partial}_\Lambda F^\Lambda(p)
            +\partial_\Lambda\alpha^\Lambda\,\Gamma^{(4,1,0)\Lambda}(k,k',q,q),
            \label{eq_app_fRG+MF: full A flow}
        \end{split}
    \end{equation}
    with $\Gamma^{(6,0,0)\Lambda}$ the 3-fermion coupling. We recall that in all the above flow equations, we have considered only the terms in which the center of mass momentum $q$ enters parametrically in the equations. This means that we have assigned to the flow equation for $\mathcal{A}^\Lambda$ only contributions in the particle-particle channel and we have neglected in all flow equations all the terms that contain a loop with the normal single scale propagator $\widetilde{\partial}_\Lambda G^\Lambda(k)$. Within a reduced model, where the bare interaction is nonzero only for $q=0$ scattering processes, the mean-field is the exact solution and one can prove that, due to the reduced phase space, only the diagrams that we have considered in our truncation of the flow equations survive~\cite{Salmhofer2004}. In order to treat the higher order couplings, $\Gamma^{(0,3,0)\Lambda}$, $\Gamma^{(2,2,0)\Lambda}$, $\Gamma^{(4,1,0)\Lambda}$  and $\Gamma^{(6,0,0)\Lambda}$, one can approximate their flow equations in order to make them integrable in way similar to Katanin's approximation for the 3-fermion coupling. The integrated results are the fermionic loop integrals schematically shown in Fig.~\ref{fig_app_fRG+MF: katanin}. Skipping any calculation, we just state that this approximation allows for absorbing the second and third terms on the right hand-side of Eqs.~\eqref{eq_app_fRG+MF: full P sigma flow},~\eqref{eq_app_fRG+MF: full h_sigma flow} and~\eqref{eq_app_fRG+MF: full A flow} into the first one just by replacing $\widetilde{\partial}_\Lambda\Pi^\Lambda_{11}$ with its full derivative $\partial_\Lambda\Pi^\Lambda_{11}$. In summary:
    \begin{subequations}
        \begin{align}
            &\partial_\Lambda m_\sigma^\Lambda(q)=\int_p h_\sigma^\Lambda(p;q)\left[\partial_\Lambda\Pi^\Lambda_{11}(p;q)\right] h_\sigma^\Lambda(p;q),\\
            &\partial_\Lambda h^\Lambda_\sigma(k;q)=\int_p\mathcal{A}^\Lambda(k,p;q)\left[\partial_\Lambda\Pi^\Lambda_{11}(p;q)\right]h^\Lambda_\sigma(p;q),\\
            &\partial_\Lambda\mathcal{A}^\Lambda(k,k';q)=\int_p\mathcal{A}^\Lambda(k,p;q)\left[\partial_\Lambda\Pi^\Lambda_{11}(p;q)\right]\mathcal{A}^\Lambda(p,k';q).
        \end{align}
    \end{subequations}
    With a similar approach, one can derive the flow equations for the transverse couplings:
    \begin{subequations}
        \begin{align}
                &\partial_\Lambda m_\pi^\Lambda(q)=\int_p h_\pi^\Lambda(p;q)\left[\partial_\Lambda\Pi^\Lambda_{22}(p;q)\right] h_\pi^\Lambda(p;q),\\
                &\partial_\Lambda h^\Lambda_\pi(k;q)=\int_p\Phi^\Lambda(k,p;q)\left[\partial_\Lambda\Pi^\Lambda_{22}(p;q)\right]h_\pi^\Lambda(p;q),\\
                &\partial_\Lambda\Phi^\Lambda(k,k';q)=\int_p\Phi^\Lambda(k,p;q)\left[\partial_\Lambda\Pi^\Lambda_{22}(p;q)\right]\Phi^\Lambda(p,k';q).
        \end{align}
    \end{subequations}
    \section{Calculation of the irreducible vertex in the bosonic formalism}
    \label{app: irr V bosonic formalism}
    In this appendix we provide a proof of Eq.~\eqref{eq_fRG+MF: irr V bosonic formalism} by making use of matrix notation. 
    If the full vertex can be decomposed as in Eq.~\eqref{eq_fRG+MF: vertex at Lambda crit}
    \begin{equation}
        V=\mathcal{Q}+\frac{h [h]^T}{m},
    \end{equation}
    we can plug this relation into the definition of the irreducible vertex, Eq.~\eqref{eq_fRG+MF: irr V fermionic}. With some algebra we obtain
    \begin{equation}
        \begin{split}
            \widetilde{V}=\left[1+V\Pi\right]^{-1}V
            =\left[1+\frac{\widetilde{h} [h]^T}{m}\Pi\right]^{-1}\left[\widetilde{\mathcal{Q}}+\frac{\widetilde{h} [h]^T}{m}\right],
        \end{split}
        \label{eq_app_fRG+MF: V tilde appendix I}
    \end{equation}
    where in the last equality we have inserted a representation of the identity,
    \begin{equation}
        1=\left[1+\mathcal{Q}\Pi\right]\left[1+\mathcal{Q}\Pi\right]^{-1},  
    \end{equation}
    in between the two matrices and we have made use of definitions~\eqref{eq_fRG+MF: reduced C tilde} and~\eqref{eq_fRG+MF: reduced yukawa}. With a bit of simple algebra, we can analytically invert the matrix on the left in the last line of Eq.~\eqref{eq_app_fRG+MF: V tilde appendix I}, obtaining
    \begin{equation}
        \left[1+\frac{\widetilde{h} [h]^T}{m}\Pi\right]^{-1}=1-\frac{\widetilde{h} [h]^T}{\widetilde{m}}\Pi,
    \end{equation}
    %
    Where $\widetilde{m}$ is defined in Eq.~\eqref{eq_fRG+MF: reduced mass P tilde}. By plugging this result into Eq.~\eqref{eq_app_fRG+MF: V tilde appendix I}, we finally obtain
    \begin{equation}
        \widetilde{V}=\widetilde{\mathcal{Q}}+\frac{\widetilde{h} [\widetilde{h}]^T}{\widetilde{m}},
    \end{equation}
    that is the result of Eq.~\eqref{eq_fRG+MF: irr V bosonic formalism}.
    \section{Algorithm for the calculation of the superfluid gap}
    The formalism described in Sec.~\ref{sec_fRG+MF: bosonic flow and integration} allows us to formulate a minimal set of closed equations required for the calculation of the gap. We drop the $\Lambda$ superscript, assuming that we have reached the final scale. The gap can be computed using the Ward identity, so we can reduce ourselves to a single self consistent equation for $\alpha$, that is a single scalar quantity, and another one for $h_\pi$, momentum dependent. The equation for $\alpha$ is Eq.~\eqref{eq_fRG+MF: alpha solution}. The transverse Yukawa coupling is calculated through Eq.~\eqref{eq_fRG+MF: h_pi}. The equations are coupled since the superfluid gap $\Delta=\alpha h_\pi$ appears in the right hand side of both.\\
    We propose an iterative loop to solve the above mentioned equations. By starting with the initial conditions $\alpha^{(0)}=0$ and $h_\pi^{(0)}(k)=0$, we update the transverse Yukawa coupling at every loop iteration $i$ according to Eq.~\eqref{eq_fRG+MF: h_pi}, that can be reformulated in the following algorithmic form:
    \begin{equation}
        h_\pi^{(i+1)}(k)=\int_{k'} \left[M^{(i)}(k,k')\right]^{-1}\,\widetilde{h}^{\Lambda_s}(k'), 
        \label{eq_fRG+MF_app: h_pi loop}
    \end{equation}
    with the matrix $M^{(i)}$ defined as 
    \begin{equation}
        M^{(i)}(k,k')=\delta_{k,k'}-\mathcal{\widetilde{Q}}^{\Lambda_s}(k,k')\,\Pi_{22}^{(i)}(k';\alpha^{(i)}),
    \end{equation}
    and the $22$-bubble rewritten as
    \begin{equation}
        \Pi_{22}^{(i)}(k;\alpha) = \frac{1}{G^{-1}(k)G^{-1}(-k)+\alpha^2\left[h_\pi^{(i)}(k)\right]^2},
    \end{equation}
    with $G(k)$ defined in Eq.~\eqref{eq_fRG+MF: G at Lambda_s}.
    Eq.~\eqref{eq_fRG+MF_app: h_pi loop} is not solved self consistently at every loop iteration $i$, because we have chosen to evaluate the r.h.s with $h_\pi$ at the previous iteration. $\alpha^{(i+1)}$ is calculated by self consistently solving
    \begin{equation}
        1=\frac{1}{\widetilde{m}^{\Lambda_s}}\int_k \widetilde{h}^{\Lambda_s}(k)\, \Pi_{22}^{(i+1)}(k;\alpha) \,h_\pi^{(i+1)}(k)
        \label{eq_fRG+MF_app: alpha loop}
    \end{equation}
    for $\alpha$. The equation above is nothing but Eq.~\eqref{eq_fRG+MF: alpha solution} where the solution $\alpha=0$ has been factorized away.
    The loop consisting of Eqs.~\eqref{eq_fRG+MF_app: h_pi loop} and~\eqref{eq_fRG+MF_app: alpha loop} must be repeated until convergence is reached in $\alpha$ and, subsequently, in $h_\pi$.
    This formulation of self consistent equations is not computationally lighter than the one in the fermionic formalism, but more easily controllable, as one can split the frequency and momentum dependence of the gap (through $h_\pi$) from the strength of the order ($\alpha$). Moreover, thanks to the fact that $h_\pi$ is updated with an explicit expression, namely Eq.~\eqref{eq_fRG+MF_app: h_pi loop}, that is in general a well behaved function of $k$, the frequency and momentum dependence of the gap is assured to be under control.

    %
    \chapter{Alternative derivation of the SBE flow equations}
    \label{app: SBE_fRG_app}
    In this Appendix, we present an alternative derivation of the flow equations presented in Chap.~\ref{chap: Bos Fluct Norm}, obtained by the introduction bosonic fields via three \emph{Hubbard-Stratonovich transformations} (HST). We re-write the bare interaction as
    \begin{equation}
        Un_{\up}n_{\down}=3Un_{\up}n_{\down} - 2 Un_{\up}n_{\down},
    \end{equation}
    and apply three different HST on each of the first three terms, one for physical channel. In formulas this can be expressed as  
    \begin{equation}
        \begin{split}
            \mathcal{Z}_\text{Hubbard}&=\int \mathcal{D}\left(\psi,\Bar{\psi}\right)e^{-\mathcal{S}_\text{Hubbard}\left[\psi,\Bar{\psi}\right]}
            =\int\mathcal{D}\boldsymbol{\Phi}\mathcal{D}\left(\psi,\Bar{\psi}\right)e^{-\mathcal{S}_\text{bos}\left[\psi,\Bar{\psi},\boldsymbol{\Phi}\right]},
        \end{split}
        \label{eq_app_SBE_fRGZ}
    \end{equation}
    where $\boldsymbol{\Phi}=(\Vec{\phi}_m,\phi_c,\phi_p,\phi_p^*)$ collects all three bosonic fields, $\mathcal{S}_\mathrm{Hubbard}$ is the bare Hubbard action, and the bosonized one is given by
    \begin{equation}
        \begin{split}
            \mathcal{S}_\text{bos}\left[\psi,\Bar{\psi},\boldsymbol{\Phi}\right] = 
            &-\int_{k,\sigma} \Bar{\psi}_{k,\sigma} \left(i\nu+\mu-\epsilon_k\right) \psi_{k,\sigma}\\
            &-\frac{1}{2}\int_{q}\phi_c(-q) \frac{1}{U} \phi_c(q)
            -\frac{1}{2}\int_{q}\Vec{\phi}_m(-q) \cdot \frac{1}{U} \Vec{\phi}_m(q)\\
            &+\int_{q}\phi^*_p(q)\frac{1}{U} \phi_p(q)
           +\int_{k,q,\sigma}\phi_c(q) \,\Bar{\psi}_{k+\frac{q}{2},\sigma}\psi_{k-\frac{q}{2},\sigma}\\
           &+\int_{k,q}\sum_{\sigma,\sigma'}\Vec{\phi}_m(q)\cdot \,\Bar{\psi}_{k+\frac{q}{2},\sigma}\Vec{\sigma}_{\sigma\sigma'}\psi_{k-\frac{q}{2},\sigma'}\\
           &+\int_{k,q}\left[\phi_p(q) \,\Bar{\psi}_{\frac{q}{2}+k,\uparrow}\Bar{\psi}_{\frac{q}{2}-k,\downarrow}+\phi^*_p(q) \,\psi_{\frac{q}{2}-k,\downarrow}\psi_{\frac{q}{2}+k,\uparrow}\right]\\       &
           -2U \int_0^\beta d\tau \sum_j n_{j,\uparrow}(\tau) n_{j,\downarrow}(\tau).
           \label{eq_app_SBE_fRG: HS action}
        \end{split}
    \end{equation}
    The remaining (not bosonized) $-2U$ term in $\mathcal{S}_\text{bos}$, avoids double counting of the bare interaction.
    
    We then introduce the RG scale via a regulator acting on the fermions. The regularized generating functional reads as
    \begin{equation}
        \begin{split}
        W^\Lambda\left[\eta,\Bar{\eta},\boldsymbol{J}\right]&=-\ln
        \int\mathcal{D}\boldsymbol{\Phi}\int \mathcal{D}\left(\psi,\Bar{\psi}\right)
        e^{-S^\Lambda_\text{bos}\left[\psi,\Bar{\psi},\boldsymbol{\Phi}\right]+\left(\Bar{\psi},\eta\right)+\left(\Bar{\eta},\psi\right) + \left(\boldsymbol{\Phi},\boldsymbol{J}\right)},
        \label{eq_app_SBE_fRG: W Lambda}
        \end{split}
    \end{equation}
    with
    \begin{equation}
        S^\Lambda_\text{bos}\left[\psi,\Bar{\psi},\boldsymbol{\Phi}\right]=S_\text{bos}\left[\psi,\Bar{\psi},\boldsymbol{\Phi}\right]+\int_{k,\sigma} \Bar{\psi}_{k,\sigma}\,R^\Lambda(k)\,\psi_{k,\sigma}.
    \label{eq_app_SBE_fRG: reg Sbos}
    \end{equation}
    The initial conditions at $\L=\Lini$ depend on the formalism used. In the plain fRG, we impose $R^{\Lambda\rightarrow\Lini}(k)\rightarrow \infty$, so that at the initial scale the effective action must equal $\mathcal{S}_\mathrm{bos}$. Differently, within the DMF\textsuperscript 2RG, the regulator must fulfill
    \begin{equation}
        R^{\Lambda_{\text{ini}}}(k) = \epsilon_{\boldsymbol{k}} - \Delta_\text{AIM}\left(\nu\right),
    \end{equation}
    so that we have
    \begin{equation}
        \mathcal{S}^{\Lini}\left[\psi,\Bar{\psi},\boldsymbol{\Phi}\right] =
        \mathcal{S}_\text{AIM}\left[\psi,\Bar{\psi},\boldsymbol{\Phi}\right],
    \end{equation}
    where $\mathcal{S}_\text{AIM}\left[\psi,\Bar{\psi}\right]$ is the action of the self-consistent AIM, where (local) bosonic fields have been introduced via HST. The initial conditions for the effective action therefore read as
    \begin{equation}
        \Gamma^{\Lini}\left[\psi,\Bar{\psi},\boldsymbol{\Phi}\right]=\Gamma_\text{AIM}\left[\psi,\Bar{\psi},\boldsymbol{\Phi}\right],
    \end{equation}
    with $\Gamma_\text{AIM}\left[\psi,\Bar{\psi},\boldsymbol{\Phi}\right]$ the effective action of the self-consistent AIM. Expanding it in terms of 1PI functions, one recovers the initial conditions given in Sec.~\ref{eq_SBE_fRG: DMF2RG initial conditions}, where the screened interactions $D^X$ and the Yukawa couplings $h^X$ at the initial scale equal their local counterpart of the AIM. 
    
    The above defined formalism allows for a straightforward inclusion of the bosonic fluctuations that, among other things, are responsible for the fulfillment of the Mermin-Wagner theorem. In fact, the present formalism can be extended by adding some boson-boson interaction terms~\cite{Strack2008,Friederich2011,Obert2013} which can suppress the divergence of the bosonic propagators at a finite scale. 

    \chapter{Details on the RPA for spiral magnets}
    \label{app: low en spiral}
    In this Appendix, we report some details on the RPA calculation of the collective excitations in spiral magnets. 
    \section{Coherence factors}
    The coherence factors entering the bare susceptibilities $\chit^0_{ab}(\bq,\omega)$ in Eq.~\eqref{eq_low_spiral: chi0 expression} are defined as
    \begin{equation}
        \mathcal{A}^{ab}_{\ell\ell'}(\bk,\bq)=\frac{1}{2}\Tr\left[\sigma^a u_\bk^\ell\sigma^b u_{\bk+\bq}^{\ell'}\right],
        \label{eq_app_spiral: coh fact def}
    \end{equation}
    with $u^\ell_\bk$ given by (see Eq.~\eqref{eq_low_spiral: ukl def})
    \begin{equation}
         u_\bk^\ell = \sigma^0 + \ell \, \frac{h_\bk}{e_\bk} \sigma^3 +
         \ell \, \frac{\Delta}{e_\bk} \sigma^1. 
    \end{equation}
    Here, $\ell$ and $\ell'$ label the quasiparticle bands, and $a$ and $b$ correspond to the charge-spin indices. Performing the trace, we get the following expression for the charge-charge coherence factor
    \begin{equation} 
         A^{00}_{\ell\ell'}(\bk,\bq) =
         1 + \ell\ell'\,\frac{h_\bk h_{\bk+\bq} + \Delta^2}{e_\bk e_{\bk+\bq}},
    \end{equation}
    while for the mixed charge-spin ones we have
    \begin{eqnarray}
         A^{01}_{\ell\ell'}(\bk,\bq) &=&
         \ell \, \frac{\Delta}{e_\bk} + \ell' \, \frac{\Delta}{e_{\bk+\bq}}, \\
         A^{02}_{\ell\ell'}(\bk,\bq) &=&
         -i \ell\ell' \, \Delta \frac{h_\bk - h_{\bk+\bq}}{e_\bk e_{\bk+\bq}}, \\
         A^{03}_{\ell\ell'}(\bk,\bq) &=&
         \ell \, \frac{h_\bk}{e_\bk} + \ell' \, \frac{h_{\bk+\bq}}{e_{\bk+\bq}} .
    \end{eqnarray}
    The diagonal spin coherence factors are given by
    \begin{eqnarray}
        A^{11}_{\ell\ell'}(\bk,\bq) &=&
         1 - \ell\ell'\,\frac{h_\bk h_{\bk+\bq} - \Delta^2}{e_\bk e_{\bk+\bq}}, \\
         A^{22}_{\ell\ell'}(\bk,\bq) &=&
         1 - \ell\ell'\,\frac{h_\bk h_{\bk+\bq} + \Delta^2}{e_\bk e_{\bk+\bq}}, \\
        A^{33}_{\ell\ell'}(\bk,\bq) &=&
         1 + \ell\ell'\,\frac{h_\bk h_{\bk+\bq} - \Delta^2}{e_\bk e_{\bk+\bq}}, \label{eq_app_spiral: A33}
    \end{eqnarray}
    and the off-diagonal ones by
    \begin{eqnarray}
         A^{12}_{\ell\ell'}(\bk,\bq) &=&
         -i\ell \, \frac{h_\bk}{e_\bk} + i\ell' \, \frac{h_{\bk+\bq}}{e_{\bk+\bq}}, \\
         A^{13}_{\ell\ell'}(\bk,\bq) &=&
         \ell\ell' \, \Delta \frac{h_\bk + h_{\bk+\bq}}{e_\bk e_{\bk+\bq}}, \\
         A^{23}_{\ell\ell'}(\bk,\bq) &=&
         -i \ell \, \frac{\Delta}{e_\bk} + i\ell' \, \frac{\Delta}{e_{\bk+\bq}}.
    \end{eqnarray}
    The remaining off-diagonal coherence factors can be easily obtained from the above expressions and the relation $\mathcal{A}^{ba}_{\ell\ell'}(\bk,\bq)=[\mathcal{A}^{ab}_{\ell\ell'}(\bk,\bq)]^*$. The $\mathcal{A}^{ab}_{\ell\ell'}(\bk,\bq)$ are purely imaginary if and only if one of the two indices equals two, and real in all other cases. Thus, the exchange of $a$ and $b$ gives
    \begin{equation}
        \mathcal{A}^{ba}_{\ell\ell'}(\bk,\bq) = p^a p^b \mathcal{A}^{ab}_{\ell\ell'}(\bk,\bq)
        \label{eq_app_spiral: coh fact exchange a,b}
    \end{equation}
    with $p^a=+1$ for $a=0,1,3$ and $p^{a=2}=-1$. 
    
    Using $\xi_\bk=\xi_{\mbk}$, one obtains $h_{\mbk-\bQ}=-h_\bk$, $g_{\mbk-\bQ}=g_\bk$, $e_{\mbk-\bQ}=e_\bk$ and $u^\ell_{\mbk-\bQ}=\sigma^1 u^\ell_{\bk}\sigma^1$. From Eq.~\eqref{eq_app_spiral: coh fact def}, one sees that
    \begin{equation}
        A^{ab}_{\ell'\ell}(\mbk-\bQ-\bq,\bq) = \frac{1}{2}\Tr\left[\widetilde{\sigma}^b u^\ell_{\bk+\bq}\widetilde{\sigma}^a u^{\ell'}_{\bk}\right],
    \end{equation}
    with $\widetilde{\sigma}^a=\sigma^1\sigma^a\sigma^1=s^a\sigma^a$, where $s^a=+1$ for $a=0,1$, and $s^a=-1$ for $a=2,3$. Using Eq.~\eqref{eq_app_spiral: coh fact exchange a,b}, one obtains
    \begin{equation}
         A^{ab}_{\ell'\ell}(\mbk-\bQ-\bq,\bq) = s^a s^b \mathcal{A}^{ba}_{\ell\ell'}(\bk,\bq)=s^{ab}\mathcal{A}^{ab}_{\ell\ell'}(\bk,\bq),
         \label{eq_app_spiral: Aq symmetry}
    \end{equation}
    where 
    \begin{equation}
        s^{ab}=s^a s^b p^a p^b = (1-2\delta_{a3})(1-2\delta_{b3}).
    \end{equation}
    \section{Symmetries of the bare susceptibilities}
    In this Section, we prove the symmetries of the bare bubbles listed in Table~\ref{tab_low_spiral: symmetries}. 
    \subsection{Parity under frequency sign change}
    We decompose expression~\eqref{eq_low_spiral: chi0 expression} into intraband and interband contributions
    \begin{equation}
        \begin{split}
             \chit_{ab}^0(\bq,z) =
             &-\frac{1}{8}\sum_\ell\int_\bk A^{ab}_{\ell\ell}(\bk,\bq)\frac{f(E^\ell_\bk)-f(E^\ell_\bkq)}{E^\ell_\bk-E^\ell_\bkq+z}\\
             &-\frac{1}{8}\sum_\ell\int_\bk A^{ab}_{\ell,\ellnot}(\bk,\bq)\frac{f(E^\ell_\bk)-f(E^\ellnot_\bkq)}{E^\ell_\bk-E^\ellnot_\bkq+z},
        \end{split}
    \end{equation}
    with $z$ a generic complex frequency. 
    Splitting the difference of the Fermi functions, and making the variable change $\bk\to\mbk-\bQ-\bq$ in the integral in the second term, we obtain for the intraband term
    \begin{equation}
        \begin{split}
            [\chit_{ab}^0(\bq,z)]_\mathrm{intra}=
            &-\frac{1}{8}\sum_\ell\int_\bk A^{ab}_{\ell\ell}(\bk,\bq)\frac{f(E^\ell_\bk)}{E^\ell_\bk-E^\ell_\bkq+z}\\
            &-\frac{1}{8}\sum_\ell\int_\bk A^{ab}_{\ell\ell}(\bk,\bq)\frac{f(E^\ellnot_{\mbk-\bQ-\bq})}{E^\ell_{\mbk-\bQ}-E^\ell_{\mbk-\bQ-\bq}-z}.
        \end{split}
    \end{equation}
    Using Eq.~\eqref{eq_app_spiral: Aq symmetry} and $E^\ell_{\mbk-\bQ}=E^\ell_{\bk}$, we obtain
    \begin{equation}
        [\chit_{ab}^0(\bq,z)]_\mathrm{intra} =
        -\frac{1}{8} \sum_\ell\int_\bk A^{ab}_{\ell\ell}(\bk,\bq)       f(E^\ell_\bk)
        \left( \frac{1}{E^\ell_\bk-E^\ell_\bkq+z} +
        \frac{s^{ab}}{E^\ell_\bk-E^\ell_\bkq-z} \right) \, .
    \end{equation}
    Similarly, we rewrite the interband term as
    \begin{equation}
        \begin{split}
             [\chit_{ab}^0(\bq,z)]_\mathrm{inter}=
             &-\frac{1}{8}\sum_\ell\int_\bk A^{ab}_{\ell,\ellnot}(\bk,\bq)
             \frac{f(E^\ell_\bk)}{E^\ell_\bk-E^\ellnot_\bkq+z} \nonumber \\
             & -\frac{1}{8}\sum_\ell\int_\bk A^{ab}_{\ellnot,\ell}(\mbk- \bq-\bQ,\bq)
             \frac{-f(E^\ell_{\mbk-\bQ})}{-(E^\ell_{\mbk-\bQ}-E^\ellnot_{\mbk-\bQ-\bq}-z)}, 
        \end{split}
    \end{equation}
    where in the second term we have made the substitution $\ell\to\ellnot$. Using again Eq.~\eqref{eq_app_spiral: Aq symmetry}, we get
    \begin{equation}
         [\chit_{ab}^0(\bq,z)]_\mathrm{inter} =
         -\frac{1}{8}\sum_\ell\int_\bk A^{ab}_{\ell,\ellnot}(\bk,\bq) f(E^\ell_\bk)
         \left( \frac{1}{E^\ell_\bk-E^\ellnot_\bkq+z} + 
         \frac{s^{ab}}{E^\ell_\bk-E^\ellnot_\bkq-z} \right).
    \end{equation}
    Summing up the interband and intraband terms, we obtain
    \begin{equation}
        \chit^0_{ab}(\bq,-z) = s^{ab}\chit^0_{ab}(\bq,z).
        \label{eq_app_spiral: frequency change symm}
    \end{equation}
    In the physical case of retarded suscecptibilities, that is, $z=\omega+i0^+$, we get
    \begin{subequations}
        \begin{align}
            &\chit^{0r}_{ab}(\bq,-\omega)=s^{ab}\chit^{0r}_{ab}(\bq,\omega),\\
            &\chit^{0i}_{ab}(\bq,-\omega)=-s^{ab}\chit^{0i}_{ab}(\bq,\omega),
        \end{align}
    \end{subequations}
    with $\chit^{0r}_{ab}$ and $\chit^{0i}_{ab}$ defined in the main text. 
    \subsubsection{Parity under momentum sign change}
    Performing the variable change $\bk\to\bk-\bq/2$ in the definition of the bare susceptibility, we get
    \begin{equation}
        \chit^0_{ab}(\bq,z) = \chit_0^{ab}(\bq,z)= -\frac{1}{8}\sum_{\ell\ell'} \int_\bk
        A^{ab}_{\ell\ell'}\left(\bk-\frac{\bq}{2},\bq\right)
        \frac{f(E^{\ell}_{\bk-\frac{\bq}{2}})-f(E^{\ell'}_{\bk+\frac{\bq}{2}})}
        {E^{\ell}_{\bk-\frac{\bq}{2}}-E^{\ell'}_{\bk+\frac{\bq}{2}}+z}.
    \end{equation}
    Using
    \begin{equation}
     A^{ab}_{\ell'\ell}\left(\bk+\frac{\bq}{2},-\bq\right) =
     A^{ba}_{\ell\ell'}\left(\bk-\frac{\bq}{2},\bq\right) = p^{a}p^{b}\,A^{ab}_{\ell\ell'}\left(\bk-\frac{\bq}{2},\bq\right) \, ,
    \end{equation}
    we immediately see that 
    \begin{equation} \label{eq_low_spiral:Aq->-q}
     \chit^0_{ab}(-\bq,-z) = p^{a}p^{b} \, \chit^0_{ab}(\bq,z).
    \end{equation}
    Combining this result with Eq.~\eqref{eq_app_spiral: frequency change symm}, we obtain
    \begin{equation}
        \chit^0_{ab}(-\bq,z) = p^{ab} \, \chit^0_{ab}(\bq,z),
    \end{equation}
    with $p^{ab}$ defined as 
    \begin{equation}
        p^{ab} = p^a p^b s^{ab} = s^a s^b =  (1-2\delta_{a2})(1-2\delta_{b2})(1-2\delta_{a3})(1-2\delta_{b3}).
    \end{equation}
    In the case of retarded susceptibilities, that is, $z=\omega+i0^+$, we get
    \begin{subequations}
        \begin{align}
            &\chit^{0r}_{ab}(-\bq,\omega) = p^{ab} \chit^{0r}_{ab}(\bq,\omega),\\
            &\chit^{0i}_{ab}(-\bq,\omega) = -p^{ab} \chit^{0i}_{ab}(\bq,\omega).
        \end{align}
    \end{subequations}
    \section{Calculation of \texorpdfstring{$\chit^0_{33}(\pm\bQ,0)$}{chitilde033(+-Q,0)}}
    In this Appendix we prove the relation \eqref{eq_low_spiral: chi33(q,0)} for $\chit^0_{33}(-\bQ,0)$. The corresponding relation for $\chit^0_{33}(\bQ,0)$ follows from the parity of $\chit^0_{33}(\bq,\omega)$ under $\bq\to-\bq$. Using the general expression~(\ref{eq_low_spiral: chi0 def}) for the bare susceptibility, and Eq.~\eqref{eq_app_spiral: A33} for the coherence factor $A^{33}_{\ell\ell'}(\bk,\bq)$, one obtains
    \begin{eqnarray}
     \chit^0_{33}(-\bQ,0) &=&
     - \frac{1}{8} \int_\bk \left[ 1 + \frac{h_\bk h_{\bk-\bQ}-\Delta^2}{e_\bk e_{\bk-\bQ}}\right]
     \left(\frac{f(E^+_\bk)-f(E^+_{\bk-\bQ})}{E^+_\bk-E^+_{\bk-\bQ}} +
     \frac{f(E^-_\bk)-f(E^-_{\bk-\bQ})}{E^-_\bk-E^-_{\bk-\bQ}}\right) \nonumber \\
     && - \frac{1}{8} \int_\bk \left[1-\frac{h_\bk h_{\bk-\bQ}-\Delta^2}{e_\bk e_{\bk-\bQ}}\right]
     \left(\frac{f(E^+_\bk)-f(E^-_{\bk-\bQ})}{E^+_\bk-E^-_{\bk-\bQ}} +
     \frac{f(E^-_\bk)-f(E^+_{\bk-\bQ})}{E^-_\bk-E^+_{\bk-\bQ}}\right) \nonumber \\[2mm]
     &=& - \frac{1}{4} \sum_{\ell=\pm}\int_\bk 
     \left\{\left[1-\frac{h_\bk h_\mbk+\Delta^2}{e_\bk e_\mbk}\right]
     \frac{f(E^\ell_\bk)}{E^\ell_\bk-E^\ell_\mbk} +
     \left[1+\frac{h_\bk h_\mbk+\Delta^2}{e_\bk e_\mbk}\right]
     \frac{f(E^\ell_\bk)}{E^\ell_\bk-E^\ellnot_\mbk}\right\} \nonumber \\[2mm]
     &=& \sum_{\ell=\pm} \int_\bk \frac{(-\ell) f(E^\ell_\bk)}{4e_\bk}
     \left\{\frac{2\ell e_\bk (g_\bk-g_\mbk)+2 h_\bk(h_\bk -h_\mbk)}
     {(E^\ell_\bk-E^\ellnot_\mbk)(E^\ell_\bk-E^\ell_\mbk)}\right\} \, .
    \end{eqnarray}
    In the second equation we have used $h_{\bk-\bQ}=-h_{\mbk}$, $e_{\bk-\bQ}=e_{\mbk}$, and $E^\pm_{\bk-\bQ}=E^\pm_{\mbk}$. It is easy to see that the linear combinations
    $\gmk = g_\bk - g_\mbk$, $h^\pm_\bk = h_\bk \pm h_\mbk$, and $e^\pm_\bk = e_\bk \pm e_\mbk$
    obey the relations $\hmk\hpk = h_\bk^2-h_\mbk^2 = e_\bk^2-e_\mbk^2 = \emk\epk$, and
    $\hmk = -\gmk$. Using these relations, we finally get
    \begin{eqnarray}
     \chit^0_{33}(-\bQ,0) &=&
      \sum_{\ell=\pm}\int_\bk \frac{(-\ell) f(E^\ell_\bk)}{4e_\bk}
      \left\{\frac{2\ell e_\bk \gmk + 2h_\bk\hmk }{(\gmk + \ell \epk)(\gmk + \ell \emk)}
      \right\} \nonumber \\
      &=& \sum_{\ell=\pm}\int_\bk \frac{(-\ell) f(E^\ell_\bk)}{4e_\bk}
      \left\{\frac{2\ell e_\bk \gmk + \emk\epk + (\gmk)^2 }{(\gmk + \ell \epk)(\gmk + \ell \emk)}
      \right\} \nonumber \\
      &=& \sum_{\ell=\pm}\int_\bk \frac{(-\ell) f(E^\ell_\bk)}{4e_\bk} =
      \int_\bk \frac{f(E^-_\bk) - f(E^+_\bk)}{4e_\bk} \, .
    \end{eqnarray}
    \section{Expressions for \texorpdfstring{$\kappa_\alpha^{30}(\bzero)$}{ka30(0)} and \texorpdfstring{$\kappa_\alpha^{31}(\bzero)$}{ka31(0)}}
    \label{app: ka30 and ka31}
    In this appendix, we report explicit expressions for the off-diagonal paramagnetic contributions to the spin stiffness, namely $\kappa_\alpha^{30}(\bzero)$, and $\kappa_\alpha^{31}(\bzero)$. 
    
    For $\kappa_\alpha^{30}(\bzero)$, we have, after having made the trace in \eqref{eq_low_spiral: paramagnetic contr Kernel} explicit,
    \begin{equation}
        \begin{split}
            &\kappa_\alpha^{30}(\bzero)=\lim_{\bq\to\bzero}K_{\mathrm{para},\alpha 0}^{31}(\bq,\bq',0)=\\&=-\frac{1}{4}\int_\bk T\sum_{\nu_n} \left\{\left[G^2_\bk(i\nu_n)+F^2_\bk(i\nu_n)\right]\gamma^\alpha_\bk-\left[\overline{G}^2_\bk(i\nu_n)+F^2_\bk(i\nu_n)\right]\gamma^\alpha_{\bk+\bQ}\right\}\delta_{\bq',\bzero}\\
            &=-\frac{1}{4}\int_\bk T\sum_{\nu_n}\left\{\partial_\bk\left[G_\bk-\overline{G}_\bk\right]+4F_\bk^2\,\partial_{k_\alpha}h_\bk\right\}\delta_{\bq',\bzero},
        \end{split}
    \end{equation}
    where we have made use of properties \eqref{eq_low_spiral: derivatives of G} in the last line. The first term vanishes when integrated by parts, while the Matsubara summation for the second yields
    \begin{equation}
        \begin{split}
            \kappa_\alpha^{30}(\bzero)=
            -\frac{\Delta^2}{4}\int_\bk \left[\frac{f(E^-_\bk)-f(E^+_\bk)}{e^3_\bk}+\frac{f^\prime(E^+_\bk)+f^\prime(E^-_\bk)}{e_\bk^2}\right](\partial_{k_{\alpha}}h_{\bk}).
        \end{split}
    \end{equation}
    For $\kappa_\alpha^{31}(\bzero)$ we have
    \begin{equation}
        \begin{split}
            \lim_{\bq\to\bzero}K_{\mathrm{para},\alpha 0}^{31}&(\bq,\bq',0)=\\
            &=-\frac{1}{4}\int_\bk T\sum_{\nu_n} \left[G_\bk(i\nu_n)F_\bk(i\nu_n)\gamma^\alpha_\bk-\overline{G}_\bk(i\nu_n)F_\bk(i\nu_n)\gamma^\alpha_{\bk+\bQ}\right]\left(\delta_{\bq',\bQ}+\delta_{\bq',-\bQ}\right).
        \end{split}
    \end{equation}
    Defining $\kappa_\alpha^{31}(\bzero)=2K_{\mathrm{para},\alpha 0}^{31}(\bzero,\bQ,0)$ (see Eq.~\eqref{eq_low_spiral: k31 def}), and performing the Matsubara sum, we obtain
    \begin{equation}
        \begin{split}
        \kappa_\alpha^{31}(\bzero)=-\frac{\Delta^2}{4}
            \int_\bk\bigg\{ \left[\frac{h_\bk}{e_\bk}(\partial_{k_\alpha}g_\bk)+(\partial_{k_\alpha}h_\bk)\right]\frac{f'(E^+_\bk)}{e_\bk}
            +\Big[&\frac{h_\bk}{e_\bk}(\partial_{k_\alpha}g_\bk)-(\partial_{k_\alpha}h_\bk)\Big]\frac{f'(E^-_\bk)}{e_\bk}\\
            &+\frac{h_\bk}{e_\bk^2}(\partial_{k_\alpha}g_\bk) \frac{f(E^-_\bk)-f(E^+_\bk)}{e_\bk}\bigg\}.
        \end{split}
    \end{equation}
    Furthermore, it is easy to see that $K_{\alpha 0}^{31}(\bzero,\pm\bQ,0)=\mp iK_{\alpha 0}^{32}(\bzero,\pm\bQ,0)$, which, together with Eq.~\eqref{eq_low_spiral: k32 def} proves $\kappa_\alpha^{31}(\bzero)=\kappa^\alpha_{32}(\bzero)$. We remark that in the N\'eel limit both $\kappa_\alpha^{30}(\bzero)$ and $\kappa_\alpha^{31}(\bzero)$ vanish as their integrands are odd under $\bk\to\bk+\bQ$.
    \section{Expressions for \texorpdfstring{$\chit_0^{-a}(Q)$}{chit0-a(Q)}}
    \label{app: chit0-a(Q)}
    We report here the RPA expressions for the off-diagonal bare susceptibilities $\chit_0^{-a}(Q)$, with $a=0,1,2$. They can all be obtained by computing the trace and the Matsubara summation in Eq.~\eqref{app: chit0-a(Q)}. We obtain
    \begin{subequations}
        \begin{align}
            &\chit_0^{-0}(Q)=-\frac{1}{16}\int_\bk \sum_{\ell,\ell'=\pm}\left[\ell\frac{\Delta}{e_\bk}+\ell'\frac{\Delta}{e_{\bk+\bQ}}+\ell\ell'\frac{\Delta(h_{\bk+\bQ}-h_\bk)}{e_\bk e_{\bk+\bQ}}\right]F_{\ell\ell'}(\bk,\bQ,0),\\
            &\chit_0^{-1}(Q)=-\frac{1}{16}\int_\bk \sum_{\ell,\ell'=\pm}\left[1+\ell\frac{h_\bk}{e_\bk}-\ell'\frac{h_{\bk+\bQ}}{e_{\bk+\bQ}}-\ell\ell'\frac{h_\bk h_{\bk+\bQ}-\Delta^2}{e_\bk e_{\bk+\bQ}}\right]F_{\ell\ell'}(\bk,\bQ,0),\\
            &\chit_0^{-2}(Q)=+\frac{i}{16}\int_\bk \sum_{\ell,\ell'=\pm}\left[1+\ell\frac{h_\bk}{e_\bk}-\ell'\frac{h_{\bk+\bQ}}{e_{\bk+\bQ}}-\ell\ell'\frac{h_\bk h_{\bk+\bQ}+\Delta^2}{e_\bk e_{\bk+\bQ}}\right]F_{\ell\ell'}(\bk,\bQ,0).
        \end{align}
    \end{subequations}
    with $F_{\ell\ell'}(\bk,\bq,\omega)$ defined as in Eq.~\eqref{eq_low_spiral: Fll def}.


\end{appendices}
\backmatter

\rhead[\fancyplain{}{\bfseries Bibliography}]{\fancyplain{}{\bfseries\thepage}}         \lhead[\fancyplain{}{\bfseries\thepage}]{\fancyplain{}{\bfseries Bibliography}} 
\nocite{apsrev42Control}
\bibliographystyle{mybst.bst}
\bibliography{main.bib}
\rhead[\fancyplain{}{\bfseries Bibliography}]{\fancyplain{}{\bfseries\thepage}}         \lhead[\fancyplain{}{\bfseries\thepage}]{\fancyplain{}{\bfseries Bibliography}} 
%
%
\rhead[\fancyplain{}{\bfseries }]{\fancyplain{}{\bfseries\thepage}}         \lhead[\fancyplain{}{\bfseries\thepage}]{\fancyplain{}{\bfseries }} 
\chapter*{Acknowledgments}
\addcontentsline{toc}{chapter}{Acknowledgments}

    \rhead[\fancyplain{}{\bfseries Acknowledgments}]{\fancyplain{}{\bfseries\thepage}}
    \lhead[\fancyplain{}{\bfseries\thepage}]{\fancyplain{}{\bfseries Acknowledgments}}
    I want to thank here all the people who have helped and supported me during the realization of this thesis.
    
    I am indebted to my supervisor, Walter Metzner for having given me the opportunity to work in his group in the Max Planck Institute for Solid State Research, which allowed me to live a wonderful scientific and human experience. I am grateful to him for sharing with me his deep knowledge on condensed matter physics and his mathematical rigor. I also thank him for his constant support along my PhD and for being available to discuss at any moment I needed it. Finally, I have appreciated how he has helped me to develop my scientific independence. Maria Daghofer is also thanked for co-refereeing this thesis. 
    
    A very special thanks goes also to my colleague, office mate and friend Demetrio Vilardi, for having accompanied me during my PhD as an "older brother" and for having shared with me his enormous experience in matter of physics, computer programming and, on top of all, life. I have enjoyed the several and long discussions that we have had along the years on topics of all kind.
    
    I also want to thank all people in the quantum many-body department for having contributed to a friendly and stimulating environment. In particular, I want to thank "the old generation" of PhD students: Moritz Hirschmann, Andreas Leonhardt, Oleksii Maistrenko, Johannes Mitscherling, Lukas Schwarz, and Jachym S\'ykora  for enlightening scientific discussions and for having welcomed me in the group. A thank goes also to the "new generation" of PhD, master, and guest students: Kirill Alpin, Sayan Banerjee, Hannes Braun, Steffen Bollmann, Lukas Debbeler, Rafael Haenel, Niclas Heinsdorf, Simon Klein, Raffaele Mazzilli, Shunsuke Nakamura, Nikolaos Parthenios, Janika Reichstetter, Kaustubh Roy, Robin Scholle, and Sida Tian, for the several interesting discussions we have had, the lunch and coffee breaks, and for the leisure time we have spent together, which made my stay in Stuttgart more pleasant after the pandemic. Furthermore, I acknowledge all the stimulating discussions I have had with the postdocs, guest scientists, and group leaders in Metzner Department and Classen and Schäfer independent research groups: Debmalya Chakraborty, Laura Classen, Lorenzo Del Re, Elio König, Andres Greco, Dirk Manske, Matteo Puviani, Pavel Ostrovsky, Thomas Schäfer, Andreas Schnyder, and Xianxin Wu. I am very grateful to our secretary Jeanette Schüller-Knapp and to Birgit King for their administrative and linguistic support.  
    
    I want to thank Alessandro Toschi from Vienna and Sabine Andergassen and her group in Tübingen, Ayman El-Eryani, Kilian Fraboulet, and Sarah Heinzelmann, for a fruitful collaboration that has allowed me to learn many things. A thanks goes also to Michael Scherer for having invited me to visit his group in Bochum.
    
    I am very thankful to all friends that I have met at both the Max Planck Institutes in Stuttgart, with whom a I have spent countless hours doing disparate activities: Gon\c{c}alo Antunes, Pol Cabanach, Karla Cordero, Alex Kossak, Gaurav Gardi, Charlotte Le Mouel, Lina Martin, Nikhilesh Murty, Matteo Puviani, Hiroto Takahashi, and Atsushi Ueda. Thank you friends for having made my time in Stuttgart enjoyable. 
    
    Last but not least, I thank my parents Luciano and Monica, my sisters Laura and Marta, all my close relatives and my family-in-law for the warm and constant support they have shown to me during my years of PhD. 
    
    Finally, I thank my wife Elisa for having been close to me in every moment and for having always had a word of comfort, even in the most difficult times. 

\chapter*{List of publications}
\addcontentsline{toc}{chapter}{List of publications}

    \rhead[\fancyplain{}{\bfseries List of publications}]{\fancyplain{}{\bfseries\thepage}}
    \lhead[\fancyplain{}{\bfseries\thepage}]{\fancyplain{}{\bfseries List of publications}}
    This thesis is based on the following publications:
        \begin{itemize}
            \item P.~M.~Bonetti\footnotemark[1], J.~Mitscherling\footnotemark[1]\footnotetext[1]{Equal contribution}, D.~Vilardi, and W.~Metzner, Charge carrier drop at the onset of pseudogap behavior in the two-dimensional Hubbard model, \href{https://link.aps.org/doi/10.1103/PhysRevB.101.165142}{Phys.~Rev.~B \textbf{101}, 165142 (2020)}.
            \item D.~Vilardi, P.~M.~Bonetti, and W.~Metzner, Dynamical functional renormalization group computation of order parameters and critical temperatures in the two-dimensional Hubbard model, \href{https://link.aps.org/doi/10.1103/PhysRevB.102.245128}{Phys.~Rev.~B \textbf{102}, 245128 (2020)}.
            \item P.~M.~Bonetti, Accessing the ordered phase of correlated Fermi systems:  Vertex bosonization and mean-field theory within the functional renormalization group, \href{https://link.aps.org/doi/10.1103/PhysRevB.102.235160}{Phys.~Rev.~B \textbf{102}, 235160 (2020)}.
            \item P.~M.~Bonetti,  A.~Toschi,  C.~Hille,  S.~Andergassen,  and  D.~Vilardi,  Single  boson exchange representation of the functional renormalization group for strongly interacting many-electron systems, \href{https://doi.org/10.1103/PhysRevResearch.4.013034}{Phys.~Rev.~Research \textbf{4}, 013034 (2022)}.
            \item P.~M.~Bonetti, and W.~Metzner, Spin stiffness, spectral weight, and Landau damping of magnons in metallic spiral magnets, \href{https://link.aps.org/doi/10.1103/PhysRevB.105.134426}{Phys.~Rev.~B \textbf{105}, 134426 (2022)}.
            \item P.~M.~Bonetti, Local Ward identities for collective excitations in fermionic systems with spontaneously broken symmetries, \href{https://doi.org/10.48550/arXiv.2204.04132}{arXiv:2204.04132}, \emph{accepted in Physical Review B} (2022). 
            \item P.~M.~Bonetti, and W.~Metzner, SU(2) gauge theory of the pseudogap phase in the two-dimensional Hubbard model, \href{https://doi.org/10.48550/arXiv.2207.00829}{arXiv:2207.00829} (2022).
        \end{itemize}
        Other publications have been produced during the period as PhD student, but they are not covered in this thesis:
        \begin{itemize}
            \item P.~M.~Bonetti, A.~Rucci, M.~L.~Chiofalo, and V.~Vuleti\'c, Quantum effects in the Aubry transition, \href{https://doi.org/10.1103/PhysRevResearch.3.013031}{Phys.~Rev.~Research \textbf{3}, 013031 (2021)}.
            \item K.~Fraboulet, S.~Heinzelmann, P.~M.~Bonetti, A.~El-Eryani, D.~Vilardi, A.~Toschi, and S.~Andergassen, Single-boson~exchange~functional renormalization group application to the two-dimensional Hubbard model at weak coupling, \href{https://doi.org/10.48550/arXiv.2206.14478}{arXiv:2206.14478} (2022).
        \end{itemize}
\end{document}